\documentclass[]{report}

\usepackage{setspace}
\usepackage[T1]{fontenc}
\usepackage{graphicx}
\usepackage{epsfig}
\usepackage{rotating}
\usepackage{amssymb}
\usepackage{cancel} 

\usepackage{subfig}
\usepackage[labelfont=bf]{caption} 

\usepackage{dsfont}
\usepackage{psfrag}
\usepackage{amsmath,euscript,array,mathrsfs}
\usepackage{bbold}
\usepackage{epsf}
\usepackage{xcolor}
\definecolor{mypink}{RGB}{255,20,147}
\definecolor{mybrown}{RGB}{139,69,19}
\definecolor{mygrey}{RGB}{106,106,106}
\definecolor{mypurple}{RGB}{128,0,128}
\definecolor{mycyan}{RGB}{0,170,255}

\usepackage{cite} 

\newcommand{\comm}[1]{} 
\newcommand{\beq}{\begin{equation}}
\newcommand{\eeq}{\end{equation}}
\newcommand{\be}{\begin{equation}}
\newcommand{\ee}{\end{equation}}
\newcommand{\beqs}{\begin{eqnarray}}
\newcommand{\eeqs}{\end{eqnarray}}
\newcommand{\nn}{\nonumber}

\def\a{\alpha}
\def\b{\beta}
\def\c{\chi}
\def\d{\delta}
\def\e{\epsilon}
\def\f{\phi}

\def\g{\gamma}

\def\k{\kappa}
\def\l{\lambda}  \def\L{\Lambda}

  \def\w{\omega}
  \def\th{\theta}
\def\r{\rho}
\def\s{\sigma}
\def\t{\tau}
  \def\U{\Upsilon}

\def\z{\zeta}

\def\ca{{\cal A}}
\def\cb{{\cal B}}

\def\cd{{\cal D}}

\def\cf{{\cal F}}

\def\ch{{\cal H}}
\def\ci{{\cal I}}

\def\ck{{\cal K}}
\def\cl{{\cal L}}
\def\cm{{\cal M}}
\def\cn{{\cal N}}
\def\co{{\cal O}}

\def\cq{{\cal Q}}
\def\car{{\cal R}}
\def\cs{{\cal S}}

\def\cv{{\cal V}}
\def\cw{{\cal W}}

\def\cz{{\cal Z}}

\def\pa{\partial} 

\def\half{\frac{1}{2}}
\def\bo{{\raise-.3ex\hbox{\large$\Box$}}} 
  
\def\di{\mbox{d}}

\def\pin{Poincar\'{e} invariance }  
\def\is{\hspace{-1mm}} 

\comm{
\title{
	Non-perturbative aspects of gauge theories from gauge-gravity dualities
}
\author{John Roughley}
\date{}
}

\begin{document}

\begin{titlepage}
\centering
	\vspace*{1cm}
{\huge\textbf{Non-perturbative aspects of gauge\textcolor{white}{l} theories from gauge-gravity\textcolor{white}{l} dualities  \hspace{2mm}}\textcolor{white}{l}  }\\		
\vspace{10mm}
\begin{center}		
{\Large\textbf{John Roughley  \hspace{2mm}} } 
\end{center}
\vfill
\begin{center}		
{\Large \textcolor{white}{l}Submitted to Swansea University in fulfilment of\newline 
\textcolor{white}{ppp} the requirements for the Degree of\newline
Doctor of Philosophy\textcolor{white}{ll} }
\end{center}
		
\vspace{8mm}
\begin{center}		
\hspace{1.8mm}\includegraphics[width=0.3\textwidth]{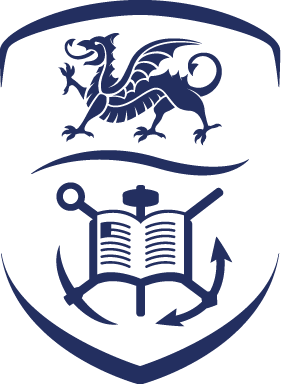}	
\end{center}	
{\large
				Department of Physics\\[-2mm]
				Swansea University\\[-2mm]
				United Kingdom\\
				August 2021
			}		
\end{titlepage}

\newpage
\begin{singlespace}	
\section*{Abstract}
In this Ph.D. Thesis we consider two specific supergravities which are well-established within the literature on holography, and which are known to provide the low-energy effective 
description of either superstring theory or M-theory: the six-dimensional half-maximal theory of Romans, and the maximal supergravity in seven dimensions.\par 
We implement their dimensional reduction by compactifying on an $S^{1}$ and $T^{2}$, respectively, to obtain a five-dimensional sigma-model coupled to gravity. Spectra of bosonic excitations are computed numerically by considering field fluctuations on background geometries which holographically realise confinement. We furthermore propose a diagnostic tool to detect mixing effects between scalar resonances and the pseudo-Nambu--Goldstone boson associated with spontaneous breaking of conformal invariance: the dilaton. This test consists of neglecting a certain component of the spin-0 fluctuation variables, effectively disregarding their back-reaction on the underlying geometry; where discrepancies arise compared to the complete calculation we infer dilaton mixing. For both theories this analysis evinces a parametrically light dilaton.\par
For each supergravity we uncover a tachyonic instability within their parameter space; motivated by these pathological findings we proceed to conduct an investigation into their respective phase structures, reasoning that there must necessarily exist some mechanism by which these instabilities are rendered physically inaccessible. We compile a comprehensive catalogue of geometrically distinct backgrounds admissible within each theory, and derive general expressions for their holographically renormalised free energy $\cf$. Another numerical 
routine is employed to systematically extract data for some special deformation parameters, and $\cf$ is plotted in units of an appropriate universal scale.\par 
Our analysis proves fruitful: each theory exhibits clear evidence of a first-order phase transition which induces the spontaneous decompactification of the shrinking circular dimension before the instability manifests, favouring instead a class of singular solutions. The aforementioned dilaton resonance appears only along a \emph{metastable} portion of the branch of confining backgrounds. 

\end{singlespace}

\newpage
\section*{Declaration}
I declare that I am the sole author of this Thesis, and that the work contained herein is the result of my own investigations except where otherwise stated in the text. This Thesis is based on work produced in collaboration, and which appears within the following publications:\\
{\small
\begin{itemize}
	\item[[1\hspace{-1.4mm}]] D.~Elander, M.~Piai, and J.~Roughley,\newline
	``Holographic glueballs from the circle reduction of Romans supergravity,''\newline
	JHEP \textbf{02}, 101 (2019), doi:10.1007/JHEP02(2019)101\newline
	[arXiv:1811.01010 [hep-th]].
	\item[[2\hspace{-1.4mm}]] D.~Elander, M.~Piai, and J.~Roughley,\newline
	``Probing the holographic dilaton,''\newline
	JHEP \textbf{06}, 177 (2020), [erratum: JHEP \textbf{12}, 109 (2020)],\newline
	doi:10.1007/JHEP06(2020)177\newline
	[arXiv:2004.05656 [hep-th]].
	\item[[3\hspace{-1.4mm}]] D.~Elander, M.~Piai, and J.~Roughley,\newline
	``Dilatonic states near holographic phase transitions,''\newline
	Phys. Rev. D \textbf{103}, 106018 (2021), doi:10.1103/PhysRevD.103.106018\newline
	[arXiv:2010.04100 [hep-th]].
	\item[[4\hspace{-1.4mm}]] D.~Elander, M.~Piai, and J.~Roughley,\newline
	``Light dilaton in a metastable vacuum,''\newline
	Phys. Rev. D \textbf{103}, no.4, 046009 (2021), doi:10.1103/PhysRevD.103.046009\newline
	[arXiv:2011.07049 [hep-th]].
\end{itemize}
}
$\text{}$\newline
\noindent All other sources and references are provided in the appended bibliography.\\ \\ 
This Thesis and the work that it comprises has not previously been accepted in substance for any other degree or qualification, and is not being concurrently submitted in candidature for any other degree or qualification.\\ \\
I hereby give consent for this Thesis, if accepted, to be made available online in the University’s Open Access Repository and for inter-library loan, and for its title and abstract to be made available to outside organisations.
\begin{flushright}
	\emph{John Roughley}\\ August 2021.
	\end{flushright}

\tableofcontents

\newpage
\section*{Acknowledgements}
I would like to thank Maurizio Piai and Biagio Lucini for acting as my doctoral advisors over the course of my candidacy, and for their helpful comments and criticisms of this Thesis during its development. I would like to express in particular my gratitude to Maurizio Piai and Daniel Elander, with whom I have collaborated in the past few years, for their immeasurable expertise and guidance throughout this period. I have thoroughly enjoyed working with you both, and I would be remiss if I did not take this opportunity to say so again.\par 
Finally I would also like to thank my family for their unwavering and unconditional support during my studies, and for providing me with the motivation to push through the setbacks which inevitably arise when conducting research.\\ \\
I have benefited from the financial support endowed by the Science and Technology Facilities Council (STFC) studentship ST/R505158/1.     

\cleardoublepage
\addcontentsline{toc}{chapter}{\listfigurename}
\listoffigures

\cleardoublepage
\addcontentsline{toc}{chapter}{\listtablename}
\listoftables

\chapter{Introduction}
\label{Chap:Intro}
\section{Background}
\label{Sec:Background}
The non-perturbative nature of strongly interacting Quantum Field Theories (QFTs) makes them particularly difficult to study, since the methods of perturbation theory are insufficient to provide reliable results; its failure in this context means that physical observables of interest cannot be computed. This is especially concerning given that the Standard Model, which to date provides our best understanding of elementary particles and their fundamental interactions, includes a Yang-Mills theory describing (at low energies) the strong-coupling physics of quarks and gluons: Quantum Chromodynamics (QCD). Ab initio calculations in this context are possible, though only by making use of lattice numerical methods which are typically challenging and resource-intensive. Complementary approaches to this problem which provide a more general understanding of similar strongly-coupled systems, without relying on expensive and time-consuming calculations on a supercomputer, would certainly be desirable.\par
A major breakthrough in our understanding of these systems came with the much celebrated and influential 1997 paper by J.~Maldacena~\cite{Maldacena:1997re} in which the so-called AdS/CFT correspondence was first proposed, providing a successful realisation of the \emph{holographic principle} earlier developed by G.~'t Hooft~\cite{tHooft:1993dmi} and L.~Susskind~\cite{Susskind:1994vu} in the context of string theory. In its original form, the AdS/CFT correspondence is a conjectured relation between two apparently dissimilar theories: type-IIB superstring theory (or its low-energy supergravity limit) formulated on the product space geometry $\text{AdS}_{5}\times S^{5}$, and the $\cn=4$ superconformal Yang-Mills (SYM) theory in four dimensions with gauge group $SU(N_{c})$. They have in common certain symmetry properties: the isometry group of $\text{AdS}_{5}$ and the conformal group of four-dimensional Minkowski space are isomorphic, both being $SO(4,2)$, and moreover there is an isomorphism $SO(6)\cong SU(4)$ between the isometry group of the five-sphere $S^{5}$ and the $R$-symmetry group of $\cn=4$ SYM~\cite{Aharony:1999ti}. The correspondence conjectures that a gravitational theory formulated on the curved $\text{AdS}_{5}$ bulk geometry, and a conformal field theory (CFT) situated at the four-dimensional boundary (flat Minkowski space), should both describe the same underlying physics; this would therefore be an example of a holographic \emph{duality}.
What makes this dual description so powerful is that it relates the strong-coupling regime on one side of the duality with the weak-coupling regime on the other, so that otherwise unfeasible non-perturbative field theory calculations at low energy scales could in principle be rendered as a comparatively simple perturbative computation in one additional dimension, with gravity. The low energy classical supergravity description of the bulk side of the duality is realised in the large $N_{c}$ limit~\cite{tHooft:1973alw} (see also reviews in Refs.~\cite{Coleman:1980nk,Manohar:1998xv}) with small string coupling $g_{s}\to 0$, holding the 't Hooft coupling $\l\equiv N_{c}\,g_{s}\equiv N_{c}\,g_{YM}^{2}$ large $\l\gg1$ and fixed; here $g_{YM}$ is the gauge coupling of the corresponding field theory.\par
The AdS/CFT correspondence saw further development shortly afterwards, starting from Refs.~\cite{Gubser:1998bc,Witten:1998qj} (see also Ref.~\cite{Banks:1998dd}), wherein the relationship between states propagating on the higher-dimensional geometry and properties of the dual CFT was more precisely defined. It was postulated, for example, that each supergravity field $\f(x^{\mu},z)$ (where $z$ is the radial coordinate and $x^{\mu}$ are the four Minkowski dimensions of the UV boundary at $z=0$) should correspond to a gauge-invariant operator $\co(x^{\mu})$ in the dual CFT, of which the scaling dimension $\Delta$ is related to the bulk mass of $\f$. Furthermore, the asymptotic value assumed by each supergravity field at the conformal boundary $\f(x^{\mu},0)\equiv\f_{0}$ should be understood to act as a source for this dual operator. An equivalence between the generating functional of the boundary field theory and the partition function of the bulk gravitation model was proposed (see Refs.~\cite{Bianchi:2001kw,Skenderis:2002wp,Papadimitriou:2004ap} for discussion in the context of holographic renormalisation, and Refs.~\cite{Petersen:1999zh,Erdmenger:2007cm,Nastase:2007kj} for more general reviews):
\begin{equation}
\cz_{\cm}[\f_{0}]=\int_{\f\sim \f_{0}}\is\is D\f\, e^{-S[\f]}
=\Big\langle e^{-\is\int_{\pa\cm}\f_{0}\co}\Big\rangle_{CFT}\,,
\end{equation}
where the left side is the supergravity partition function (here written in the Euclidean signature) with boundary conditions (BCs) imposed on each bulk field $\f$ at the UV boundary of the space $\pa\cm$, and the right side is the CFT generating functional with corresponding sources $\f_{0}$ and operators $\co$. Using this prescription, it became possible to compute important field theoretic quantities such as correlation functions and condensates, by taking functional derivatives of the supergravity action with respect to the appropriate sources. These innovations laid the foundations for what is now referred to as the \emph{holographic dictionary}, a more general catalogue of associated quantities and parameters on either side of the duality.\par 
Despite the context of its original construction, the applicability of the AdS/CFT correspondence is not restricted to holographic systems for which the higher-dimensional geometry is $\text{AdS}_{5}$. The complete classification of possible supergravities---based on their underlying superalgebra---which admit supersymmetric $\text{AdS}_{D}$ solutions was provided by Nahm~\cite{Nahm:1977tg} (see also Refs.~\cite{Kac:1977em,DeWitt:1981wm}); although no such solutions exist for $D>7$, their non-supersymmetric counterparts may yet be discovered in higher dimensions (see for example Ref.~\cite{Cordova:2018eba}). Hence, the correspondence can be generalised to include geometries with other numbers of non-compactified dimensions.   
Furthermore, since its inception the correspondence has been developed in order to be applicable to a broader class of holographic systems; these include, for example, models in which the higher-dimensional geometry deviates from Anti-de Sitter space as one travels radially away from the UV boundary (breaking conformal invariance in the dual field theory), in addition to models which preserve different amounts of supersymmetry (SUSY). It is due to these developments that AdS/CFT correspondences have come to be known more generically as \emph{gauge/gravity} dualities, 
although these names are typically understood to be synonymous and are often used interchangeably.\par 
Gauge/gravity dualities have also been proposed as a tool with which one may holographically model a four-dimensional field theory which exhibits \emph{confinement} at low energies, and the dictionary has been extended to facilitate the calculation of appropriately renormalised 2-point functions~\cite{Bianchi:2001kw,Skenderis:2002wp,Papadimitriou:2004ap} of relevance to computing composite state (glueball) mass spectra. In the context of QCD, the term `confinement' refers to the phenomenon that quarks, antiquarks, and gluons (which belong to the fundamental, antifundamental, and adjoint representations of the gauge group $SU(3)_{c}$, respectively) cannot be isolated below a certain energy threshold, and must form colour-neutral composite states which transform as singlets under the colour gauge group: hadrons (including glueballs, which are examples of \emph{exotic} mesons). Furthermore, in real-world QCD, it requires increasingly more energy to separate a quark-antiquark pair, and above a certain length scale it becomes energetically favourable to instead generate a new $q\bar{q}$ pair from the vacuum (so-called \emph{hadronisation}).\par 
An alternative definition of confinement exists for more general QFTs which do not exhibit this hadronisation at low energies, and it is this definition which we shall adopt throughout this Thesis. For a given QFT we can study the interaction of two source particles by considering the static (time-invariant) potential between them; in QED this is the Coulomb potential between two electric charges, and in QCD it is the potential between two colour charges (a quark-antiquark pair for example). For a generic strongly interacting QFT, confinement manifests as a static potential between a quark-antiquark pair that increases linearly with separation, which can be deduced by studying the behaviour of \emph{Wilson loops}. A Wilson loop which encloses the flux tubes between the $q\bar{q}$ pair scales with the \emph{area} of the contour as separation is increased; this is the ``area law'' of confinement. Conversely, for a QFT which does not exhibit confinement---for example Quantum Electrodynamics---the Wilson loop scales instead with the \emph{perimeter} of the loop contour.
In gauge/gravity correspondences, the expectation value of rectangular Wilson loops of area $L\times T$ (space$\times$time)---which are localised at the UV boundary of the bulk---can be computed using a standard holographic prescription~\cite{Rey:1998ik,Maldacena:1998im} (see also Refs.~\cite{Kinar:1998vq,Brandhuber:1999jr,
	Avramis:2006nv,Nunez:2009da,Faedo:2013ota}): in the uplifted ten-dimensional geometry one hangs an open string with endpoints fixed to the loop contour at the boundary, and which is allowed to explore the bulk geometry along the radial dimension. By minimising the classical action of this configuration in the $T\to\infty$ limit, one can compute the energy of the system as a function of the quark-antiquark separation $L$ and recover the expected linear behaviour for the static potential of a confining theory.\par
On the gravity side of the duality, confinement is manifested as a geometric property of the bulk spacetime manifold, and there are known to exist at least two realisations. As originally suggested by Witten in Ref.~\cite{Witten:1998zw}, one such method is via the toroidal compactification of a supergravity which admits $\text{AdS}_{D}$ background configurations, in such a way that one internal circle $S^{1}$ of the torus smoothly shrinks to zero volume at a finite value of the radial coordinate; the resultant tapering of the bulk manifold naturally introduces a low-energy limit in the dual field theory which lives on the four-dimensional boundary, which in turn may be intuitively interpreted as the confinement scale of composite states (see Ref.~\cite{Brower:2000rp} for an early example of glueball spectra computed in this way).\par 
The alternative method to toroidal compactification is related to what is known in the literature as the \emph{conifold}~\cite{Candelas:1989js,Klebanov:1998hh,Klebanov:2000nc,PandoZayas:2000ctr,Klebanov:2000hb,Maldacena:2000yy,Chamseddine:1997nm,Butti:2004pk,Nunez:2008wi} (see also Refs.~\cite{Acharya:2004qe,Strominger:1995cz}). Briefly, one can consider a product space geometry of the form~\cite{Klebanov:2000nc} $\cm^{10}=\text{AdS}_{5}\times T^{1,1}$,
 where $T^{1,1}$ describes the five-dimensional base of a special type of six-dimensional manifold containing a conical singularity (a \emph{conifold}), and which has certain properties which make it particularly interesting to the holographic study of four-dimensional field theories with unbroken supersymmetry~\cite{Atiyah:2001qf}. The base space $T^{1,1}$ of this cone is topologically equivalent to $S^{2}\times S^{3}$, and the conical singularity at the cone apex can be smoothed (or `repaired') by allowing either the 2-sphere or 3-sphere to maintain a finite non-zero volume at the end of space in the radial direction~\cite{Candelas:1989js}: the former case is referred to as the \emph{resolved} conifold and the latter as the \emph{deformed} conifold, and both exhibit the same UV asymptotic behaviour as the singular conifold. 
 It was demonstrated by I.~Klebanov and M.~Strassler in Ref.~\cite{Klebanov:2000hb} that the type-IIB supergravity solution  propagating on the deformed conifold geometry exhibits certain behaviour near to the IR end of space which, in the dual $\cn=1$ non-conformal gauge theory, can be interpreted as confinement. A similar solution was constructed shortly afterwards by J.~Maldacena and C.~Nu\~{n}ez in Ref.~\cite{Maldacena:2000yy}, and it was later shown that these are two limits of a one-parameter family of solutions referred to as the \emph{baryonic branch}~\cite{Butti:2004pk}. 
 It should be clarified that we mention the conifold only for the sake of completeness, and much of the important underlying physics and several technical aspects of this geometry have been omitted here; for our purposes going forward we will always be referring to toroidal compactification when discussing confinement within holography.\par
In this Thesis we will primarily focus our attention on two specific supergravity theories, both of which have been extensively studied in the context of \emph{top-down} holography (i.e. starting from a rigorously defined higher-dimensional theory of quantum gravity), and each of which is known to represent the low-energy limit of either superstring theory or M-theory. What makes them particularly interesting candidates for further investigation is that they are relatively simple supergravity models, which nevertheless provide an interesting framework within which to study the phenomenology of strongly-coupled field theories. Both of them admit AdS background solutions, and may be toroidally compactified in order to geometrically realise confinement in the dual field theory. Furthermore, as we shall see, both theories admit unique supersymmetric fixed point solutions which may be deformed to generate several physically distinct classes of background configurations.\par 
The first of these theories is the six-dimensional half-maximal $\cn=(2,2)$ gauged supergravity with $F_{4}$ superalgebra and $SU(2)$ gauge group, the existence of which was originally predicted in Ref.~\cite{DeWitt:1981wm}, and which was first constructed and written explicitly by Romans~\cite{Romans:1985tw}. It is known to be obtainable from massive type-IIA supergravity~\cite{Romans:1985tz} via the reduction of the ten-dimensional geometry on a \emph{warped} four-sphere: $\cm^{10}\to \text{AdS}_{6}\times S^{4}$, which preserves an $SO(4)$ isometry of the compactified space (the warp factor appearing in the lift to ten dimensions has a non-trivial dependence on one of the angles which parametrises the $S^{4}$, so that the internal geometry is topologically a foliation of 3-spheres, and hence only the $SO(4)\subset SO(5)$ isometry is preserved) and breaks half of the supersymmetry~\cite{Brandhuber:1999np,Cvetic:1999un}. Alternative lifts to type-IIB supergravity are known, see for example Refs.~\cite{Hong:2018amk,Jeong:2013jfc}. The isomorphism $SO(4)\cong SU(2)\times SU(2)$ contains two copies of the $SU(2)$ subgroup: one of these manifests the $R$-symmetry group of the dual theory living on the boundary of the $\text{AdS}_{6}$ space, while the other provides the supergravity gauge group. This theory, widely known as \emph{Romans supergravity}, is an illustrative example of an interesting theory which presents a rich topic for exploration despite its relative simplicity, and it has been applied for many different purposes in a variety of contexts in the literature~\cite{
	Hong:2018amk,Jeong:2013jfc,Wen:2004qh,Kuperstein:2004yf,Elander:2013jqa,DAuria:2000afl,Andrianopoli:2001rs,Seiberg:1996bd,Morrison:1996xf,Intriligator:1997pq,Douglas:1996xp,Ganor:1996pc,Nishimura:2000wj,Ferrara:1998gv,Gursoy:2002tx,Nunez:2001pt,Karndumri:2012vh,Lozano:2012au,Karndumri:2014lba,Chang:2017mxc,Gutperle:2018axv,Suh:2018tul,Suh:2018szn,Kim:2019fsg,Chen:2019qib}. These include, for example, to study somewhat atypical strongly-coupled field theories in five dimensions via the $\text{AdS}_{6}$/$\text{CFT}_{5}$ correspondence~\cite{Seiberg:1996bd,Morrison:1996xf,Intriligator:1997pq,Douglas:1996xp,Ganor:1996pc,Nishimura:2000wj,Ferrara:1998gv}, to investigate non-trivial renormalisation group (RG) flows using holography~\cite{Gursoy:2002tx,Nunez:2001pt,Karndumri:2012vh}, and to compute the spectra of glueball masses in a four-dimensional confining field theory by compactifying the dual of a five-dimensional field theory on a circle~\cite{Wen:2004qh,Kuperstein:2004yf,Elander:2013jqa}. As a final comment, and in anticipation of Section~\ref{Sec:Formalism6Dmodel} where we shall introduce the formalism of Romans supergravity, we here mention that it is known in the literature that the scalar manifold of half-maximal non-chiral supergravities in $D=6$ dimensions can be extended by introducing vector multiplets which couple to the pure theory~\cite{DAuria:2000afl,Andrianopoli:2001rs} (see also Refs.~\cite{Freedman:2012zz,Tanii:2014gaa}). However, for our purposes we shall neglect to include these additional multiplets, and will instead consider only the minimal bosonic field content of the theory: one real scalar coupled to gravity, four vectors, and a 2-form.\par
The second theory which we shall study is the seven-dimensional maximal $\cn=4$ gauged supergravity with gauge group $Sp(4)\cong SO(5)$, which was originally constructed in Refs.~\cite{Pernici:1984xx,Pernici:1984zw} by compactifying the maximally supersymmetric eleven-dimensional supergravity on a four-sphere $\cm^{11}\to\text{AdS}_{7}\times S^{4}$; the isometry group of the compact space manifests the $SO(5)$ gauge group in $D=7$ dimensions. There is known to exist a related compactification---first predicted in Ref.~\cite{Freund:1980xh}, see also Refs.~\cite{deWit:1986oxb,deWit:1981sst,deWit:1982bul,Page:1983mke,Duff:1990xz,Nicolai:2011cy}---on a 7-sphere $\cm^{11}\to\text{AdS}_{4}\times S^{7}$ which also retains maximal supersymmetry and for which the isometry group of the $S^{7}$ realises an $SO(8)$ gauge group, but we will not be studying this theory. It was shown in Ref.~\cite{Nastase:1999cb} that the spectrum of the $\text{AdS}_{7}\times S^{4}$ system may be \emph{consistently} truncated to neglect the massive Kaluza-Klein states of the compact $S^{4}$, retaining only the graviton supermultiplet (see Ref.~\cite{Aharony:1999ti} for a review). The lift of the $D=7$ supergravity back to eleven dimensions is known to simplify if one further truncates the theory to retain only a single scalar $\f$ field~\cite{Lu:1999bc}, and it is this truncated model which we shall be investigating. The scalar potential of this simplified system admits two distinct critical point solutions with exactly $\text{AdS}_{7}$ background geometry, in addition to more general solutions which interpolate between them~\cite{Campos:2000yu}. A further dimensional reduction to $D=5$ can be performed by compactifying two of the external dimensions of the AdS space on a torus $T^{2}=S^{1}\times S^{1}$, which introduces two additional spin-0 fields to the model; solutions to the classical equations of motion (EOMs) may be constructed in such a way that one of these circles always maintains a non-zero volume, and the corresponding scalar which parametrises this volume can be interpreted as the dilaton field of the uplifted type-IIA supergravity. For the critical point solution $\f=0$ this toroidally reduced system was first proposed by Witten~\cite{Witten:1998zw} in the context of studying confining field theories using holography, and it has been used elsewhere in the literature, for example as the dual of quenched  QCD~\cite{Sakai:2004cn,Sakai:2005yt} to geometrically realise spontaneously broken chiral symmetry. We will introduce the necessary formalism for this model, and perform the dimensional reduction on a torus, in Chapter~\ref{Chap:SpectraWitten}.\par 
We conclude this introductory section by briefly describing the two main objectives of the work which constitutes this Thesis. Firstly, we shall exploit gauge/gravity dualities in order to compute the mass spectra of composite states in four-dimensional strongly-coupled field theories, by considering fluctuations of the higher dimensional supergravity fields about their background configurations. We will calculate these spectra numerically for the two theories introduced above: Romans six-dimensional half-maximal supergravity compactified on a circle, and the seven-dimensional maximal supergravity compactified on a torus, for background solutions which admit a dual interpretation in terms of confining field theories. For the spin-0 states, we will then proceed to repeat the computation using what we henceforth refer to as the \emph{probe approximation}, a diagnostic `tool' with which it is possible to determine to what extent the physical scalar states mix with the dilaton of the theory; we will provide a proper introduction for these concepts in the following chapter. For both supergravity models, in the process of extracting the spectra we will uncover the existence of a tachyonic instability within a certain region of parameter space, which appears as a result of a runaway direction in the scalar potential.\par
Secondly, motivated by our discovery of these instabilities, we will conduct an energetics analysis for each of the models by computing the (appropriately renormalised) free energy as a function of a set of universal deformation parameters, for several geometrically distinct classes of background solutions. In both cases we will uncover the existence of a first-order phase transition which ensures that the unstable branch of solutions is never energetically favoured, and we will demonstrate that---beyond a certain critical value of a control parameter---the system prefers to spontaneously decompactify the dimension wrapped on the shrinking $S^{1}$ to restore the maximum $(D-1)$-dimensional Poincar\'{e} invariance.

\section{Motivation: \emph{Conformality Lost}}
\label{Sec:Motivation}
It was proposed in Ref.~\cite{Kaplan:2009kr}, and further discussed in Refs.~\cite{Gorbenko:2018ncu,Gorbenko:2018dtm}, that one possible marker of the transition from a conformal regime to a non-conformal regime within a QFT (of relevance to the study of the QCD \emph{conformal window}) is the merging of two beta function fixed points, resulting in the complexification of the scaling dimension of a field theory operator; in the context of the AdS/CFT correspondence this is realised on the gravity side of the duality by a scalar field acquiring a mass which falls below the Breitenlohner-Freedman (BF) stability bound~\cite{Breitenlohner:1982jf}. As a brief reminder of this mass bound, recall that in Refs.~\cite{Gubser:1998bc,Witten:1998qj} it was shown that the AdS/CFT correspondence predicts that the scaling dimension $\Delta$ of the gauge-invariant boundary operator dual to a scalar field $\f$ propagating on $\text{AdS}_{d+1}$ (with curvature radius $R$) is given by: 
\begin{equation}
\label{Eq:Delta}
m^{2}R^{2}=\Delta(\Delta-d)\,,
\end{equation}
which as a quadratic equation in $\Delta$ admits two distinct solutions for the scaling dimension:
\begin{equation}
\Delta_{\pm}=\frac{1}{2}\Big[d\pm\sqrt{d^{2}+4m^{2}R^{2}}\Big]\,.
\end{equation} 
We see that the necessary condition for $\Delta_{\pm}$ to be real is given by the bound 
\begin{equation}
\label{Eq:BFbound}
m^{2}\geqslant m_{BF}^{2}\equiv-\frac{d^{2}}{4R^{2}}\,,
\end{equation}
which was originally derived for a massive free scalar field in Ref.~\cite{Breitenlohner:1982jf}, by determining which asymptotic boundary conditions for the scalar ensure that the system has finite energy. It is worth noting that the BF bound defined in Eq.~(\ref{Eq:BFbound}) does not forbid the existence of fields with negative mass squared; while in Minkowski space such states would indicate the presence of a classical instability in the theory, for scalar fields propagating on AdS geometries the energetic stability requirement is instead given by a lower bound on admissible $m^{2}<0$ values~\cite{Witten:1998qj}.\par 
In a more recent study~\cite{Pomarol:2019aae}, this idea of relating the BF bound to conformal invariance was tested within the framework of bottom-up holography, using a simple $\text{AdS}_{5}$ model with a ``hard-wall'' IR brane used to introduce by hand an end of space and to generate a mass gap in the putative dual field theory. The authors explored the dynamics of this system, and in particular investigated the spectrum of resonances in the boundary theory as parameters were dialled to approach the BF bound in the bulk. They observed that the dilaton was indeed the lightest state in the scalar spectrum, although they furthermore concluded that this state was not \emph{parametrically} light and hence its mass could not be tuned to be arbitrarily small compared to the other resonances.\par 
Motivated by these findings we will adopt an alternative approach to studying dilaton phenomenology, though instead in the context of supergravity within the established framework of top-down holography. 
We shall pursue a line of investigation which generalises the notion of proximity to the BF bound as a signal of broken conformal invariance (applicable only to AdS geometries), and will instead consider proximity to regions within the system parameter space which introduce a tachyonic instability in the spectrum of fluctuations; in this way we will be able to conduct a study of the spectra of resonances analogous to that of Ref.~\cite{Pomarol:2019aae}, but for background solutions which model confinement by departing from Anti-de Sitter space. The objective of our research will nevertheless be similar: to ascertain to what degree the scalar states of the boundary theory may be identified with the dilaton, and to deduce whether or not the dilaton is the (parametrically) lightest resonance.

\section{Thesis overview}
\label{Sec:Overview}
The Thesis is organised as follows. In Chapter~\ref{Chap:GeneralFormalism} we shall introduce all of the general formalism which is applicable to both of the supergravity models that we will be studying, and briefly discuss the underlying methods of our investigation. In Sec.~\ref{Sec:ComputingSpectra} we present the five-dimensional holographic formalism required to compute the spectra of gauge-invariant states in a dual four-dimensional field theory, and provide a descriptive outline of the numerical routine which we employ to extract values for the physical masses. In Sec.~\ref{Sec:DilatonFormalism} we discuss dilaton phenomenology in the context of holography, and describe in more detail our implementation of the probe approximation. We then provide a schematic overview of our investigation into the energetics and phase structure of the two supergravity theories in Sec.~\ref{Sec:EnergeticsFormalism}.\par 
Part~\ref{PartOne}---which comprises Chapters~\ref{Chap:nTorus}, \ref{Chap:SpectraRomans}, and \ref{Chap:SpectraWitten}---is dedicated to presenting and discussing the results of our numerical spectra computations.
In Chapter~\ref{Chap:nTorus} we extract the spectrum of excitations for an example toy model (the reduction of a generic $\text{AdS}_{D}$ system on an $n$-torus), and then proceed to demonstrate the aforementioned probe approximation by applying it first to this holographic model as a proof of concept. 
Then, in Chapters~\ref{Chap:SpectraRomans} and~\ref{Chap:SpectraWitten} we present in detail all of the necessary formalism required to describe the two supergravity theories, perform the dimensional reduction of each via toroidal compactification, show explicitly the class of background solutions which geometrically realise confinement, derive the equations satisfied by the gauge-invariant field fluctuations, compute the spectra of massive excitations, and then finally compare these results to those obtained from the corresponding probe approximation. Chapter~\ref{Chap:SpectraRomans} focuses exclusively on Romans six-dimensional supergravity, while Chapter~\ref{Chap:SpectraWitten} is dedicated to the maximal seven-dimensional theory.\par 
Part~\ref{PartTwo}---which comprises Chapters~\ref{Chap:EnergeticsRomans} and~\ref{Chap:EnergeticsWitten}---is dedicated to presenting and discussing our exploration of the phase structure of the two supergravity theories that we are considering, by conducting a numerical energetics analysis in each case.
In Chapter~\ref{Chap:EnergeticsRomans} we present this study for Romans six-dimensional supergravity, introducing a classification of several geometrically distinct types of background solutions, and providing a thorough derivation of the holographically renormalised free energy. We furthermore introduce a scale setting scheme which is necessary in order to compare these different classes of solutions, and then provide an outline of the numerical routine used to extract data for the required parameters. We conclude by presenting the free energy as a function of these parameters for each class of background within our catalogue, alongside a few other interesting quantities. Chapter~\ref{Chap:EnergeticsWitten} repeats this entire analysis for the seven-dimensional supergravity.\par
Finally for the main body of the document, Chapter~\ref{Chap:Discussion} is dedicated to summarising our key findings and conclusions; we furthermore comment on some interesting (and rather surprising) phenomenological similarities between the two compactified supergravity theories.\ We then conclude with a discussion on some issues that our work leaves unresolved, and suggest some related topics---building upon this Thesis---which future work may potentially seek to address.\par  
In Appendix~\ref{App:2-Forms} we derive three equivalent formulations of a particular Lagrangian density which is of relevance to our spectra computation for the six-dimensional supergravity, and in Appendix~\ref{App:CritSpectra} we present tabulated numerical masses that are obtained by considering fluctuations about some special background solutions. In Appendix~\ref{App:LambdaSpectra} we implement an alternative normalisation of the spectra plots for the two supergravity theories, by making use of a universal energy scale which is introduced as part of our phase structure analysis. In Appendix~\ref{App:GravInvariants} we provide expressions for some gravitational invariants which may be derived using the metric ans\"{a}tze that we adopt, and present some plots to demonstrate explicitly the geometric differences between some of the solution classes that we consider. Finally, Appendix~\ref{App:ParamPlots} contains some additional plots which highlight the non-trivial implicit relations between the various parameters which are introduced in our energetics analysis of the two supergravity theories.

\chapter{General formalism}
\label{Chap:GeneralFormalism}
\section{Computing spectra}
\label{Sec:ComputingSpectra}
\subsection{Holographic formalism in \emph{D=}5 dimensions}
\label{Subsec:HolographicFormalism}
According to the dictionary of gauge-gravity dualities, the mass spectrum of composite bound states in a ($D-1$)-dimensional strongly-coupled field theory can be computed by studying the spectrum of small fluctuations around an asymptotically-AdS background configuration in the corresponding dual $D$-dimensional supergravity model. For the purposes of this Thesis we are interested in obtaining the spectra of massive states for confining four-dimensional field theories, and we are hence required to examine the bosonic field excitations of their weakly-coupled five-dimensional gravitational duals.\par 
We shall be considering the dimensional reduction of two well-known supergravities, and so we dedicate this introductory chapter to defining all of the general holographic formalism which is common to each system that we study. To start with, consider the five-dimensional geometry described by the following line element ansatz:
\begin{equation}
\label{Eq:5Dmetric}
\di s^2_5 \,=\, e^{2A} \di x^2_{1,3} \,+\,\di r^2 \,,
\end{equation}  
where $\di x_{1,3}^2$ is the four-dimensional Minkowski metric with ``mostly plus'' signature $\eta_{\mu\nu}\equiv {\rm diag}\,(-1\,,\,1\,,\,1\,,\,1)$, $A$ is the metric warp factor, and $r$ is the holographic coordinate which parametrises the radial dimension. In computing the spectra, we constrain the holographic coordinate to take values in the closed interval $r\in[r_{1},r_{2}]$, where $r_{1}$ is the infrared (IR) boundary and $r_{2}$ is the ultraviolet (UV) boundary; these boundaries have no physical meaning, and are introduced as regulators of the dual field theory with the understanding that the physical spectrum results are recovered only after taking appropriate limits. We define indices to run over $\mu,\nu\in\{0,1,2,3\}$ and $M,N\in\{0,1,2,3,5\}$, so that the determinant of the five-dimensional metric is given by $\text{det}(g_{MN})\equiv g_{5}=-e^{8A}$. Finally, to ensure that Poincar\'{e} invariance is manifestly preserved along the Minkowski $x^{\mu}$ directions, we demand that all fields of the gravitational model are functions \emph{only} of the holographic coordinate $r$, including the warp factor $A=A(r)$.\par
We next introduce the conventions which we shall adopt for classical gravity. The Christoffel symbols (metric connection) are given by
\begin{equation}
\Gamma^P_{\,\,\,\,MN}\,\equiv\,\frac{1}{2}g^{PQ}\left(\frac{}{}\partial_Mg_{NQ}+\partial_Ng_{QM}-\partial_Qg_{MN}\right)\,,
\end{equation}
so that the Riemann tensor is
\begin{equation}
\mathcal{R}_{MNP}^{\,\,\,\,\,\,\,\,\,\,\,\,\,\,\,\,\,Q}\,\equiv\,\partial_N\Gamma^Q_{\,\,\,\,MP}-\partial_M\Gamma^Q_{\,\,\,\,NP}+\Gamma^S_{\,\,\,\,MP}\Gamma^Q_{\,\,\,\,SN}
-\Gamma^S_{\,\,\,\,NP}\Gamma^Q_{\,\,\,\,SM}\,,
\end{equation}
the Ricci tensor can be written as
\begin{equation}
\mathcal{R}_{MN}\,\equiv\,\mathcal{R}_{MPN}^{\,\,\,\,\,\,\,\,\,\,\,\,\,\,\,\,\,P}\,,
\end{equation}
and the Ricci curvature scalar as
\begin{equation}
\mathcal{R}_{5}\,\equiv\,\mathcal{R}_{MN}g^{MN}=-8A''-20(A')^{2}\,.
\end{equation}
The radial coordinate $r$ parametrises a bounded segment of the five-dimensional bulk manifold, and hence an induced metric $\tilde{g}_{MN}$ is needed on each of the two four-dimensional boundaries. We introduce the five-vector $n_{M}=(0,0,0,0,1)$ which is defined to be orthonormal to the boundaries:
\begin{equation}
1= g_{MN}n^{M}n^{N}\,,~~~~~~~~0=\tilde{g}_{MN} n^{M}\,,
\end{equation}
so that the induced metric is given by
\begin{equation}
\tilde{g}_{MN}\,\equiv\, g_{MN}-n_{M}n_{N}=\,{\rm diag}\,(e^{2A}\eta_{\mu\nu}\,,\,0)\,.
\end{equation}
We next define the covariant derivative acting on a generic $(1,1)$-tensor (which may be generalised for tensors of any rank):
\begin{equation}
\nabla_M T^{P}_{\,\,\,\,N}\,\equiv\,\partial_MT^{P}_{\,\,\,\,N}+\Gamma^P_{\,\,\,\,MQ}T^{Q}_{\,\,\,\,N}-\Gamma^Q_{\,\,\,\,MN}T^{P}_{\,\,\,\,Q}\,,
\end{equation}
which allows us to define the extrinsic curvature $\mathcal{K}$ in terms of the following symmetric tensor:
\begin{equation}
\mathcal{K}_{MN}\,\equiv\,\nabla_M n_N\,=\,\pa_M n_N-\Gamma^Q_{\,\,\,\,MN}n_Q\,,
\end{equation}
so that the extrinsic curvature scalar is $\mathcal{K}\equiv g^{MN}\mathcal{K}_{MN}=4\pa_{r}A$.\par
Now that we have characterised the underlying geometry of the five-dimensional gravitational model, we next introduce the formalism (following the notation of Refs.~\cite{Bianchi:2003ug,Berg:2005pd,Berg:2006xy,Elander:2009bm,Elander:2010wd}) necessary to describe a sigma-model of $n$ scalars coupled to gravity in five dimensions. The general action may be written as
\begin{align}
\label{Eq:5Daction}
\mathcal{S}_{5} &=\int\is\di^5x \bigg\{\sqrt{-g_5}\bigg[\frac{\mathcal{R}_{5}}{4}-\frac{1}{2}G_{ab}g^{MN}\pa_M\Phi^a\pa_N\Phi^b-\mathcal{V}(\Phi^a)\bigg]\nonumber \\
&\hspace{10mm}+\sum_{i=1,2}\delta(r-r_i)(-)^i\sqrt{-\tilde{g}}\bigg[\frac{\mathcal{K}}{2}+\lambda_{i}(\Phi^a)\bigg]\bigg\},
\end{align} 
where $g_{5}$ is the determinant of the metric defined by Eq.~(\ref{Eq:5Dmetric}), $\tilde{g}$ is the determinant of the induced metric, $\mathcal{R}_{5}$ is the Ricci curvature scalar, and $\mathcal{K}$ is the extrinsic curvature. The $n$ scalars coupled to gravity are denoted by $\Phi^a$ (with $a=1,\ldots,n$), $\mathcal{V}$ is the sigma-model scalar potential, and $\lambda_{i}$ are boundary-localised potentials (see Ref.~\cite{Elander:2010wd} for details). The $(-1)^i$ term in the action manifests an antiparallel orientation of the orthonormal vector $n_{M}$ at the IR boundary. The kinetic term for the scalar fields is contracted with $G_{ab}$, the sigma-model metric, with which we may define quantities analogous to those describing the spacetime geometry; the sigma-model connection on the $n$-dimensional scalar manifold may be written as
\begin{equation}
\mathcal{G}^{a}_{bc} \,\equiv\, \frac{1}{2} G^{ad} \left( \partial_b G_{cd} + \partial_ c G_{db} - \partial_d G_{bc} \right)\, ,
\end{equation} 
while the sigma-model Riemann tensor is
\begin{equation}
\mathcal{R}^{a}_{\,\,\,\,bcd} \,\equiv\, \partial_c \mathcal{G}^{a}_{\,\,\,\,bd} - \partial_d \mathcal{G}^{a}_{\,\,\,\,bc} + \mathcal{G}^{a}_{\,\,\,\,ce} \mathcal{G}^{e}_{\,\,\,\,bd} - \mathcal{G}^{a}_{\,\,\,\,de} \mathcal{G}^{e}_{\,\,\,\,bc}\, .
\end{equation} 
Finally, for a generic field which carries a sigma-model index $X^{a}$, we define the sigma-model covariant derivative
\begin{equation}
D_b X^a \,\equiv\, \pa_b X^a + \mathcal{G}^{a}_{\,\,\,\,bc} X^c\, ,
\end{equation}
and the background covariant derivative
\begin{equation}
\mathcal{D}_{M} X^a \,\equiv\, \pa_{M} X^a + \mathcal{G}^{a}_{\,\,\,\,bc} \pa_{M} \Phi^b X^c\, ,
\end{equation}
and for the sigma-model derivative we adopt the notation $\mathcal{V}^a\equiv G^{ab}\pa_b \mathcal{V}$ with  
$\pa_b \cv\equiv \frac{\pa \cv}{\pa \Phi^b}$. Note that for our purposes it is sufficient to only consider the background covariant derivative $\mathcal{D}_{M}$ acting on a sigma-model $(1,0)$-tensor $X^{a}$, as we will not encounter any dynamical fields carrying more than one sigma-model index (the one exception being $G_{ab}$). With our conventions established, and recalling that we will henceforth assume that the bulk fields and warp factor are functions \emph{only} of the radial coordinate $r$, we can write down the classical equations of motion which are derived from the variation of the general five-dimensional action in Eq.~(\ref{Eq:5Daction})~\cite{Bianchi:2003ug,Elander:2010wd}:
\begin{align}
0&=\pa_r^2\Phi^a\,+\,4\pa_rA\pa_r\Phi^a\,+\,{\cal G}^a_{\,\,\,\,bc}\pa_r\Phi^b\pa_r\Phi^c\,-\,\mathcal{V}^a\,, \label{Eq:ScalarEOM}\\
0&=3\pa_r^{2}A\,+\,6(\pa_{r}A)^2\,+\,G_{ab}\pa_r\Phi^a\pa_r\Phi^b\,+\,2\mathcal{V}\,, \label{Eq:Einstein1}\\
0&=6(\pa_r A)^2\,-\,G_{ab}\pa_r\Phi^a\pa_r\Phi^b\,+\,2\mathcal{V}\,, \label{Eq:Einstein2}
\end{align}
where Eqs.~(\ref{Eq:ScalarEOM}) are the equations of motion for the scalars, while Eq.~(\ref{Eq:Einstein1}) and Eq.~(\ref{Eq:Einstein2}) come from the Einstein field equations.\par 
As a brief aside, we here note that the task of finding background solutions to these equations of motion is somewhat simplified for cases in which we are able to identify a superpotential $\mathcal{W}(\Phi^{a})$, so that the scalar potential $\mathcal{V}(\Phi^{a})$ in $D$ dimensions satisfies the following condition~\cite{Berg:2005pd}:
\begin{equation}
{\cal V}=\frac{1}{2}G^{ab}{\cal W}_{a}{\cal W}_{b}-\frac{D-1}{D-2}{\cal W}^2\,,
\label{Eq:Superpotential}
\end{equation}
where $G_{ab}$ is the sigma-model metric and $\mathcal{W}_{a}\equiv \pa_{a}\mathcal{W}$, and where we adopt a domain-wall (DW) metric ansatz of the form
\begin{equation}
\label{Eq:RomansSUSYmetric}
\di s^2_D =e^{2{\cal A}}\di x_{1,D-2}^2+\di r^2\,,
\end{equation}
with warp factor ${\cal A}$ and radial coordinate $r$. In such a case where these criteria are met, one may solve instead the following system of first-order equations:
\begin{align}
\pa_{r}\Phi^{a}&=G^{ab}\pa_{b}{\cal W}\,,  \label{Eq:FirstOrderPhi}\\
\pa_{r}{\cal A}&=-\frac{2}{D-2}{\cal W}\,, \label{Eq:FirstOrderA}
\end{align}
to find a special set of solutions which are guaranteed to also satisfy the original second-order equations of motion. We will make use of this superpotential formalism in later chapters. To conduct our numerical analysis of the glueball spectra, it is convenient to employ the gauge-invariant formalism developed in Refs.~\cite{Bianchi:2003ug,Berg:2005pd,Berg:2006xy,Elander:2009bm,Elander:2010wd}, which we shall briefly review here.\par 
Having identified a background solution to the classical equations of motion, we proceed to expand the scalar fields as
\begin{equation}
\Phi^a(x^\mu,r) = \bar{\Phi}^{a}(r) + \varphi^{a}(x^\mu,r) \,,
\end{equation} 
where $\varphi^{a}(x^\mu,r)$ represent small fluctuations about the background solution $\bar{\Phi}^{a}(r)$. We furthermore parametrise the fluctuations of the metric by decomposing the tensor $g_{MN}$ according to the Arnowitt-Deser-Misner (ADM) prescription~\cite{Arnowitt:1959ah}: we consider the foliation of the five-dimensional bulk manifold into four-dimensional hypersurfaces along the radial dimension, rewriting the $D=5$ metric as follows:
\begin{align}
\label{EQ:ADM1}
\di s^2_5&= e^{2A} \Big(\eta_{\mu\nu}+h_{\mu\nu}\Big) \di x^{\mu} \di x^{\nu} +2\nu_{\mu}\di x^{\mu}\di r + \Big((1+\nu)^2+\nu_{\sigma}\nu^{\sigma}\Big)\di r^2\,,\\
h^{\mu}_{\,\,\,\,\nu}&=\big(h^{TT}\big)^{\mu}_{\,\,\,\,\nu}
+\pa^{\mu}\epsilon_{\nu}+\pa_{\nu}\epsilon^{\mu}+\frac{\pa^{\mu}\pa_{\nu}}{\Box}H+\frac{1}{3}\delta^{\mu}_{\,\,\,\,\nu}h\,,\nn\\
\label{EQ:ADM2}
&=\big(h^{TT}\big)^{\mu}_{\,\,\,\,\nu}
+iq^{\mu}\epsilon_{\nu}+iq_{\nu}\epsilon^{\mu}+\frac{q^{\mu}q_{\nu}}{q^{2}}H+\frac{1}{3}\delta^{\mu}_{\,\,\,\,\nu}h\,,
\end{align}   
where $\Box\equiv\eta^{\mu\nu}\pa_{\mu}\pa_{\nu}$ is the d'Alembert operator, $h^{TT}$ is the transverse and traceless component of the metric fluctuation, and $\epsilon_{\mu}$ is transverse. The linearised equations of motion for the field fluctuations $\{\varphi^a,\nu,\nu^{\mu},(h^{TT})^{\mu}_{\,\,\,\,\nu},h,H,\epsilon^{\mu}\}$ may then be equivalently reformulated in terms of the following gauge-invariant (under infinitesimal diffeomorphisms) physical variables:
\begin{align}
\label{Eq:a}
\mathfrak{a}^a&\equiv\varphi^a\,-\,\frac{\pa_r\bar{\Phi}^a}{6\pa_r A} h\,,\\
\mathfrak{b}&\equiv\nu\,-\,\pa_r\left(\frac{h}{6\pa_r A}\right)\,,\\
\mathfrak{c}&\equiv e^{-2A}\pa_{\mu}\nu^{\mu}\,-\,\frac{e^{-2A}q^2 }{6\pa_r A}\,h\,-\,\frac{1}{2}\pa_r H\,,\\
\label{Eq:d}
\mathfrak{d}^{\mu}&\equiv e^{-2A} \Pi^{\mu}_{\,\,\,\,\nu}\nu^{\nu} \,-\,\pa_r \epsilon^{\mu}\,,\\
\label{Eq:e}
\mathfrak{e}^{\mu}_{\,\,\,\,\nu}&\equiv(h^{TT})^{\mu}_{\,\,\,\,\nu}\,,
\end{align}
with $ \Pi^{\mu}_{\,\,\,\,\nu}\equiv \delta^{\mu}_{\,\,\,\,\nu}-\frac{q^{\mu}q_{\nu}}{q^2}$ the transverse momentum projector satisfying $\Pi^{\mu\nu}q_{\mu}=0$, so that the equations of motion for the field fluctuations decouple into sectors according to spin. The equation of motion for the spin-1 field $\mathfrak{d}^{\mu}$ is algebraic and hence is not used to compute the spectra of vector composite states; the equations of motion for $\mathfrak{b}$ and $\mathfrak{c}$ are also algebraic, and their solutions may be written in terms of $\mathfrak{a}^{a}$~\cite{Elander:2010wd}. We are therefore left with the equations of motion for two independent spin sectors. Defining $M^{2}\equiv -q^{2}$ as the four-dimensional mass of the fluctuations, the tensorial fluctuations $\mathfrak{e}^{\mu}_{\,\,\,\,\nu}(M,r)$ satisfy the bulk equation
\begin{equation}
\label{Eq:TensorFluct}
0=\big[\pa_{r}^2+4\pa_r A \pa_r +e^{-2A}M^2\big] \mathfrak{e}^{\mu}_{\,\,\,\,\nu}\,,
\end{equation}
and are subject to Neumann boundary conditions:
\begin{equation}
\label{Eq:TensorFluctBC}
0=\pa_r \mathfrak{e}^{\mu}_{\,\,\,\,\nu}\Big|_{r_i}\,.
\end{equation}    
Likewise, the equations of motion for the spin-0 fluctuations $\mathfrak{a}^{a}(M,r)$ may be written as 
\begin{align}
\label{Eq:ScalarFluct}
0 &= \Big[ \cd_{r}^{2} + 4\pa_{r}A\cd_{r} + e^{-2A} M^2 \Big] \mathfrak{a}^a \,+\,\\ \nonumber
& - \Big[ \cv^a{}_{|c} - \car^a{}_{bcd} \pa_{r}\bar\Phi^b \pa_{r}\bar\Phi^d + \frac{4 (\pa_{r}\bar\Phi^a \cv^b + \cv^a 
	\pa_{r}\bar\Phi^b) G_{bc} }{3\pa_{r} A} 
+ \frac{16\cv\pa_{r}\bar\Phi^a \pa_{r}\bar\Phi^b G_{bc}}
{9 (\pa_{r}A)^2} \Big] \mathfrak{a}^c\,,
\end{align}
where we have introduced the notation $\cv^a_{\,\,\,\,|b}\equiv \pa_b \cv^a + \mathcal{G}^a_{\,\,\,\,bc}\cv^c$, with corresponding boundary conditions
\begin{align}
\label{Eq:ScalarFluctBC}
\pa_r\bar{\Phi}^a\pa_r\bar{\Phi}_{b}\cd_{r} \mathfrak{a}^{b}\Big|_{r_i}=
\bigg[-\frac{3}{2}\pa_{r}A\frac{M^{2}}{e^{2A}}\delta^a_{\,\,\,\,b}+\pa_r \bar{\Phi}^{a}\bigg(\frac{4\mathcal{V}}{3\pa_r A}\pa_r \bar{\Phi}_{b}+\mathcal{V}_b\bigg)
\bigg]\mathfrak{a}^b\Big|_{r_i}\,.
\end{align} 
This procedure and the resulting equations of motion for the field fluctuations are quite general, and for any physical system of interest which can be similarly modelled as an $n$-scalar sigma-model coupled to gravity in five dimensions, it is possible to compute the mass spectra for the spin-0 and spin-2 glueball sectors of its corresponding dual strongly-coupled field theory, with some caveats \cite{Elander:2010wd}. The analogous formalism for an arbitrary number of dimensions is discussed in Section 4 of Ref.~\cite{Elander:2020csd}.\par
We will make use of this same formalism throughout Chapters~\ref{Chap:nTorus}\,-\,\ref{Chap:SpectraWitten} while conducting our numerical study of the spectra for three distinct holographic systems. In Chapter~\ref{Chap:SpectraRomans}---where we consider the six-dimensional supergravity---we will generalise this procedure by supplementing the sigma-model action with terms accounting for the contributions of 1- and 2-forms, and in so doing we will be able to extract the complete spectra of bosonic excitations for the theory.

\subsection{Numerical implementation} 
\label{Subsec:NumericalImp}  
We conclude Section~\ref{Sec:ComputingSpectra} by briefly outlining the procedure used to compute the mass spectra using the formalism of the previous chapter, and describe the qualitative structure of a numerical routine which would allow one to most easily employ these techniques in practice.\par
It is first necessary to identify background profiles for the scalar fields $\Phi^{a}$ and warp factor $A$ which solve the classical equations of motion, subject to the simplifying assumption that the profiles are functions only of the radial coordinate; let us assume that the background solutions are generated over the domain $r\in[r_{1}^{\text{num}},r_{2}^{\text{num}}]$, where the superscript label denotes that these values are the numerical endpoints of the backgrounds. Once the background solutions have been obtained, we proceed to solve the fluctuation equations by employing the mid-determinant method. For a chosen trial value of the mass $M$, we impose independently the boundary conditions for the fluctuations in the IR and UV---at $r_{1}$ and $r_{2}$, respectively---and use the bulk fluctuation equation(s) to evolve these solutions towards an intermediate value of the radial coordinate $r_{*}$ with $r_{1}^{\text{num}}<r_{1}<r_{*}<r_{2}<r_{2}^{\text{num}}$ (note that the computation of the spin-0 spectrum for an $n$-scalar sigma-model requires that we solve a system of $n$ fluctuation equations, and hence it is necessary to evolve $n$ linearly independent solutions towards the intermediate $r_{*}$). We then construct the $2\times 2$ matrix $\mu(M^2)$ using the evolved solutions and their radial derivatives, evaluated at the midpoint $r_{*}$ (for the fluctuations of $n$ sigma-model scalars this matrix would instead have dimensions $2n\times 2n$). Schematically, for a generic field fluctuation $f(M,r)$ we have
\begin{equation}
\mu(M^{2}) \equiv \begin{pmatrix}
f_{I}(r_{*})& f_{U}(r_{*})\\
f'_{I}(r_{*}) & f'_{U}(r_{*}) 
\end{pmatrix}\,,
\end{equation}          
where the subscript $I$ denotes a solution generated by setting up boundary conditions at the IR regulator, the subscript $U$ represents a solution which evolves backwards from the UV regulator, and primes here denote differentiation with respect to the radial coordinate $r$. We compute the determinant of $\mu$ and repeat this process by varying the trial (squared) mass $M^2$, obtaining for each iteration a single data point $\{M^{2},\,\text{det}(\mu)\}$. The mass spectrum---for the particular choice of the two regulators---is then given by the discrete set of $M^2$ values for which $\text{det}(\mu)=0$, i.e. the set of trial mass values for which we can construct linearly dependent fluctuations which evolve from the IR and UV and smoothly connect at some intermediate midpoint (see for example Ref.~\cite{Berg:2006xy} for an outline of this \emph{midpoint determinant} method).\par
Given the numerical nature of this algorithm, it is worth commenting on a couple of technicalities which arise when computing the spectrum in this way. We first remind the Reader that the parameters $r_{1}$ and $r_{2}$, which are introduced as holographic regulators of the dual field theory, are unphysical, and that the spectrum of physical masses would be obtained only in the limit in which the effect of these regulators is removed: $r_{1}\to r_{o}$ and $r_{2}\to\infty$, where $r_{o}$ is the physical end of the bulk geometry which sets the scale of confinement at low energies. In practice, it is not numerically feasible to take these limits when solving the equations of motion for the fluctuations, and so care should be taken to ensure that the values assigned to the regulators are sufficiently low (high) in the IR (UV) that the extracted tower of states is not subject to any spurious cutoff effects or numerical artefacts. Typically the regulators should be chosen as close to the numerical endpoints of the backgrounds as possible, though this may be limited by the numerical precision being used. Similarly, it is also necessary to check the convergence of the spectrum as a function of the midpoint $r_{*}$, as this parameter may need to be tuned in order to optimise the numerics. The second technicality concerns identifying the zeros of the mass matrix determinant; $\det(\mu(M^2))$ is a function which oscillates around zero, and can only be obtained numerically using a finite number of trial mass values, limited by time and computational resources. As a result, in practice we approximate the zeros of the function by the points at which the determinant changes sign, and hence the accuracy of the spectrum scales with the number of trial masses that we iterate over.

\section{Identifying the dilaton}
\label{Sec:DilatonFormalism}
Nambu--Goldstone Bosons (NGBs) are massless scalar particles that appear in QFT models which exhibit a spontaneously broken continuous symmetry, where the model would otherwise be exactly invariant under these symmetry transformations. For cases in which this spontaneously broken symmetry is not exact (an \emph{approximate} symmetry which is explicitly broken by the Lagrangian of the theory) the corresponding spin-0 particles which are generated are referred to as \emph{pseudo}-Nambu--Goldstone Bosons (pNGBs), and have small non-zero masses. 
As a specific example, the \emph{dilaton} is a hypothetical scalar particle which appears in models which manifest the spontaneous breaking of (approximate) scale invariance, and is the pNGB associated with the breakdown of \emph{dilatation} invariance; in addition to invariance under the Poincar\'{e} and special conformal transformations, dilatation invariance is a necessary requirement for a CFT. The mass of the dilaton is controlled by the degree to which scale invariance is explicitly broken, and is completely suppressed (i.e. the dilaton becomes massless) in the limit in which the spontaneously broken scale symmetry is exact at the level of the Lagrangian.\par 
The dilaton has been studied in many different contexts, and work dedicated to understanding its phenomenology has produced a significant number of papers in the literature: these include, for example, early attempts to describe the dilaton in terms of a low-energy effective field theory~\cite{Migdal:1982jp,Coleman:1985rnk}, work on modelling dynamical electroweak symmetry breaking with a composite dilaton~\cite{Leung:1985sn,Bardeen:1985sm,Yamawaki:1985zg}, studies of near-conformal field theories and lattice data~\cite{Matsuzaki:2013eva,
	Golterman:2016lsd,Kasai:2016ifi,Golterman:2016hlz,Hansen:2016fri,Golterman:2016cdd,Appelquist:2017wcg,
	Appelquist:2017vyy,Golterman:2018mfm,Cata:2019edh,Appelquist:2019lgk,Cata:2018wzl,
	Brown:2019ipr}, and extensions to the standard model which contain a composite Higgs particle~\cite{Goldberger:2008zz,Hong:2004td,Dietrich:2005jn,Hashimoto:2010nw,
	Appelquist:2010gy,Vecchi:2010gj,Chacko:2012sy,
	Bellazzini:2012vz,Bellazzini:2013fga,Abe:2012eu,Eichten:2012qb,Hernandez-Leon:2017kea}. The dilaton has also featured in the context of model building using bottom-up holography~\cite{Elander:2011aa,Kutasov:2012uq,Lawrance:2012cg,Elander:2012fk,Goykhman:2012az,
	Evans:2013vca,Megias:2014iwa,Elander:2015asa}, including within braneworld systems which implement the Goldberger-Wise moduli stabilisation mechanism~\cite{Goldberger:1999uk,
	DeWolfe:1999cp,Goldberger:1999un,Csaki:2000zn,ArkaniHamed:2000ds,Rattazzi:2000hs,Kofman:2004tk}; other papers have instead explored dilaton phenomenology in the context of top-down holography, in studies of certain special confining field theories from the conifold~\cite{ Elander:2009pk,Elander:2012yh,Elander:2014ola,
	Elander:2017cle,Elander:2017hyr}. The wide variety of papers on this topic is in part explained by the relative simplicity of computing spectra in these models: the relevant low-energy features of a system which is known to descend from superstring theory are often retained when one instead considers a sigma-model coupled to gravity, in which high-energy degrees of freedom are neglected and the spectra of fluctuations about the supergravity backgrounds may be calculated rigorously.\par
Despite this relative abundance of papers on the topic, however, the important question of how to actually deduce whether a light scalar state in a given model is indeed a dilaton presents a non-trivial technical difficulty; the spectrum of spin-0 states is typically sourced both by the field theory operators dual to the scalar supergravity fields, in addition to the dilatation operator itself. These physical states may arise as a result of mixing effects between these operators, and it is hence natural to ask how one may distinguish a dilaton (or a mass eigenstate which results from significant mixing with the dilaton) from other generic states with the same quantum numbers. To begin to address this issue, we will next formally introduce the \emph{probe approximation} which we earlier alluded to, and which is discussed and implemented as a test for a variety of models in Ref.~\cite{Elander:2020csd}.\par
We start by reminding the Reader of the gauge-invariant scalar combination $\mathfrak{a}^a(M,r)$ which was introduced in Eq.~(\ref{Eq:a}) of Sec.~\ref{Subsec:HolographicFormalism}:
\begin{align}
\mathfrak{a}^{a}(M,r)&\equiv\varphi^{a}(M,r)\,-\,\frac{\pa_{r}\bar{\Phi}^{a}(r)}{6\pa_{r} A(r)} h(M,r)\notag\\
\label{Eq:aProbe}
&\equiv\varphi^{a}(M,r)\,-\,\Gamma^{a}(r) h(M,r)\,,
\end{align}
where $\varphi^a(M,r)$ are the leading order scalar field fluctuations about the background solutions $\bar{\Phi}^a(r)$, $h(M,r)$ is the (four-dimensional) trace of the tensor component of the five-dimensional ADM-decomposed metric, and we have reinstated the explicit dependences on $M$ and the radial coordinate $r$. According to the holographic dictionary, the bulk fields $\varphi^{a}$ are associated with the scalar operators which define the dual theory, while the metric perturbation $h$ couples to the trace of the stress-energy tensor of the boundary theory and hence sources the dilatation operator. We can therefore predict that a test intended to determine the extent to which each spin-0 state mixes with the dilaton should equivalently measure the mixing effects between the fluctuations of both the sigma-model scalars and the metric.\par
Hence, the diagnostic tool which we propose (and which we call the probe approximation) consists of computing the spectrum of scalar states for each background solution in two separate ways: firstly, by making use of the fluctuation equations and boundary conditions presented in Eqs.~(\ref{Eq:ScalarFluct},\,\ref{Eq:ScalarFluctBC}), which are satisfied by the complete gauge-invariant scalar variables $\mathfrak{a}^{a}$ and which preserve any dilaton admixture which may be present, and secondly by then implementing the approximation $h=0$, which has the effect of decoupling the field fluctuations $\varphi^{a}$ from the dilaton. This second calculation essentially `switches off' any back-reaction which the scalar fluctuations may induce on the bulk geometry, and completely neglects the contribution to the mass eigenstates coming from the metric perturbation. Any spin-0 states which are unaffected by this probe approximation therefore cannot be interpreted as resulting from mixing with the dilaton, since by definition the approximation should only be valid for the fluctuations of the sigma-model scalar fields. By contrast,  if when comparing the spectra for the two computations we observe significant discrepancies between one or more states, then we may infer that the metric perturbation component of the gauge-invariant variable $\mathfrak{a}^{a}$ is \emph{not} negligible, and furthermore that these states are (at least partially) identifiable as the dilaton. For future reference, we shall use the phrase \emph{approximate dilaton} to refer to any state which is determined to be a significant admixture with the dilaton, even when the mass is not necessarily light compared to other states in the spectrum.\par
The bulk equations and boundary conditions which are satisfied by the probe states $\mathfrak{a}^a|_{h=0}\equiv\mathfrak{p}^{a}$ may be obtained by considering the series expansion of $\mathfrak{a}^{a}$ in powers of the vanishingly small parameter $\Gamma^{a}(r)\ll 1$ as defined in Eq.~(\ref{Eq:aProbe}), which to leading order gives
\begin{equation}
\label{Eq:ProbeScalarFluct}
0=\Big[\cd_{r}^{2} + 4\pa_{r} A\,\cd_{r}+e^{-2A} M^{2}\Big]\mathfrak{p}^a-\Big[\cv^{a}_{\,\,\,\,|c}
-\car^{a}_{\,\,\,\,bcd}\pa_{r}\bar{\Phi}^b\pa_{r} \bar{\Phi}^d\Big]\mathfrak{p}^c\,,
\end{equation}
for the bulk fluctuations, while the boundary conditions reduce to the simple Dirichlet form:
\begin{equation}
\label{Eq:ProbeScalarFluctBC}
0=\mathfrak{p}^a\Big|_{r_i}\,.
\end{equation}
We conclude this section with a brief but nevertheless important clarification. Although the probe approximation, as we have defined it, will prove to be an invaluable tool when attempting to detect the presence of (partially) dilatonic scalar states, we emphasise that the approximation is merely a convenient diagnostic test which does not by itself provide any meaningful information about a given model, when removed from this context. The utility of this tool relies on the comparison of its results to those obtained from the proper computation of the complete gauge-invariant spectrum, and it does not otherwise provide us with any further physical insight.

\section{Energetics analysis of phase structure}
\label{Sec:EnergeticsFormalism}
We propose an investigation into the phase structure of two particular supergravity models, chosen for their relative simplicity in the context of top-down holography as examples of gravitational models which admit classical solutions with confining four-dimensional field theories as their boundary duals: the circle compactifiction of Romans six-dimensional half-maximal supergravity~\cite{Romans:1985tw}, and the toroidal compactification of the seven-dimensional maximal supergravity~\cite{Nastase:1999cb,Pernici:1984xx,Pernici:1984zw,Lu:1999bc,Campos:2000yu} admitting a background configuration which provides the holographic description of confinement in four dimensions, as proposed by Witten~\cite{Witten:1998zw}.\par  
Each of these supergravity models admits several distinct classes of background solutions---with correspondingly different bulk geometries---and the question arises of how to use these various solutions in order to explore the phase structure of the model and to ascertain the existence of a phase transition. As we shall see in Chapters~\ref{Chap:SpectraRomans} and~\ref{Chap:SpectraWitten}, our computation of the glueball mass spectra reveals the presence of a classical instability in both systems (in the form of a tachyonic scalar state), and hence we anticipate the existence of a phase transition by necessity: the models which we consider are well-defined and established supergravities, and so there must be some mechanism by which the unphysical region of parameter space containing the instability is separated from the physical region, and is not itself physically realised.\par  
To this end, and with our motivation established, we intend to conduct an energetics analysis of the various classes of solutions within these two theories, predicting that the branch of solutions which contains the tachyonic state must prove to be energetically disfavoured for choices of parameters which bring the system in proximity of the instability. More specifically, we will compute the holographically renormalised free energy density $\cf$ of the system, taking care to use a prescription which allows us to legitimately and unambiguously compare background solutions belonging to different classes.\par 
A detailed derivation of the free energy density for the two models will be provided in their respective Chapters~\ref{Chap:EnergeticsRomans} and~\ref{Chap:EnergeticsWitten}, together with an explanation of the various parameters upon which $\cf$ depends; here we provide only a schematic definition for the free energy, starting with the classical action of a $D$-dimensional system containing a single scalar field coupled to gravity, with two boundaries situated at regulated values of the holographic coordinate $\r=\r_{i}$. This action is given by
\begin{equation}
\label{Eq:GenActionF}
\cs=\cs_D
+\sum_{i=1,2}(-)^{i}\is\int\is\di^{D-1}x \,\sqrt{-\tilde{\hat{g}}}\left[
\frac{\ck}{2}+\l_{i}
\right]_{\r=\r_i},
\end{equation}         
where $\cs_{D}$ is the classical bulk action which contains the $D$-dimensional Ricci scalar $\cal{R}_{D}$ and the kinetic and potential terms for the scalar field, $\tilde{\hat{g}}$ denotes the determinant of the pullback of the metric induced at each boundary, $\ck$ is the extrinsic curvature term of the Gibbons-Hawking-York (GHY) action which we are required to include due to the presence of boundaries, and $\l_{i}$ are boundary-localised potentials which we presently neglect to specify for simplicity. We will be more explicit in later chapters, but here it suffices to state that the classical equations of motion obtained from the bulk action, together with the large-$\r$ asymptotic expansions for the scalar field and metric warp factor, may be used to derive the required expression for the free energy:   
\begin{equation}
\label{Eq:FreeEnergyDef}
F\equiv - \lim_{\r_1\rightarrow \r_o}\lim_{\r_2\rightarrow +\infty}{\cal S}\equiv\int\is\text{d}^{D-1}x\,\cf\, ,
\end{equation}
where $\r_o$ is the physical end of the geometry with $\r_o < \r_{1}$, $\cs$ is the complete (appropriately renormalised) on-shell action, and $\cf$ is the free energy density. We will later show that $\cf$ may be formulated as a function of a special set of deformation parameters which characterise the UV (large $\r$) asymptotic field expansions and that, in order to obtain any meaningful results from this analysis, it is necessary to employ a numerical routine to extract physical values for these parameters. This necessity arises because the evolution of the non-linear classical equations of motion into the bulk geometry, combined with the imposition of boundary conditions in the IR, yields non-trivial implicit functional relations between the UV parameters; this encodes the non-perturbative dynamics of the dual field theory. \par 
Briefly, this numerical process is as follows: for each class of solutions within the supergravity models, we use their IR (small $\r$) field expansions to generate a family of numerical backgrounds which solve the classical equations of motion, and then systematically match each of these backgrounds (and their derivatives) to the general UV expansions, solving for each parameter in turn. In this way we are able to extract a complete table of values for the various UV parameters, with each single set of values  unambiguously identifying a unique numerical background within the family. We will provide a more comprehensive explanation of this procedure in Chapters~\ref{Chap:EnergeticsRomans} and~\ref{Chap:EnergeticsWitten}.\par       
As we earlier alluded to, it is insufficient to simply plot the free energy density using this acquired UV parameter data, as there exists a further subtlety which must first be addressed. We will later demonstrate that it is necessary to introduce an appropriate scale setting procedure by which we rescale all physical quantities of interest to our study, including the free energy density and its arguments. Only then are we able to plot the free energy density $\cf$ as a function of the numerically obtained UV parameter data for each branch of solutions, and explore the phase structure for the two supergravity models.

\part{Spectra of composite states}
\label{PartOne}

\chapter{Example application: $T^{n}$-compactification of $\textrm{AdS}_{n+5}$ system  }
\label{Chap:nTorus}
\section{Formalism of the \emph{D}-dimensional model}
\label{Sec:FormalismAdSDmodel}
As discussed in Chapter~\ref{Chap:Intro} this Thesis will focus primarily on the study of two specific supergravities, both of which are well-known in the literature on top-down holography; these are the six-dimensional half-maximal theory first written by Romans~\cite{Romans:1985tw} which we compactify on an $S^{1}$, and the seven-dimensional maximal theory~\cite{Nastase:1999cb,Pernici:1984xx,Pernici:1984zw,Lu:1999bc,Campos:2000yu} compactified on a torus $T^2$, which admits the $\text{AdS}_{7}$ background solution constructed by Witten~\cite{Witten:1998zw}.
Although it is known that no supersymmetric $\text{AdS}_{D}$ supergravity solutions exist for $D>7$ (see Refs.~\cite{Nahm:1977tg,Kac:1977em,DeWitt:1981wm}), recent work has uncovered the existence of non-supersymmetric $\text{AdS}_{8}$ solutions within type-IIA supergravity~\cite{Cordova:2018eba}; similar solutions which do not fall under the exhaustive classification of Nahm~\cite{Nahm:1977tg} may yet be discovered. Furthermore, higher-dimensional models have also proven to be of phenomenological interest in the context of the \emph{clockwork mechanism}~\cite{Choi:2015fiu,Kaplan:2015fuy,Giudice:2016yja}, where the compactification of a large number of dimensions may be used to generate mass hierarchies without relying on the introduction of other small parameters~\cite{Teresi:2018eai}. We therefore find it instructive to begin by investigating a generic $D$-dimensional system described by the Einstein--Hilbert action, supplemented by a constant potential term $\cv_{D}<0$. Such a model admits background solutions which realise an $\text{AdS}_{D}$ bulk geometry. We discuss this model mainly as an example application of the probe approximation introduced in Sec.~\ref{Sec:DilatonFormalism}, and we do not provide any further justification of its inclusion here; the results of our analysis will nevertheless motivate a brief remark when we come to apply the same techniques to our spectra computations for the two supergravity theories.\par 
The simple pure gravity action in $D=5+n$ dimensions which we shall adopt is defined as follows:   
\begin{equation}
\label{Eq:DAction}
\cs_D=\int\is\di^{D}x\, \sqrt{-\hat{g}_D} \left(\frac{\car_{D}}{4}-\cv_{D}\right)\,,
\end{equation}
where $\hat{g}_{D}$ is the determinant of the D-dimensional metric tensor, $\car_{D}\equiv \hat{g}^{\hat{M}\hat{N}}R_{\hat{M}\hat{N}}$ with $\hat{M},\hat{N}\in\{0,1,2,3,5,\ldots,D-1,D\}$ is the corresponding Ricci curvature scalar, and $\mathcal{V}_{D}$ is the constant potential. The $\text{AdS}_{D}$ geometry has a radius of curvature given by
\begin{equation}
R^{2}\equiv-\frac{1}{4\cv_{D}}(D-1)(D-2)\,,
\end{equation}
which can be fixed to unity by defining our potential as
\begin{equation}
\cv_{D}=-\frac{1}{4}(D-1)(D-2)=-\frac{1}{4}(n+4)(n+3)\,.
\end{equation}

\section{Toroidal reduction to \emph{D=}5 dimensions}
\label{Sec:nToroidalReduction}
\subsubsection{The metric}
We start by assuming that $n$ internal dimensions of the geometry each wrap around a separate $S^{1}$---together describing an $n$-torus $T^{n}$---and reduce the system to five dimensions by compactifying on this torus. We furthermore assume that the individual volumes of the $n$ circles are parametrised by two scalar fields only, so that the $n-2$ additional scalars which would be introduced when $D\geqslant8$ have been (consistently) truncated. Our adopted ansatz for the $D$-dimensional line element (for $n\geqslant 2$) may hence be written as follows: 
\begin{equation} \label{Eq:Dmetric}
\di s^{2}_{D}=e^{-2\d\bar{\c}}\di s_{5}^{2} 
+e^{\frac{6}{n}\d \bar{\c}}\left(\sum_{i=1}^{n-1}
e^{\sqrt{\frac{8}{n(n-1)}}\bar{\w}}\di \theta_{i}^{2}
+e^{-\sqrt{\frac{8(n-1)}{n}}\bar{\w}} \di \theta_{n}^{2}\right)\,,
\end{equation}
where $\di s_{5}^{2}$ is the metric for the five-dimensional domain-wall geometry as defined in Eq.~(\ref{Eq:5Dmetric}), and $0\leqslant\theta_i < 2\pi$ for $i=1,\ldots,n$ are the periodic coordinates which parametrise the $T^{n}$. The two aforementioned scalars are $\bar{\c}$ and $\bar{\w}$, where the latter is here associated with the generator of the $U(1)^{n}\simeq SO(2)^{n}$ symmetry of the $T^{n}$. A natural choice for the free parameter $\d=\d(n)$ will become apparent in the process of dimensionally reducing the system. From this metric ansatz, and assuming that the scalars and warp factor are dependent \emph{only} on the holographic coordinate $r$, we obtain
\begin{equation}
\sqrt{-\hat{g}_{D}}=e^{4A-2\d\bar{\c}}=e^{-2\d\bar{\c}}\sqrt{-g_{5}}\,,
\end{equation}
which we note is independent of $\bar{\w}$. For the Ricci scalar we derive the following expression:
\begin{align}
\car_{D}&=-2e^{2\d\bar{\c}}\Big[4A''-\d\bar{\c}''+10\big(A'\big)^{2}
-4\d A'\bar{\c}'+\tfrac{3(D-2)}{2(D-5)}\d^{2}\big(\bar{\c}'\big)^{2} +\big(\bar{\w}'\big)^{2}   \Big]\nn\\
=&-2e^{2\d\bar{\c}}\Big[4A''-\d\bar{\c}''+10\big(A'\big)^{2}
-4\d A'\bar{\c}'+\tfrac{3(n+3)}{2n}\d^{2}\big(\bar{\c}'\big)^{2} +\big(\bar{\w}'\big)^{2}   \Big]\,,
\end{align} 
so that the following useful relation is satisfied:
\begin{align}
\sqrt{-\hat{g}_{D}}\car_{D}&=\sqrt{-g_{5}}\Big[\car_{5} +2\d\bar{\c}''
+8\d A'\bar{\c}' -\tfrac{3(D-2)}{D-5}\d^{2}\big(\bar{\c}'\big)^{2} -\big(\bar{\w}'\big)^{2}  \Big]\nn\\
&=\sqrt{-g_{5}}\Big[\car_{5} +2\d\bar{\c}''
+8\d A'\bar{\c}' -\tfrac{3(n+3)}{n}\d^{2}\big(\bar{\c}'\big)^{2} -\big(\bar{\w}'\big)^{2}  \Big]\,.  \label{Eq:nTorusRelation}
\end{align}
We conclude this subsection by observing that for background solutions which have vanishing $\bar{\w}=0$, the bulk geometry preserves an $n$-dimensional rotational symmetry within the space spanned by the toroidal coordinates; in such a case the metric takes the following form:
\begin{equation}\label{Eq:DmetricOmega0}
\di s^{2}_{D}=\di\r^{2} +e^{2(A-\d\bar{\c})}\di x^{2}_{1,3}
+e^{\frac{6}{n}\d \bar{\c}}\sum_{i=1}^{n}\di\theta^{2}_{i}\,,
\end{equation}
where we have also introduced the convenient reparametrisation of the radial coordinate via $\di r\equiv e^{\d\bar{\c} }\di\r$. For backgrounds which further satisfy the identification $An=(n+3)\d\bar{\c}$, Poincar\'{e} invariance is locally preserved within the $(n+4)$-dimensional subspace spanned by the Minkowski and toroidal coordinates, and the metric ansatz simplifies to
\begin{equation}\label{Eq:DmetricSymmetric}
\di s^{2}_{D}=\di\r^{2} +e^{2\ca} \left(
\di x^{2}_{1,3} +\sum_{i=1}^{n}\di\theta^{2}_{i} \right)\,,
\end{equation}
with $\ca\equiv\frac{3}{n+3}A=\frac{9}{2n}\d^{2}A$. We shall briefly return to this remark in Sec.~\ref{Sec:nTorusEOMs}.

\subsubsection{The action}
Having characterised the underlying geometry of the model, we next turn our attention to reducing the action $\cs_{D}$ to five dimensions by compactifying on the $n$-torus $T^{n}$; in performing this reduction we assume that none of the fields have any dependence on the torus angles. By direct substitution of Eq.~(\ref{Eq:nTorusRelation}) we find that Eq.~(\ref{Eq:DAction}) may be rewritten as follows:
\begin{align}
\cs_{D}=\int\prod_{i=1}^{n}\di\theta_{i}
\int\is\di^{5}x\,\sqrt{-g_{5}}\bigg(
\frac{\car_{5}}{4} +\half\d\bar{\c}''&+2\d A'\bar{\c}' -\tfrac{3(n+3)}{4n}\d^{2}\big(\bar{\c}'\big)^{2} \nn\\ &\hspace{8.8mm}-\half\big(\bar{\w}'\big)^{2} -e^{-2\d\bar{\c}}\cv_{D}
\bigg)\,, \label{Eq:DActionSubR}
\end{align}
where primes denote differentiation with respect to $r$. This can then be reformulated solely in terms of five-dimensional dynamical quantities by postulating equivalence to an expression of the form 
\begin{equation}
\cs_{D}=\int\prod_{i=1}^{n}\di\theta_{i}
\left\{\tilde{\cs}_{5} +\pa\cs \right\}
=(2\pi)^n\left\{\tilde{\cs}_{5} + \pa\cs \right\}\,, \label{Eq:DActionTotalDeriv}
\end{equation}
where the integrand measure simply gives the total volume of the $n$ circles internal to the $T^{n}$ as a prefactor. Here $\tilde{\cs}_{5}$ is the general five-dimensional action presented in Eq.~(\ref{Eq:5Daction}) (neglecting the boundary-localised contributions):
\begin{equation}
\tilde{\cs}_{5}=\int\is\di^{5}x\,\sqrt{-g_{5}}
\bigg(\frac{\car_{5}}{4}-\half G_{ab} g^{MN}\pa_{M}\Phi^{a}\pa_{N}\Phi^{b} -\cv \bigg)\,,   \label{Eq:nTorus5Daction}
\end{equation}
with $\Phi^{a}=\{\bar{\c},\,\bar{\w}\}$, while the total derivative term $\pa S$ is given by
\begin{equation}
\pa\cs=\frac{\d}{2}\int\is\di^{5}x \,\pa_{M}\Big(
\sqrt{-g_5} \,g_{5}^{MN} \pa_{N}\bar{\c} \Big)\,.
\end{equation}
By comparing Eqs.~(\ref{Eq:DActionSubR}) and~(\ref{Eq:DActionTotalDeriv}) we therefore deduce that $\cv$ must be related to the constant potential appearing in $\cs_{D}$ by the relation
\begin{equation}
\cv=e^{-2\d \bar{\c}}\cv_{D}\,,
\end{equation}
and we furthermore find that $G_{\bar{\w}\bar{\w}}=1$. The kinetic term for the scalar $\bar{\c}$ may be canonically normalised if we also choose $G_{\bar{\c}\bar{\c}}=1$ and hence fix the free parameter $\d$ to be
\begin{equation}
\label{Eq:delta}
\d^{2} = \frac{2n}{3(3+n)}\,,
\end{equation}
so that the sigma-model metric of the dimensionally reduced system is simply the identity matrix $G_{ab}=\d_{ab}$. The Ricci scalar simplifies to
\begin{equation}
\car_{D}=\tfrac{2}{3}\Big[10\cv_{D} +n(n+4)\bar{\c}'\Big]\,.
\end{equation}

\section{Equations of motion and solutions}
\label{Sec:nTorusEOMs}
\subsubsection{Equations of motion}
The classical equations of motion which follow from the toroidal reduction to five-dimensions are derived from the general results shown in Eqs.~(\ref{Eq:ScalarEOM}\,-\,\ref{Eq:Einstein2}) of Section~\ref{Subsec:HolographicFormalism}; recalling that we assume field dependence \emph{only} on the radial coordinate (and hence no dependence on the Minkowski and torus coordinates), these EOMs are given by:
\begin{align}
\pa^{2}_{r}\bar{\c} + 4\pa_{r}\bar{\c}\pa_{r}A &= \frac{\pa\cv}{\pa\bar{\c}}\,,\\
\pa^{2}_{r}\bar{\w} + 4\pa_{r}\bar{\w}\pa_{r}A &= 0\,,\\
3\pa^{2}_{r}A +6(\pa_{r}A)^{2} +(\pa_{r}\bar{\c})^{2} +(\pa_{r}\bar{\w})^{2}&= -2\cv\,,\\
6(\pa_{r}A)^{2} -(\pa_{r}\bar{\c})^{2} -(\pa_{r}\bar{\w})^{2} &= -2\cv\,.
\end{align}
This system of equations may then be conveniently rewritten in terms of the $D$-dimensional potential $\cv_{D}$ by implementing the change of radial coordinate $\pa_{r}\equiv e^{-\d\bar{\c}}\pa_{\r}$ defined just after Eq.~(\ref{Eq:DmetricOmega0}), so that we equivalently have
\begin{align}
\pa^{2}_{\r}\bar{\c}
+\big(4\pa_{\r}A-\d\pa_{\r}\bar{\c}\big)\pa_{\r}\bar{\c}&=
-2\d\cv_{D}\, ,\label{Eq:nTorusEOM1}\\
\pa^{2}_{\r}\bar{\w}+\big(4\pa_{\r}A-\d\pa_{\r}\bar{\c}\big)
\pa_{\r}\bar{\w}&=0
\, ,\label{Eq:nTorusEOM2}\\
3\pa^{2}_{\r}A +6(\pa_{\r}A)^{2} +(\pa_{\r}\bar{\c})^{2} +(\pa_{\r}\bar{\w})^{2} -3\d\pa_{\r}A\pa_{\r}\bar{\c} &=
-2\cv_{D}\, ,\label{Eq:nTorusEOM3}\\
6(\pa_{\r}A)^{2}-(\pa_{\r}\bar{\c})^{2}-(\pa_{\r}\bar{\w})^{2}&=
-2\cv_{D}\, .\label{Eq:nTorusEOM4}
\end{align}
As an aside we notice that the combined summation of Eq.~(\ref{Eq:nTorusEOM1}), $-\tfrac{\d}{2}\times$Eq.~(\ref{Eq:nTorusEOM3}), and $-\tfrac{\d}{2}\times$Eq.~(\ref{Eq:nTorusEOM4}) gives a vanishing quantity, and correspondingly that
\begin{equation}
\label{Eq:nTorusVanish}
\tfrac{3\d}{2}A'' -\bar{\c}'' + 6\d(A')^{2} +\d(\bar{\c}')^{2}
-\left(4+\tfrac{3\d^2}{2}\right)A'\bar{\c}' = 0\,.
\end{equation}
This expression may be reformulated as a total derivative with respect to $\r$, so that
\begin{equation}
\label{Eq:nTorusC}
e^{4A-\bar{\c}}\Big(\tfrac{3\d}{2}\pa_{\r}A -\pa_{\r}\bar{\c} \Big)=C
\end{equation}
represents a conserved quantity at all energy scales, for some background-dependent constant $C$. We hence observe that by simultaneously imposing the constraints $An=(n+3)\d\bar{\c}$ and $\bar{\w}=An-(n+3)\d\bar{\c}$, all equations of motion are satisfied and we recover the maximally symmetric geometry which locally preserves $(D-1)$-dimensional \pin as described by the metric in Eq.~(\ref{Eq:DmetricSymmetric}). We mention this observation here merely in passing, though we shall later see that analogous conserved quantities also exist for the two supergravity theories, and these will play an important role in our energetics analysis of their respective phase structures.    

\subsubsection{Confining solutions}
In order to holographically compute the spectra of gauge-invariant fluctuations $\mathfrak{a}^{a}$ and $\mathfrak{e}^{\mu}_{\,\,\,\,\nu}$ as defined in Eqs.~(\ref{Eq:a}) and~(\ref{Eq:e}), respectively, we require that our dimensionally reduced model is able to geometrically realise a low-energy scale of confinement within the dual field theory. This motivates us to here introduce a class of background solutions for which one of the internal circles of the $n$-torus (parametrised by $\theta_{n}$) shrinks to a point at some finite value of the radial coordinate $\r=\r_{o}$ in the deep IR, so that the bulk geometry tapers and smoothly closes off. As discussed in Chapter~\ref{Chap:Intro} we may naturally interpret this geometric property as an intrinsic low-energy limit in the boundary theory, and the spectra of gauge-invariant fluctuations about these background profiles as physical states which exhibit confinement. Conversely, in the large-$\r$ limit we asymptotically recover the $\text{AdS}_{D}$ geometry with unit curvature.\par 
The family of analytical solutions to the classical equations of motion presented in Eqs.~(\ref{Eq:nTorusEOM1}\,-\,\ref{Eq:nTorusEOM4}) which meet these requirements may be written as~\cite{Elander:2020csd}  
\begin{align}
\label{Eq:nTorusConfChi}
\bar{\c}(\r)&=\bar{\c}_{I}+\sqrt{\tfrac{n+3}{6n(n+4)^2}}\bigg\{
n(n+4)\ln\Big[\tfrac{2}{n+4}\Big]
+n\ln\Big[\tfrac{1}{2}\sinh\big((n+4)\r\big)\Big]\nn\\
&\hspace{57mm}
-4\ln\Big[\coth\big(\tfrac{1}{2}(n+4)\r\big)\Big]
\bigg\}\,,\\
\label{Eq:nTorusConfOmega}
\bar{\w}(\r)&=\bar{\w}_{I}-\sqrt{\tfrac{n-1}{2n}}\ln\Big[
\tanh\Big(\tfrac{1}{2}(n+4)\r\Big)\Big]\,,\\
\label{Eq:nTorusConfA}
A(\r)&=\tfrac{n+3}{3(n+4)}\ln\Big[\tfrac{1}{2}\sinh\big((n+4)\r\big)\Big]
+\tfrac{1}{3(4+n)}\ln\Big[\tanh\big(\tfrac{1}{2}(n+4)\r\big)\Big]\,,
\end{align}
where an integration constant $\r_{o}$ which fixes the end of space has been set to zero without loss of generality, and we have exploited the fact that an additive shift to $A(\r)$ leaves the EOMs invariant to also set another constant $A_{I}$ to zero. A third integration constant $\bar{\c}_{I}$ is not a free parameter however, and is constrained by the requirement that the $D$-dimensional geometry does not contain a conical singularity. To demonstrate this point explicitly let us first consider series expanding the exact solutions in proximity to the end of space at $\r=\r_{o}=0$, which yields the following IR expansions:  
\begin{align}
\label{Eq:nTorusConfChiExp}
\bar{\c}(\r)&=\bar{\c}_{I}+\sqrt{\tfrac{n+3}{6n}}\bigg[
(n-1)\ln(2) +6\ln(\r) -\tfrac{1}{6}(n-2)(n+4)\r^{2} +\ldots\bigg]\,,\\
\label{Eq:nTorusConfOmegaExp}
\bar{\w}(\r)&=\bar{\w}_{I}-\sqrt{\tfrac{n-1}{2n}}\bigg[
\ln\left(\tfrac{n+4}{2}\right) +\ln(\r) -\tfrac{1}{12}(n+4)^{2}\r^{2}
+\ldots\bigg]\,,\\
\label{Eq:nTorusConfAExp}
A(\r)&=\frac{1}{3}\bigg[
\ln\left(\tfrac{n+4}{2}\right) +\ln(\r) +\tfrac{1}{12}(n+4)(2n+5)\r^{2} +\ldots\bigg]\,,
\end{align} 
where the unwritten subsequent terms are of order $\co\big(\r^4\big)$. We restrict our attention to the two-dimensional subspace spanned by $\r$ and $\theta_{n}$ (which has the topology of a cylinder), and examine its behaviour when we impose that the $S^{1}$ parametrised by $\theta_{n}$ shrinks to zero volume by directly substituting in these expansions. The following expression is obtained for the line element:
\begin{align}
\di\tilde{s}_{2}^{2}&=\di\r^2 + e^{\frac{6}{n}\d\bar{\c}
	-\sqrt{\frac{8(n-1)}{n}}\bar{\w}}\di\theta_{n}^{2}\nn\\
&=\di\r^2 + e^{\sqrt{\frac{24}{n(n+3)}}\bar{\c}_{I}
	-\sqrt{\frac{8(n-1)}{n}}\bar{\w}_{I} }\di\theta_{n}^{2}\nn\\
&=\di\r^{2} + \r^{2}\di\theta_{n}^{2}\,,
\end{align} 
where in going from the second line to the third we have made the necessary identification
\begin{equation}
\bar{\c}_{I}=\Big[\tfrac{1}{3}(n-1)(n+3)\Big]^{\half}\bar{\w}_{I}\,,
\end{equation} 
to ensure that we recover the standard metric of the plane in polar coordinates and hence avoid an angular deficit. The remaining integration constant $\bar{\w}_{I}$ may otherwise be freely assigned, and for simplicity we choose to fix $\bar{\w}_{I}=\bar{\c}_{I}=0$ henceforth.

\subsubsection{Hyperscaling violating solutions}
If we instead consider the large-$\r$ limit of the analytical solutions in Eqs.~(\ref{Eq:nTorusConfChi}\,-\,\ref{Eq:nTorusConfA}) in proximity of the UV boundary, we obtain the following exact solutions:
\begin{align}
\bar{\c}_{hv}(\r)&=\Big[\tfrac{1}{6}n(n+3)\Big]^{\half}\r\,,\\
\bar{\w}_{hv}(\r)&=0\,,\\
A_{hv}(\r)&=\frac{1}{3}(n+3)\r\,,
\end{align}
which satisfy the relations $\bar{\w}=An-(n+3)\d\bar{\c}=0$ discussed just after Eq.~(\ref{Eq:nTorusC}), corroborating our statement that the confining backgrounds asymptotically realise an $\text{AdS}_{D}$ geometry in the far UV. They correspond to the \emph{hyperscaling violating} (HV) solutions studied in Refs.~\cite{Elander:2015asa,Teresi:2018eai} (see also Ref.~\cite{Dong:2012se} for a general review of hyperscaling violation in the context of holography), and we can briefly demonstrate this behaviour---following the notation of Ref.~\cite{Elander:2015asa}---by defining 
\begin{equation}
\a(n)\equiv\frac{1}{3}(n+3) \quad,\quad \g(n)\equiv\Big[\tfrac{1}{6}n(n+3)\Big]^{\half}\,,
\end{equation}
so that $A_{hv}=\a\r$ and $\bar{\c}_{hv}=\g\r$, and furthermore by introducing the reparametrisation of the holographic coordinate $\r\to z$ defined via
\begin{equation}
\di\r \equiv \big(\g\d-\a\big)^{-1}z^{-1}\di z = -z^{-1}\di z\,. 
\end{equation}
After substituting in for $A_{hv}$ and $\bar{\c}_{hv}$ and implementing this change of coordinate, the five-dimensional metric may be reformulated as follows:
\begin{align}
\di s_{5}^{2}&=e^{2A_{hv}(\r)}\di x_{1,3}^{2} + e^{2\d\bar{\c}_{hv}(\r)}\di\r^{2} \nn\\
&=z^{-2-\frac{2n}{3}}\Big(\di x_{1,3}^{2} + \di z^{2}  \Big)\,,
\end{align}
which we see transforms as $\di s_{5}^{2}\to \s^{-\frac{2n}{3}}\di s_{5}^{2}$ under a generic coordinate rescaling $x^{\mu}\to\s x^{\mu}$ and $z\to\s z$, exhibiting a hyperscaling coefficient dependent on the dimensionality of the $n$-torus.\par 
In the next section we shall use these HV solutions---which approximate the analytical confining backgrounds in the large-$n$ limit---to highlight some interesting properties of the fluctuation equations for the gauge-invariant scalars $\mathfrak{a^{a}}$. Although the mass spectra are computed using the exact solutions in Eqs.~(\ref{Eq:nTorusConfChi}\,-\,\ref{Eq:nTorusConfA}) we find that the simpler HV solutions nevertheless capture some important qualitative features of the results; in particular they provide an effective estimate of an upper bound on the dimensionality of the $n$-torus, above which the $\mathfrak{p}^{\bar{\c}}$ probe states acquire a spurious dependence on the imposed boundary conditions.   

\subsubsection{Skewed solutions}
As a concluding remark, we observe that the classical equations of motion presented in Eqs.~(\ref{Eq:nTorusEOM1}\,-\,\ref{Eq:nTorusEOM4}) are invariant under the transformation $\bar{\w}(\r)\to-\bar{\w}(\r)$; the system hence admits an additional class of backgrounds which are related to, but geometrically distinct from, the solutions which exhibit confinement. Their behaviour at the end of space can be determined by again examining the two-dimensional line element parametrised by $\r$ and $\theta_{n}$:
\begin{align}
\di\tilde{s}_{2}^{2}&=\di\r^2 + e^{\frac{6}{n}\d\bar{\c}
	-\sqrt{\frac{8(n-1)}{n}}\bar{\w}}\di\theta_{n}^{2}\nn\\
&=\di\r^2 +\Big(\tfrac{2}{n+4}\Big)^{4-\frac{4}{n}}\r^{-2+\frac{4}{n}}
\,\di\theta_{n}^{2}\,,
\end{align}
from which we deduce that this geometry does \emph{not} smoothly close off in the deep IR. Instead, the volume of the $S^{1}$ spanned by $\theta_{n}$ converges to a non-zero constant if the $T^{n}$ saturates the dimensionality lower bound ($n=2$) imposed by our metric ansatz, and diverges in the $\r\to\r_{o}=0$ limit if $n>2$; the name \emph{skewed} is chosen to reflect this geometric property. Although these backgrounds are mentioned here merely as an interesting aside, we will find that similar types of solutions exist within the two compactified supergravity theories; in each case the equations of motion are found to be invariant under a transformation which flips the sign of a linear combination of fields, while simultaneously leaving another combination unchanged.

\section{Fluctuation equations}
\label{Sec:nTorusFluctEqns}
\subsubsection{Gauge-invariant states}
To compute the spectra of physical resonances in the toroidally reduced system---and ultimately to test our probe state analysis for detecting a dilaton admixture---we consider fluctuations about the confining backgrounds introduced in Sec.~\ref{Sec:nTorusEOMs}, and solve the corresponding equations presented in Eqs.~(\ref{Eq:TensorFluct}\,-\,\ref{Eq:ScalarFluctBC}) using the numerical procedure described in Sec.~\ref{Subsec:NumericalImp}. As earlier anticipated, in this subsection we shall use the hyperscaling violating solutions instead to discuss some important qualitative features of these spectra.\par 
We start by considering the gauge-invariant field 
$\mathfrak{e}^{\mu}_{\,\,\,\,\nu}(M,\r)$ associated with the tensor fluctuations of the ADM-decomposed metric, which satisfy the bulk equation:
\begin{align}
\label{Eq:nTorusTensorFluct}
0&=\Big[\pa_{\r}^2 +\big(4\pa_{\r}A-\d\bar{\c}\big)\pa_{\r} +e^{2(\d\bar{\c}-A)}M^2\Big] \mathfrak{e}^{\mu}_{\,\,\,\,\nu}\\
&=\Big[\pa_{\r}^2 +(n+4)\pa_{\r} +e^{-2\r}x^{2}M^{2}\Big] \mathfrak{e}^{\mu}_{\,\,\,\,\nu}\,,  \label{Eq:nTorusTensorFluctHV}
\end{align}
where the second line follows from the direct substitution of the HV backgrounds, with $\d=\d(n)$ as defined in Eq.~(\ref{Eq:delta}).\ The parameter $x\equiv e^{-A_{I}}$ has been introduced to absorb any arbitrary constant which may be added to $A_{hv}(\r)$.\ The simplified Eq.~(\ref{Eq:nTorusTensorFluctHV}) admits a general solution which comprises linear combinations of Bessel functions $J_{\a}$ and $Y_{\a}$, given by
\begin{equation}
\mathfrak{e}^{\mu}_{\,\,\,\,\nu}(M,\r)=e^{-(n+4)\frac{\r}{2}}\Big[
c_{J}J_{2+\frac{n}{2}}(xMe^{-\r}) + c_{Y}Y_{2+\frac{n}{2}}(xMe^{-\r})
\Big]\,,  \label{Eq:nTorusTensorSolution}
\end{equation} 
where $c_{J}$ and $c_{Y}$ are constants. By imposing the required Neumann boundary conditions at $\r=0$ and $\r\to\infty$ (the latter necessitating $c_{Y}$=0), we determine that solutions in the large-$n$ limit correspond to the zeros of $J_{1+\frac{n}{2}}(xM)$. Hence, as discussed in Ref.~\cite{Elander:2020csd} using approximations from Ref.~\cite{Bessel}, the spectrum of spin-2 states asymptotes to a \emph{gapped continuum} as the number of circles diverges (see Fig.~\ref{Fig:nTorusSpectrum}).\par
Turning our attention now to the gauge-invariant spin-0 fields $\mathfrak{a}^{a}(M,\r)$, constructed from the fluctuations of the sigma-model scalars and the trace of the tensor component of the ADM-decomposed metric, we find that the (coupled) bulk equations may be rewritten as follows: 
\begin{align}
\label{Eq:nTorusScalarFluct}
0 &= \Big[\pa_{\r}^{2} +\big(4\pa_{r}A-\d\pa_{\r}\bar{\c} \big)\pa_{\r} + e^{2(\d\bar{\c}-A)} M^2 \Big] \mathfrak{a}^a \nn\\
& -e^{2\d\bar{\c}}\Big[\d^{a\bar{\c}}\pa_{\bar{\c}}^{2}\cv 
+\tfrac{4}{3\pa_{\r} A}\pa_{\bar{\c}}\cv
\Big(\pa_{\r}\bar\Phi^a +\d^{a\bar{\c}}\pa_{\r}\bar{\c}\Big)
 +\tfrac{16\cv}{9(\pa_{\r}A)^2}\pa_{\r}\bar\Phi^a \pa_{\r}\bar{\c} \Big] \mathfrak{a}^{\bar{\c}}\nn\\
 & -e^{2\d\bar{\c}}\Big[\tfrac{4}{3\pa_{\r} A}\d^{a\bar{\c}}\pa_{\bar{\c}}\cv\pa_{\r}\bar{\w}
 +\tfrac{16\cv}{9(\pa_{\r}A)^2}\pa_{\r}\bar\Phi^a \pa_{\r}\bar{\w} \Big] \mathfrak{a}^{\bar{\w}}\,.
\end{align}
These equations are simplified significantly after making the replacements using the HV solutions, and in particular the gauge-invariant scalars $\mathfrak{a}^{\bar{\c}}$ and $\mathfrak{a}^{\bar{\w}}$ completely decouple. We are left with: 
\begin{equation}
 0=\Big[\pa_{\r}^2 +(n+4)\pa_{\r} +e^{-2\r}x^{2}M^{2}\Big] \mathfrak{a}^{a}\,,   \label{Eq:nTorusScalarFluctHV}
\end{equation}
which is identical to the corresponding equation for the tensor modes in Eq.~(\ref{Eq:nTorusTensorFluctHV}). The boundary conditions satisfied by the scalar fluctuations were introduced in Eq.~(\ref{Eq:ScalarFluctBC}), and are rewritten in a more convenient form below:
\begin{align}
\label{Eq:nTorusScalarFluctBC}
0&=\Big[\tfrac{3}{2}M^{2}e^{-2A}\pa_{\r}A\Big]
\mathfrak{a}^{a}\Big\rvert_{\r_i}
-\pa_{\r}\bar{\Phi}^{a}\Big[
\pa_{\bar{\c}}\cv -e^{-2\d\bar{\c}}\pa_{\r}\bar{\c}\pa_{\r}
+\tfrac{4\cv}{3\pa_{\r}A}\pa_{\r}\bar{\c} \Big]   \mathfrak{a}^{\bar{\c}}\Big\rvert_{\r_i}\nn\\
&\hspace{36.7mm}+\pa_{\r}\bar{\Phi}^{a}\pa_{\r}\bar{\w}\Big[
e^{-2\d\bar{\c}}\pa_{\r} -\tfrac{4\cv}{3\pa_{\r}A}  \Big]   
\mathfrak{a}^{\bar{\w}}\Big\rvert_{\r_i}\\
&=\Big[\tfrac{1}{2}(n+3)x^{2}M^{2}e^{-2\r}\Big]
\mathfrak{a}^{a}\Big\rvert_{\r_i}
 +\sqrt{\tfrac{n}{6}(n+3)}
 \pa_{\r}\bar{\Phi}^{a}\pa_{\r}\mathfrak{a}^{\bar{\c}}\Big\rvert_{\r_i}\,,
\end{align}
where once again the second equality follows from the substitution of the HV solutions. We therefore see that $\mathfrak{a}^{\bar{\w}}$ satisfies Dirichlet boundary conditions (recall that $\bar{\w}_{hv}(\r)=0$), while $\mathfrak{a}^{\bar{\c}}$ instead obeys the following (Robin) boundary conditions:
\begin{equation}  \label{Eq:nTorusChiFluctBC}
0=\bigg[\frac{n}{3}\pa_{\r} +x^{2}M^{2}e^{-2\r} \bigg]\mathfrak{a}^{\bar{\c}}\Big\rvert_{\r_i}\,.
\end{equation}
The general solution for $\mathfrak{a}^{\bar{\w}}(M,\r)$ takes the same form as that in Eq.~(\ref{Eq:nTorusTensorSolution}), though the required Dirichlet BCs mean that solutions are instead given by the zeros of $J_{2+\frac{n}{2}}(xM)$; in the large-$n$ limit this tower of states hence becomes degenerate with the continuum spectrum of the spin-2 $\mathfrak{e}^{\mu}_{\,\,\,\,\nu}$ states. The Robin BCs obeyed by the $\mathfrak{a}^{\bar{\c}}(M,\r)$ fluctuations yield the same results in the large-$n$ limit, albeit with the presence of an additional isolated state with mass $M<1$ (as shown in Fig.~\ref{Fig:nTorusSpectrum}).  

\subsubsection{Probe states}  
Let us conclude this qualitative discussion by examining the equations and boundary conditions satisfied by the probe states $\mathfrak{p}^{a}$; we remind the Reader that the probe approximation is implemented by neglecting the component of $\mathfrak{a^{a}}$ that is proportional to the metric fluctuation $h$, which is equivalent to assuming that $\Gamma^{a}(\r)=\frac{\pa_{\r}\bar{\Phi}^{a}(\r)}{\pa_{\r}A(\r)}\ll1$. The general fluctuation equation is presented in Eq.~(\ref{Eq:ProbeScalarFluct}), and for the purposes of our toroidally reduced sigma-model we find that it may be written as follows:  
\begin{equation}
0 = \Big[\pa_{\r}^{2} +\big(4\pa_{r}A-\d\pa_{\r}\bar{\c} \big)\pa_{\r} + e^{2(\d\bar{\c}-A)} M^2 \Big] \mathfrak{p}^a 
-e^{2\d\bar{\c}}G^{a\bar{\c}}\big(\pa_{\bar{\c}}^{2}\cv\big)
\mathfrak{p}^{\bar{\c}}\,,
\end{equation}
where we have taken advantage of the fact that $\pa_{\bar{\w}}\cv=0$. Notice that the probe approximation decouples the two scalars (since $G_{ab}=\d_{ab}$), but also introduces an asymmetry; although the fluctuations of $\mathfrak{p}^{\bar{\w}}$ obey the same equation as previously seen with the full gauge-invariant states in Eq.~(\ref{Eq:nTorusScalarFluctHV}), the corresponding equation for the $\mathfrak{p}^{\bar{\c}}$ states contains an additional contribution proportional to the potential $\cv$ of the dimensionally reduced model. We see that implementing the probe approximation greatly simplifies Eq.~(\ref{Eq:nTorusScalarFluct}), but at the cost of spoiling the exact cancellation between the three terms in the second line.\par 
After substituting in once more for the hyperscaling violating backgrounds we find that the $\mathfrak{p}^{\bar{\c}}$ equation is modified to read:
\begin{equation}
0=\Big[\pa_{\r}^2 +(n+4)\pa_{\r} +e^{-2\r}x^{2}M^{2}
+\tfrac{2}{3}n(n+4)\Big] \mathfrak{p}^{\bar{\c}}\,,   \label{Eq:nTorusChiProbeHV}
\end{equation}
which admits the following general solution of Bessel functions:
\begin{align}  \label{Eq:nTorusChiProbeSolution}
\mathfrak{p}^{\bar{\c}}(M,\r)&=e^{-(n+4)\frac{\r}{2}}\Big[
c_{J}J_{\a^{\bar{\c}}}(xMe^{-\r}) + c_{Y}Y_{\a^{\bar{\c}}}(xMe^{-\r})
\Big]\,,\nn\\
&\hspace{10mm}\text{with}\quad \a^{\bar{\c}}\equiv
\sqrt{\tfrac{(12-5n)(n+4)}{12}}\,.
\end{align}
By imposing the required Dirichlet BCs at $\r=0$ and $\r\to\infty$, we therefore find that solutions for $\mathfrak{p}^{\bar{\c}}$ are given by the zeros of $J_{\a^{\bar{\c}}}(xM)$. More specifically, \emph{real} solutions exist only for $n\leqslant\frac{12}{5}$, and we can predict that within theories obtained by compactifying on higher-dimensional tori $T^{n\geqslant3}$, the probe approximation will completely fail to capture the physical $\mathfrak{a}^{\bar{\c}}$ states. For clarity, we remind the Reader that the HV solutions are obtainable from the analytical confining solutions by taking the $\r\to\infty$ limit of the latter. We hence anticipate that this dimensionality bound should provide a reasonable estimate of the maximum number of compactified circles our model permits, before a subset of the probe states (those governed by $\bar{\c}$) acquire a spurious dependence on the UV boundary conditions.

\section{Mass spectra}
\label{Sec:nTorusMassPlots}
\subsubsection{Gauge-invariant states}
\begin{figure}[h!]
	\begin{center}
		\makebox[\textwidth]{\includegraphics[
			width=0.56\paperwidth
			]
			{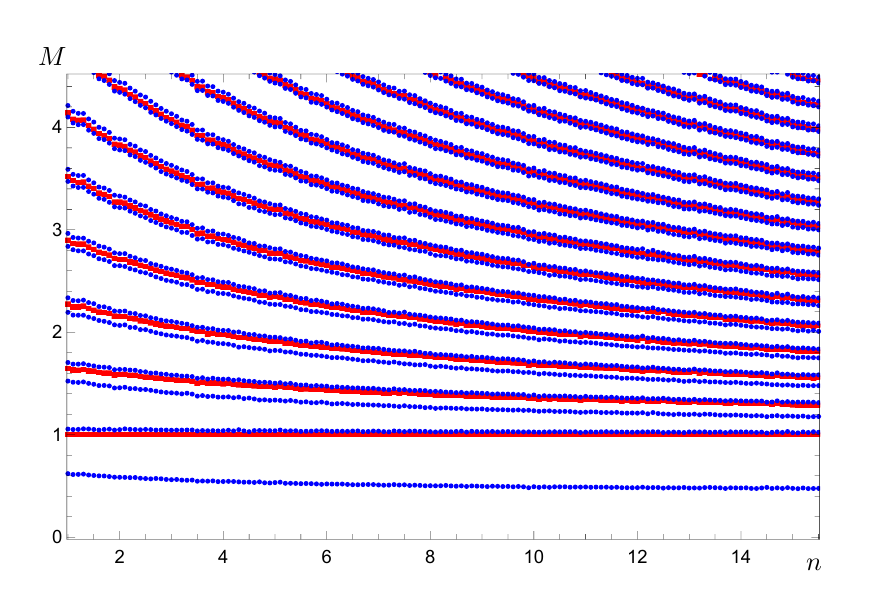}}
	\end{center}
	\vspace{-6mm}
	\caption[Mass spectra for an example application]{The spectra of masses $M$ as a function of the number of $n$-torus dimensions compactified on circles, $n\in[1,\,15.5]$. The action $\tilde{\cs}_{5}$ is obtainable from the toroidal compactification of a higher-dimensional pure gravity theory for integer $n\geqslant2$ only, though we analytically continue our numerical computation to permit all $n\geqslant1$. The blue disks represent the gauge-invariant scalar fluctuations $\mathfrak{a}^{\bar{\c}}$ and $\mathfrak{a}^{\bar{\w}}$, while the red squares denote the tensor states $\mathfrak{e}^{\mu}_{\,\,\,\,\nu}$. All states are normalised in units of the lightest tensor mass, and the spectra were computed using regulators $\r_{1}=10^{-3}$ and $\r_{2}=8$.}
	\label{Fig:nTorusSpectrum}
\end{figure}

Let us now return to the confining backgrounds. The results of our numerical spectrum computation for the complete gauge-invariant scalar fluctuations $\mathfrak{a}^{a}$, together with the tensor modes $\mathfrak{e}^{\mu}_{\,\,\,\,\nu}$, are presented in Fig.~\ref{Fig:nTorusSpectrum}; they are represented by the blue disks and red squares, respectively. Our five-dimensional sigma-model is obtainable via the toroidal compactification of a higher-dimensional gravity theory for integer $n\geqslant2$ only (recall from Eq.~(\ref{Eq:Dmetric}) our $D$-dimensional metric ansatz); nevertheless we find it worthwhile to simply extend our computation to permit all $n\geqslant1$, with the understanding that these additional backgrounds do not admit a sensible interpretation in terms of a lift to the higher-dimensional pure gravity theory.\par 
As anticipated in Sec.~\ref{Sec:nTorusFluctEqns} we see that both of the physical spin sectors produce a spectrum which gradually approaches that of a continuum as the number of compactified dimensions is increased, so that in the $n\to\infty$ limit we would expect to obtain an infinitely dense band of resonances. This observation is supplemented by the caveat that we also find an additional scalar state which appears to remain light and separated from the other modes; this isolated state (associated with $\bar{\c}$) was not detected by the computation presented in Fig.~2 of Ref.~\cite{Teresi:2018eai}, though the spin-0 results are otherwise qualitatively similar.\par 
The complete set of scalar modes comprises two separate towers of states, one for each of the sigma-model scalars appearing in the dimensionally reduced theory.\ As inferred from our analysis of the fluctuation equations using the hyperscaling violating solutions---which we remind the Reader approximate the confining solutions for large $n$---these two towers eventually both become degenerate with the spin-2 resonances as the dimensionality of the $T^{n}$ increases.        
\\
Conversely, when the system contains relatively few circles the two scalar towers are more easily distinguished; the ratio of the $\mathfrak{a}^{\bar{\w}}$ and $\mathfrak{e}^{\mu}_{\,\,\,\,\nu}$ masses is always of order $\sim1$, while the $\mathfrak{a}^{\bar{\w}}$ states are slightly lighter and separated from them both. This effect is more pronounced with the lightest states in the spectrum, and in particular the very lightest resonance for $\mathfrak{a^{\bar{\c}}}$ has mass $M<1$ due to the Robin boundary conditions shown in Eq.~(\ref{Eq:nTorusChiFluctBC}). 

\subsubsection{Probe states}
\begin{figure}[h!]
	\begin{center}
		\makebox[\textwidth]{\includegraphics[
			width=0.56\paperwidth
			]
			{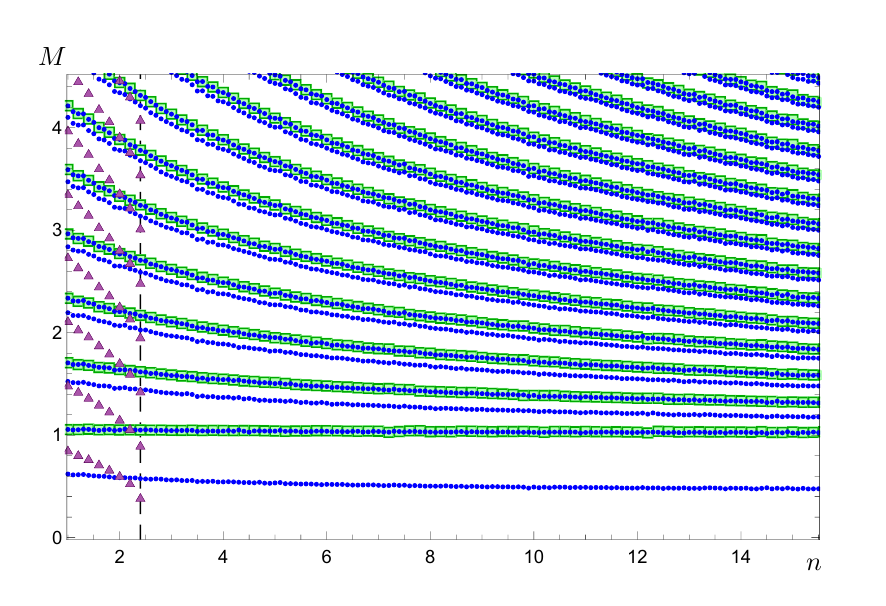}}
	\end{center}
	\vspace{-6mm}
	\caption[Probe spectra for an example application]{The spectra of masses $M$ as a function of the number of $n$-torus dimensions compactified on circles, $n\in[1,\,15.5]$. The action $\tilde{\cs}_{5}$ is obtainable from the toroidal compactification of a higher-dimensional pure gravity theory for integer $n\geqslant2$ only, though we analytically continue our numerical computation to permit all $n\geqslant1$. The blue disks represent the gauge-invariant scalar fluctuations $\mathfrak{a}^{\bar{\c}}$ and $\mathfrak{a}^{\bar{\w}}$ (as shown also in Fig.~\ref{Fig:nTorusSpectrum}). 
		We here additionally include the results of our probe computation for the $\mathfrak{p}^{\bar{\c}}$ states
		(purple triangles) and $\mathfrak{p}^{\bar{\w}}$ (green squares), with the former shown only for $n\leqslant\frac{12}{5}$ (denoted by the vertical dashed line, see the discussion in Sec.~\ref{Sec:nTorusFluctEqns} for details). All states are normalised in units of the lightest tensor mass, and the spectra were computed using regulators $\r_{1}=10^{-3}$ and $\r_{2}=8$. }
	\label{Fig:nTorusProbeSpectrum}
\end{figure}

Having discussed our spectra results for the physical fluctuations of the toroidally reduced sigma-model, we now turn our attention to the corresponding results for the probe computation. In Fig.~\ref{Fig:nTorusProbeSpectrum} we reproduce the same tower of $\mathfrak{a}^{a}$ scalar resonances as shown in Fig.~\ref{Fig:nTorusSpectrum} (denoted again by the blue disks), supplemented by the probe states $\mathfrak{p^{\bar{\w}}}$ (green squares) and $\mathfrak{p^{\bar{\c}}}$ (purple triangles). The analysis has been analytically continued as before to permit all $n\geqslant1$, though for $n<2$ the five-dimensional system is not obtainable from the compactification of the pure gravity theory on a $T^{n}$; we do not attempt to provide a physically realistic motivation for choices $n\notin\mathbb{Z}$.\par  
Let us start by considering the probe resonances associated with the scalar $\bar{\w}$, which we observe approximate very effectively the tower of physical $\mathfrak{a}^{\bar{\w}}$ resonances for all values of $n$. While this behaviour was predicted when the number of compactified dimensions is large---we determined that the bulk equation and boundary conditions for $\mathfrak{a}^{\bar{\w}}$ and $\mathfrak{p}^{\bar{\w}}$ are in agreement in this limit---the universal success of the probe approximation (including at lower values of $n$) implies that the gauge-invariant excitations associated with $\bar{\w}$ contain a negligible $h$ component. Since the boundary value of the field fluctuation $h$ is identified via the holographic dictionary with the source of the dilatation operator in the dual field theory, we infer that this tower of states is not the result of any significant dilaton mixing effects.\par 
As anticipated in the paragraph following Eq.~(\ref{Eq:nTorusChiProbeSolution}), the corresponding $\mathfrak{p}^{\bar{\c}}$ probe states are plotted only up to the dimensionality upper bound identified from the Bessel function index $\a^{\bar{\c}}$ (at $n=12/5$, here denoted by the dashed line), and they unambiguously fail to capture the gauge-invariant $\mathfrak{a}^{\bar{\c}}$ resonances. In particular, the lightest excitation in the physical spectrum is a spin-0 state of mass $M<1$ associated with the fluctuations of the sigma-model scalar $\bar{\c}$. This state evidently contains a significant $h$ component, in addition to the contribution coming from the scalar field fluctuation $\varphi^{\bar{\c}}$, and it is hence reasonable to interpret this excitation as being (at least partially) identifiable with the dilaton; a similar observation leads us to conclude that in fact the entire tower of $\mathfrak{a}^{\bar{\c}}$ resonances, including the heavier states, are to some degree dilatonic.\par 
The cause of this phenomenon, and of the modified bulk equation for $\mathfrak{p}^{\bar{\c}}$ as shown in Eq.~(\ref{Eq:nTorusChiProbeHV}), is the nature of the confining solutions themselves; we notice that the analytical expressions for $\bar{\c}(\r)$ and the warp factor $A(\r)$ which are introduced in Eqs.~(\ref{Eq:nTorusConfChi},\,\ref{Eq:nTorusConfA}) yield a ratio which is always of order $\pa_{\r}\bar{\c}/\pa_{\r}A\sim1$. Hence the probe approximation---which is implemented by assuming this ratio to be negligibly small---is never justified, and the $\mathfrak{p}^{\bar{\c}}$ probe states will necessarily fail to capture the excitations of the complete gauge-invariant computation. The field fluctuations of the scalar $\bar{\c}$ and of the (trace of the) ADM-decomposed metric governed by $h$ are inseparable, and dilaton mixing is intrinsic to their corresponding gauge-invariant combination $\mathfrak{a}^{\bar{\c}}$. An equivalent statement is that the back-reaction induced in the underlying geometry by the fluctuations of $\bar{\c}$ is always a considerable effect.\par 
We conclude by observing another interesting feature of the $\mathfrak{p}^{\bar{\c}}$ probe states, when the number of compactified dimensions is small. Despite the fact that these unphysical resonances provide in general a very poor approximation of the complete scalar states associated with $\bar{\c}$, we nevertheless notice that there exists a privileged choice for $n$ at which the two calculations are coincidentally in agreement; for $n\approx2$ the towers of $\mathfrak{a}^{\bar{\c}}$ and $\mathfrak{p}^{\bar{\c}}$ excitations briefly intersect, and any appreciable dilaton mixing effects are hence suppressed. The relevance of this observation will become apparent in Chapter~\ref{Chap:SpectraWitten} when we discuss the spectra of composite states within field theories holographically dual to the seven-dimensional supergravity discussed in Chapter~\ref{Chap:Intro}, compactified on a $T^{n=2}=S^{1}\times S^{1}$.               
\clearpage

\chapter{Six-dimensional half-maximal supergravity}
\label{Chap:SpectraRomans}
\section{Formalism of the six-dimensional model}
\label{Sec:Formalism6Dmodel}
\subsubsection{The action in $D=6$ dimensions}
We start by presenting the six-dimensional action for the model, which describes 32 bosonic degrees freedom (d.o.f.) organised within the following field content: one real scalar $\f$ ($1\times1$), one vector $A_{\hat{M}}$ ($1\times 4$) transforming as a $U(1)$ gauge boson, three vectors $A^{i}_{\hat{M}}$ ($3\times 4$) in the \textbf{3} representation of $SU(2)$, one
2-form $B_{\hat{M}\hat{N}}$ ($1\times 6$), and the six-dimensional metric
tensor $\hat{g}_{\hat{M}\hat{N}}$ ($1\times 9$); we have put in parentheses the degrees of freedom carried by each individual field. Hatted uppercase Latin indices $\hat{M}\in\{0,1,2,3,5,6\}$ represent the coordinates of the six-dimensional spacetime, while $i\in\{1,2,3\}$ is the gauge index of $SU(2)$. The complete action may be written as~\cite{Romans:1985tw}
{\small
\begin{align}
\label{Eq:6DAction}
\mathcal {S}_{6}=\int\is\text{d}^{6}x\,\sqrt{-\hat g_{6}}
\bigg(\frac{\mathcal{R}_{6}}{4}&- \hat g^{\hat{M}\hat{N}}\pa_{\hat{M}}\phi\pa_{\hat{N}}\phi -\mathcal{V}_{6}(\phi)  
-\frac{1}{4}e^{-2\phi}\hat g^{\hat{M}\hat{R}}
\hat g^{\hat{N}\hat{S}}\sum_{i}\hat F_{\hat{M}\hat{N}}^{i}\hat F_{\hat{R}\hat{S}}^{i}\, \notag \\
&\hspace{-14mm}-\frac{1}{4}e^{-2\phi}\hat g^{\hat{M}\hat{R}}
\hat g^{\hat{N}\hat{S}}\hat{\mathcal{H}}_{\hat{M}\hat{N}}\hat{\mathcal{H}}_{\hat{R}\hat{S}}
-\frac{1}{12}e^{4\phi}\hat g^{\hat{M}\hat{R}}
\hat g^{\hat{N}\hat{S}}\hat g^{\hat{T}\hat{U}}\hat G_{\hat{M}\hat{N}\hat{T}}\hat G_{\hat{R}\hat{S}\hat{U}} \bigg)\,,
\end{align} }%
where $\hat{g}_{6}$ is the determinant of the six-dimensional metric, $\mathcal{R}_{6}\equiv \hat{g}^{\hat{M}\hat{N}}\car_{\hat{M}\hat{N}}$ is the corresponding Ricci curvature scalar, and $\mathcal{V}_{6}(\f)$ is the potential for the real scalar field $\f$. Summation over repeated indices is implied. The tensors are defined as
\begin{align}
\hat F_{\hat{M}\hat{N}} &\equiv \partial_{\hat{M}}A_{\hat{N}}-
\partial_{\hat{N}}A_{\hat{M}}\,,\\
\hat F_{\hat{M}\hat{N}}^{i} &\equiv \partial_{\hat{M}}A_{\hat{N}}^{i}-
\partial_{\hat{N}}A_{\hat{M}}^{i}+g\epsilon^{ijk}A_{\hat{M}}^{j}A_{\hat{N}}^{k}\,,\\
\hat{\mathcal{H}}_{\hat{M}\hat{N}} &\equiv \hat F_{\hat{M}\hat{N}} + mB_{\hat{M}\hat{N}}\,,\\
\hat G_{\hat{M}\hat{N}\hat{T}}&\equiv 3\partial_{[\hat{M}}B_{\hat{N}\hat{T}]}=
\partial_{\hat{M}}B_{\hat{N}\hat{T}}+
\partial_{\hat{N}}B_{\hat{T}\hat{M}}+
\partial_{\hat{T}}B_{\hat{M}\hat{N}}\,,
\end{align}
where $\epsilon^{ijk}$ is the three-dimensional Levi-Civita symbol. We follow the same conventions as in Ref.~\cite{Jeong:2013jfc}, by fixing the gauge coupling as $g=\sqrt{8}$ with the mass parameter given by $m=\frac{g}{3}$. We furthermore adopt the conventional definition for the complete anti-symmetrisation of a generic $(0,3)$-tensor $X_{ABC}$ using
\begin{equation}
X_{[ABC]}\equiv\frac{1}{3!}\big(X_{ABC}+X_{BCA}+X_{CAB}-X_{ACB}-X_{BAC}-X_{CBA}\big)\,,
\end{equation}  
which may easily be generalised for a tensor of any order.

\subsubsection{Critical points of the $D=6$ potential}
\begin{figure}[t]
	\begin{center}
		\includegraphics[width=10cm]{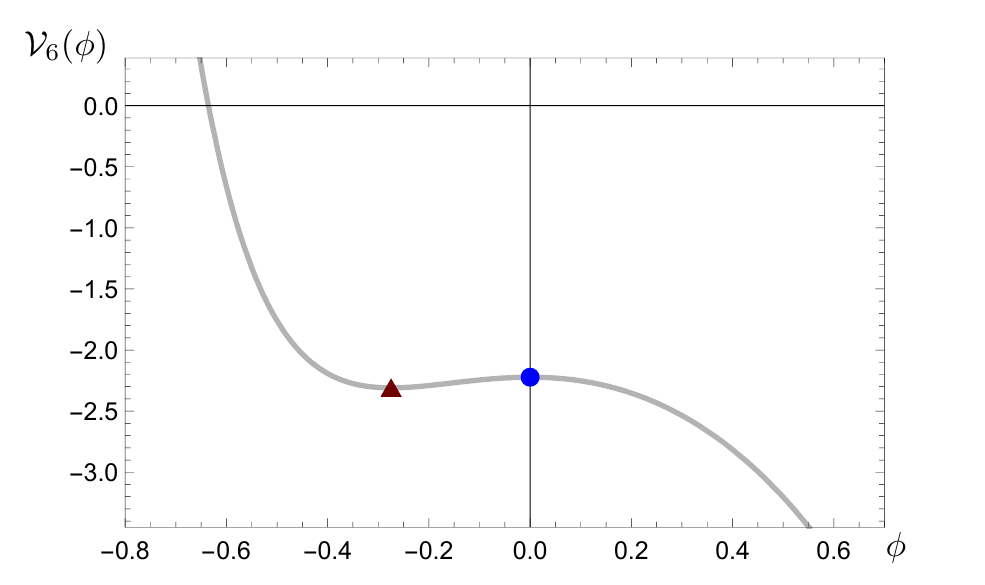}
		\caption[Scalar potential of the six-dimensional supergravity]{The potential ${\cal V}_6(\phi)$ as a function of the scalar $\phi$ in the sigma-model coupled to gravity in $D=6$ dimensions. The blue disk denotes the supersymmetric critical point $\phi=\phi_{UV}=0$, while the dark-red triangle represents the non-supersymmetric critical point  $\f=\f_{IR}=-\frac{1}{4}\ln(3)$.}
		\label{Fig:ScalarPotential6D}
	\end{center}
\end{figure}

The scalar potential $\mathcal{V}_{6}(\f)$ appearing in the six-dimensional action is given by
\begin{equation}
\label{Eq:V6}
\mathcal{V}_{6}(\phi)=\frac{1}{9}\bigg(e^{-6\phi}-9e^{2\phi}
-12e^{-2\phi}\bigg) \,,
\end{equation}
which admits two critical points, each corresponding to a distinct five-dimensional conformal field theory; these special values for $\f$ are given by:
\begin{align}
\f_{UV}&=0 \hspace{19mm} \bigg(\mathcal{V}_{6}(\f_{UV})=-\frac{20}{9}\bigg)\,,\\
\f_{IR}&=-\frac{1}{4}\ln(3) \hspace{8mm}\bigg(\mathcal{V}_{6}(\f_{IR})=-\frac{4}{\sqrt{3}}\bigg)\,.
\end{align}
We label the critical points with subscripts to reflect the fact that one can construct numerical solutions which interpolate between the two, which in the dual theory corresponds to a renormalisation group (RG) flow between two fixed points, from high energies (short distance scales) at $\f_{UV}$ to low energies (long distance scales) at $\f_{IR}$. The former critical point is a global maximum of the potential which preserves supersymmetry and is dual to a $D=5$ superconformal gauge theory~\cite{Ferrara:1998gv}, while the latter is a global minimum which breaks supersymmetry~\cite{Gursoy:2002tx}; a plot of $\mathcal{V}_{6}(\f)$ is shown in Fig.~\ref{Fig:ScalarPotential6D}.

For each of the two critical point values of the scalar field $\f(r)$, the six-dimensional bulk geometry is that of $\text{AdS}_{6}$; following the same conventions as in Ref.~\cite{Elander:2013jqa}, the curvature radii are given by~\cite{Karndumri:2012vh} 
\begin{align}
R_{UV}^{2}&\equiv-5\big[\mathcal{V}_{6}(\f_{UV})\big]^{-1}
=\frac{9}{4}\, ,\\
R_{IR}^{2}&\equiv-5\big[\mathcal{V}_{6}(\f_{IR})\big]^{-1}=\frac{5\sqrt{3}}{4}\label{Eq:RomansRIR}\, .
\end{align} 
We can extract the mass of the scalar field $\f$ by considering small perturbations about each critical point, reading off the mass in each case as the coefficient of the term quadratic in $\f$ in the expansion of the potential. We obtain 
\begin{align}
\mathcal{V}_{6}(\f_{UV})&=-\frac{20}{9}-\frac{8\f^{2}}{3}
+\mathcal{O}\big(\f^{3}\big),\\
\mathcal{V}_{6}(\f_{IR})&=-\frac{4}{\sqrt{3}}
+\frac{8}{\sqrt{3}}\big(\f-\f_{IR}\big)^{2}
+\mathcal{O}\big((\f-\f_{IR})^{3}\big)\,,
\end{align} 
and hence
\begin{align}
m_{UV}^{2}&=-\frac{8}{3}\quad \rightarrow \quad m_{UV}^{2}R_{UV}^{2}=-6\,,\\
m_{IR}^{2}&=\frac{8}{\sqrt{3}}\quad \rightarrow \quad m_{IR}^{2}R_{IR}^{2}=10\,.
\end{align}
Finally, we can use the mass of the scalar supergravity field $\f$ and the curvature radius of the $\text{AdS}$ geometry to compute the scaling dimension $\Delta$ of the gauge-invariant boundary operator which is dual to $\f$. From the required relation shown in Eq.~(\ref{Eq:Delta}), we determine $\Delta$ separately for each of the two critical points: 
\begin{equation}
\label{Eq:RomansDeltas}
\Delta_{UV}=3 \quad , \quad \Delta_{IR}=\frac{1}{2}\Big(5+\sqrt{65}\Big)\,.
\end{equation}

\section{Circle reduction to \emph{D=}5 dimensions}
\label{Sec:CircleReduction}
\subsubsection{The metric}
To compute the spectra of composite states in a four-dimensional confining field theory, it is first necessary to dimensionally reduce the gravitational model to five dimensions. To this end we compactify one of the external dimensions---parametrised by the coordinate $\eta\in[0,2\pi)$---on a circle $S^{1}$, by making use of the five-dimensional metric introduced in Eq.~(\ref{Eq:5Dmetric}) and the following ansatz:     
\begin{align}\label{Eq:6Dmetric}
\text{d}s_{6}^{2}&=e^{-2\c}\text{d}s_{5}^{2} + e^{6\c}\big(\text{d}\eta+V_{M}\text{d}x^{M}\big)^{2}\notag\\
&=e^{-2\c}\big(e^{2A(r)}\text{d}x_{1,3}^{2} +\text{d}r^{2}\big)+ e^{6\c}\big(\text{d}\eta+V_{M}\text{d}x^{M}\big)^{2}\notag\\
&=e^{-2\c}\big(e^{2A(\r)}\text{d}x_{1,3}^{2} +e^{2\c}\text{d}\r^{2}\big)+ e^{6\c}\big(\text{d}\eta+V_{M}\text{d}x^{M}\big)^{2}\, ,
\end{align}
where the index $M\in\{0,1,2,3,5\}$ labels the coordinates of the $D=5$ system, $V_{M}$ is a covariant five-vector (the graviphoton, or gravivector) coming from the decomposition of the $D=6$ metric tensor, and in going from the second to the third line we have introduced a convenient redefinition of the radial coordinate via $\pa_{r}\equiv e^{-\c}\pa_{\r}$ ($\text{d}r=e^{\c}\text{d}\r$). The dynamical field $\c$ is introduced to parametrise the volume of the circle, and appears as an additional scalar in the sigma-model coupled to five-dimensional gravity; a further three sigma-model scalars $\pi^{i}$ result from the decomposition of the $SU(2)$ adjoint vectors as $A_{\hat{M}}^{i}=\{A_{\mu}^{i}, A_{5}^{i}, \pi^{i}\}$, where  $\mu\in\{0,1,2,3\}$ labels the four Minkowski spacetime coordinates.\par 
We remind the Reader of our assumption that the scalars $\{\f,\c\}$ and the warp factor $A$ are functions only of the holographic coordinate $\r$, and we emphasise that to ensure local Poincar\'{e} invariance is manifestly preserved along the Minkowski directions for this metric ansatz, we must additionally assume that \emph{only} these three fields are permitted to acquire non-zero radial profiles; furthermore, we observe that Poincar\'{e} invariance is extended to include the $\eta$ direction if we enforce the identification $A=4\c$.\par 
Finally, we note the following useful relations which may be derived given the metric ansatz in Eq.~(\ref{Eq:6Dmetric}):
\begin{align}
\hat{g}^{MR}&=e^{2\c}g^{MR}\,,\\ 
\hat{g}^{6M}&=-e^{2\c}g^{MN}V_{N}\,,\\ \hat{g}^{66}&=e^{-6\c}+e^{2\c}g^{MN}V_{M}V_{N}\,,
\end{align}
where $\hat{g}$ denotes the $D=6$ metric, $g^{MN}$ is the (inverse of the) $D=5$ metric defined in Eq.~(\ref{Eq:5Dmetric}), and uppercase Latin indices again run over $M\in\{0,1,2,3,5\}$. The determinants of the five- and six-dimensional metric tensors are related by $\sqrt{-\hat{g}_6} = e^{-2\c}\sqrt{-g_5}$. The six-dimensional Ricci scalar, written in terms of the holographic coordinate $\r$, is given by
\begin{equation}
\car_{6}=-2\Big(4A''-\c''+10(A')^{2}+7(\c')^{2}-8A'\c'\Big)\,.
\end{equation} 

\subsubsection{The action} 
We next turn our attention to  the six-dimensional action written in Eq.~(\ref{Eq:6DAction}) by compactifying on a circle and decomposing its constituent fields and tensors, reformulating the action in terms of their lower-dimensional analogues. We will make the assumption that each supergravity field assumes a constant value along the $S^{1}$-compact dimension $x^{6}$, so that $\pa_{6}f=0$ for any generic field or tensor $f$, and we hence neglect fluctuations along the circle to retain only its zero modes.\par
We start by considering only the contributions to the six-dimensional action which are independent of the $U(1)$ fields $A_{\hat{M}}$ and $B_{\hat{M}\hat{N}}$, which amounts to neglecting the final two terms in Eq.~(\ref{Eq:6DAction}), and noting that with this simplification the action $\mathcal{S}_{6}$ may be conveniently rewritten as       

\begin{equation}
\label{Eq:6DActionTotalDeriv}
\mathcal{S}_{6}=\int\is\text{d}\eta\,
\bigg\{\tilde{\mathcal{S}}_{5} + \half\int\is\text{d}^{5}x\,
\pa_{M}\Big(\sqrt{-g_{5}}\,g^{MN}\pa_{N}\c\Big)\bigg\} 
\,,
\end{equation}
where the five-dimensional action $\tilde{\mathcal{S}}_{5}$ is given by
\begin{align}
\label{Eq:5DActionNoU1}
\tilde{\mathcal{S}}_{5}=\int\is\text{d}^{5}x\,\sqrt{-g_{5}}
\bigg(\frac{\mathcal{R}_{5}}{4}&-\half G_{ab} g^{MN}\partial_{M}\Phi^{a}\partial_{N}\Phi^{b} -\mathcal{V}(\phi,\chi)\notag\\  
&-\frac{1}{4}H_{AB}g^{MR}g^{NS}F_{MN}^{A}F_{RS}^{B} \bigg)\,.
\end{align}
Here the index $a\in\{1,2,3\}$ represents the scalar fields of the sigma-model coupled to five-dimensional gravity so that $\Phi^a = \{ \phi, \chi, \pi^i \}$, the indices $A,B\in\{1,2\}$ denote the field strength tensors for the vector fields $\{V_{M},A^{i}_{M}\}$, and the new scalar potential $\mathcal{V}$ is related to the original six-dimensional potential by $\mathcal{V}(\phi,\chi) = e^{-2\chi}\mathcal{V}_{6}(\phi)$.
The sigma-model metric $G_{ab}$ and the metric $H_{AB}$ for the vector field strengths are given by 
\begin{align}
G_{ab} &= \text{diag} \Big(2, 6, e^{-6\chi-2\f}\Big) \,, \\
\label{Eq:RomansHABmetric}
H_{AB} &= \text{diag}\Big(\tfrac{1}{4}e^{8\c},e^{2\c-2\f}\Big) \,,
\end{align}
while the vector field strengths $F^{A}=\{F^{V},F^{i}\}$ are defined as follows:
\begin{align}
F_{MN}^{V}&\equiv \partial_{M}V_{N}-\partial_{N}V_{M}\,,\\
F_{MN}^{i}&\equiv \partial_{M}A_{N}^{i}-\partial_{N}A_{M}^{i}
+g\epsilon^{ijk}A_{M}^{j}A_{N}^{k}
+(V_{M}\partial_{N}\pi^{i}-V_{N}\partial_{M}\pi^{i})\,.
\end{align} 
The total derivative contribution to Eq.~(\ref{Eq:6DActionTotalDeriv}) does not affect the equations of motion and hence we disregard it, so that the dimensionally reduced action---neglecting the $U(1)$ fields $\{A_{\hat{M}},\, B_{\hat{M}\hat{N}} \}$---is given by Eq.~(\ref{Eq:5DActionNoU1}).\par
We now consider the complementary action containing only the contributions of the $U(1)$ fields:
{\small\begin{align}
\mathcal{S}_{6}^{U(1)}=\int\is\text{d}^{6}x\,\sqrt{-\hat g_{6}}
\bigg(&-\frac{1}{4}e^{-2\phi}\hat g^{\hat{M}\hat{R}}
\hat g^{\hat{N}\hat{S}}\hat{\mathcal{H}}_{\hat{M}\hat{N}}\hat{\mathcal{H}}_{\hat{R}\hat{S}}\notag\\
&-\frac{1}{12}e^{4\phi}\hat g^{\hat{M}\hat{R}}
\hat g^{\hat{N}\hat{S}}\hat g^{\hat{T}\hat{U}}\hat G_{\hat{M}\hat{N}\hat{T}}\hat G_{\hat{R}\hat{S}\hat{U}} \bigg)\,,
\end{align} }
which can be decomposed in terms of five-dimensional quantities and rewritten:
{\small\begin{align}
	\label{Eq:5DActionU1}
	\mathcal{S}_{6}^{U(1)}=\int\is\text{d}\eta\,\text{d}^{5}x\,
	\sqrt{-g_{5}}\bigg\{
	&-\frac{1}{4}H^{(2)}g^{MR}g^{NS}\mathcal{H}_{MN}\mathcal{H}_{RS}\notag\\
	&-\frac{1}{12}K^{(2)}g^{MR}g^{NS}g^{TU}G_{MNT}G_{RSU}\,
	-\frac{1}{2}G^{(1)}g^{NS}\mathcal{H}_{6N}\mathcal{H}_{6S}\notag\\
	&-\frac{1}{4}H^{(1)}g^{NS}g^{TU}G_{6NT}G_{6SU}
	\bigg\}\,,
	\end{align}}%
where the prefactors are given by $H^{(2)}=e^{2\chi-2\phi}$, $K^{(2)}=e^{4\chi+4\phi}$, $G^{(1)}=e^{-6\chi-2\phi}$, and $H^{(1)}=e^{-4\chi+4\phi}\,$; the superscript numbers describe the field content of each term, with a $^{(2)}$ denoting an expression containing the 2-form $B_{MN}$, and $^{(1)}$ denoting an expression containing only 1-forms, independent of $B_{MN}$. The hatted tensors decompose into their five-dimensional analogues according to the following definitions:     
{\small\begin{align}
\mathcal{H}_{MN}& \equiv \hat F_{MN}+mB_{MN} + \left( V_M \partial_N A_6 - V_N \partial_M A_6 \right) + m \left( B_{6M} V_N - B_{6N} V_M \right)\,,\\
\mathcal{H}_{6N}& \equiv \hat{\mathcal{H}}_{6N} = \partial_{6}A_{N}-\partial_{N}A_{6}+mB_{6N}=-\partial_{N}A_{6}+mB_{6N}\,,\\
G_{MNT}& \equiv 3\partial_{\small[M}B_{NT\small]}  - 6V_{\small[M} \partial_N B_{T\small]6}\,,\\
G_{6NT}&\equiv \hat G_{6NT} =
\partial_{6}B_{NT}-
\partial_{N}B_{6T}+
\partial_{T}B_{6N}=\partial_{T}B_{6N}-\partial_{N}B_{6T}\,.
\end{align} }
By combining the contributions coming from the two reduced actions in Eq.~(\ref{Eq:5DActionNoU1}) and Eq.~(\ref{Eq:5DActionU1}) we finally obtain our complete five-dimensional action, written as follows:
{\small\begin{align}
	\label{Eq:5DActionComplete}
	\mathcal{S}_{5}=\int\is\text{d}^{5}x\,\sqrt{-g_{5}}
	\bigg(\frac{\mathcal{R}_{5}}{4}-\half G_{ab} g^{MN}\partial_{M}\Phi^{a}\partial_{N}\Phi^{b} &-\mathcal{V}(\phi,\chi)  
	-\frac{1}{4}H_{AB}g^{MR}g^{NS}F_{MN}^{A}F_{RS}^{B}\,\notag\\
	-\frac{1}{4}e^{2\chi-2\phi}g^{MR}g^{NS}\mathcal{H}_{MN}\mathcal{H}_{RS}
	&-\frac{1}{12}e^{4\chi+4\phi}g^{MR}g^{NS}g^{TU}G_{MNT}G_{RSU}\,\notag\\
	-\frac{1}{2}e^{-6\chi-2\phi}g^{NS}\mathcal{H}_{6N}\mathcal{H}_{6S}
	&-\frac{1}{4}e^{-4\chi+4\phi}g^{NS}g^{TU}G_{6NT}G_{6SU}
	\bigg)\,,
	\end{align}}%
where the original 32 physical degrees of freedom contained within Eq.~(\ref{Eq:6DAction}) are now carried by the following five-dimensional fields: six scalars $\{\phi,\chi,\pi^{i},A_{6}\}$ ($6\times1$), six vectors $\{A_{M},A^{i}_{M},B_{6N},V_{M}\}$ ($6\times3$), one 2-form $B_{MN}$ ($1\times3$), and the metric tensor $g_{MN}$ ($1\times5$).

\section{Equations of motion and confining solutions}
\label{Sec:RomansEOMs}
\subsubsection{Equations of motion}
Admissible background configurations for the model are found by solving the classical equations of motion derived from the circle-reduced five-dimensional action $\mathcal{S}_{5}$ written in Eq.~(\ref{Eq:5DActionComplete}). Using the general results presented in Eqs.~(\ref{Eq:ScalarEOM}\,-\,\ref{Eq:Einstein2}) of Section~\ref{Subsec:HolographicFormalism}, and recalling that all of the supergravity fields are assumed to be functions only of the holographic coordinate, we obtain the following:
\begin{align}
\pa^{2}_{r}\f + 4\pa_{r}\f\pa_{r}A &= \half\frac{\pa\cv}{\pa\f}\,,\\
\pa^{2}_{r}\c + 4\pa_{r}\c\pa_{r}A &= \frac{1}{6}\frac{\pa\cv}{\pa\c}\,,\\
3\pa^{2}_{r}A +6(\pa_{r}A)^{2} +2(\pa_{r}\f)^{2} +6(\pa_{r}\c)^{2}
 &= -2\cv\,,\\
3(\pa_{r}A)^{2} -(\pa_{r}\f)^{2} -3(\pa_{r}\c)^{2} &= -\cv\,.
\end{align}
This system of equations may be reformulated in terms of the six-dimensional scalar potential $\cv_{6}(\f)=e^{2\c}\cv(\f,\c)$ by implementing the change of radial coordinate $r\to\r$ defined just after Eq.~(\ref{Eq:6Dmetric}), so that we have
\begin{align}
\pa^{2}_{\r}\f+(4\pa_{\r}A-\pa_{\r}\c)\pa_{\r}\f&=
\frac{1}{2}\frac{\pa\cv_{6}}{\pa\f}\, ,\label{Eq:RomansEOM1}\\
\pa^{2}_{\r}\c+(4\pa_{\r}A-\pa_{\r}\c)\pa_{\r}\c&=
-\frac{\cv_{6}}{3}\, ,\label{Eq:RomansEOM2}\\
3(\pa_{\r}A)^{2}-(\pa_{\r}\f)^{2}-3(\pa_{\r}\c)^{2}&=
-\cv_{6}\, ,\label{Eq:RomansEOM3}\\
3\pa^{2}_{\r}A +6(\pa_{\r}A)^{2} +2(\pa_{\r}\f)^{2} +6(\pa_{\r}\c)^{2} -3\pa_{\r}A\pa_{\r}\c &=
-2\cv_{6}\, .\label{Eq:RomansEOM4}
\end{align}
We note that only the first three of these equations are independent; Eq.~(\ref{Eq:RomansEOM4}) may alternatively be obtained by differentiating Eq.~(\ref{Eq:RomansEOM3}) with respect to $\r$ and substituting for the two scalars $\f(\r)$ and $\c(\r)$ using their respective equations of motion. 
Furthermore, we may compactly rewrite these equations by introducing the following convenient quantities:
\begin{align}
\a\equiv 4A-\c\quad &, \quad \b\equiv A-4\c  \\
\bigg(
\Rightarrow
\quad 
\c=\tfrac{1}{15}\big(\a-4\b\big)\quad &, \quad A=\tfrac{1}{15}\big(4\a-\b\big)    \bigg)\, ,\nn
\end{align}
so that we have
\begin{align}
\pa^{2}_{\r}\f+\pa_{\r}\a\pa_{\r}\f&=
\frac{1}{2}\frac{\pa\cv_{6}}{\pa\f}\, ,\label{Eq:EOMalphaphi}\\
\pa^{2}_{\r}\a+(\pa_{\r}\a)^{2}&=
-5\cv_{6}\, ,\label{Eq:EOMalpha}\\
(\pa_{\r}\a)^{2}-(\pa_{\r}\b)^{2}-5(\pa_{\r}\f)^{2}
&=-5\cv_{6}\,,\\
\pa^{2}_{\r}\b+\pa_{\r}\a\pa_{\r}\b&=0\label{Eq:EOMbeta}\,.
\end{align} 
Reformulated in this way, we can make two important observations: firstly that the equations of motion are invariant under the sign change $\b\to-\b$ (or equivalently $A-4\c \to 4\c-A$) while holding $\a\to\a$ invariant, and secondly that Eq.~(\ref{Eq:EOMbeta}) may be rewritten as a vanishing total derivative, so that 
\begin{equation}
\label{Eq:C}
-e^{\a}\,\pa_{\r}\b=C \quad\Longleftrightarrow\quad e^{4A-\c}\Big(4\pa_{\r}\c-\pa_{\r}A\Big)=C\, ,
\end{equation}    
where $C$ is some background-dependent integration constant; we will return to these useful observations in Chapter~\ref{Chap:EnergeticsRomans} when we investigate the phase structure for this supergravity model, but we note them here for convenience. We observe that the warp factor constraint $A=4\c$ which was introduced in Sec.~\ref{Sec:CircleReduction} to ensure the local preservation of five-dimensional Poincar\'{e} invariance simply corresponds to $\b=0$, and that the equations of motion satisfied by background solutions on such a domain-wall geometry may be rewritten as 
\begin{align}
4\partial^{2}_{\rho}\phi+15\partial_{\rho}A\partial_{\rho}\phi&=2\frac{\partial\mathcal{V}_{6}}{\partial\phi}\, ,\label{Eq:EOMc1}\\
3\partial^{2}_{\rho}A + 4(\partial_{\rho}\phi)^{2}
&=0\, ,\label{Eq:EOMc2}\\
45(\partial_{\rho}A)^{2}-16(\partial_{\rho}\phi)^{2}&=-16\mathcal{V}_{6}\,, \label{Eq:EOMc3}
\end{align} 
or equivalently
\begin{align}
\pa^{2}_{\r}\f+\pa_{\r}\a\pa_{\r}\f
&=\frac{1}{2}\frac{\pa\mathcal{V}_{6}}{\pa\f}\, ,\\
\pa^{2}_{\r}\a+(\pa_{\r}\a)^{2}&=-5\mathcal{V}_{6}\, ,\\
(\pa_{\r}\a)^{2}-5(\pa_{\r}\f)^{2}
&=-5\mathcal{V}_{6}\,.
\end{align} 
Finally, we note for future reference that we can solve Eq.~(\ref{Eq:EOMc3}) algebraically for $A'(\r)$ and substitute back into Eq.~(\ref{Eq:EOMc1}) to obtain the following second order differential equation written solely in terms of the scalar $\f$:
\begin{equation}
\label{Eq:ParametricPhi}
0=3\f''+\sqrt{5}\f'
\Big[\big(3\f'\big)^{2}+\g^{-3}\Big(9\g^{4}+12\g^{2}-1\Big)\Big]^{\frac{1}{2}}
+\g\Big(3-4\g^{-2}+\g^{-4}\Big)\,,
\end{equation}
where primes denote derivatives with respect to $\r$, and we have here defined $\g\equiv e^{2\f(\r)}$. We will return to this useful equation later in order to visualise how the various domain-wall solutions parametrically flow away from the unique supersymmetric fixed point $\f=0$.  

\subsubsection{Confining solutions}
We are interested in computing the spectra of bosonic composite states (glueballs) in a confining field theory dual to our reduced five-dimensional effective model, and hence we require that the geometric configuration on the gravity side is able to incorporate confinement as a physical low-energy limit; with this in mind, we here introduce a class of background solutions which we shall refer to as \emph{confining}. \par 
The defining property of this class of solutions is that at some small but finite value of the holographic coordinate $\r=\r_o$, the $S^{1}$-compact dimension (parametrised by the coordinate $\eta$, and with size governed by the scalar $\c$) shrinks to a point so that the bulk geometry smoothly closes off and ends; we use this tapering property of the manifold to realise a physical low-energy limit at which the corresponding dual four-dimensional field theory may be probed, and interpret the spectra of fluctuations about these background profiles as physical states which exhibit confinement at a certain low-energy threshold.\par 
When $\f$ is equal to either of the two critical point values $\{\f_{IR},\,\f_{UV}\}$ of the six-dimensional potential, there exist exact analytical solutions for the scalar field $\c$ and the metric warp factor $A$ which satisfy the equations of motion in five dimensions; defining $\f_{p}$ to be either one of these critical points and $v\equiv\mathcal{V}_{6}(\f_{p})$, these solutions are given by~\cite{Wen:2004qh,Kuperstein:2004yf,Elander:2013jqa}:
\begin{align}
\f&=\f_{p}\,,\\
\c(\r)&=\c_{I}-\frac{1}{5}\ln\bigg[
\cosh\bigg(\tfrac{{\sqrt{-5v}}}{2}(\r-\r_{o}) \bigg) \bigg]\notag\\
&\hspace{47mm}+\frac{1}{3}\ln\bigg[
\sinh\bigg(\tfrac{{\sqrt{-5v}}}{2}(\r-\r_{o})\bigg)\bigg]
\,,\label{Eq:RomansConfChi}\\
A(\r)&=A_{I}-\frac{4}{15}\ln(2)+\frac{4}{15}\ln\bigg[
\sinh\bigg(\sqrt{-5v}(\r-\r_{o}) \bigg) \bigg]\notag\\
&\hspace{47mm}+\frac{1}{15}\ln\bigg[
\tanh\bigg(\tfrac{{\sqrt{-5v}}}{2}(\r-\r_{o})\bigg)\bigg]
\,,\label{Eq:RomansConfA}
\end{align}    
where $\c_{I}$, $A_{I}$, and $\r_{o}$ are three integration constants. We note that $\r_{o}$ may be freely chosen to fix the physical end of space for the solutions at small $\r$, and unless otherwise specified we will always set this constant to zero. The equations of motion in Eqs.~(\ref{Eq:RomansEOM1}\,-\,\ref{Eq:RomansEOM4}) are invariant under an additive shift to the warp factor $A(\r)$, and hence we are also free to choose $A_{I}=\c_{I}$. The remaining integration constant $\c_{I}$, however, is not a free parameter and must be chosen to ensure that the six-dimensional geometry is regular at the end of space $\r=\r_{o}$. As stated earlier, the scalar field $\c(\r)$ is introduced in the dimensional reduction procedure as an additional degree of freedom which controls the size of the circular sixth dimension parametrised by the coordinate $\eta$; this coordinate is periodic (with period $2\pi$), and hence we must fix $\c_{I}$ to avoid a conical singularity at the end of space. In proximity of this point the geometry resembles a two-dimensional plane described by the following line element: 
\begin{align}
\label{Eq:2Dmetric}
\di\tilde{s}_{2}^{2}&= \di\r^{2}+e^{6\c}\di\eta^{2}\\
&=\di\r^{2}-\frac{5}{4}ve^{6\c_{I}}(\r-\r_{o})^2\di\eta^{2}
+\ldots\,,
\end{align}  
where in going from the first line to the second line we have directly substituted in for $\c$ using Eq.~(\ref{Eq:RomansConfChi}). To put this metric into the form of standard Euclidean polar coordinates it is hence necessary to make the identification 
\begin{equation}
\c_{I}=\frac{1}{6}\ln\bigg(\frac{-4}{5v}\bigg)\,.
\end{equation} 
More generally, it is also possible to numerically construct background solutions for cases in which $\f$ is not equal to either of the two critical point values, but rather smoothly interpolates between them. These solutions may be obtained by solving the classical equations of motion and imposing boundary conditions on the bulk fields using the following IR (small $(\r-\r_o)$) expansions:
{\small\begin{align}
	\phi(\r) &= \phi_{I}-\frac{1}{12}e^{-6\phi_{I}} \left(1-4 e^{4\phi_{I}}+3 e^{8\phi_{I}}\right)(\r-\r_{o})^2 \nn\\
	&\hspace{20mm}-\frac{1}{324} e^{-12\phi_{I}}\left(4-28 e^{4\phi_{I}}+51 e^{8\phi_{I}}-27 e^{16\phi_{I}}\right)(\r-\r_{o})^4 \nn\\
	&\hspace{20mm}+\mathcal{O}\left((\r-\r_o)^6\right)\, ,\\
	\c(\r)&=\c_{I}+\frac{1}{3}\ln\left(\frac{5}{3}\right)
	+\frac{1}{3}\ln (\r-\r_{o})
	-\frac{1}{27} e^{-2\phi_{I}} \Big(2+\sinh\big(4\phi_{I}
	+\ln(3)\big)\Big)(\r-\r_{o})^2\,  \nn\\
	&\hspace{20mm}+\frac{5}{486}e^{-4\phi_{I}}\Big(2+\sinh\big(4\phi_{I}
	+\ln(3)\big)\Big)^2(\r-\r_{o})^4 \nn\\
	&\hspace{20mm}+\mathcal{O}\left((\r-\r_o)^6\right)\, ,\label{Eq:RomansChiIRExpansion}\\
	A(\r)&= A_{I}+\frac{1}{3}\ln\left(\frac{5}{3}\right)
	+\frac{1}{3}\ln(\r-\r_{o})
	-\frac{7}{324}e^{-6\phi_{I}}\left(1-12 e^{4\phi_{I}}-9 e^{8\phi_{I}}\right)(\r-\r_{o})^2\, \nn\\
	&\hspace{6mm}+\frac{1}{17496}\left(108-67 e^{-12\phi_{I}}+636 e^{-8\phi_{I}}-2124 e^{-4\phi_{I}}-1053 e^{4\phi_{I}}\right)(\r-\r_{o})^4\nn\\
	&\hspace{20mm}+\mathcal{O}\left((\r-\r_o)^6\right)\, ,\label{Eq:RomansAIRExpansion}
	\end{align}}%
where the additional variable $\f_{I}$ here parametrises an entire family of possible background solutions. The special choice $\f_{I}=\f_{p}$ produces a solution equivalent to one of the analytical solutions, while $\f_{I}\in(\f_{IR},\,\f_{UV})$ ensures that the resultant generated background $\f(\r)$ interpolates between the two critical points, with the exact choice determining at what energy scale the transition between the two distinct CFTs occurs in the dual boundary theory. When computing the spectra of excitations we will permit $\f_{UV} < \f_{I}$, however we will furthermore impose the constraint $\f_{IR} \leqslant \f_{I}$ so that $\f(\r)$ is bounded from below. Finally, we note that by substituting the $\c$ IR expansion into Eq.~(\ref{Eq:2Dmetric}), the constraint on $\c_{I}$ to avoid a conical singularity at the end of space becomes 
\begin{equation}
\c_{I}=\frac{1}{3}\ln\bigg(\frac{3}{5}\bigg)\,.
\end{equation}

\section{Physical mass spectra: graviton modes}
\label{Sec:RomansMassPlots}
This section is dedicated to presenting the results of our numerical mass spectra computation, restricting attention to the field fluctuations which descend from the six-dimensional graviton of the supergravity multiplet; the complementary results obtained for the various other (bosonic) fields which comprise the theory are presented separately in Sec.~\ref{Sec:MoreBosons}, while the probe state analysis will be discussed in Sec.~\ref{Sec:RomansProbePlots}.\par
As per our line element ansatz in Eq.~(\ref{Eq:6Dmetric}), after dimensionally reducing the system by compactifying on a circle, the $9$ degrees of freedom carried by the metric tensor $\hat{g}_{\hat{M}\hat{N}}$ of the six-dimensional supergravity action $\cs_{6}$ may be decomposed into a graviton $g_{MN}$ (propagating 5 d.o.f.), a graviphoton $V_{M}$ (3 d.o.f.), and a (gravi)scalar $\hat{g}_{66}$ (1 d.o.f.) which governs the volume of the $S^{1}$. The fluctuations of these fields may be reformulated in terms of the physical gauge-invariant variables $\mathfrak{e}^{\mu}_{\ \nu}$ (5 d.o.f.), $V_{\mu}$ (2 d.o.f.), and $\mathfrak{a}^{a}$ (2$\times$1 d.o.f.), for which we then compute the spectra of excitations.\par 
Although we have already introduced the bulk equations and boundary conditions which are satisfied by the tensor and sigma-model scalar modes in Eqs.~(\ref{Eq:TensorFluct}\,-\,\ref{Eq:ScalarFluctBC}), we have not yet provided the corresponding equations for the graviphoton; we shall instead derive these (Eqs.~(\ref{Eq:VmuFluct},\,\ref{Eq:VmuFluctBC})) explicitly in Sec.~\ref{Sec:MoreBosons}, when we consider generalised supplementary actions to describe the remaining bosonic fields of the supergravity multiplet in five dimensions. As we shall see these derivations are rather non-trivial, and we therefore find it convenient to simply present the graviphoton spectrum here first alongside the other constituents of the $D=6$ graviton.\par
\begin{figure}[h!]
	\begin{center}
		\makebox[\textwidth]{\includegraphics[width=0.76\paperwidth]
			{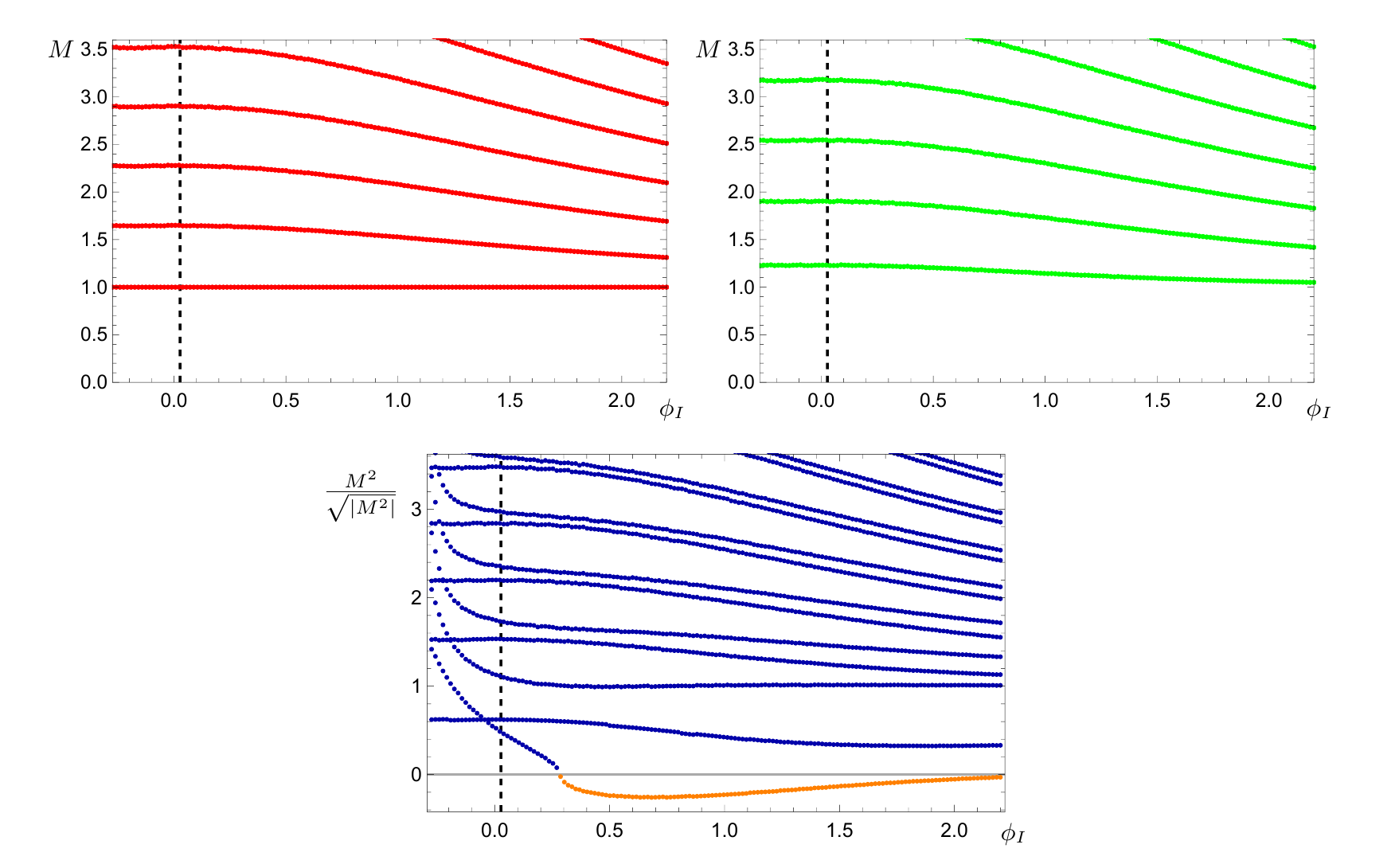}}
	\end{center}
\vspace{-5mm}
	\caption[Mass spectra of graviton resonances for the six-dimensional supergravity]{The spectra of masses $M$ as a function of the one free parameter which characterises the class of confining solutions, $\f_{I}\in[\f_{IR},2.2]$. From top to bottom, left to right: 
	the spectra of fluctuations for the tensors 
	$\mathfrak{e}^{\mu}_{\ \nu}$ (red), 
	the graviphoton $V_{\mu}$ (green), and the two scalars 
		$\mathfrak{a}^{a}$ (blue). The orange disks in the scalar 
 spectrum represent masses for which $M^{2}<0$, and hence denote a tachyonic state. The vertical dashed lines mark a critical value of the IR parameter $\f_{I}= \f_{I}^{*}>0$, which we shall formally introduce in Sec.~\ref{Sec:RomansPhaseStruct}. All states are normalised in units of the lightest tensor mass, and the spectra were computed using regulators $\r_{1}=10^{-4}$ and $\r_{2}=12$. }
	\label{Fig:Spectra1}
\end{figure}

The results of our numerical analysis, normalised in units of the lightest spin-2 state, are shown in the three panels of Fig.~\ref{Fig:Spectra1}. We here widen the scope of our previous computation in Ref.~\cite{Elander:2018aub} by permitting $\f_{I}>0$, corresponding to backgrounds that are driven away from the non-supersymmetric critical point solution and which `roll' down the runaway direction of the potential $\cv_{6}$ (see Fig.~\ref{Fig:ScalarPotential6D}) as they evolve towards the end of the geometry in the deep IR. As a
\\
consistency check, we verified that for $\f_{I}\in[\f_{IR},\f_{UV}]$ our results are in agreement with Ref.~\cite{Elander:2018aub}.
The numerical masses obtained by considering fluctuations about the backgrounds which admit constant $\f(\r)=\f_{p}\forall\r$ are tabulated in Appendix~\ref{App:CritSpectra} for convenience.\par
We here briefly remind the Reader that by varying the IR expansion parameter $\f_{I}$ one may generate an entire family of inequivalent confining background solutions, with the specific choice of $\f_{I}$ determining the energy scale at which the transition between the critical point solutions occurs; for $\f_{I}\in(\f_{IR},\,\f_{UV})$ the $S^{1}$-compactified dimension contracts to a point and the geometry smoothly closes off before the $\f(\r)$ background has had sufficient time to reach $\f_{IR}$. 
Still restricting our attention to the $\f_{I}\leqslant0$ region of the plots in Fig.~\ref{Fig:Spectra1}, we observe that the complete towers of spin-2 and spin-1 states, in addition to a subset of the spin-0 states (which are associated with fluctuations of $\c$), exhibit a \emph{universal} nature; these resonances appear to be independent of the scale at which the aforementioned interpolation occurs. The only source of scale-dependence which characterises these physical excitations---and moreover which is common to all three spin sectors---is that of the geometric confinement mechanism. That is to say, these states are sensitive only to the energy scale at which the dual field theory confines, and are otherwise insensitive to the specific details of the background profiles which are being fluctuated; this phenomenon has previously been discussed in Ref.~\cite{Elander:2013jqa}, and we confirm that our scalar spectrum (for both $\mathfrak{a}^{\f}$ and $\mathfrak{a}^{\c}$) agrees with the results of Figure~4 and Table~1 therein.\par 
Similar investigations in the context of top-down holography are known to exist already in the literature, and we now devote some time to quantitatively comparing (where possible) our results to these previous studies.\par 
In Ref.~\cite{Wen:2004qh} C.~Wen and H.~Yang conducted an early glueball spectrum analysis of (large-$N_{c}$) $\text{QCD}_{4}$ by considering the bosonic fluctuations of a dual $\text{AdS}_{6}$-Schwarzschild black hole geometry, raising the `temperature' of a circle-compactified \emph{thermal} coordinate to holographically realise confinement. The authors computed the spectra of glueball states which derive from fluctuations of the higher-dimensional metric, however their study considered only the critical $\f=\f_{UV}$ background solution; hence, they did not allow for perturbations of the supergravity field $\f$ which correspond to deforming the boundary CFT to realise an RG flow towards the IR fixed point. Nevertheless, we may compare our results to the data presented in Table~1 of their paper---normalising in units of the lightest tensor state---and we observe very good agreement with all three of our universal towers $\{\mathfrak{e}^{\mu}_{\ \nu},\,V_{\mu},\,\mathfrak{a}^{\c}\}$ (see Table~\ref{Tbl:RomansUV} in Appendix~\ref{App:CritSpectra}).\par  
Another investigation was conducted shortly after by S.~Kuperstein and J.~Sonnenschein in Ref.~\cite{Kuperstein:2004yf}, wherein the authors considered the fluctuations of supergravity fields propagating on a similar $S^{1}$-compactified $\text{AdS}_{6}$ background geometry (again restricting attention to the trivial $\f=0$ solution). Their analysis supplemented the spectrum computation for excitations of the $D=6$ graviton with those derived also from a Ramond--Ramond 1-form, which yielded two additional towers of states compared to Ref.~\cite{Wen:2004qh}; we shall postpone discussion of these two towers until Sec.~\ref{Sec:MoreBosons}. For the modes which may be identified with our universal states in Fig.~\ref{Fig:Spectra1}, we once again infer excellent agreement by comparing our $\mathfrak{e}^{\mu}_{\ \nu}$ data to that presented in Table~2, $V_{\mu}$ to Table~3, and our $\mathfrak{a}^{\c}$ data to that in Table~4 (after normalising each in units of the lightest spin-2 excitation). There does exist, however, one significant discrepancy between our results and those of Ref.~\cite{Kuperstein:2004yf}; the authors find an additional set of heavy scalar states in their study which do not agree with our tower of $\mathfrak{a}^{\f}$ excitations, and---since $\f$ is the only spin-0 field which appears in the six-dimensional supergravity---it is hence not clear how these states should be interpreted in this context.\par 
As we have explained, our spectra calculations for the field fluctuations of the six-dimensional supergravity are predicated on the application of gauge-gravity dualities, and in particular the geometric implementation of confinement; we interpret the masses extracted from our analysis as physical states in the dual four-dimensional model, which at low energies resembles a typical Yang-Mills theory. This correspondence lends itself to comparison with lattice studies of gauge theories extrapolated to large-$N_{c}$, and in Refs.~\cite{Lucini:2010nv,Bennett:2017kga} the authors perform numerical computations on a cubic lattice for models admitting $SU(N\to\infty)$ and $Sp(4)\cong SO(5)$ gauge groups, respectively. Our results presented in Fig.~\ref{Fig:Spectra1} for the background-dependent modes show qualitative agreement with these two studies, and we direct the Reader's attention to the discussion in Ref.~\cite{Elander:2018aub} for further details and clarification of some technical subtleties associated with the comparison.\par 
We conclude this section by commenting on the complementary $\f_{I}>0$ region of the spectra in Fig.~\ref{Fig:Spectra1}, which extends our earlier results from Ref.~\cite{Elander:2018aub}. Firstly, we notice that the universal background-independence phenomenon described for solutions which dualise RG flows between the two fixed points does not continue to manifest as the transition parameter $\f_{I}$ is dialled higher. Indeed, we clearly see that the spectra for all three gauge-invariant modes which descend from the $D=6$ graviton become increasingly densely packed. In the large-$\f_{I}$ limit we deduce that the heavier excitations become asymptotically degenerate, while a few of the lightest scalar states remain comparatively light and discrete; this gapped continuum phenomenon has also been observed in other contexts in the literature---see for example Refs.~\cite{Elander:2017cle,Elander:2017hyr} and~\cite{Brandhuber:2000fr,Brandhuber:2002rx}---and we shall encounter a similar effect for the other fluctuation spectra presented in Sec.~\ref{Sec:MoreBosons}.\par
Secondly, and perhaps most significantly, we notice that the lightest resonance in the spin-0 spectrum acquires a negative squared mass as the IR parameter $\f_{I}$ is dialled above the trivial critical point solution, and hence corresponds to a tachyonic instability. Since we are studying the field fluctuations of a model obtained by dimensionally reducing an established and well-defined supergravity theory from a top-down holographic perspective, we predict by necessity the existence of a phase transition which would prevent these pathological backgrounds from being realised. It is this observation which crucially motivates our exploration of the theory phase space in later sections, by cataloguing several geometrically distinct classes of solutions which are admitted by the $S^{1}$-compactified supergravity, and then proceeding to systematically compute their (appropriately renormalised) free energy. We shall postpone further discussion of this exercise until Chapter~\ref{Chap:EnergeticsRomans}.             

\section{Probe spectrum analysis}
\label{Sec:RomansProbePlots}
As discussed in Sec.~\ref{Sec:Motivation}, we are interested in generalising the study of dilaton phenomenology as presented in Ref.~\cite{Pomarol:2019aae} to be applicable to non-AdS geometries. Having now computed the mass spectrum of gauge-invariant scalar fluctuations in Sec.~\ref{Sec:RomansMassPlots} (in addition to the spectra of the two other universal graviton modes), we next present the results of our probe state analysis using the formalism introduced in Sec.~\ref{Sec:DilatonFormalism}.\par 
As a brief review, our dilaton investigation is based primarily on the comparison of two spectrum calculations: the first is for the complete, gauge-invariant scalar fluctuations $\mathfrak{a}^{a}$ as defined in Eq.~(\ref{Eq:a}), while the second is for those same states evaluated instead using the probe approximation $\mathfrak{a}^{a}|_{h=0}\equiv\mathfrak{p}^{a}$. In the latter case, the component of $\mathfrak{a}^{a}$ which corresponds to the scalar fluctuations of the metric $h$ (the boundary value of which sources the dilatation operator of the dual field theory) is switched off by hand. We anticipate that for any physical states which differ appreciably between the two computations, the contribution of the metric perturbation $h$ is not insignificant, and those states exhibit non-trivial mixing with the dilaton.\par
\begin{figure}[h!]
	\begin{center}
		\makebox[\textwidth]{\includegraphics[
			width=0.56\paperwidth
			]
			{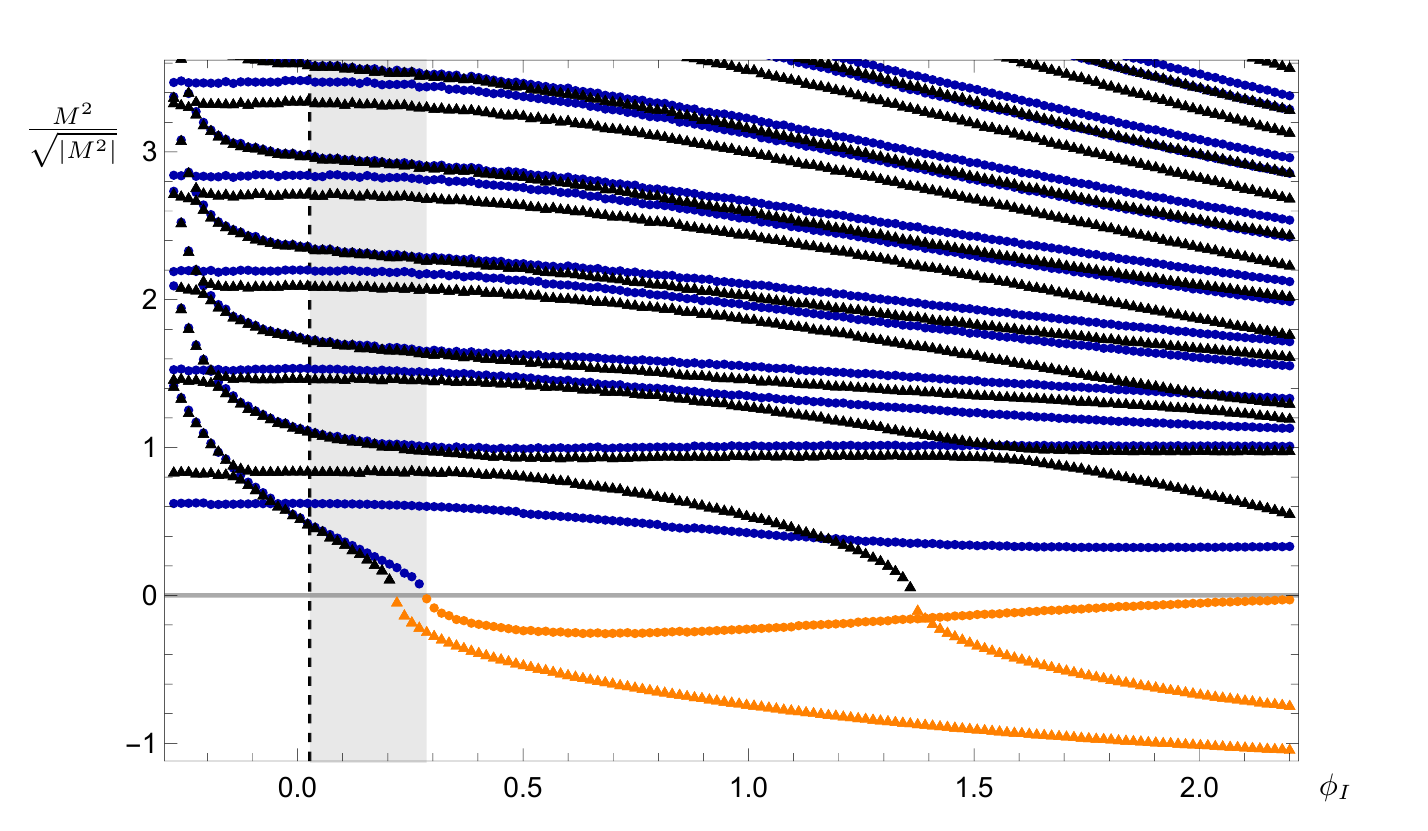}}
	\end{center}
\vspace{-5mm}
	\caption[Probe spectra for the six-dimensional supergravity]{The spectra of masses $M$ as a function of the one free parameter which characterises the class of confining solutions, $\f_{I}\in[\f_{IR},2.2]$. All states are normalised in units of the lightest tensor mass, and the spectrum was computed using regulators $\r_{1}=10^{-4}$ and $\r_{2}=12$. As in Fig.~\ref{Fig:Spectra1},
		the blue disks represent the two scalars of the model $\phi$ and $\chi$, while the orange disks denote the tachyon. 
		We here additionally include the results of our mass spectrum computation using the probe approximation for $M^{2}>0$
		(black triangles) and $M^{2}<0$ (orange triangles). 
	The vertical dashed line marks a critical value of the IR parameter $\f_{I}= \f_{I}^{*}>0$, while the shaded grey region denotes the region of parameter space for which the confining solutions are metastable; we shall elaborate on these points in Sec.~\ref{Sec:RomansPhaseStruct}.}
	\label{Fig:RomansSpectrumProbe}
\end{figure}

\begin{figure}[h!]
	\begin{center}
		\makebox[\textwidth]{\includegraphics[width=0.56\paperwidth]
			{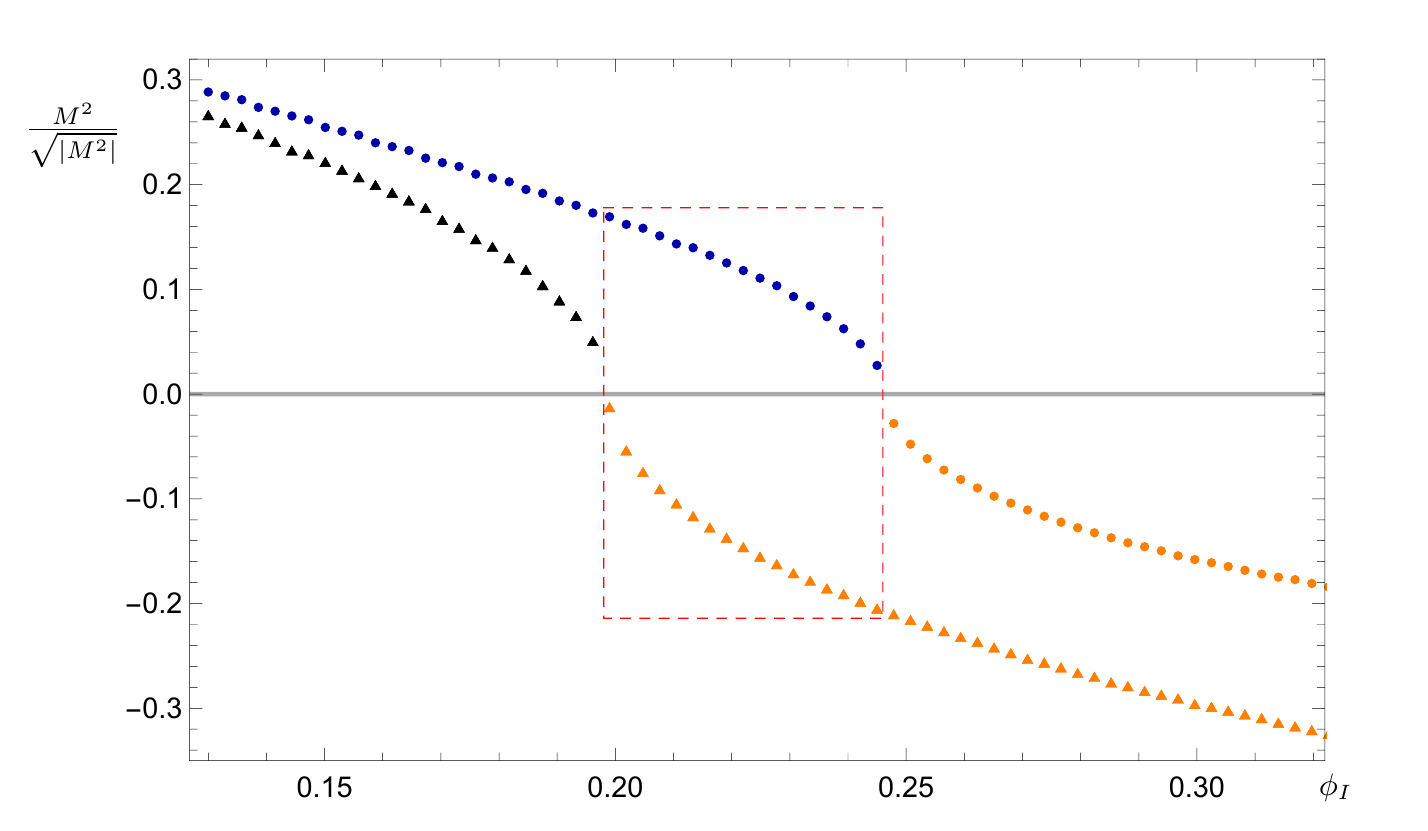}}
	\end{center}
\vspace{-5mm}
	\caption[Magnification of the probe spectra for the six-dimensional supergravity]{A magnification of the plot shown in Fig.~\ref{Fig:RomansSpectrumProbe}. The states are normalised in units of the lightest tensor mass, and are computed with regulators $\r_{1} =10^{-9}$ and $\r_{2}=15$. We focus in particular on the lightest state of the spectrum, in the plot region where the tachyonic states first appear. The dashed red box is intended to enclose an important feature of the full spectrum, namely a region of $\f_{I}$ parameter space wherein the probe approximation completely disagrees with the full gauge-invariant scalar computation. There exists a finite range of values for the IR parameter $\f_{I}$ for which the squared masses $M^{2}$ of the physical scalars $\mathfrak{a}^{b}$ and the probes $\mathfrak{a}^{b}\rvert_{h=0}\equiv\mathfrak{p}^{b}$ differ by a minus sign, and hence the probe approximation unambiguously fails.}
	\label{Fig:RomansSpectrumProbeZoom}
\end{figure}

 
The results of our probe state analysis are presented in Figures~\ref{Fig:RomansSpectrumProbe} and~\ref{Fig:RomansSpectrumProbeZoom} (see also Fig.~4 of Ref.~\cite{Elander:2020csd}), and from even a cursory examination it is clear that the probe approximation does not accurately capture the physical spectrum for any value of the IR parameter $\f_{I}$; this is especially true for the two lightest states. We see from Fig.~\ref{Fig:RomansSpectrumProbe} that for negative values of $\f_{I}$ in proximity of the IR critical point $\f_{I}=\f_{IR}$, the lightest state is completely missed by the probe calculation while the second lightest state is well approximated. In this region of parameter space we may infer that the former mass eigenstate has a significant dilaton contribution, while the latter does not.\par 
As we increase $\f_{I}$ to approach the UV fixed point at $\f_{I}=0$, we notice a reversal of this tendency: the probe approximation now begins to effectively capture the lightest state, while the next-to-lightest state instead exhibits dilatonic behaviour. Further along still, within the shaded region of the plot, we make another important observation: just before the appearance of the tachyon at $\f_{I}=\f_{I}^{\t}\sim 0.25$ the probe approximation again fails to capture the lightest state, the mass of which becomes parametrically light compared to the rest of the spectrum.                 
From this observation we conclude that---at least in proximity to the appearance of the tachyonic instability---we may legitimately identify the lightest state as an approximate dilaton; we provide a magnified view of this plot region in Fig.~\ref{Fig:RomansSpectrumProbeZoom} for emphasis. 
We furthermore notice that even the heavier states of the spectrum are not always captured well by the probe approximation, and as $\f_{I}$ is dialled higher we indeed see evidence of the probe states becoming lighter, and eventually tachyonic. This highlights the fact that mixing effects between the gauge-invariant spin-0 mass eigenstates and the dilaton are not trivial, and are not restricted to the lightest states in the tower.\par 
To summarise then, we have uncovered the existence of a tachyonic instability in the mass spectrum for a class of regular background solutions to Romans $D=6$ supergravity. Moreover, we have provided evidence that this tachyonic mode contains a significant component coming from the scalar fluctuation of the five-dimensional metric, and hence exhibits significant mixing with the approximate dilaton of the theory. As we discussed in Sec.~\ref{Sec:EnergeticsFormalism}, these findings will motivate our energetics analysis later in Chapter~\ref{Chap:EnergeticsRomans}, since we anticipate by necessity that a phase transition exists to prohibit the system from reaching the unstable region of the theory parameter space. Our investigation into the phase structure of the model will also provide us with some useful parameter relations, which we shall use to elucidate the nature of the dilaton; we will return to these spectra results later on in Sec.~\ref{Sec:RomansProbeRevisit}, and re-examine them in this context.        

\section{Physical mass spectra: \emph{p}-form modes }
\label{Sec:MoreBosons}
As we have discussed, the spectra of composite states are obtained by fluctuating the supergravity fields about their classical background configurations, and determining which values of the mass parameter $M$ allow for the fluctuation equations to be simultaneously satisfied over the entire domain of the holographic coordinate $\r$. In Sec.~\ref{Sec:ComputingSpectra} we presented the equations and boundary conditions which are obeyed by the gauge-invariant variables constructed from the fluctuations of the metric and the sigma-model scalars, and in this section we shall derive the corresponding equations for the various other fields which define $\cs_{6}$. We shall proceed
by considering generic actions to describe gauge-invariant 1- and 2-forms in five dimensions---supplementary to the action of a sigma-model coupled to five-dimensional gravity---and then decomposing the fields into their constituent four-dimensional components. The required fluctuation equations and boundary conditions are then obtained by demanding that an infinitesimal variation with respect to each field should vanish on-shell.\par
Before continuing however, we will first address an important point concerning our treatment of the various $p$-forms and fields. The five-dimensional action of the reduced system, shown in Eq.~(\ref{Eq:5DActionComplete}), is invariant under a number of gauge transformations: there are $U(1)$ invariances associated with the graviphoton $V_{M}$, the vector $B_{6N}$, and the pseudo-scalar $A_{6}$, an $SU(2)$ invariance of the vectors $A_{N}^{i}$ and the pseudo-scalars $\pi^{i}$, gauge-invariances for the 2-form $B_{MN}$ and the vector $A_{M}$, in addition to diffeomorphisms of the fluctuations of the sigma-model scalars and the metric tensor, which were considered in Sec.~\ref{Subsec:HolographicFormalism}. We here emphasise the fact that we may treat these various gauge-invariances separately, since we assume that \emph{only} the metric and the sigma-model scalars $\f$ and $\c$ acquire non-trivial background profiles (we fluctuate every other field about their trivial configuration), and furthermore that to compute the mass spectra it is sufficient to retain only terms which are up to second order in the field fluctuations.  

\subsubsection{Vectors (1-forms) in $D=5$ dimensions}
Let us start by recalling the terms in the five-dimensional action of Eq.~(\ref{Eq:5DActionComplete}) that contain a $U(1)$-invariant vector (1-form) contribution, which combine to give: 
\begin{align}
\mathcal{S}_{5}^{(1f)}=\int\is\text{d}^{5}x\,\sqrt{-g_{5}}
\bigg(&-\frac{1}{4}H_{11}g^{MR}g^{NS}F_{MN}^{V}F_{RS}^{V}
-\frac{1}{2}G^{(1)}g^{NS}\mathcal{H}_{6N}\mathcal{H}_{6S}\notag\\
&-\frac{1}{4}H^{(1)}g^{NS}g^{TU}G_{6NT}G_{6SU}
\bigg)\,.
\end{align}
After substituting in for the tensors $F_{MN}^{V}$, $\mathcal{H}_{6N}$, and $G_{6NT}$, this action may be rewritten as follows:
\begin{align}
\mathcal{S}_{5}^{(1f)}=\int\is\text{d}^{5}x\,\sqrt{-g_{5}}
\bigg(&-\frac{1}{4}H_{11}g^{MR}g^{NS}\big(\pa_{M}V_{N}-\pa_{N}V_{M}\big)\big(\pa_{R}V_{S}-\pa_{S}V_{R}\big)\notag\\
&-\frac{1}{2}G^{(1)}g^{NS}\big(\pa_{N}A_{6}-mB_{6N}\big)\big(\pa_{S}A_{6}-mB_{6S}\big)
\notag\\
&-\frac{1}{4}H^{(1)}g^{NS}g^{TU}\big(\pa_{T}B_{6N}-\pa_{N}B_{6T}\big)\big(\pa_{U}B_{6S}-\pa_{S}B_{6U}\big)
\bigg)\,.
\end{align}
We shall find that after decomposing the five-vectors which comprise this sub-system into four-dimensional quantities, we will ultimately derive the fluctuation equations and BCs satisfied by the set of fields $\{V_{\mu},\,A_{\mu}^{i},\,B_{6\mu},\,X(A_{6},B_{65})\}$. More generally, let us instead consider the following action describing a spontaneously broken $U(1)$ theory in five dimensions:
\begin{align}
{\cal S}_5^{(1)} =\int\is\di^{4}x\di r\, \sqrt{-g_5}\bigg\{ 
&-\frac{1}{4}H \,g^{MR}\,g^{NS}\, F_{MN}F_{RS} \nn\\
&-\frac{1}{2}G \, g^{MN}\Big(\pa_M \pi+m A_M\Big)\Big(\pa_N \pi+m A_N\Big)
\bigg\}\,,
\label{Eq:RomansGeneralU1}
\end{align}
where $G=G(\Phi^{a})$ and $H=H(\Phi^{a})$ are sigma-model geometric factors which generalise $G^{(1)}$ and $H^{(1)}$, $F_{MN}\equiv 2\partial_{[M}A_{N]}$ is the field strength for the generic 1-form $A_{M}$, $\pi$ is a pseudo-scalar (0-form), and the mass $m$ is a symmetry-breaking parameter; gauge-invariance of the terms $\pa_M \pi+m A_M$ manifests $\forall m$ via the transformations $A_{M}\rightarrow A_{M}- \pa_{M}\a$ and $\pi\rightarrow\pi+m\a$, where $\a$ is a function of the spacetime coordinates $x^{M}$. Similarly to our treatment in Sec.~\ref{Subsec:HolographicFormalism} of the metric using the ADM decomposition, we will rewrite the fields of the action in terms of their constituent four-vectors.\par
As a brief aside, we here remind the Reader that the number of on-shell degrees of freedom $f_{m=0}$ for a massless $p$-form in $D$ dimensions is given by~\cite{Nastase:2011aa} 
\begin{equation}
f_{m=0} = \begin{pmatrix}
D-2\\
p
\end{pmatrix} = \frac{(D-2)!}{(D-2-p)!\,p!}\,,
\end{equation}  
while for a massive $p$-form in $D$ dimensions the number of on-shell degrees of freedom $f_{m\neq 0}$ is
\begin{equation}
f_{m\neq 0} = \begin{pmatrix}
D-1\\
p
\end{pmatrix} = \frac{(D-1)!}{(D-1-p)!\,p!}\,,
\end{equation} 
so that in $D=5$ dimensions a massless (massive) 1-form has 3 (4) on-shell degrees of freedom. The massless scalar fields $A_{5}$ and $\pi$ present in Eq.~(\ref{Eq:RomansGeneralU1}) both behave as Goldstone bosons, with the former coming from the Kaluza-Klein dimensional reduction of the five-vector $A_{M}$, and the latter resulting from the spontaneous symmetry breaking of the global $U(1)$ in $D=5$; a combination of these two fields provides the additional degree of freedom required for the longitudinal components of a infinite tower of massive four-vectors $A_{\mu}$ (analogous to the Higgs mechanism, here the vectors acquire mass after ``eating'' the Goldstone bosons), while another combination forms the massive pseudo-scalars $X$. We shall return to this point later.\par
After decomposing the fields and rewriting the action $\mathcal{S}_{5}^{(1)}$ in terms of four-vectors and (pseudo-)scalars, we Fourier transform according to the following conventions:   
\begin{align}
\label{Eq:Ftransf1}
\psi(x^{\mu})&\equiv\int\is\frac{\di^{4}q}{(2\pi)^2}\,
e^{i q_{\mu}x^{\mu}}\tilde{\psi}(q_{\mu})\,,\\
\label{Eq:Ftransf2}
\delta^{(4)}(q_{\mu})&\equiv\int\is\frac{\di^{4}x}{(2\pi)^4}\,
e^{i q_{\mu}x^{\mu}}\,,
\end{align}
and make use where necessary of the symmetrisation condition
\begin{equation}
\int\is\di^{4}q\, \psi(q)\varphi(-q)\equiv \int\is\di^{4}q\,\half\Big[\psi(q)\varphi(-q)+\psi(-q)\varphi(q)\Big]\,,
\end{equation}
where $\psi$ and $\varphi$ are generic fields. After some algebra and integration by parts---and neglecting to show tildes on transformed fields---we find that the general action of Eq.~(\ref{Eq:RomansGeneralU1}) may be written as follows:
\beqs
{\cal S}_5^{(1)} &=&\int\is \di^{4}q\,\di r \left\{ 
-\frac{1}{2}H\,A_{\mu}(-q)\,q^2 P^{\mu\nu} A_{\nu}(q)\,
-\frac{1}{2}H e^{2A} q^2 A_{5}(-q)A_5(q)
\,\right.\nonumber\\
&&\nonumber\left.
-\frac{1}{2}A_{\mu}(-q)\eta^{\mu\nu}\Big[-\pa_r\left(H e^{2A} \pa_r A_{\nu}(q)\right)\Big]
\,\right.\\
&&\nonumber\left.
+\sum_{i=1,2}(-)^i\delta(r-r_i)\left[-\frac{1}{2} H e^{2A} A_{\mu}(-q) \eta^{\mu\nu} \pa_r A_{\nu}(q)\right]
\,\right.\\
&&\nonumber\left.
-\frac{1}{2}\Big[iq^{\mu}A_{\mu}(-q)\pa_r\left( H e^{2A} A_5(q)\right)\,+\,(q\leftrightarrow -q)\Big]
\,\right.\\
&&\nonumber\left.
+\sum_{i=1,2}(-)^i\delta(r-r_i)\left[\frac{1}{2}i H e^{2A} q^{\mu}A_{\mu}(-q) A_5(q)\,+\,(q\leftrightarrow -q)\right]
\,\right.\\
&&\nonumber\left.
-\frac{1}{2}m^2 G e^{4A} A_5(-q)A_5(q)
-\frac{1}{2}\pi(-q)\pa_r\left[- G e^{4A} \pa_r \pi(q)\right]
\,\right.\\
&&\nonumber\left.
+\sum_{i=1,2}(-)^i\delta(r-r_i)\left[-\frac{1}{2}G e^{4A} \pi(-q) \pa_r \pi(q) \right]
\,\right.\\
&&\nonumber\left.
-\frac{1}{2}\pi(-q) \pa_r\Big[-m G e^{4A} A_5(q)\Big]
\,\right.\\
&&\nonumber\left.
-\frac{1}{2}A_5(-q) \Big[m G e^{4A} \pa_r \pi(q)\Big]
\,\right.\\
&&\nonumber\left.
+\sum_{i=1,2}(-)^i\delta(r-r_i)\left[-\frac{1}{2}m G e^{4A} \pi(-q) A_5(q)\right]
\,\right.\\
&&\nonumber\left.
-\frac{1}{2}G e^{2A}\Big[q^2 \pi(-q)\pi(q)+m^2 \eta^{\mu\nu}A_{\mu}(-q)A_{\nu}(q)\Big]
\,\right.\\
&&\left.\, -\frac{1}{2}m G e^{2A} \,
\Big[-iq_{\mu}\pi(-q)  \eta^{\mu\nu}A_{\nu}(q)\,+\, (q\rightarrow -q)\Big]
\right\}\,,
\eeqs
where $P^{\mu\nu}(q^2)\equiv\eta^{\mu\nu}-\frac{q^{\mu}q^{\nu}}{q^2}$ is the transverse momentum projector satisfying $P^{\mu\nu}q_{\mu}=0$ (with $\frac{q^{\mu}q^{\nu}}{q^2}$ the longitudinal momentum projector). We have neglected to show explicitly the radial coordinate dependences (i.e.\ $A_{\mu}(q)$ should be understood to mean $A_{\mu}(q,r)$), and terms containing a delta function $\d(r-r_{i})$ are contributions which are localised to the two endpoints of the radial segment parametrised by $r$. The presence of these boundaries motivates our inclusion of the following generic boundary-localised kinetic terms for the vector $A_{\mu}$:
\begin{align}
\cs_{D}^{(1)} &= \int\is \di^{4}x\,\di r \sum_{i=1,2} (-)^i
\delta(r-r_i)\sqrt{-g_5}D_i \left\{-\frac{1}{4}\tilde{g}^{MN}\tilde{g}^{RS}F_{MR}F_{NS}\right\}\,\nn\\
&=\int\is \di^{4}q\,\di r \sum_{i=1,2}(-)^i\delta(r-r_i)
\left\{-\frac{1}{2} D_{i}q^{2}A_{\mu}(-q) P^{\mu\nu} A_{\nu}(q) \right\}\,,
\end{align}
and for the pseudo-scalar $\pi$:
\begin{align}
{\cal S}_C^{(1)}&=\int\is\di^{4}x\,\di r \sum_{i=1,2}(-)^{i}\delta(r-r_i)\sqrt{-g_5}\bigg\{
-\frac{1}{2} C_{i} \Big[\pa_{\mu}\pi+m A_{\mu}\Big]\tilde{g}^{\mu\nu} \Big[\pa_{\nu}\pi+ m A_{\nu}\Big] \bigg\}\,\nn\\
&=\int\is\di^{4}q\,\di r \sum_{i=1,2}(-)^{i}\delta(r-r_i)\bigg\{
-\frac{1}{2} C_i e^{2A}\Big[q_{\mu}\pi(-q)+i m A_{\mu}(-q)\Big]\nn\\
&\hspace{56mm}\times\eta^{\mu\nu} 
\Big[q_{\nu}\pi(q)-i m A_{\nu}(q)\Big] \bigg\}\,.
\end{align}
The four constants $D_{i}$ and $C_{i}$ are in general introduced as counter-terms in the process of holographic renormalisation, however for our purposes it is justified to simply set them both to zero (see Appendix B.2 of Ref.~\cite{Elander:2018aub} for details), which we shall do after we have derived the fluctuation equations.\par
We notice that the action $\mathcal{S}_{5}^{(1)}$ contains unphysical mixing terms between the vector $A_{\mu}$ and the Goldstone pseudo-scalar $\pi$, which must be removed by hand. For this purpose we introduce a general $R_{\xi}$ gauge-fixing bulk action, in addition to separate boundary expressions, so that each action contains the terms required to exactly cancel the undesired contributions; the bulk term is given by   
\begin{align}
{\cal S}^{(1)}_{\xi}
&=\int\is\di^{4}q\,\di r \bigg\{-\frac{H}{2\xi}
\Big[q^{\mu}A_{\mu}(-q)+ i\frac{\xi}{H} mG e^{2A} \pi(-q)
+ i \frac{\xi}{H} \pa_r\Big(H e^{2A} A_5(-q)\Big)\Big] \nn\\
&\hspace{15mm}\times
\Big[q^{\nu}A_{\nu}(q) - i\frac{\xi}{H} mG e^{2A} \pi(q) -i\frac{\xi}{H} \pa_r\Big(H e^{2A} A_5(q)\Big)\Big]
\bigg\}\,, 
\end{align}
while the boundary-localised gauge-fixing terms are
\begin{align}
{\cal S}_M^{(1)}
&=\int\is\di^{4}q\,\di r \sum_{i=1,2}(-)^{i}\delta(r-r_i)\bigg\{
-\frac{1}{2M_i}\nn
\Big[q^{\mu}A_{\mu}(-q)- i M_i H e^{2A} A_5(-q) \nn\\
&\hspace{70mm}+ i m M_i C_i e^{2A} \pi(-q) \Big] \nn\\
&\hspace{20mm}\times
\Big[q^{\nu}A_{\nu}(q)+ i M_i H e^{2A} A_5(q) - i m M_i C_i e^{2A} \pi(q) \Big]
\bigg\}\,.
\end{align}
In general there is no reason to assume that the gauge-fixing parameter $\xi$ is independent of the radial segment parametrised by the holographic coordinate $r$; in the case that $\xi=\xi(r)$ the $U(1)$ symmetry of the five-dimensional model would generate an infinite number of spontaneously broken $U(1)$ gauge theories in $D=4$ dimensions, with each admitting an independent gauge-fixing choice. We shall make the simplifying assumption that $\xi$ is a scale-invariant constant, however.\par
We are now ready to compute the equations for the field fluctuations of this sub-system. For the sake of brevity we will show explicitly the derivation of the $A_{\mu}$ equation only, with the understanding that the other equations are obtained in much the same way. Our starting point is the total action ${\cal S}_{\text{Tot}}^{(1)}\equiv {\cal S}_{5}^{(1)}+{\cal S}_{D}^{(1)}+{\cal S}_{C}^{(1)}+{\cal S}_{\xi}^{(1)}+{\cal S}_{M}^{(1)}$, from which we collect terms containing an $A_{\mu}$ contribution: 
\begin{align}
{\cal S}_{A_{\mu}}^{(1)}
&=\int\is\di^{4}q\,\di r \bigg\{-\half H q^{2}P^{\mu\nu}A_{\mu}(-q)A_{\nu}(q) +\half\eta^{\mu\nu}A_{\mu}(-q)\pa_{r}\Big(He^{2A}\pa_{r}A_{\nu}(q)\Big)\nn\\ &\hspace{20mm}-\half m^{2}Ge^{2A}\eta^{\mu\nu}A_{\mu}(-q)A_{\nu}(q) -\frac{1}{2\xi}Hq^{\mu}q^{\nu}A_{\mu}(-q)A_{\nu}(q)
\bigg\}\nn\\ 
&+\int\is\di^{4}q\,\di r \sum_{i=1,2}(-)^{i}\delta(r-r_i)\bigg\{
-\half He^{2A}\eta^{\mu\nu}A_{\mu}(-q)A_{\nu}(q)\nn\\
&\hspace{20mm}-\half D_{i} q^{2}P^{\mu\nu}A_{\mu}(-q)A_{\nu}(q)-\half m^{2}C_{i}e^{2A}\eta^{\mu\nu}A_{\mu}(-q)A_{\nu}(q)\nn\\
&\hspace{20mm}-\frac{1}{2M_{i}}q^{\mu}q^{\nu}A_{\mu}(-q)A_{\nu}(q) 
\bigg\}\,,
\end{align}  
and we remind the Reader that $\{C_{i},\,D_{i},\,M_{i}\}$ are constants, but $H$ and $G$ are functions of the sigma-model scalars (and hence are $r$-dependent). The equations for the fluctuations are derived by considering an infinitesimal variation of the action with respect to its constituent fields, and requiring that this variation vanishes on-shell. 
For contributions to the bulk action which generically take the form $\g(r)\pa_{r}\big[\z(r)\d A_{\mu}(-q)\big]$ (i.e. for terms in which the field variation appears within a derivative) it is convenient to rewrite them using partial integration:
\begin{equation}
\g(r)\pa_{r}\Big[\z(r)\d A_{\mu}(-q)\Big] = \pa_{r}\Big[\d A_{\mu}(-q)\g(r)\z(r)\Big]-\z(r)\d A_{\mu}(-q)\pa_{r}\g(r)\,,
\end{equation}
and then recasting the total derivative piece instead as a boundary term. After taking the variation of the action, we find that the condition for $\d{\cal S}_{A_{\mu}}^{(1)}$ to vanish is that its integrand must be trivial. From the bulk contribution to the action we therefore have
\begin{align}
0&=\d A_{\mu}(-q)\bigg[-\half H q^{2}P^{\mu\nu}A_{\nu}(q) +\half\eta^{\mu\nu}\pa_{r}\Big(He^{2A}\pa_{r}A_{\nu}(q)\Big)\nn\\ &\hspace{20mm}-\half m^{2}Ge^{2A}\eta^{\mu\nu}A_{\nu}(q) -\frac{1}{2\xi}Hq^{\mu}q^{\nu}A_{\nu}(q)\bigg]\,,
\end{align}
and from the boundary contribution:  
\begin{align}
0&=\d A_{\mu}(-q)\bigg[
-\half He^{2A}\eta^{\mu\nu}A_{\nu}(q)-\half D_{i} q^{2}P^{\mu\nu}A_{\nu}(q)\nn\\
&\hspace{20mm}-\half m^{2}C_{i}e^{2A}\eta^{\mu\nu}A_{\nu}(q)-\frac{1}{2M_{i}}q^{\mu}q^{\nu}A_{\nu}(q)\bigg]\bigg|_{r=r_i}\,.
\end{align}
We remind the reader that $\eta^{\mu\nu}=P^{\mu\nu}+\frac{q^{\mu}q^{\nu}}{q^2}$, where $P^{\mu\nu}$ and $\frac{q^{\mu}q^{\nu}}{q^2}$ are the transverse and longitudinal momentum projectors, respectively, and hence we obtain distinct fluctuation equations for the two polarisations of $A_{\mu}(q)$. The boundary-localised gauge-fixing parameters $M_{i}$ are independent of the bulk dynamics and may be assigned an arbitrary value; we will conveniently choose to fix $M_{i}=\frac{\xi}{D_{i}}$ so that the equations satisfied by the longitudinal component of the spin-1 fields are identical to those for the transverse component, after making the replacement $q^2\to\frac{q^2}{\xi}$. We shall only compute the spectrum for the transverse polarisation of $A_{\mu}$, the longitudinal polarisation retaining an unphysical gauge dependence. Finally then, we obtain the following bulk equations and boundary conditions for the field fluctuations of the four-vectors $A_{\mu}(q,r)$: 
\begin{align}
\label{Eq:1formVectorEqn}
0&=\left[q^2 H  -\pa_r \left( H e^{2A} \pa_r \right) + m^2 G e^{2A}\right] P^{\mu\nu}A_{\mu}(q,r)\,,\\
\label{Eq:1formVectorBC}
0&=\left[\frac{}{}H e^{2A} \pa_r +q^2 D_i + m^2 C_i e^{2A} \right] P^{\mu\nu}A_{\nu}(q,r)\Big|_{r=r_i}\,,\\
0&=\left[\frac{q^2}{\xi} H  -\pa_r \left( H e^{2A} \pa_r \right) + m^2 G e^{2A}\right] \frac{q^{\mu}q^{\nu}}{q^2}A_{\mu}(q,r)\,,\\
0&=\left[H e^{2A} \pa_r +\frac{q^2}{\xi} D_i + m^2 C_i e^{2A} \right] \frac{q^{\mu}q^{\nu}}{q^2}A_{\nu}(q,r)\Big|_{r=r_i}\,.
\end{align}
The corresponding equations for the pseudo-scalars $A_{5}$ and $\pi$ are similarly obtained by varying the necessary contributions to ${\cal S}_{\text{Tot}}^{(1)}$, the result of which provides the following conditions: 
\begin{align}
0&=\d A_{5}(-q)\bigg[-\half He^{2A} q^{2}A_{5}(q)-\half m^{2}Ge^{4A}A_{5}(q)\nn\\ 
&\hspace{20mm}-\half mGe^{4A}\pa_{r}\pi(q)
+\half m\xi He^{2A}\pa_{r}\left(\frac{G}{H}e^{2A}\pi(q)\right)\nn\\
&\hspace{20mm}+\half\xi He^{2A}\pa_{r}\left(\frac{1}{H}\pa_{r}\Big(He^{2A}A_{5}(q)\Big)\right)
\bigg]\,,
\end{align} 
\begin{align}
0&=\d \pi(-q)\Big[-\half Ge^{2A} q^{2}\pi(q)-\half m\xi\frac{G}{H}e^{2A}\pa_{r}\Big(He^{2A}A_{5}(q)\Big)\nn\\
&\hspace{20mm}+\half\pa_{r}\Big(Ge^{4A}\pa_{r}\pi(q)\Big)-\half m^{2}\xi\frac{G^2}{H}e^{4A}\pi(q)\nn\\
&\hspace{20mm}+\half m \pa_{r}\Big(Ge^{4A}A_{5}(q)\Big)
\Big]\,,
\end{align} 
\begin{align}
0&=\d A_{5}(-q)\bigg[-\half m\xi G e^{4A}\pi(q)
-\half\xi e^{2A}\pa_{r}\Big(He^{2A}A_{5}(q)\Big)\nn\\ 
&\hspace{20mm}-\half M_{i}H^{2}e^{4A}A_{5}(q)
+\half mC_{i}M_{i} He^{4A}\pi(q)\bigg]\bigg|_{r=r_i}\,,
\end{align}
\begin{align}
0&=\d \pi(-q)\bigg[-\half G e^{4A}\pa_{r}\pi(q)
-\half C_{i}q^{2} e^{2A}\pi(q)\nn\\ 
&\hspace{20mm}+\half mC_{i} M_{i}He^{4A}A_{5}(q)
-\half mGe^{4A}A_{5}(q)\nn\\
&\hspace{20mm}-\half m^{2}C_{i}^{2}M_{i}e^{4A}\pi(q)
\bigg]\bigg|_{r=r_i}\,.
\end{align}
These expressions appear to be considerably more complicated than those for the vector polarisations due to the mixing between $A_{5}$ and $\pi$, however by introducing the following convenient reparametrisations:  
\begin{align}
A_5&\equiv\frac{1}{m}\left(\frac{m X}{e^{4A} G} - \partial_r \pi\right)\,,\\
\pi&\equiv Y + \frac{m \partial_r X}{q^2 e^{2A} G}\,,
\end{align}
and after some algebraic manipulation, we find that the equations and boundary conditions for the two new scalars $X$ and $Y$ decouple, and we have the following equivalent formulation:
\begin{align}
\label{Eq:1formScalarEqn}
0&=\left[\pa_r^2  -\left(2 \pa_r A +\frac{\pa_r G}{G}\right)\pa_r -\left( e^{-2A} q^2 +m^{2}\frac{G}{H}\right)\right]X(q,r)\,,\\
\label{Eq:1formScalarBC}
0&=\Big[C_i \pa_r  \,+\, G\,\Big] X(q,r) \Big|_{r=r_i}\,,\\
0&=\left[\pa_r^2  +\left(2 \pa_r A +\frac{\pa_r H}{H}\right)\pa_r  
-\left(e^{-2A} \frac{q^2}{\xi} +m^{2}\frac{G}{H}\right)\right]Y(q,r)\,,\\
0&=\left[He^{2A}\pa_r \,+\,\left(\frac{D_i}{\xi} q^2 + m^2 C_ie^{2A}\right)\right] Y(q,r)\Big|_{r=r_i}\,.
\end{align}
We see that the equations for $Y(q,r)$ are gauge-dependent, and hence we will only compute the spectrum for the gauge-invariant scalar combination $X(q,r)=\frac{G}{m}e^{4A}\big(mA_{5}+\pa_{r}\pi\big)$. We furthermore notice that the equations for this unphysical scalar $Y$ are identical to those satisfied by the longitudinal polarisation of the vectors $\frac{q^{\mu}q^{\nu}}{q^2}A_{\mu}(q,r)$, which manifestly corroborates our earlier claim that although one combination of the pseudo-scalars $A_{5}$ and $\pi$ generates a physical tower of massive spin-0 states, another combination $Y$ instead supply the additional degrees of freedom necessary to form the longitudinal components of a tower of massive (but gauge-dependent) spin-1 states.\par 
To summarise then, given the bulk equations and boundary conditions satisfied by the transverse polarisation of a generic vector $A_{\mu}$ shown in Eqs.~(\ref{Eq:1formVectorEqn}) and (\ref{Eq:1formVectorBC}), we obtain the following fluctuation equations for the set of fields $\{V_{\mu},\,A_{\mu}^{i}\,,B_{6\mu}\}$:
\begin{align}
\label{Eq:VmuFluct}
0&= P^{\mu\nu}\left[e^{-\c}\partial_{\r}\left(\frac{}{}e^{2A+7\c}\partial_{\r}V_{\nu}\right)+M^{2}e^{8\c} V_{\nu}\right]\, ,\\
\label{Eq:VmuFluctBC}
0&=P^{\mu\nu}\pa_{\r}V_{\nu}\Big|_{\rho_{i}}\, ,
\end{align}
where we have replaced $H= H_{11}=\frac{1}{4}e^{8\c}$ from the field strength metric shown in Eq.~(\ref{Eq:RomansHABmetric}) (and with $G=0$), 
\begin{align}
0&= P^{\mu\nu}\left[e^{-\c}\partial_{\r}
\Big(e^{2A+\c-2\f}\pa_{\r}A^{i}_{\nu}\Big)
+M^{2}e^{2\c-2\f} A^{i}_{\nu}\right]\, ,\\
0&=P^{\mu\nu}\pa_{\r}A_{\nu}^{i}\Big|_{\rho_{i}}\, ,
\end{align}
for $H= H_{22}=\frac{1}{4}e^{2\c-2\f}$ (and again with $G=0$), and then for $H= H^{(1)}=e^{-4\c+4\f}$ with $G= G^{(1)}=e^{-6\c-2\f}$ we have
\begin{align}
0&=
\left[M^{2}+e^{3\c-4\f}
\pa_{\r}(e^{2A-5\c+4\f}\pa_{\r})-\frac{8}{9}e^{2A-2\c-6\f}\right]
P^{\mu\nu}B_{6\nu}\, ,\\
0&=P^{\mu\nu}\pa_{\r}B_{6\nu}\Big|_{\rho_{i}}\, .
\end{align}
In each expression we have reintroduced the convenient change of radial coordinate $\pa_{r} = e^{-\c}\pa_{\r}$, we have set $C_{i}=D_{i}=0$, and we remind the Reader that with our conventions $q^{2}=-M^{2}$ and $m^2 = \frac{8}{9}$. Finally, from Eqs.~(\ref{Eq:1formScalarEqn}) and (\ref{Eq:1formScalarBC}), with the replacements $A_{5}\to B_{65}$ and $\pi\to A_{6}$ and with $H= H^{(1)}=e^{-4\c+4\f},\, G= G^{(1)}=e^{-6\c-2\f}$, we obtain the equations satisfied by the fluctuations of the physical spin-0 combination $X(M,r)$: 
\begin{align}
0&=\pa^{2}_{\r}X+\Big(5\pa_{\r}\c-2\pa_{\r}A+2\pa_{\r}\f\Big)\pa_{\r}X
+\left(M^{2}e^{-2A+2\c}-\frac{8}{9}e^{-6\f}\right)X\, ,\\
0&=X\Big|_{\rho_{i}}\, .
\end{align}

\subsubsection{2-forms in $D=5$ dimensions}
Analogously to our treatment of the 1-forms in the previous section we start by considering a general action describing 2-forms in $D=5$ dimensions, which is supplementary to the action for a sigma-model coupled to gravity. To construct such an action we first define the 3-form field-strength tensor for a generic 2-form $B_{MN}$ as follows:  
\begin{equation}
G_{MNT}\equiv 3\pa_{[M}B_{NT]}\,=\,\pa_MB_{NT}+\pa_NB_{TM}+\pa_TB_{MN}\,,
\end{equation}
which is invariant under the gauge transformation
\begin{equation}
B_{MN}\to B_{MN}-2\pa_{[M}\a_{N]} = B_{MN}-(\pa_{M}\a_{N}-\pa_{N}\a_{M})\,,
\end{equation}
for some arbitrary five-vector $\a_{M}$ that depends on the coordinates of the five-dimensional spacetime. By then introducing a 1-form $A_{M}$ which transforms up to an additive shift:  
\begin{equation}
A_{M}\to A_{M} + m\a_{M}\,,
\end{equation}
with $m$ a constant, we see that one can build the following gauge-invariant 2-form:
\begin{equation}
{\cal H}_{MN} \equiv \pa_{M}A_{N} - \pa_{N}A_{M} + m B_{MN}\equiv F_{MN}+ m B_{MN}\,.
\end{equation} 
Using the field strength tensor $G_{MNT}$ and the 2-form $\mathcal{H}_{MN}$, we may therefore construct the desired gauge-invariant action (see Appendix~\ref{App:2-Forms} for details):  
\begin{align}
{\cal S}_5^{(2)} =\int\is\di^{4}x\,\di r\, \sqrt{-g_5}\bigg\{ 
&-\frac{1}{4}H\,g^{MR} g^{NS}\,{\cal H}_{MN} {\cal H}_{RS}\nn\\
&-\frac{1}{12}K\, g^{MR}\,g^{NS}\,g^{TU} G_{MNT}G_{RSU}\bigg\}\,,
\label{Eq:2formAction}
\end{align}
where $H=H(\Phi^{a})$ and $K=K(\Phi^{a})$ are general sigma-model geometric factors.\ As with our treatment of the 1-form sub-system, we will decompose these fields into their four-dimensional constituents, and will eventually obtain the fluctuation equations and boundary conditions obeyed by the set of fields\newline $\{B_{\mu\nu},\,X_{\mu}(A_{\mu},B_{5\mu}),\,A_{5}\}$. 
We proceed as before by Fourier transforming ${\cal S}_5^{(2)}$ according to Eqs.~(\ref{Eq:Ftransf1}\,-\,\ref{Eq:Ftransf2}), so that written out explicitly we have the following action formulated in momentum-space:
\beqs
{\cal S}_5^{(2)} &=&\int\is\di^{4}q\,\di r \left\{
-\frac{1}{2} H e^{2A}\Big[\pa_{r}A_{\mu}(-q)
+ m B_{5\mu}(-q)\Big]\eta^{\mu\nu}\Big[\pa_{r}A_{\nu}(q)+ m B_{5\nu}(q)\Big]\right.\nonumber\\
&&
\left.
-\frac{1}{2}H q^2 e^{2A} A_5(-q)A_5(q)
\right.\nonumber\\
&&\nonumber\left.
-\frac{1}{2}H e^{2A}\Big[i A_5(-q)\Big(q^{\mu}\pa_{r}A_{\mu}(q)+m q^{\mu}B_{5\mu}(q)\Big)\,+\,(q\leftrightarrow -q)\Big]
\right.\\
&&\nonumber\left.
-\frac{1}{2}H A_{\mu}(-q)\,q^2 P^{\mu\nu} A_{\nu}(q)
\right.\\
&&\nonumber\left.
-\frac{1}{4}H m^2 B_{\mu\nu}(-q) \,\eta^{\mu\r}\eta^{\nu\s}\, B_{\r\s}(q)
\right.\\
&&\nonumber\left.
-\frac{1}{2}H\eta^{\mu\nu}\Big[i m q^{\r}B_{\r\mu}(-q)A_{\nu}(q)\,+\,(q\leftrightarrow -q)\Big]
\right.\\
&&\nonumber\left.
-\frac{1}{4} B_{\mu\nu}(-q)\,\eta^{\mu\r}\eta^{\nu\s}\left[-\pa_{r}\Big(K \pa_{r}B_{\r\s}(q)\Big)\right]
\right.\\
&&\nonumber\left.
+\sum_{i=1,2}(-)^i\d(r-r_i)\left[-\frac{1}{4}K B_{\mu\nu}(-q)\,\eta^{\mu\r}\eta^{\nu\s}\,\pa_{r}B_{\r\s}(q)\right]
\right.\\
&&\nonumber\left.
-\frac{1}{2}K \,B_{5\mu}(-q)\,q^2P^{\mu\nu}\,B_{5\nu}(q)
\right.\\
&&\nonumber\left.
-\frac{1}{2} K\eta^{\mu\nu}\Big[-i q^{\r}\pa_{r}B_{\r\mu}(-q)\, \,B_{5\nu}(q)\,+\,(q\leftrightarrow -q)\Big]
\right.\\
&&\left.
-\frac{1}{4}K e^{-2A} B_{\mu\nu}(-q)\,q^2 \,P^{\mu\r} P^{\nu\s}\,B_{\r\s}(q)
\frac{}{}\right\}\,.
\eeqs
where again $P^{\mu\nu}(q^2)\equiv\eta^{\mu\nu}-\frac{q^{\mu}q^{\nu}}{q^2}$ is the projector onto the transverse momentum polarisation, and we have neglected to show explicitly field dependences on the radial coordinate. Due to the presence of boundaries we supplement the above action with generic boundary-localised kinetic terms for the 3- and 2-forms, which are written as follows:
{\small
	\begin{align}
	{\cal S}_{E}^{(2)}
	&=\int\is\di^{4}x\,\di r \sum_{i=1,2}(-)^{i} \delta(r-r_i)\sqrt{-g_5}\bigg\{
	-\frac{1}{12}\,E_{i}\, K\, \tilde{g}^{\mu\sigma}\tilde{g}^{\nu\tau}\tilde{g}^{\rho\omega}G_{\mu\nu\rho}G_{\sigma\tau\omega}
	\bigg\}\nn\\
	&=\int\is\di^{4}q\,\di r \sum_{i=1,2}(-)^{i} \delta(r-r_i)\bigg\{
	-\frac{1}{4}e^{-2A}\,E_{i}\,K\,
	B_{\mu\nu}(-q)\,q^2\,P^{\mu\r}P^{\nu\s}\,B_{\r\s}(q)
	\bigg\}\,,\\
	{\cal S}_{D}^{(2)}&=\int \di^4x\,\di r \sum_{i=1,2}(-)^{i} \d(r-r_i)\sqrt{-g_5}\bigg\{
	-\frac{1}{4}D_{i}\, H\,  \tilde{g}^{\mu\s}\tilde{g}^{\nu\t} 
	\ch_{\mu\nu}\ch_{\s\t}
	\bigg\}\nn\\
	&=\int\is\di^{4}q\,\di r \sum_{i=1,2}(-)^{i} \delta(r-r_i)\bigg\{
	-\frac{1}{4} D_{i} H 
	\Big[q_{\mu}A_{\nu}(-q)-q_{\nu}A_{\mu}(-q)+ i m B_{\mu\nu}(-q)
	\Big]\nn\\
	&\hspace{34mm}\times\eta^{\mu\r}\eta^{\nu\s}
	\Big[q_{\r}A_{\s}(q)-q_{\s}A_{\r}(q)- i m B_{\r\s}(q)\Big]
	\bigg\}\,.
	\end{align}
}%
where the constants $E_{i}$ and $D_{i}$ take the role of counter-terms in the process of holographic renormalisation (analogously to $D_{i}$ and $C_{i}$ in our treatment of the 1-forms), and we note for convenience that
\begin{equation}
P^{\mu\rho} P^{\nu\sigma}
=\Big(\eta^{\mu\rho} \eta^{\nu\sigma}-2 \frac{q^{\mu}q^{\rho}}{q^2}\eta^{\nu\sigma}\Big)\, .
\end{equation} 
The bulk action ${\cal S}_5^{(2)}$ contains non-physical mixing terms between forms of different order, and hence it is necessary to introduce appropriately chosen gauge-fixing terms to remove them by hand. To cancel mixing terms between the 2-form and the 1-forms $B_{5\nu}$ and $A_{\nu}$ we add to ${\cal S}_5^{(2)}$ the following expressions:  
{\small
	\begin{align}
	{\cal S}_{\Xi,2}^{(2)}
	&=\int\is\di^{4}q\,\di r \bigg\{-\frac{K}{2\Xi}e^{2A}
	\Big[e^{-2A}q^{\r} B_{\r\mu}(-q)+i\frac{\Xi}{K}  \pa_{r} \Big(KB_{5\mu}(-q)\Big)
	+i\frac{\Xi}{K}mH A_{\mu}(-q)\Big]\nn\\
	&\hspace{10mm}\times \eta^{\mu\nu} 
	\Big[e^{-2A}q^{\s} B_{\s\nu}(q)-i\frac{\Xi}{K}  \pa_{r} \Big(KB_{5\nu}(q)\Big)
	-i\frac{\Xi}{K} m H A_{\nu}(q)\Big]
	\,\bigg\}\,,\\
	{\cal S}_{N,2}^{(2)}	
	&=\int\is\di^{4}q\,\di r \sum_{i=1,2}(-)^{i} \delta(r-r_i)\bigg\{\nn\\
	&\hspace{4mm}-\frac{K}{2N_i}e^{2A}\eta^{\mu\nu}
	\Big[e^{-2A}q^{\r}B_{\r\mu}(-q) -iN_{i}B_{5\mu}(-q)
	+im\frac{N_i}{K}D_{i}H A_{\mu}(-q)\Big]\nn\\
	&\hspace{20mm}\times\Big[e^{-2A}q^{\s}B_{\s\nu}(q)+iN_iB_{5\nu}(q)
	-i m \frac{N_i}{K}D_{i}H A_{\nu}(q)\Big]\bigg\}\,.
	\end{align} 
}%
The gauge-fixing parameter $\Xi$ could in principle be dependent on the radial coordinate $r$, however for simplicity we will assume that it is a constant. We also make the convenient choice to fix the boundary-localised parameter $N_i=\frac{\Xi}{E_i}$. 
To remove the terms which mix the 1-forms with the 0-form $A_{5}$, we furthermore add the following expressions: 
{\small
	\begin{align}
	{\cal S}^{(2)}_{\xi,1}
	&=\int\is\di^{4}q\,\di r\bigg\{
	-\frac{K}{2\xi}\Big[q^{\mu}B_{5\mu}(-q)-\frac{i\xi}{K}mHe^{2A} A_{5}(-q)\Big]
	\Big[q^{\nu}B_{5\nu}(q)+\frac{i\xi}{K}mHe^{2A} A_{5}(q)\Big]\nn\\
	&\hspace{10mm}-\frac{H}{2\xi}\Big[q^{\mu}A_{\mu}(-q)+\frac{i\xi}{H}
	\pa_{r}\Big(He^{2A}A_5(-q)\Big)\Big]\nn\\
	&\hspace{40mm}\times\Big[q^{\nu}A_{\nu}(q)-\frac{i\xi}{H}
	\pa_r\Big(He^{2A}A_5(q)\Big)\Big]
	\bigg\},\\
	{\cal S}^{(2)}_{M,1}
	&=\int\is\di^{4}q\,\di r \sum_{i=1,2}(-)^{i}\delta(r-r_i)
	\bigg\{
	-\frac{H}{2M_i}\Big[q^{\mu}A_{\mu}(-q) - i M_{i}e^{2A} A_5(-q)\Big]\nn\\
	&\hspace{54mm}\times
	\Big[q^{\nu}A_{\nu}(q) + i M_{i}e^{2A} A_5(q)\Big]
	\bigg\}\,.
	\end{align} 
}%
where the constants $\xi$ and $M_{i}$ here reprise their roles as bulk and boundary-localised gauge-fixing parameters, respectively.\par 
Following the same procedure as with the system of 1-forms, we derive the equations for the field fluctuations by summing together all contributions in the bulk and at the boundaries: 
\begin{equation*}
{\cal S}_{\text{Tot}}^{(2)}\equiv {\cal S}_{5}^{(2)}+{\cal S}_{E}^{(2)}+{\cal S}_{D}^{(2)}+{\cal S}_{\Xi,2}^{(2)}+{\cal S}_{N,2}^{(2)}+{\cal S}^{(2)}_{\xi,1}+{\cal S}^{(2)}_{M,1}\,, 
\end{equation*}
and then taking an infinitesimal variation of this total action to obtain $\d{\cal S}_{\text{Tot}}^{(2)}$. The fluctuation equations satisfied by the fields which comprise this sub-system are derived in the same manner as previously: by demanding that this variation vanishes on-shell, which amounts to ensuring that its integrand is equal to zero. For the transverse components of the generic 2-form $B_{\r\s}$ we hence find the following results:  
\begin{align}
\label{Eq:2formEqn}
0&=\Big[Kq^2 e^{-2A}-\pa_r\Big(K \pa_r\Big)+H m^2\Big] P^{\mu\r} P^{\nu\s} B_{\r\s}(q,r)\,,\\
\label{Eq:2formBC}
0&=\Big[K\,E_i q^2 e^{-2A}+ K \pa_r +D_i H m^2\Big] P^{\mu\r} P^{\nu\s} B_{\r\s}(q,r)\Big|_{r=r_i} \,,
\end{align}
and we note that our convenient choice $N_i=\frac{\Xi}{E_i}$ ensures that the corresponding equations for the longitudinal polarisation of $B_{\r\s}$ may be obtained by simply replacing $q^{2}\to\frac{q^2}{\Xi}$, although we will only compute the gauge-invariant spectrum of states for the transverse component.\par 
The contributions of the vector fields $B_{5\mu}$ and $A_{\mu}$ couple non-trivially, and we treat their polarisations separately. For the transverse components $P^{\mu\nu}B_{5\mu}$ and $P^{\mu\nu}A_{\mu}$ we define a generalised $U(1)$ gauge-invariant field $X_{\mu}$ and its complementary field $Y_{\mu}$ via the relations     
\begin{align}
B_{5\mu}&\equiv \frac{1}{m}\left(\frac{m X_{\mu}}{e^{2A} H} - \pa_r A_{\mu}\right)\,,\\
P_\mu{}^\nu A_\nu&\equiv Y_{\mu}+\frac{m \partial_r X_{\mu}}{q^2  H}\,,
\end{align}
so that after some algebra we find that $X_{\mu}$ and $Y_{\mu}$ are completely decoupled in the reformulated equations, and satisfy the following:
\begin{align}
0&=\left[\pa_r^2 -
\frac{\pa_r H}{H}\pa_r
-\left(q^{2}e^{-2 A} + m^2 \frac{H}{K} \right) 
\right] X_{\mu}(q,r)\,,\\
0&=\left[\pa_r  +\frac{1}{D_i} \right] X_{\mu}(q,r)\Big|_{r=r_i}\,,\\
0&=\left[\pa_r^2 +\frac{\pa_r K}{K}\pa_r
-\left(\frac{q^2}{\Xi}e^{-2 A} + m^2 \frac{H}{K} \right) 
\right]Y_{\mu}(q,r)\,,\\
0&=\left[\pa_r  +\frac{q^2}{N_i}e^{-2A} + D_i m^2 \frac{H}{K}\right] Y_{\mu}(q,r)\Big|_{r=r_i}\,.
\end{align}
We make the important observation that with our choice $N_i=\frac{\Xi}{E_i}$ the bulk and boundary equations for the four-vector $Y_{\mu}$ are manifestly gauge-dependent, and hence the fluctuations of the transverse component of $Y_{\mu}$ do not generate a physical tower of states; moreover, we notice that these same equations are identical to those for the transverse component of the 2-form shown in Eqs.~(\ref{Eq:2formEqn}\,-\,\ref{Eq:2formBC}) after making the replacement $q^{2}\to\frac{q^2}{\Xi}$. This evinces an underlying Higgs-like mechanism analogous to that in our treatment of the 1-form sector, where the spontaneous $U(1)$ symmetry breaking caused a combination of 0-forms to act as Goldstone bosons, consequently generating a mass for the longitudinal polarisations of a gauge-dependent 1-form. In this sub-system the two degrees of freedom carried by the transverse component of the 1-form $Y_{\mu}$ are ``Higgsed'' into the massless 2-form $B_{\mu\nu}$ (which is dual to a scalar, both having 1 degree of freedom), transmuting it into a massive 2-form (dual to a massive vector, both with three degrees of freedom).\par
Next let us consider the longitudinal components of the vectors $B_{5\mu}^{L}$ and $A_{\mu}^{L}$, which again mix non-trivially. We therefore introduce the longitudinally-polarised fields $X_{\mu}^{L}$ and $Y_{\mu}^{L}$ by defining 
\begin{align}
B_{5\mu}^L&\equiv \frac{1}{m}\left(\frac{m X_{\mu}^L}{e^{2A} H} - \pa_r A_{\mu}^L\right)\,,\\
A_{\mu}^L&\equiv Y_{\mu}^L+\xi\frac{m \pa_r X_{\mu}^L}{q^2  H}\,,
\end{align}
so that we obtain the following decoupled equations:
\begin{align}
0&=\left[\pa_{r}^{2} -\frac{\pa_r H}{H}\pa_r 
-\left(\frac{q^2}{\xi}e^{-2 A} + m^2 \frac{H}{K} \right) 
\right] X_{\mu}^L(q,r)\,,\\
0&=\left[\pa_r +\frac{1}{D_i} \right]
X_{\nu}^L(q,r)\Big|_{r=r_i}\,,\\
0&=\left[\pa_{r}^{2} +\frac{\pa_r K}{K}\pa_r   
-\left(\frac{q^2}{\xi \Xi}e^{-2 A} + m^2 \frac{H}{K} \right) 
\right] Y_{\mu}^L(q,r)\,,\\
0&=\left[\pa_r  +\frac{q^2}{\xi N_i}e^{-2A} + D_i m^2
\frac{H}{K}\right]
Y_{\mu}^L(q,r)\Big|_{r=r_i}\,,
\end{align}
where we have conveniently fixed the boundary-localised constant $M_{i}=\frac{\xi}{D_{i}}$ so that these equations are identical to those for the transverse vectors $X_{\mu}$ and $Y_{\mu}$, up to the replacement $q^{2}\to\frac{q^2}{\xi}$. Hence we see that none of the longitudinal vectors are physical (being manifestly dependent on the gauge-fixing parameters $\xi$ and $\Xi$), and that the only physical spin-1 mass spectrum coming from the decomposition of the generic five-dimensional 2-form action is that of the transverse field $X_{\mu}(q,r)=\frac{H}{m}e^{2A}\big(mB_{5\mu}+\pa_{r}A_{\mu}\big)$.\par 
It only remains to consider the last degree of freedom within this sub-system, which is carried by the scalar $A_{5}$. The bulk equations and boundary conditions for this field are completely decoupled from the other spin sectors, and are written as follows:   
\begin{align}
\label{Eq:RemnantEqn}
0&=\left(\frac{q^2}{\xi} + m^2 e^{2A} \frac{H}{K}\right)A_5(q) - \pa_r \left[\frac{1}{H}\pa_r \left(H e^{2A} A_5(q)\right)\right]\,,\\
\label{Eq:RemnantBC}
0&=\frac{e^{2A}}{D_i} A_5(q)+\frac{1}{H}\pa_r \Big[H e^{2A}A_5(q)\Big]\Big|_{r=r_i}\,,
\end{align}
where again we have replaced $M_{i}=\frac{\xi}{D_{i}}$, and hence note that the fluctuations of $A_{5}$ are gauge-dependent. A final noteworthy observation is that these equations may be reparametrised in terms of the new scalar $\tilde{A}_{5}\equiv He^{2A}A_{5}$, after which they agree exactly with those equations for the longitudinal vector $X_{\mu}^{L}$; this is the same mechanism present for the 1-form sub-system, wherein a gauge-dependent 0-form assumed the role of a Goldstone Boson to provide the additional degrees of freedom required to generate a mass for an otherwise massless longitudinally-polarised spin-1 field.\par 
This decomposition procedure, which we have described in detail for 1- and 2-forms in $D=5$ dimensions, generalises the Higgs mechanism to generic $p$-forms in an arbitrary number of dimensions, and ultimately results in a physical tower of massive states for the subset of fields which exhibit gauge-independence. Conversely no physical spectrum exists for those fields which exhibit a spurious gauge-dependence, and which are merely remnants of the Higgs mechanism in the generic $R_{\xi}$ gauge.\par    
To summarise then, after choosing to set $D_{i}=E_{i}=0$, implementing once more the change of coordinate defined via $\pa_{r} = e^{-\c}\pa_{\r}$, and substituting $q^{2}=-M^{2}$ and $m^2 = \frac{8}{9}$, we have the following final results for the equations satisfied by the fluctuations of the fields $B_{\r\s}(M,r)$ and $X_{\mu}(M,r)$:
\begin{align}
0&=\Big[M^{2}e^{-2A}+e^{-5\c-4\f}\pa_{\r}\Big(e^{3\c+4\f}\pa_{\r} \Big) 
-\frac{8}{9}e^{-2\c-6\f} \Big]P^{\mu\r}P^{\nu\s}B_{\r\s}\,,\\
0&= P^{\mu\tau} P^{\nu\s} \pa_\r B_{\tau\s} \Big|_{\r_i}\,\\
0&=P^{\mu\nu}\Big[\pa_{\r}
\Big(e^{-\c}\pa_{\r}X_{\nu} \Big)
-2e^{-\c}\Big(\pa_{\r}\c-\pa_{\r}\f\Big)\pa_{\r}X_{\nu}\nn\\
&\hspace{50mm}+e^{\c}\Big(e^{-2A}M^{2}-\frac{8}{9}e^{-2\c-6\f} \Big)X_{\nu} \Big]\,,\\
0&= P^{\mu\nu} X_{\nu} \Big|_{\r_i}\,.
\end{align}

\subsubsection{Mass spectra for $p$-forms}
In Sec.~\ref{Sec:RomansMassPlots} we presented the results of our numerical spectra computation for the supergravity field fluctuations which descend from the six-dimensional graviton. Having now derived the equations and boundary conditions which are satisfied by the fluctuations of generic 1- and 2-forms in $D=5$ dimensions, we can proceed to present the corresponding spectra plots for the remaining fields of the six-dimensional supergravity; we remind the Reader that these are the $SU(2)$ adjoint six-vectors $A_{\hat{M}}^{i}$, the $U(1)$ six-vector $A_{\hat{M}}$, and the $U(1)$ 2-form $B_{\hat{M}\hat{N}}$.\par
\begin{figure}[h!]
	\begin{center}
		\makebox[\textwidth]{\includegraphics[width=0.76\paperwidth]
			{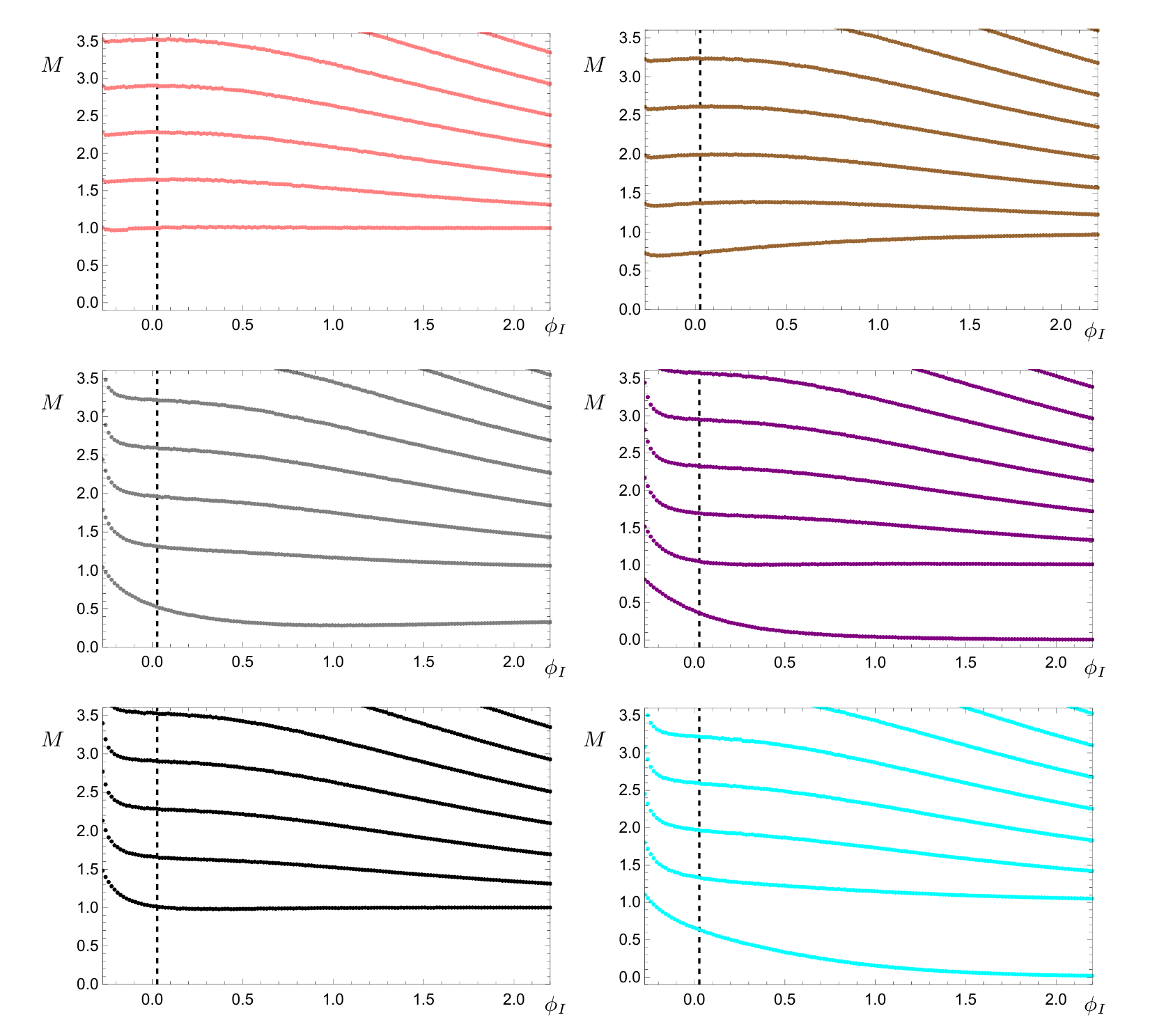}}
	\end{center}
	\vspace{-5mm}
	\caption[Mass spectra of $p$-form resonances for the six-dimensional supergravity]{The spectra of masses $M$ as a function of the one free parameter which characterises the class of confining solutions, $\f_{I}\in[\f_{IR},2.2]$. From top to bottom, left to right: the spectra of fluctuations 
		of the $SU(2)$ adjoint pseudo-scalars $\pi^{i}$ (pink), $SU(2)$ adjoint vectors $A^{i}_{\mu}$ (brown), 
		$U(1)$ scalar combination $X$ (grey), 
		$U(1)$ transverse vector $B_{6\mu}$ (purple), 
		$U(1)$ transverse vector combination $X_{\mu}$ (black), and 
		the $U(1)$ 2-form $B_{\mu\nu}$ (cyan). The vertical dashed lines mark a critical value of the IR parameter $\f_{I}= \f_{I}^{*}>0$, which we shall formally introduce in Sec.~\ref{Sec:RomansPhaseStruct}. All states are normalised in units of the lightest tensor mass. The spectra were computed using regulators $\r_{1}=10^{-4}$ and $\r_{2}=12$ with the exception of the $U(1)$ scalar combination $X$, for which the choice $\r_{1}= 10^{-7}$ was used instead to minimise numerical cutoff effects which were present for the lightest state at large values of $\f_{I}$. }
	\label{Fig:Spectra2}
\end{figure}

The decomposition of these bosonic fields into their constituent components---and the subsequent reformulation necessary to decouple a subset of the four-vectors and scalars---is non-trivial, and for convenience we provide a schematic overview: 
\begin{align*}
&\hspace{20.5mm}\overbrace{\big\uparrow\hspace{53mm}\big\uparrow}^{\textcolor{mygrey}{X} }
\\[-2mm]
\hspace{-4mm}A_{\hat{M}}^{i}\to\big\{\textcolor{mybrown}{A_{\mu}^{i}},\,\cancel{A_{5}^{i}},\,\textcolor{mypink}{\pi^{i}}\big\}
\quad;\quad 
A_{\hat{M}}&\to\big\{A_{\mu},\,\cancel{A_{5}},\,A_{6}\big\}
\quad;\quad B_{\hat{M}\hat{N}}\to\big\{\textcolor{mycyan}{B_{\mu\nu}},\,B_{5\mu},\,\textcolor{mypurple}{B_{6\mu}},\,B_{65}\big\}
\\[-2mm]
&\hspace{7.5mm}\underbrace{\big\downarrow\hspace{50mm}\big\downarrow}_{X_{\mu}}
\end{align*}
where the coloured highlighting is intended to facilitate identification with the corresponding spectra in Fig.~\ref{Fig:Spectra2}. The strikethrough applied to the fifth components of each 1-form represents the fact that---according to our analysis in the previous section, in particular the general results derived in Eqs.~(\ref{Eq:RemnantEqn}\,-\,\ref{Eq:RemnantBC})---these scalars do not produce physical (gauge-invariant) towers of states. Rather, the one degree of freedom carried by each of these scalars is Higgsed into a massless vector (2 d.o.f.) to provide its longitudinal polarisation with a mass; hence we disregard them.\par 
We emphasise that the bulk equation and boundary conditions which are satisfied by $\mathfrak{a}^{\pi}$---associated with the triplet of $SU(2)$ pseudo-scalars $\pi^{i}$---are identical to those obeyed by the other spin-0 fluctuations $\{\mathfrak{a}^{\f},\mathfrak{a}^{\c}\}$ presented in Eqs.~(\ref{Eq:ScalarFluct}\,-\,\ref{Eq:ScalarFluctBC}), although there exists a slight technical subtlety which we shall now address. We remind the Reader that these equations exhibit non-trivial coupling between the three gauge-invariant fluctuations $\mathfrak{a}^{a}$, and that in general the physical spectra must be extracted by identifying the zeros of a $6\times6$ matrix (see Sec.~\ref{Subsec:NumericalImp} for details). However we furthermore remind the Reader that we permit only the metric and the sigma-model scalars $\f$ and $\c$ to acquire non-vanishing background profiles, and hence after substituting for $\pi^{i}(\r)=0$ we find that the equations conveniently separate into two decoupled subsystems for $\{\mathfrak{a}^{\f},\mathfrak{a}^{\c}\}$ and $\mathfrak{a}^{\pi}$; the spectrum resulting from the former subsystem was presented in Fig.~\ref{Fig:Spectra1}. From Eq.~(\ref{Eq:RomansHABmetric}) we note that the $G_{\pi\pi}$ component of the sigma-model metric $G_{ab}(\Phi^{c})$ is dependent on the other two scalars $\f$ and $\c$, and hence the corresponding $\mathfrak{a}^{\pi}$ computation is somewhat complicated by the fact that $\car^{a}_{\,\,\,\,bcd}$ (the Riemann tensor describing the scalar manifold of the dimensionally reduced model) contains non-trivial components.\par 
Let us now turn our attention to the spectra which descend from the 1- and 2-form fields of the six-dimensional supergravity, presented in Fig.~\ref{Fig:Spectra2}. All states are again normalised in terms of the lightest tensor mass, and for the $\f_{I}\leqslant0$ region of the plots we confirm agreement with our previous computation in Ref.~\cite{Elander:2018aub} (see also Tables~\ref{Tbl:RomansUV} and~\ref{Tbl:RomansIR}). Our first observation pertains to the spectra of the two modes which descend from the $SU(2)$ vectors $A_{\hat{M}}^{i}$, the four-vectors $A_{\mu}^{i}$ and the pseudo-scalars $\pi^{i}$, which are shown in the top panels of Fig.~\ref{Fig:Spectra2}. We notice that of the six fields, these two towers are distinctive in the sense that they exhibit the same universal (background-independent) behaviour for negative $\f_{I}$ as was also encountered with the graviton modes in Fig.~\ref{Fig:Spectra1} of Sec.~\ref{Sec:RomansMassPlots}.\par
The spectrum of pseudo-scalar excitations is further distinguished by the interesting observation that it is approximately degenerate with the tower of tensor $\mathfrak{e}^{\mu}_{\ \nu}$ modes shown in Fig.~\ref{Fig:Spectra1}, which is clearly demonstrated---at least for the two critical point solutions---in Tables~\ref{Tbl:RomansUV} and~\ref{Tbl:RomansIR}. Recall that in Sec.~\ref{Sec:RomansMassPlots} we compared to Ref.~\cite{Kuperstein:2004yf} the results of our numerical computation for modes which descend from the $D=6$ graviton, with which we found good agreement. The authors of this paper furthermore considered the fluctuations of an RR 1-form, from which they extracted two additional towers corresponding to pseudo-scalar and spin-1 resonances; we verify that these results are in agreement with two of our spectra, namely the $\pi^{i}$ and $A_{\mu}^{i}$ modes which descend from the triplet of $SU(2)$ 1-forms. Perhaps more significantly, we notice that the aforementioned degeneracy between the pseudo-scalars and spin-2 metric excitations present in our results is not a novel phenomenon, and was also observed in Ref.~\cite{Kuperstein:2004yf}. Our computation refines this finding by demonstrating that it is not restricted to the trivial supersymmetric solution, and indeed applies to an entire class of confining backgrounds that are distinguished by the scale at which their duals exhibit an RG flow.\par
Let us conclude this discussion by considering the remaining plots of Fig.~\ref{Fig:Spectra2}. These are the modes $\{X,B_{6\mu},X_{\mu},B_{\mu\nu}\}$, which all descend from the $U(1)$ fields $A_{\hat{M}}$ and $B_{\hat{M}\hat{N}}$ of the six-dimensional supergravity, and prior to the work in Refs.~\cite{Elander:2018aub,Elander:2020ial} the computation of their spectra had not been attempted in the literature. We start by observing that these four towers of states are dissimilar to those resulting from the decomposition of $\hat{g}_{\hat{M}\hat{N}}$ and $A_{\hat{M}}^{i}$, in that they do not exhibit universal background-independence for $\f_{I}\leqslant0$. The resonances of these modes instead become appreciably heavier as the IR parameter is dialled lower towards $\f_{I}=\f_{IR}$, so that within this subset of solutions the lightest states are extracted by fluctuating the fields evaluated on the trivial $\f_{I}=0$ background (see also Fig.~3 of \cite{Elander:2018aub}).\par 
Our final observation further divides the set of $\{X,B_{6\mu},X_{\mu},B_{\mu\nu}\}$ modes into two subsets, based on their behaviour in the large-$\f_{I}$ limit. The two spectra which descend from the triplet of $SU(2)$ six-vectors exhibit a mass gap, and in the $\f_{I}\to\infty$ limit we anticipate that each will asymptotically approach a gapped continuum; we remind the Reader that this phenomenon was also observed for the three graviton modes presented in Fig.~\ref{Fig:Spectra1} (if we disregard the tachyonic state). Of the four spectra which descend from the six-dimensional $U(1)$ fields, we notice that only the composite modes $X(A_{6},B_{65})$ and $X_{\mu}(A_{\mu},B_{5\mu})$ show the same behaviour, with their lightest masses converging to a finite non-zero value. 
Conversely, for the other two fields $\{B_{6\mu},B_{\mu\nu}\}$ which descend from the 2-form (the latter being dual to a massive vector propagating three d.o.f.), we observe that the lightest mass in each case appears to be parametrically suppressed in the large-$\f_{I}$ limit; it is not obvious whether this phenomenon is indicative of any significant underlying physical effects, and we mention it here merely as an interesting feature of the spectra.       
\clearpage

\chapter{Seven-dimensional maximal supergravity}
\label{Chap:SpectraWitten}
\section{Formalism of the seven-dimensional model}
\label{Sec:Formalism7Dmodel}
\subsubsection{The action in $D=7$ dimensions}
As anticipated in Section~\ref{Sec:Background}, the second theory which we shall be investigating is the seven-dimensional maximal supergravity originally constructed in Refs.~\cite{Pernici:1984xx,Pernici:1984zw}, which we truncate to retain only a single real scalar field $\f$ coupled to gravity. We start by defining hatted uppercase Latin indices $\hat{M}\in\{0,1,2,3,5,6,7\}$ to here represent the coordinates of the seven-dimensional spacetime, so that the truncated action we will adopt may be written as follows~\cite{Elander:2020csd,Elander:2013jqa} (see also
Refs.~\cite{Pernici:1984xx,Pernici:1984zw}):
\begin{equation}
\label{Eq:7DAction}
\cs_{7}=\int\is\di^{7}x\, \sqrt{-\hat{g}_7}\bigg(\frac{{\cal R}_7}{4}-\frac{1}{4}\hat{g}^{\hat{M}\hat{N}}\pa_{\hat{M}}
\f\pa_{\hat{N}}\f -\cv_{7}(\f)\bigg),
\end{equation}
where $\hat{g}_{7}$ is the determinant of the seven-dimensional metric tensor, $\mathcal{R}_{7}\equiv \hat{g}^{\hat{M}\hat{N}}R_{\hat{M}\hat{N}}$ is the corresponding Ricci curvature scalar, $\mathcal{V}_{7}(\f)$ is the scalar potential, and summation over repeated indices is implied.

\subsubsection{Critical points of the $D=7$ potential}
\begin{figure}[h!]
	\begin{center}
		\makebox[\textwidth]{\includegraphics[width=0.42\paperwidth]
			{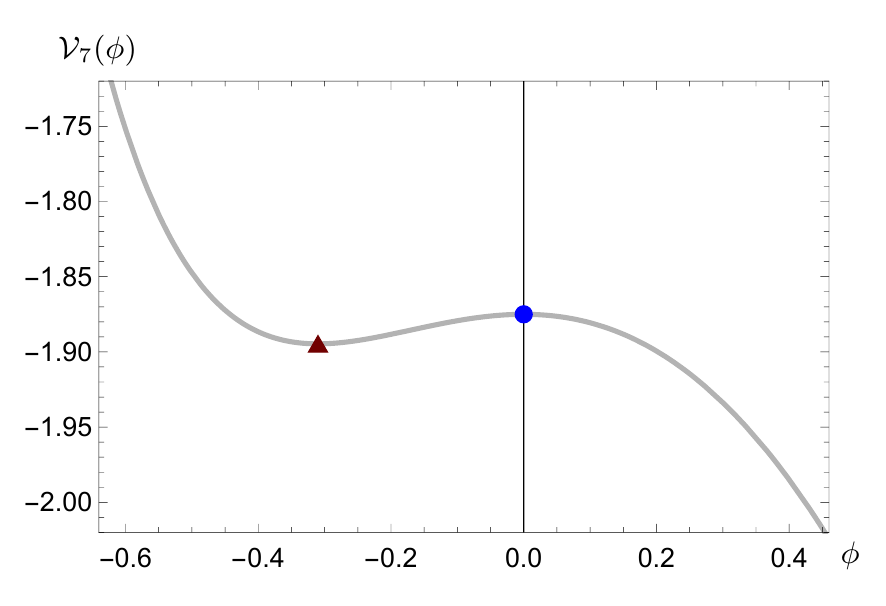}}
	\end{center}
\vspace{-5mm}
	\caption[Scalar potential of the seven-dimensional supergravity]{The potential $\cv_{7}(\f)$ as a function of the scalar $\f$ in the sigma-model coupled to gravity in $D=7$ dimensions. The blue disk denotes the supersymmetric critical point $\f=\f_{UV}=0$, while the dark-red triangle represents the non-supersymmetric critical point  $\f=\f_{IR}=-\frac{\ln(2)}{\sqrt{5}}$.}
	\label{Fig:ScalarPotential7D}
\end{figure}

The potential of the seven-dimensional sigma-model, which is plotted in Fig.~\ref{Fig:ScalarPotential7D}, may be written as
\begin{equation}
\label{Eq:V7}
\cv_{7}(\f)=\half\bigg(\frac{1}{4}e^{-\frac{8}{\sqrt{5}}\f}-2e^{-\frac{3}{\sqrt{5}}\f} -2e^{\frac{2}{\sqrt{5}}\f}\bigg)\,,
\end{equation}
and admits two stationary point solutions for the scalar field $\f$. Each of these critical points realises a distinct $\text{AdS}_{7}$ geometry in the gravitational theory, corresponding to two separate six-dimensional dual CFTs living on the boundary. These critical values are as follows:  
\begin{align}
\f_{UV}&=0 \hspace{18.5mm} \bigg(\mathcal{V}_{7}(\f_{UV})=-\frac{15}{8}\bigg)\,,\\
\f_{IR}&=-\frac{1}{\sqrt{5}}\ln(2) \hspace{8mm}\bigg(\mathcal{V}_{7}(\f_{IR})=-\frac{5}{2^{7/5}}\bigg)\,,
\end{align}
where, as with Romans six-dimensional supergravity, we have adopted the subscript labels to reflect the fact one may construct numerical background solutions which interpolate between the two extrema, realising a holographic RG flow from the $\f=\f_{UV}$ supersymmetric field theory at high energies to the (perturbatively unstable) $\f=\f_{IR}$ non-supersymmetric theory at low energies.
\\
By adopting the same conventions as in Ref.~\cite{Elander:2013jqa}, though noting that we have here defined $\cs_{7}$ (and hence also $\cv_{7}$) with an additional factor of one half, the (squared) curvature radii for the two $\text{AdS}_{7}$ geometries are given by 
\begin{align}
R_{UV}^{2}&\equiv-15\big[2\cv_{7}(\f_{UV})\big]^{-1}=4\, ,\\
R_{IR}^{2}&\equiv-15\big[2\cv_{7}(\f_{IR})\big]^{-1}=3\cdot 2^{2/5}\, . \label{Eq:WittenRIR}
\end{align} 
As before, the mass of the scalar field $\f$ propagating on the two inequivalent $\text{AdS}_{7}$ geometries may be extracted by considering small perturbations about each extrema of the potential, with the coefficient of the term quadratic in $\f$ providing $m^{2}/4$. We find that 
\begin{align}
\mathcal{V}_{7}(\f_{UV})&=-\frac{15}{8}-\frac{\f^{2}}{2}
+\mathcal{O}\big(\f^{3}\big),\\
\mathcal{V}_{7}(\f_{IR})&=-\frac{5}{2^{7/5}}
+\frac{1}{2^{2/5}}\big(\f-\f_{IR}\big)^{2}
+\mathcal{O}\big((\f-\f_{IR})^{3}\big)\,,
\end{align} 
and we therefore obtain the well-known results (see for example Ref.~\cite{Campos:2000yu}):
\begin{align}
m_{UV}^{2}&=-2\quad \rightarrow \quad m_{UV}^{2}R_{UV}^{2}=-8\,,\\
m_{IR}^{2}&=2^{8/5}\quad \rightarrow \quad m_{IR}^{2}R_{IR}^{2}=12\,.
\end{align}
Then, using Eq.~(\ref{Eq:Delta}) we can compute the scaling dimension $\Delta$ of the boundary operator dual to $\f$ for each of the two critical points, keeping in each case the largest quadratic root: 
\begin{equation}
\label{Eq:WittenDeltas}
\Delta_{UV}=4 \quad , \quad \Delta_{IR}=3+\sqrt{21}\,.
\end{equation}

\section{Toroidal reduction to \emph{D=}5 dimensions}
\label{Sec:ToroidalReduction}
\subsubsection{The metric}
We reduce the system to five dimensions by compactifying two of the external directions, parametrised by $\eta$ and $\zeta$, on a torus $T^{2}\equiv S^{1}\times S^{1}$; the volumes of the two circles in the torus are controlled by the additional sigma-model scalars $\c$ and $\w$.
Using the $D=5$ line element as defined in Eq.~(\ref{Eq:5Dmetric}), we adopt the following ansatz for the seven-dimensional metric:     
\begin{align}\label{Eq:7Dmetric}
\di s_{7}^{2}&=e^{-2\c}\di s_{5}^{2} + e^{3\c}\Big(e^{-2\w}\di\eta^{2}+e^{2\w}\di\zeta^{2}\Big)\notag\\
&=e^{-2\c}\Big(e^{2A(r)}\di x_{1,3}^{2} +\di r^{2}\Big)+ e^{3\c}\Big(e^{-2\w}\di\eta^{2}+e^{2\w}\di\zeta^{2}\Big)\notag\\
&=e^{-2\c}\Big(e^{2A(\r)}\di x_{1,3}^{2} +e^{2\c}\di\r^{2}\Big)+ e^{3\c}\Big(e^{-2\w}\di\eta^{2}+e^{2\w}\di\zeta^{2}\Big)\, ,
\end{align}
where in the third line we have introduced the convenient change of coordinate defined via $\di r\equiv e^{\c}\di\r \Leftrightarrow \pa_{r}=e^{-\c}\pa_{\r}$. We assume that the background profiles for the scalar fields $\{\f,\c,\w\}$ and the warp factor $A$ are dependent \emph{only} on the holographic coordinate $\r$, and are hence independent of the periodic $T^{2}$ coordinates $0\leqslant\eta,\zeta<2\pi$.\par
We will mainly be considering background geometries in which the compactified dimension parametrised by $\zeta$ always maintains a non-zero volume, while the behaviour of the other circle (parametrised by $\eta$) depends on the specific solution in question. For backgrounds which provide the holographic realisation of confinement, the $\eta$ circle contracts to a point and vanishes at the end of space so that the bulk geometry smoothly closes off; conversely, there exist classes of \emph{domain-wall} backgrounds for which neither of the circles within the torus shrink to zero size. These DW solutions (locally) preserve six-dimensional Poincar\'{e} invariance within the space of the Minkowski and toroidal dimensions, while the tapered geometry inherent to the confining solutions reduces this symmetry to five-dimensional Poincar\'{e} invariance within the subspace parametrised by the coordinates $\{x^{\mu},\zeta\}$. We observe that by making the identification $A=\frac{5}{2}\c+\w$, the metric takes a form which manifestly shows this latter symmetry:  
\begin{equation}\label{Eq:7DmetricConstrained}
\di s_{7}^{2}=\di\r^{2}+ e^{2\ca}\Big(\di x_{1,3}^{2}+ e^{-4\w}\di\eta^{2}+\di\zeta^{2}\Big)\, ,
\end{equation}
where the new warp factor is given by $\ca=\frac{3}{2}\c+\w=A-\c$. Poincar\'{e} invariance is extended to include the $\eta$ dimension if we further impose $\w=0$. Since the less restrictive five-dimensional Poincar\'{e} symmetry will apply to all background solutions that we consider, we shall henceforth always assume that the constraint $ A-\frac{5}{2}\c=\w$ is satisfied, with the subset of solutions which admit the domain-wall geometry satisfying $\w=0$. We conclude this section by noting that the five- and seven-dimensional metric determinants are related via $\sqrt{-\hat{g}_7} = e^{-2\c}\sqrt{-g_5}$, and that the seven-dimensional Ricci scalar is given by
\begin{equation}
\label{Eq:7DRicci}
\car_{7}=-2\Big(4A''-\c''+10(A')^{2}+\tfrac{19}{4}(\c')^{2}
+(\w')^{2}-8A'\c'\Big)\,,
\end{equation}
where primes denote derivatives taken with respect to the radial coordinate $\r$.

\subsubsection{The action}
As we saw with Romans supergravity in Sec.~\ref{Sec:CircleReduction}, the defining action of the theory may be dimensionally reduced to a $D=5$ sigma-model coupled to gravity, with one additional scalar introduced for each $S^{1}$-compactified direction to parametrise their volume. We apply the same process here for the seven-dimensional supergravity so that, after some algebra, it can be shown that the action presented in Eq.~(\ref{Eq:7DAction}) may be conveniently rewritten as  
\begin{equation}
\label{Eq:7DActionTotalDeriv}
\cs_{7}=\int\is\di\eta\,\di\zeta\,
\bigg\{\tilde{\cs}_{5} + \half\int\is\di^{5}x\,
\pa_{M}\Big(\sqrt{-g_{5}}\,g^{MN}\pa_{N}\c\Big)\bigg\} 
\,,
\end{equation}
where the five-dimensional action $\tilde{\mathcal{S}}_{5}$ takes a similar form to Eq.~(\ref{Eq:5DActionNoU1}):
\begin{equation}
\label{Eq:7Dto5DAction}
\tilde{\cs}_{5}=\int\is\di^{5}x\,\sqrt{-g_{5}}
\bigg(\frac{\car_{5}}{4}-\half G_{ab} g^{MN}\pa_{M}\Phi^{a}\pa_{N}\Phi^{b} -\cv(\f,\c) \bigg)\,.
\end{equation}
The index $a\in\{1,2,3\}$ labels the scalar fields of the sigma-model so that $\Phi^a = \{\f,\c,\w\}$, while the scalar potential $\cv$ is related to the seven-dimensional potential via $\cv(\f,\c) = e^{-2\c}\cv_{7}(\f)$. The sigma-model metric $G_{ab}$ is given by
\begin{equation}
G_{ab}={\rm diag}\,\left(\half,\frac{15}{4},1\right)\,.
\end{equation}

\section{Equations of motion and confining solutions}
\label{Sec:WittenEOMs}
\subsubsection{Equations of motion}
The classical equations of motion which follow from the toroidally reduced five-dimensional action $\tilde{\cs}_{5}$ are derived using the general results presented in Eqs.~(\ref{Eq:ScalarEOM}\,-\,\ref{Eq:Einstein2}) of Section~\ref{Subsec:HolographicFormalism}. From the equations for the three scalars, and from the two Einstein field equations, we obtain the following:
\begin{align}
\pa^{2}_{r}\f + 4\pa_{r}\f\pa_{r}A &= 2\frac{\pa\cv}{\pa\f}\,,\\
\pa^{2}_{r}\c + 4\pa_{r}\c\pa_{r}A &= \frac{4}{15}\frac{\pa\cv}{\pa\c}\,,\\
\pa^{2}_{r}\w + 4\pa_{r}\w\pa_{r}A &= 0\,,\\
12\pa^{2}_{r}A+24(\pa_{r}A)^{2}+2(\pa_{r}\f)^{2}+15(\pa_{r}\c)^{2}
+4(\pa_{r}\w)^{2} &= -8\cv\,,\\
24(\pa_{r}A)^{2}-2(\pa_{r}\f)^{2}-15(\pa_{r}\c)^{2}
-4(\pa_{r}\w)^{2} &= -8\cv\,.
\end{align}
By implementing the radial coordinate change defined just after Eq.~(\ref{Eq:7Dmetric}), the system of equations may be rewritten in terms of the seven-dimensional potential:
\begin{align}
\pa^{2}_{\r}\f + \big(4\pa_{\r}A-\pa_{\r}\c\big)\pa_{\r}\f &= 2\frac{\pa\cv_{7}}{\pa\f}\,,  \label{Eq:1}\\
\pa^{2}_{\r}\c + \big(4\pa_{\r}A-\pa_{\r}\c\big)\pa_{\r}\c &= -\frac{8}{15}\cv_{7}\,, \label{Eq:2}\\
\pa^{2}_{\r}\w + \big(4\pa_{\r}A-\pa_{\r}\c\big)\pa_{\r}\w &=0
\,,\label{Eq:3}\\
3\pa^{2}_{\r}A + 6(\pa_{\r}A)^{2} - 3\pa_{\r}A\pa_{\r}\c 
+\Sigma &= -2\cv_{7}\,,  \label{Eq:4}\\
6(\pa_{\r}A)^{2}-\Sigma &= -2\cv_{7}\,,  \label{Eq:5}
\end{align}
where we have conveniently collected some terms together by introducing
\begin{equation}
\Sigma\equiv G_{ab}\pa_{\r}\Phi^{a}\pa_{\r}\Phi^{b}= \tfrac{1}{2}(\pa_{\r}\f)^{2}+\tfrac{15}{4}(\pa_{\r}\c)^{2}
+(\pa_{\r}\w)^{2}\,.
\end{equation}
We notice that summing the combination $-3\times$Eq.~(\ref{Eq:2}), $\tfrac{2}{5}\times$Eq.~(\ref{Eq:4}), and $\tfrac{2}{5}\times$Eq.~(\ref{Eq:5}) gives the following vanishing quantity:
\begin{equation}
\label{Eq:WittenVanish}
2\pa^{2}_{\r}A - 5\pa^{2}_{\r}\c + 8(\pa_{\r}A)^{2} + 5(\pa_{\r}\c)^{2}
-22\pa_{\r}A\pa_{\r}\c = 0\,,
\end{equation}
which is satisfied by the constraint $A-\frac{5}{2}\c=\w$. By substituting for $A\to\frac{5}{2}\c+\w$ in Eq.~(\ref{Eq:3}) and Eq.~(\ref{Eq:WittenVanish}) we obtain identical expressions, which demonstrates that our system of equations is overdetermined; we may therefore omit one of the equations to remove this redundancy, leaving us with the following:
\begin{align}
\pa^{2}_{\r}\f + \big(4\pa_{\r}A-\pa_{\r}\c\big)\pa_{\r}\f &= 2\frac{\pa\cv_{7}}{\pa\f}\,,  \label{Eq:WittenEOM1}\\
\pa^{2}_{\r}\c + \big(4\pa_{\r}A-\pa_{\r}\c\big)\pa_{\r}\c &= -\frac{8}{15}\cv_{7}\,, \label{Eq:WittenEOM2}\\
\pa^{2}_{\r}\w + \big(4\pa_{\r}A-\pa_{\r}\c\big)\pa_{\r}\w &=0
\,,  \label{Eq:WittenEOM3}\\
6(\pa_{\r}A)^{2}-\Sigma &= -2\cv_{7}\,.  \label{Eq:WittenEOM4}
\end{align}
We furthermore notice that Eq.~(\ref{Eq:WittenVanish}) may be reformulated as a vanishing total derivative with respect to $\r$, so that
\begin{equation}
\label{Eq:CC}
e^{4A-\c}\Big(5\pa_{\r}\c-2\pa_{\r}A\Big)=-2e^{4A-\c}\pa_{\r}\w=C
\end{equation}
represents a conserved quantity at all energy scales, for some constant $C$ which depends on the specific background solution. We will make use of this observation in Sec.~\ref{Sec:FderivationWitten} when we derive the free energy density for this supergravity theory. 
It is convenient to reformulate Eqs.~(\ref{Eq:WittenEOM1}\,-\,\ref{Eq:WittenEOM4}) in terms of the following linear combinations of the scalar $\c$ and warp factor $A$:
\begin{align}
\label{Eq:AlphaUpsilon}
\a\equiv 4A-\c \quad &, \quad \U\equiv A-\tfrac{5}{2}\c\, ,\\
\bigg(
\Rightarrow
\quad 
\c=\tfrac{1}{9}\big(\a-4\U\big)\quad &, \quad A=\tfrac{1}{18}\big(5\a-2\U\big)    \bigg)\, ,\nn
\end{align}
so that the equations of motion take the more compact form:
\begin{align}
\pa^{2}_{\r}\f+\pa_{\r}\a\pa_{\r}\f&=2\frac{\pa\cv_{7}}{\pa\f}\, ,\\
\pa^{2}_{\r}\a+(\pa_{\r}\a)^{2}&=-\frac{24}{5}\cv_{7}\, ,\\
20(\pa_{\r}\U)^{2}-5(\pa_{\r}\a)^{2}+6(\pa_{\r}\f)^{2}
&=24\cv_{7}\,,\\
\pa^{2}_{\r}\U+\pa_{\r}\a\pa_{\r}\U&=0\,,
\end{align}
and the conserved quantity identified in Eq.~(\ref{Eq:CC}) is given by $C=-2e^{\a}\,\pa_{\r}\U$.  
Rewritten as such, we observe that the system of equations is unchanged under the sign flip $\U\to-\U$, while holding $\a\to\a$ invariant; as we shall see in Sec.~\ref{Sec:WittenClasses}, this symmetry actually implies the existence of an additional branch of solutions which are related to the class of regular backgrounds, though which are geometrically distinct.\par
As earlier mentioned, those solutions which realise a domain-wall geometry and hence which preserve six-dimensional Poincar\'{e} invariance within the $\{x^{\mu},\eta,\zeta\}$ subspace satisfy $\U=0$; the equations of motion therefore simplify for these classes: 
\begin{align}
5\pa^{2}_{\r}\f+18\pa_{\r}A\pa_{\r}\f&=
10\frac{\pa\cv_{7}}{\pa\f}\, ,\label{Eq:WitEOMc1}\\
3\pa^{2}_{\r}A +(\pa_{\r}\f)^{2}&=0\, ,\label{Eq:WitEOMc2}\\
54(\pa_{\r}A)^{2}-5(\pa_{\r}\f)^{2}&=-20\cv_{7}\,, \label{Eq:WitEOMc3}
\end{align} 
or in terms of the linear field combination $\a$ we equivalently have
\begin{align}
\pa^{2}_{\r}\f+\pa_{\r}\a\pa_{\r}\f
&=2\frac{\pa\cv_{7}}{\pa\f}\, , \label{Eq:WitEOMc4}\\
\pa^{2}_{\r}\a+(\pa_{\r}\a)^{2}&=-\frac{24}{5}\cv_{7}\, , \label{Eq:WitEOMc5}\\
5(\pa_{\r}\a)^{2}-6(\pa_{\r}\f)^{2}&=-24\cv_{7}\,.  \label{Eq:WitEOMc6}
\end{align} 
We conclude this section by observing that Eq.~(\ref{Eq:WitEOMc6}) may be solved algebraically for $\a'(\r)$, and that this expression may subsequently be substituted into Eq.~(\ref{Eq:WitEOMc4}) to derive a 
second-order non-linear differential equation written in terms of the scalar $\f(\r)$ only. We obtain the following:
\begin{align}
0=5\f'' &+\sqrt{15}\f'
\Big[2\big(\f'\big)^{2}+\g^{-\frac{8}{5}}\Big(8\g +8\g^{2}-1\Big)\Big]^{\half}\nn\\
&+\sqrt{20}\g^{-\frac{8}{5}}\Big(1-3\g +2\g^{2}\Big) \,,
\label{Eq:WittenParametricPhi}
\end{align}
where primes denote differentiation with respect to $\r$, and we have here defined $\g\equiv e^{\sqrt{5}\f(\r)}$.\ As with the analogous expression presented in Eq.~(\ref{Eq:ParametricPhi}) for Romans supergravity, we shall later use Eq.~(\ref{Eq:WittenParametricPhi}) to produce a parametric plot of the underlying vector field governing $\f$ for DW background solutions which preserve six-dimensional Poincar\'{e} invariance.

\subsubsection{Confining solutions}
The mass spectra of bosonic composite states in a four-dimensional strongly-coupled field theory can be computed holographically by considering field fluctuations on backgrounds in the dual higher-dimensional theory which geometrically realise confinement. In this section we introduce a class of such solutions, for which one of the internal circles of the torus $T^{2}=S^{1}\times S^{1}$ shrinks to a point at some finite value of the radial coordinate $\r=\r_{o}$, and consequently for which the bulk spacetime smoothly closes off and ends; this geometric property naturally introduces a low-energy limit in the dual field theory living at the boundary, which we remind the Reader may be interpreted as the scale of confinement.\par 
In Sec.~\ref{Sec:Formalism7Dmodel} we presented the two critical points of the seven-dimensional scalar potential $\cv_{7}(\f)$, each of which realises a distinct $\text{AdS}_{7}$ geometry. When the scalar field $\f$ is equal to either of these constant values $\f=\f_{p}$, with $\f_{p}\in\{\f_{UV},\f_{IR}\}$, there are known to exist analytical solutions for the warp factor and the two additional scalars which are introduced to control the volumes of the compactified dimensions in the toroidally reduced model; these special confining solutions may be written as follows~\cite{Elander:2013jqa}:
\begin{align}
\f&=\f_{p}\,,\\
\c(\r)&=\frac{1}{3}\ln\bigg[
\frac{2}{x}\sinh(y)\cosh^{-\frac{1}{3}}(y) \bigg]\,,\label{Eq:WittenConfChi}\\
\w(\r)&=-\half\ln\bigg[
\frac{2}{x}\tanh(y) \bigg]\,,\label{Eq:WittenConfOmega}
\end{align}
where $x\equiv\left(-\frac{12}{5}v\right)^{\half}$, $v\equiv\mathcal{V}_{7}(\f_{p})$, and $y\equiv\frac{x}{2}(\r-\r_{o})$; we shall not require these exact solutions when computing the mass spectra for this theory, though we nevertheless include them here for completeness.\par
As with the analogous class of confining backgrounds for the six-dimensional supergravity it is possible to generalise the above solutions to allow for $\f$ profiles which are $\r$-dependent, in which $\f$ interpolates between the two stationary point values $\f_{UV}$ and $\f_{IR}$. The fields may be series expanded in proximity of the end of space where the $\eta$-circle collapses to a point, so that the generalised solutions are obtained by solving the EOMs subject to boundary conditions guided by the following IR expansions~\cite{Elander:2020fmv,Elander:2013jqa}:
\begin{align}
\label{Eq:WittenPhiIRExpansion}
\f(\r)&=\f_{I}-\tfrac{1}{2\sqrt{5}}e^{-\frac{8\f_{I}}{\sqrt{5}}} 
\left(1-3 e^{\sqrt{5}\f_{I}}+2 e^{2\sqrt{5}\f_{I}}\right)(\r-\r_{o})^{2}\nn\\
-&\tfrac{1}{80\sqrt{5}}e^{-\frac{16\f_{I}}{\sqrt{5}}} 
\left(9 -44 e^{\sqrt{5}\f_{I}} +57e^{2 \sqrt{5}\f_{I}} 
+2e^{3\sqrt{5}\f_{I}} -24e^{4\sqrt{5}\f_{I}}\right)(\r-\r_{o})^{4}\nn\\
&\hspace{20mm}+\co\left((\r-\r_o)^6\right)\,, \\
\label{Eq:WittenChiIRExpansion}
\c(\r)&=\c_{I}+\tfrac{1}{3}\ln(\r-\r_{o})\nn\\
-&\tfrac{1}{6000}e^{-\frac{16\f_{I}}{\sqrt{5}}} 
\left(7 -32e^{\sqrt{5}\f_{I}} +56e^{2\sqrt{5}\f_{I}} 
-224e^{3\sqrt{5}\f_{I}} -32e^{4\sqrt{5}\f_{I}}\right)(\r-\r_{o})^{4}\nn\\
&\hspace{20mm}+\co\left((\r-\r_o)^6\right)\,,\\
\label{Eq:WittenOmegaIRExpansion}
\w(\r)&=\w_{I}-\tfrac{1}{2}\ln(\r-\r_{o})
-\tfrac{1}{40}e^{-\frac{8\f_{I}}{\sqrt{5}}} 
\left(1 -8e^{\sqrt{5}\f_{I}} -8e^{2\sqrt{5}\f_{I}}\right)(\r-\r_{o})^{2}\nn\\
-&\tfrac{1}{8000}e^{-\frac{16\f_{I}}{\sqrt{5}}} 
\left(
31-8\Big(32e^{\sqrt{5}\f_{I}} -81e^{2\sqrt{5}\f_{I}}
-76e^{3\sqrt{5}\f_{I}}-68e^{4\sqrt{5}\f_{I}}\Big)
\right)(\r-\r_{o})^{4}\nn\\
&\hspace{20mm}+\co\left((\r-\r_o)^6\right)\,,
\end{align}
where $\r_{o}$ is chosen to fix the end of space along the radial direction, and $\f_{I}$ is the free parameter which determines the energy scale at which the transition from one fixed point to the other occurs. For the choice $\f_{I}=\f_{p}$ we recover the exact analytical solutions presented in Eqs.~(\ref{Eq:WittenConfChi}) and (\ref{Eq:WittenConfOmega}). For choices $\f_{I}\in(\f_{IR},\,\f_{UV})$ we generate a family of backgrounds which interpolate from $\f=\f_{UV}$ at large $\r$ towards $\f=\f_{IR}$ as one approaches the end of space, however in these cases the profile $\f(\r)$ does not have sufficient time to reach the IR fixed point before the geometry closes off and the solution terminates. Choosing $\f_{I}>\f_{UV}$ is perfectly acceptable---and we will allow for such values when computing the mass spectra in the next section---though we shall impose that $\f_{IR} \leqslant \f_{I}$ to ensure that $\f(\r)$ is bounded from below by the IR fixed point solution.\par
The integration constant $\c_{I}$ appearing in the IR expansion for the scalar $\c$ is a constrained quantity; the $S^{1}$-compactified dimension parametrised by $\eta$ is periodically identified, and as a result $\c_{I}$ is fixed by the requirement that we avoid a conical singularity at the end of space. By restricting our attention to the two-dimensional subspace spanned by $\r$ and $\eta$ in the deep IR, and furthermore by substituting in for $\c$ and $\w$ using their respective small-$\r$ expansions, we find that the bulk geometry is described by the following line element:
\begin{align}
\label{Eq:2DmetricWitten}
\di\tilde{s}_{2}^{2}&= \di\r^{2}+e^{3\c-2\w}\di\eta^{2}\\
&=\di\r^{2}+e^{3\c_{I}-2\w_{I}}(\r-\r_{o})^2\di\eta^{2}
+\ldots\,,
\end{align}
so that the necessary condition to avoid an angular deficit is given by $\c_{I}=\frac{2}{3}\w_{I}$. The value assigned to the constant $\w_{I}$ may otherwise be freely chosen.

\section{Physical mass spectra}
\label{Sec:WittenMassPlots}
In this section we present and discuss the numerical results of our spectra computation for the seven-dimensional gauged supergravity, focusing solely on the physical modes corresponding to the spin-2 fluctuations $\mathfrak{e}^{\mu}_{\ \nu}$ of the graviton, and the scalar variables $\mathfrak{a}^{a}$ constructed from the fluctuations of the sigma-model fields $\varphi^{a}$ and the spin-0 component $h$ of the ADM-decomposed metric; the results of our probe approximation analysis will be discussed separately in Sec.~\ref{Sec:WittenProbePlots}.\par 
As a brief digression, let us first comment on our choice not to retain the higher $p$-form fields of the supergravity multiplet in our investigation. We remind the Reader that in the context of superstring theory and M-theory there exists the well-known \emph{self-duality in odd dimensions} phenomenon~\cite{Townsend:1983xs}, wherein a $(2k+1)$-form field strength on a $(4k+2)$-dimensional Riemannian manifold is self-dual when acted upon by the Hodge star operator.
As explained in (for example) Refs.~\cite{Townsend:1995af,Witten:1996hc,Belov:2006jd}, this geometric property gives rise to the notion of so-called \emph{chiral} $2k$-forms, and the self-duality constraint is associated with difficulties in constructing an appropriate covariant theory action (and hence also partition function); typical terms of the form $C\wedge\star C$ are vanishing under the constraint $\star C=C$.\par
Of particular relevance in the context of $\text{AdS}_{7}$/$\text{CFT}_{6}$ holography is the maximal eleven-dimensional supergravity, in which the six-dimensional world-volume of the M5-brane contains a chiral 2-form whose field strength tensor is self-dual (see for example Refs.~\cite{Witten:1996hc,Bandos:1997ui,Pasti:1997gx} and references therein, and also earlier work in Refs.~\cite{Perry:1996mk,Schwarz:1997mc,Aganagic:1997zq}). As discussed in Sec.~\ref{Sec:Background} the compactification of this theory on an $S^{4}$ yields the desired gauged supergravity in seven dimensions, whose bosonic sector contains a self-dual  3-form~\cite{Nastase:1999cb,Lu:1999bc}. The requirement that we additionally impose the self-duality constraint on this field would somewhat complicate the spectra computation, and for this reason we choose to instead consider the consistently truncated theory comprising one sigma-model scalar coupled to gravity; this simplification is further justified by the fact that we are interested primarily in studying dilaton phenomenology, which only requires that we retain the dynamical scalar $\f$.\par
\begin{figure}[!h]
	\centering
	\subfloat{\makebox[0.2\paperwidth]
		{\includegraphics[width=0.38\paperwidth]
			{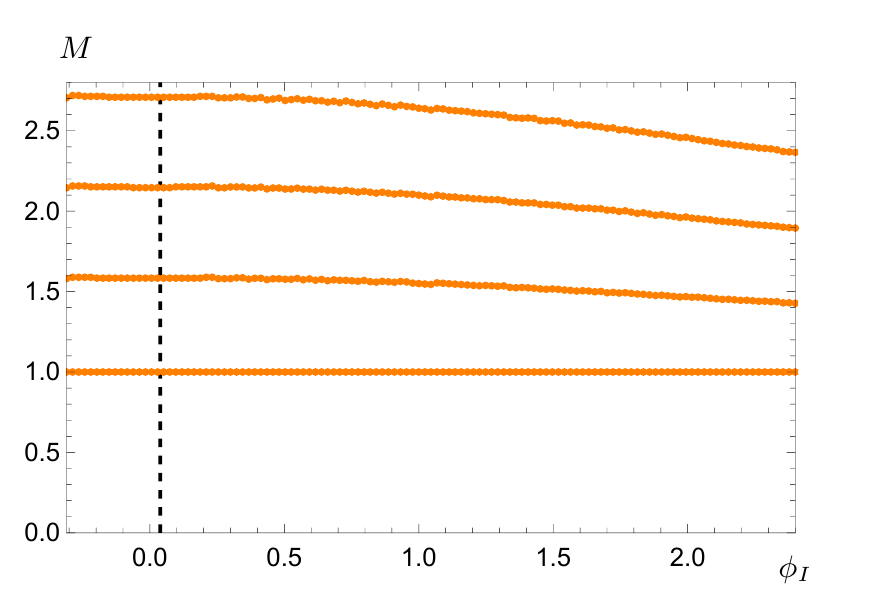}}  }
	\hfill
	\subfloat{\makebox[0.2\paperwidth]
		{\includegraphics[width=0.38\paperwidth]
			{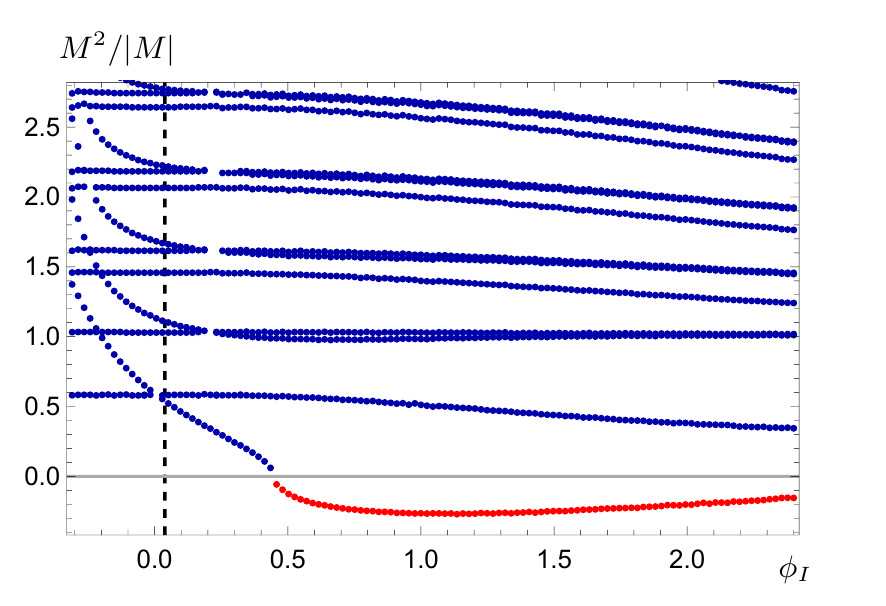}}  }
	\caption[Mass spectra of graviton resonances for the seven-dimensional supergravity]{The spectra of masses $M$ as a function of the one free parameter which characterises the class of confining solutions, $\f_{I}\in[\f_{IR},2.4]$. The left plot shows 
		the spectra of tensor fluctuations 
		$\mathfrak{e}^{\mu}_{\ \nu}$ (orange), while the right plot shows the mass eigenstates of the scalar fluctuations $\mathfrak{a}^{a}$ associated with 
		$\{\f,\c,\w\}$ (blue). The red disks in the scalar 
		spectrum represent masses for which $M^{2}<0$, and hence denote a tachyonic state. The vertical dashed lines represent the critical value of the IR parameter $\f_{I}= \f_{I}^{*}>0$ at a first-order phase transition, which we shall encounter in Sec.~\ref{Sec:WittenPhaseStruct}. All states are normalised in units of the lightest tensor mass, and were computed using regulators $\r_{1}=10^{-4}$ and $\r_{2}=12$. We acknowledge the existence of some small gaps in the scalar spectrum; these are regions where the resonances were so close to degenerate in mass that the numerical routine was unable to resolve and identify them separately, and are hence not of any physical significance.}
	\label{Fig:WittenSpectrum}
\end{figure}

 
With this clarification out of the way, let us now turn our attention to the results of our numerical computation presented in Fig.~\ref{Fig:WittenSpectrum}. As with the analogous exercise for the six-dimensional supergravity in Chapter~\ref{Chap:SpectraRomans}, we compute the spectra as a function of the one parameter $\f_{I}$ which characterises the confining backgrounds in the deep IR region of the geometry (after fixing the end of space); this parameter encodes information about the ratio of two energy scales: that at which $\f(\r)$ interpolates between the two critical point solutions in the gravitational model, and the scale at which the dual field theory exhibits confinement. We here extend the analysis of Ref.~\cite{Elander:2020csd} to include backgrounds which admit $\f>0$---corresponding to solutions which explore a runaway direction of the scalar potential $\cv_{7}(\f)$---and as a consistency check we confirm that the spectra are in agreement for the $\f\in[\f_{IR},\f_{UV}]$ region of the parameter space common to the two computations.\par
For those backgrounds with $\f_{I}\leqslant0$, which include the two critical points of the seven-dimensional potential and the family of solutions which interpolate between them, we observe in Fig.~\ref{Fig:WittenSpectrum} the same \emph{universal} behaviour as was previously encountered in Chapter~\ref{Chap:SpectraRomans} for the half-maximal theory: the tower of tensor states ($\mathfrak{e}^{\mu}_{\ \nu}$) and a subset of the scalar states ($\mathfrak{a}^{\c}$) show no dependence on the choice of background being fluctuated. Phrased another way, these massive resonances do not care about specific details of the dual RG flow on the boundary, and are affected only by the scale at which confinement is implemented geometrically in the gravitational model. We again refer the Reader to Ref.~\cite{Elander:2013jqa}, wherein the spin-0 spectrum for this class of interpolating backgrounds has previously been obtained and the same phenomenon has been discussed; we verify that our results for $\f_{I}\leqslant0$ agree with those shown in Fig.~10 and Table.~2 of this paper (after normalising appropriately), and in Table~\ref{Tbl:WittenUVandIR} we present the numerical masses that our computation yields for the critical point backgrounds $\f\in\{\f_{IR},\f_{UV}\}$ to facilitate comparison.\par
Earlier studies of the glueball spectra for $\text{QCD}_{4}$ using a top-down holographic approach from similar supergravity backgrounds exist in the literature (see for example Refs.~\cite{Csaki:1998qr,Hashimoto:1998if,Constable:1999gb}). In Ref.~\cite{Brower:2000rp} the authors started by considering M-theory formulated on the eleven-dimensional product space $\text{AdS}_{7}\times S^{4}$---dual to the six-dimensional superconformal field theory living on $N_{c}$ coincident M5-branes---and compactified two of the dimensions on a $T^{2}$ as prescribed by Witten in Ref.~\cite{Witten:1998zw}. By raising the temperature of the thermal circle internal to this torus and imposing appropriate boundary conditions they realised the so-called $\text{AdS}_{7}$ black hole geometry, which provides a holographic description of $\text{QCD}_{4}$.\par 
The authors' investigation primarily differs from our own numerical study in that they retained additional fields in their truncation, identifying six independent towers of states: five modes which descend from the graviton and 3-form of the supergravity in $D=11$ dimensions, supplemented by spin-0 excitations of the $S^{4}$ metric; we refer the Reader to the paper for further details. Two of these massive towers are comparable to the gauge-invariant fluctuations that we consider; those states which the authors refer to as $T_{4}$ and $S_{4}$ correspond respectively to the $\mathfrak{e}^{\mu}_{\ \nu}$ and $\mathfrak{a}^{a}$ resonances of our $T^{2}$-compactified model, with the caveat that they restrict attention to the trivial supersymmetric solution $\f=0$. By comparing the results of our numerical computation to those in Table~2 of Ref.~\cite{Brower:2000rp} and normalising appropriately, we find excellent agreement for the spin-2 tower and a subset of the spin-0 states (those corresponding to fluctuations of $\c$); we again refer the Reader to Table~\ref{Tbl:WittenUVandIR} of Appendix~\ref{App:CritSpectra} in which we provide numerical values for the masses that are extracted from the critical point solutions.\par 
Let us conclude this section by commenting on two significant features of the spectra presented in Fig.~\ref{Fig:WittenSpectrum}, both of which were also observed with the analogous computation for Romans six-dimensional supergravity. We first notice that the universal background-independence of the two graviton modes is restricted to the $\f_{I}\leqslant0$ region of the parameter space, corresponding to solutions which interpolate between the two critical points. As the IR parameter $\f_{I}$ is dialled higher and the scalar $\f$ is permitted to explore further along the runaway direction of the potential, we see that both towers of masses start to converge and the spectra become increasingly dense; in the large-$\f_{I}$ limit we expect to find the `gapped continuum' behaviour discussed earlier in Sec.~\ref{Sec:RomansMassPlots}.\par
Our second observation pertains to the spin-0 spectrum, in which another important phenomenon appears within the $\f_{I}>0$ region of the plot. As was encountered with the corresponding computation for the six-dimensional supergravity, at some special value of $\f_{I}=\f_{I}^{\t}$ ($\sim0.45$ in this case) we notice that the lightest scalar resonance becomes tachyonic, and hence we infer that our spectrum analysis uncovers an instability in the theory parameter space. This is potentially problematic, as we are studying the field fluctuations of a consistent truncation of an established supergravity which is known to be obtainable from compactifications of superstring and M-theory in ten and eleven dimensions~\cite{Pernici:1984xx,Pernici:1984zw,Nastase:1999cb,Lu:1999bc}; indeed, the presence of such an instability necessitates the existence of a phase transition which would prevent the system from realising these pathological backgrounds. It is this physical requirement which motivates our exploration of the theory phase space in Chapter~\ref{Chap:EnergeticsWitten}, where we shall catalogue several classes of geometrically distinct background solutions which are admitted by the $T^{2}$-reduced supergravity and systematically compute their free energy. The scalar spectrum in Fig.~\ref{Fig:WittenSpectrum} signals that a phase transition should be uncovered, and that another branch of backgrounds must become energetically favoured before the system is able reach the tachyonic instability along the branch of confining solutions. We shall return to this discussion in Chapters~\ref{Chap:EnergeticsRomans} and~\ref{Chap:EnergeticsWitten}.     
\clearpage

\section{Probe spectrum analysis}
\label{Sec:WittenProbePlots}
As discussed in Sections~\ref{Sec:Motivation}, \ref{Sec:DilatonFormalism}, and~\ref{Sec:RomansProbePlots}, the dilaton is the (massless) Nambu--Goldstone Boson associated with the spontaneous breaking of exact dilatation invariance. When this symmetry is not explicitly preserved either, for example if the CFT is deformed by switching on a source, the dilaton acquires a small non-zero mass and is instead referred to as a \emph{pseudo}-Nambu--Goldstone Boson.\ Our investigation into the phenomenology of this scalar field within the framework of top-down holography generalises the methods employed in Ref.~\cite{Pomarol:2019aae}---wherein the breaking of conformal invariance is inferred from proximity to the BF stability bound---and is predicated instead on the comparison of two distinct spectra. We consider fluctuations about backgrounds which smoothly close off in the deep IR, and hence our dilaton study is necessarily also applicable to geometries which depart from AdS.\par
We briefly remind the Reader that our diagnostic test for detecting a dilaton admixture is referred to as the \emph{probe approximation}, and that it consists of two steps: we first use Eqs.~(\ref{Eq:ScalarFluct}) and (\ref{Eq:ScalarFluctBC}) to numerically compute the spectra of complete gauge-invariant scalar fluctuations $\mathfrak{a}^{a}$ as defined in Eq.~(\ref{Eq:a}), and we then compare these results to those obtained for the corresponding probe states $\mathfrak{p}^{a}$ using Eqs.~(\ref{Eq:ProbeScalarFluct}) and (\ref{Eq:ProbeScalarFluctBC}). In the latter case we switch off by hand the metric fluctuation $h$, the scalar supergravity field dual to the dilatation operator in the boundary theory. Where discrepancies emerge between the two computations we infer that the contribution of $h$ to the mass eigenstates is not negligible, and that those states are at least partially identifiable with the dilaton.\par
\begin{figure}[t!]
	\begin{center}
		\makebox[\textwidth]{\includegraphics[
			width=0.55\paperwidth
			]
			{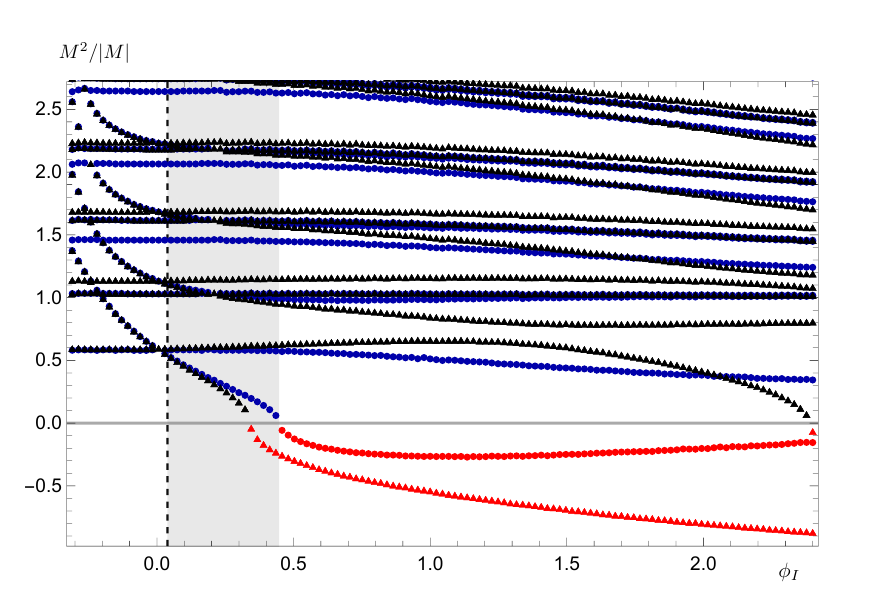}}
	\end{center}
	\vspace{-5mm}
	\caption[Probe spectra for the seven-dimensional supergravity]{The spectra of masses $M$ as a function of the one free parameter which characterises the class of confining solutions, $\f_{I}\in[\f_{IR},2.4]$. All states are normalised in units of the lightest tensor mass, and the spectrum was computed using regulators $\r_{1}=10^{-4}$ and $\r_{2}=12$. As in Fig.~\ref{Fig:WittenSpectrum},
		the blue disks represent the mass eigenstates for the three scalars of the model $\{\f,\c,\w\}$, while the red disks denote the tachyon. 
		We here additionally include the results of our mass spectrum computation using the probe approximation for $M^{2}>0$
		(black triangles) and $M^{2}<0$ (red triangles). 
		The vertical dashed line marks a critical value of the IR parameter $\f_{I}= \f_{I}^{*}>0$, while the shaded grey region denotes the region of parameter space for which the confining solutions are metastable; we shall elaborate on these points in Sec.~\ref{Sec:WittenPhaseStruct}.}
	\label{Fig:WittenSpectrumProbe}
\end{figure}

  
In Fig.~\ref{Fig:WittenSpectrumProbe} we present a direct comparison of the gauge-invariant scalar fluctuation $\mathfrak{a}^{a}$ spectrum shown in the rightmost panel of Fig.~\ref{Fig:WittenSpectrum}, to the new results obtained from our probe state $\mathfrak{p}^{a}$ computation (see also Fig.~5 of Ref.~\cite{Elander:2020csd}). There are both similarities and differences when compared to the analogous plot for the six-dimensional supergravity in Fig.~\ref{Fig:RomansSpectrumProbe}, though once again it is evident that the probe approximation fails to ever completely capture the complete tower of physical states for any value of the tunable IR parameter $\f_{I}$.    
\\
\indent Let us start by examining the $\f_{I}<0$ region of the plot in proximity of the IR critical point solution $\f=\f_{IR}$, where we notice that---contrary to Fig.~\ref{Fig:RomansSpectrumProbe}---the lightest resonance (associated with the scalar $\c$) is here well approximated by the probe analysis; moreover this is also true for half of the heavier excitations that show no discernible dependence on the choice of background, which correspond to the $\mathfrak{a}^{\w}$ fluctuation associated with $\w$ (with masses $M\approx1.03,1.61,2.18$). The other heavy background-independent states correspond to the excitations of $\c$, and these are \emph{not} well approximated by the probes (see Table~\ref{Tbl:WittenUVandIR}). To understand this observation, let us recall the results of our investigation for the toroidal compactification of a generic $\text{AdS}_{D}$ system shown in Figure~\ref{Fig:nTorusProbeSpectrum} of Chapter.~\ref{Chap:nTorus}. There we noticed that the probe states $\mathfrak{p}^{\bar{\c}}$ associated with the scalar $\bar{\c}$ (analogous to $\c$ here) coincidentally provided an effective approximation of the corresponding $\mathfrak{a}^{\bar{\c}}$ fluctuations, specifically for models obtained by compactifying $n\sim2$ external dimensions on circles. Since we are considering fluctuations about background geometries resulting from the reduction of a higher-dimensional theory on a $T^{2}$, the success of the probe approximation here is perhaps not entirely unexpected.\par 
As the parameter $\f_{I}$ is dialled higher to approach the UV critical point solution at $\f=0$ we observe that the probes effectively approximate both the lightest and next-to-lightest  gauge-invariant states, however this is not the case for the heavier resonances within the $\mathfrak{a}^{\c}$ tower (those with $M\approx1.45,2.06,2.64$); we hence infer that only the latter excitations result from significant dilaton mixing effects.\par 
Backgrounds which lie within the shaded grey region of the plot are \emph{metastable} (we shall elaborate on this point in Sec.~\ref{Sec:WittenPhaseStruct}), and as with the six-dimensional supergravity we notice that the probe approximation begins to deviate from the proper computation; this effect becomes more pronounced just before the lightest scalar $\mathfrak{a}^{a}$ turns tachyonic at $\f_{I}=\f_{I}^{\t}\sim 0.45$. This implies that, at least in proximity of the instability where the lightest state may be rendered parametrically light, it is legitimate to identify the lightest scalar excitation as being dilatonic. As with Fig.~\ref{Fig:RomansSpectrumProbe} we here too notice that the failure of the probe approximation is not restricted to the lightest states in the spectrum, and that the heavier excitations must therefore also contain non-negligible contributions from the field which sources the dilatation operator at the boundary. This is most apparent at large values of the IR parameter $\f_{I}$ where, as we found with the various spectra in Figs.~\ref{Fig:Spectra1} and~\ref{Fig:Spectra2} for the Romans theory, the physical states asymptotically converge to a gapped continuum; we see early evidence that the corresponding probes instead become lighter and eventually turn tachyonic.\par 
To summarise the results of our spectra calculations for the toroidally reduced seven-dimensional supergravity, we have uncovered the existence of a tachyonic instability which may be approached by dialling the one free parameter that labels a class of regular backgrounds. Furthermore, from our probe approximation analysis we have determined that this tachyonic state contains a significant contribution coming from the metric fluctuation $h$, which sources the dilatation operator in the dual field theory. Motivated by these findings, in Chapter~\ref{Chap:EnergeticsWitten} we shall conduct an investigation into the phase space of the theory by computing the holographically renormalised free energy density $\cf$ for several distinct classes of background solutions, with the expectation that a phase transition must exist to prevent the unstable region of parameter space from being accessed. In the process we will uncover some useful parameter relations, which will allow us to more closely examine the nature of the putative dilatonic states; we postpone further discussion on this topic until Section~\ref{Sec:WittenProbeRevisit}.

\part{Phase structure analysis}
\label{PartTwo}

\chapter{Six-dimensional half-maximal supergravity}
\label{Chap:EnergeticsRomans}
\section{Classes of solutions}
\label{Sec:RomansClasses}
There are several distinct classes (or \emph{branches}, we will use these two terms interchangeably) of background solutions which satisfy the equations of motion presented in Eqs.~(\ref{Eq:RomansEOM1}\,-\,\ref{Eq:RomansEOM4}), 
with various geometric properties and boundary-dual interpretations. We have already encountered the class referred to as \emph{confining} in Sec.~\ref{Sec:RomansEOMs}, where we used the fact that the circle-compactified dimension smoothly shrinks to a point at the end of space to model the gravitational dual of a strongly-coupled confining field theory, and to then compute the mass spectra of glueball states in such a field theory.\par 
For the purposes of this chapter, in which we conduct an investigation into the phase structure of Romans six-dimensional supergravity and (as we shall see) uncover evidence for the existence of a first-order phase transition, it will be insufficient to consider only these regular solutions; to obtain a proper understanding of the physical phase space it is necessary to study the energetics for all branches and moreover to compare them in an appropriate way. In this chapter we provide a comprehensive classification of solution types, several of which were previously unidentified prior to the work in Ref.~\cite{Elander:2020ial}.  

\subsubsection{UV asymptotic expansions}
The classes of solutions of interest to our investigation differ primarily in their geometric properties at small values of the holographic coordinate, and are classified according to their behaviour in proximity to the bulk end of space. However, due to the fact that each class represents some form of parametric deformation of the unique supersymmetric trivial fixed point $\f=0$, they all show the same convergent behaviour at large $\r$ (corresponding to UV energies in the dual field theory); more precisely, we asymptotically approach the geometry of $\text{AdS}_{6}$ (locally) in the limit $\r\to\infty$, irrespective of which specific backgrounds we are considering.\par
Therefore, after introducing a convenient new radial coordinate defined via $z\equiv e^{-2\r/3}$, we are able to write down a set of general asymptotic expansions for $\f$, $\c$, and $A$ which are valid at large $\r$ (small $z$) near the UV boundary and which are universally applicable to all branches of solutions. We present these expansions below:
{\small\begin{align}
	\label{Eq:RomansUVexpPhi}
\hspace{-11mm}\f(z)=\f_{2}z^2 &+\f_{3}z^3-6\f_{2}^{2} z^4 -4(\f_{2}\f_{3})z^5+\left(\frac{29 \f_{2}^3}{2}-\f_{3}^2\right)z^6+\frac{339}{20} \f_{2}^2 \f_{3}z^7 \nn\\
	&+\left(\frac{77 \f_{2}
		\f_{3}^2}{10}-\frac{146 \f_{2}^4}{3}\right)z^8 + \left(\frac{19 \f_{3}^3}{12}-\frac{8497 \f_{2}^3 \f_{3}}{105}\right)z^9 \nn\\  
	&+\left(\frac{6752 \f_{2}^5}{35}-\frac{1986 \f_{2}^2 \f_{3}^2}{35}\right)z^{10}+ \left(\frac{4127161 \f_{2}^4 \f_{3}}{10080}-\frac{3427 \f_{2} \f_{3}^3}{180}\right)z^{11}\nn\\
	&+\mathcal{O}\Big(z^{12}\Big)\, ,
	\end{align} }
\vspace{-8mm}
{\small\begin{align}
	\label{Eq:RomansUVexpChi}
	\chi(z)=\chi_{U}&-\frac{\ln(z)}{3}-\frac{\f_{2}^2 z^4}{12}+\chi_{5} z^5+ \left(\frac{8 \f_{2}^3}{9}-\frac{\f_{3}^2}{12}\right)z^6+\frac{32}{21} \f_{2}^2 \f_{3}z^7+ \left(\frac{3
		\f_{2} \f_{3}^2}{4}-\frac{77 \f_{2}^4}{16}\right)z^8 \nn\\
	&+\left(-\frac{1072 \f_{2}^3 \f_{3}}{135}+\frac{25 \chi_{5} \f_{2}^2}{36}+\frac{4 \f_{3}^3}{27}\right)z^9 \nn\\
	&+\left(-\frac{15
		\chi_{5}^2}{64}+\frac{172 \f_{2}^5}{9}-\frac{3181 \f_{2}^2 \f_{3}^2}{600}+\frac{9 \chi_{5} \f_{2} \f_{3}}{8}\right)z^{10} \nn\\
	&+\left(\frac{44776 \f_{2}^4 \f_{3}}{1155}-\frac{200
		\chi_{5} \f_{2}^3}{33}-\frac{96 \f_{2} \f_{3}^3}{55}+\frac{25 \chi_{5} \f_{3}^2}{44}\right)z^{11}+\mathcal{O}\Big(z^{12}\Big)\, ,
	\end{align} }
\vspace{-8mm}
{\small\begin{align}
	\label{Eq:RomansUVexpA}
	A(z)= A_{U}&-\frac{4 \ln(z)}{3}-\frac{\f_{2}^2 z^4}{3}+ \left(\frac{\chi_{5}}{4}-\frac{3 \f_{2} \f_{3}}{5}\right)z^5+ \left(\frac{32 \f_{2}^3}{9}-\frac{\f_{3}^2}{3}\right)z^6+\frac{128}{21}
	\f_{2}^2 \f_{3}z^7 \nn\\
	&+\left(3 \f_{2} \f_{3}^2-\frac{77 \f_{2}^4}{4}\right)z^8+\frac{1}{2160} \bigg(-69508 \f_{2}^3 \f_{3}+375 \chi_{5} \f_{2}^2+1280 \f_{3}^3\bigg)z^9 \nn\\ 
	&+\frac{1}{3600} \bigg(-3375 \chi_{5}^2+275200 \f_{2}^5-78936 \f_{2}^2 \f_{3}^2\bigg)z^{10} \nn\\ 
	 &+\frac{1}{18480} \bigg(2932864 \f_{2}^4 \f_{3}-28000 \chi_{5} \f_{2}^3-135324
	\f_{2} \f_{3}^3+2625 \chi_{5} \f_{3}^2\bigg)z^{11} \nn\\
	&+\mathcal{O}\Big(z^{12}\Big)\, ,
	\end{align} }%
and we also show explicitly the corresponding expansions for the two useful linear combinations $\a=4A-\c$ and $\b=A-4\c$ which were introduced in Section.~\ref{Sec:RomansEOMs}: 
{\small\begin{align}
	\a(z)=4A_{U}-\c_{U}&-5\ln(z)-\frac{5\f_{2}^{2}z^4}{4}-\frac{12}{5}\f_{2}\f_{3}z^5+\left(\frac{40\f_{2}^3}{3}-\frac{5\f_{3}^2}{4}\right)z^{6}+\frac{160}{7}\f_{2}^{2}\f_{3}z^{7} \nn\\
	&+\left(\frac{45\f_{2}\f_{3}^{2}}{4}-\frac{1155 \f_{2}^{4}}{16}\right)z^{8}+\left(\frac{20 \f_{3}^{3}}{9}-\frac{1087 \f_{2}^{3}\f_{3}}{9}\right)z^{9} \nn\\
	&+\left(-\frac{225 \c_{5}^{2}}{64}+\frac{860 \f_{2}^5}{3}-\frac{16481 \f_{2}^2\f_{3}^2}{200}-\frac{9\c_{5}\f_{2}\f_{3}}{8}\right)z^{10} \nn\\
	&+\left(\frac{45896\f_{2}^4\f_{3}}{77}-\frac{303\f_{2}\f_{3}^3}{11}\right)z^{11}+\mathcal{O}\Big(z^{12}\Big)\, ,
	\end{align} }
\vspace{-8mm}
{\small\begin{align}\hspace{-16mm}
	\b(z)=A_{U}-4\c_{U}&+\left(-\frac{15\c_{5}}{4}-\frac{3 \f_{2} \f_{3}}{5}\right)z^{5}+\left(-\frac{5\f_{2}^3\f_{3}}{12}-\frac{125\c_{5}\f_{2}^2}{48}\right)z^{9} \nn\\
	&+\left(-\frac{18}{25} \f_{2}^2 \f_{3}^2-\frac{9\c_{5}\f_{2}\f_{3}}{2}\right)z^{10} \nn\\
	&+\left(\frac{40\f_{2}^4\f_{3}}{11}+\frac{250\c_{5} \f_{2}^3}{11}-\frac{15\f_{2}\f_{3}^3}{44}-\frac{375\c_{5} \f_{3}^2}{176}\right)z^{11} \nn\\
	&+\mathcal{O}\Big(z^{12}\Big)\,.
	\end{align} }%
We observe that these general asymptotic expansions---which govern the UV behaviour of a wide variety of distinct background configurations---may be formulated in terms of only five UV parameters $\{\f_{2},\f_{3},\chi_{5},\chi_{U},A_{U}\}$. As discussed earlier in Sec.~\ref{Sec:EnergeticsFormalism} these parameters play a vital role in our energetics analysis, and in Sec.~\ref{Sec:RomansScaleSetting} we will describe in detail the numerical routine which we employ to extract meaningful data for each class of solutions using these expansions.

\subsubsection{Supersymmetric (SUSY) solutions}
The first class of solutions we will consider are obtained by making use of the superpotential formalism which was introduced in Sec.~\ref{Subsec:HolographicFormalism}, and for convenience we here reproduce the system of first-order equations that must be solved:  
\begin{align*}
\pa_{\r}\Phi^{a}&=G^{ab}\pa_{b}\cw\,,\\
\pa_{\r}\ca&=-\frac{2}{D-2}\cw\,,
\end{align*}
where we have made the coordinate replacement $r\to\r$. We remind the Reader that these solutions have as a prerequisite the condition that the bulk geometry takes the form of the domain-wall metric ansatz given in Eq.~(\ref{Eq:RomansSUSYmetric}), so that Poincar\'{e} invariance is locally preserved within the five-dimensional subspace described by the Minkowski and $\eta$ directions; as previously mentioned, this geometric property translates into the requirement that the warp factor constraint $A=4\c$ (or equivalently $\ca=\frac{3}{4}A=3\c$) is satisfied.\par
In $D=6$ dimensions the scalar potential $\mathcal{V}_{6}$ presented in Eq.~(\ref{Eq:V6}) is a function solely of $\f$, and hence the first-order equations reduce to
\begin{align}
\label{Eq:SUSYeom1}
\pa_{\r}\f&=G^{\f\f}\pa_{\f}\cw=\frac{1}{2}\pa_{\f}\cw\,,\\
\label{Eq:SUSYeom2}
\pa_{\r}{\cal A}&=-\frac{1}{2}\cw\,,
\end{align}
where in the second line we have substituted in for the $\f$ component of the (diagonal) sigma-model metric. These equations admit as a solution~\cite{Gursoy:2002tx} the superpotential $\cw=\cw_{1}$ given by:
\begin{equation}
\cw_{1}=-e^{\f}-\frac{1}{3}e^{-3\f}\,,
\label{Eq:RomansFirstOrder1}
\end{equation}
with which we have
\begin{align}
\pa_{\r}\f&=\frac{1}{2}\Big(e^{-3\f}-e^{\f}\Big)\,,
\label{Eq:RomansFirstOrder2}\\
\pa_{\r}\ca&=\frac{1}{6}\Big(3e^{\f}+e^{-3\f}\Big)\,.
\label{Eq:RomansFirstOrder3}
\end{align}  
We therefore see that the trivial supersymmetric solution with constant $\f=0$, for which the bulk geometry is that of $\text{AdS}_{6}$, has a metric warp factor which scales linearly with the radial coordinate: $\ca=\frac{2}{3}\r$.\par 
Let us here make a brief, though important, observation. In addition to the exact superpotential $\cw_{1}$ presented above, the first-order equations presented in Eqs.~(\ref{Eq:SUSYeom1}) and (\ref{Eq:SUSYeom2}) also admit a second choice for $\cw$, albeit one that we are only able to write as a power expansion in $\f$ for small perturbations about the supersymmetric fixed point $\f=0$. This expansion is given by
\begin{equation}
\label{Eq:W2}
\cw_2(\f)=-\frac{4}{3}-\frac{4}{3}\f^2+\frac{16}{3}\f^3+\frac{86}{3}
\f^4+\frac{848}{3}\f^5+\frac{988658}{315}\f^6\,+\,{\cal O}(\f^7)\,,
\end{equation}
and---as we shall later demonstrate---it will play a crucial role in our energetics analysis by providing the exact counter-terms required in our computation of the holographically renormalised free energy.\par 
More generally, there exist solutions to this same system of first-order equations for which $\f=\f(\r)$ is not a constant, but rather evolves monotonically from the trivial supersymmetric fixed point in the UV towards a \emph{good} singularity $\f\to\infty$ at the end of space. As an aside, we here clarify that when referring to background configurations in which the scalar field $\f$ exhibits singular (non-convergent) behaviour, we shall adopt the prescription of Gubser in Ref.~\cite{Gubser:2000nd}: classifying singularities as either \emph{good} or \emph{bad} depending on whether the supergravity scalar potential (evaluated on the singular solution) is bounded from above, or not, respectively.\par 
We proceed by introducing the convenient change of coordinate $\pa_{\r}\equiv e^{-\f}\pa_{\tau}$, so that Eqs.~(\ref{Eq:RomansFirstOrder2},\,\ref{Eq:RomansFirstOrder3}) may be rewritten as
\begin{align}
\pa_{\tau}\f&=-\sinh(2\f)\,,\\
\pa_{\tau}\ca&=\frac{1}{6}\Big(3e^{2\f}+e^{-2\f}\Big)\,,
\end{align} 
which admit the following exact solutions~\cite{Gursoy:2002tx}:  
\begin{align}
\f(\t)&={\rm arccoth}\left(e^{2(\t-\t_{o})}\right)\,,
\label{Eq:RomansSUSY1}\\
\ca(\t)&=\ca_{o} +\frac{1}{3}\ln\Big[\sinh(2\big(\t-\t_{o})\big)\Big]
+\frac{1}{6}\ln\Big[\tanh(\t-\t_{o})\Big]\,,
\label{Eq:RomansSUSY2}
\end{align}
where $\ca_o$ and $\tau_o$ are integration constants. By series expanding these analytical solutions we obtain the following IR (small $\t$) expansions:
\begin{align}
\f(\r)&=
-\half\ln(\t-\t_{o}) +\frac{1}{6}(\t-\t_{o})^{2}-\frac{7}{180}(\t-\t_{o})^{4} +\ldots\,,\\
\ca(\r)&=
\ca_{o} +\half\ln(\t-\t_{o}) +\frac{1}{90}(\t-\t_{o})^{4} +\ldots\,,
\end{align} 
and by making use of the $\tau\to\r$ coordinate change defined above, we therefore also find
\begin{align}
\label{Eq:RomansSUSYphi}
\f(\r)&=\ln(2) -\ln(\r-\r_{o})+\frac{1}{80}(\r-\r_{o})^{4}+\ldots\, ,\\
\ca(\r)&= \ca_{I} + \ln(\r-\r_{o})+ \frac{1}{120}(\r-\r_{o})^{4}+\ldots\, ,
\end{align}
where $\ca_{I}=\ca_{o}-\ln(2)$ and $\r_{o}$ are two new integration constants, the latter representing the radial position of the singularity in the deep IR. Recalling that $A=\frac{4}{3}\ca=4\c$, we hence equivalently have for $\c$ and $A$:
\begin{align}
\chi(\r)&=\chi_{I} + \frac{1}{3}\ln(\r-\r_{o})+ \frac{1}{360}(\r-\r_{o})^{4}+\ldots\, ,\\
A(\r)&= A_{I} + \frac{4}{3}\ln(\r-\r_{o})+ \frac{1}{90}(\r-\r_{o})^{4}+\ldots\, ,
\end{align}
where $A_{I}=\frac{4}{3}\big(\ca_{o}-\ln(2)\big)$ and $\c_{I}=\frac{A_{I}}{4}$.

\subsubsection{IR-conformal (IRC) solutions}
As with the SUSY solutions, the second class of solutions locally preserve five-dimensional Poincar\'{e} invariance by obeying the domain-wall constraint $A=4\c\Leftrightarrow \b=0$. However they are not singular, nor are they supersymmetric, instead interpolating between the two critical point solutions $\f=\f_{UV}$ and $\f=\f_{IR}$ for which the bulk background geometry is exactly $\text{AdS}_{6}$; this class therefore provides the gravitational dual description of a renormalisation group flow between two distinct five-dimensional CFTs. The circle-compact dimension parametrised by $\eta$ maintains a non-zero volume at all scales (in contrast to the confining solutions), and hence the field theories dual to this class do not exhibit a physical low-energy limit; the name \emph{IR-conformal} reflects this fact.\par 
The IR expansions for this branch are conveniently formulated in terms of the quantity $e^{-(5-\Delta_{IR})\frac{\r}{R_{IR}} }$ which is vanishingly small in the $\r\to-\infty$ limit (recall from Eq.~(\ref{Eq:RomansDeltas}) that $5-\Delta_{IR}<0$), and are given by~\cite{Elander:2020ial,Elander:2013jqa}
\begin{align}
\f(\r)&=\f_{IR} + \big(\f_{I}-\f_{IR} \big)e^{-(5-\Delta_{IR})\frac{\r}{R_{IR}} }\,+\ldots\, ,\\
\c(\r)&=\c_I+\frac{\r}{3R_{IR}} - \frac{1}{12} \big(\f_{I}-\f_{IR} \big)^{2} e^{-2(5-\Delta_{IR})\frac{\r}{R_{IR}} }\,+\ldots\, ,\\
A(\r)&=A_I+\frac{4\r}{3R_{IR}} - \frac{1}{3} \big(\f_{I}-\f_{IR} \big)^{2} e^{-2(5-\Delta_{IR})\frac{\r}{R_{IR}} }\,+\ldots\, ,
\end{align}
where $R_{IR}$ is the curvature radius of the $\text{AdS}_{6}$ geometry associated with the IR critical point solution (see Eq.~(\ref{Eq:RomansRIR})), and the integration constants $\c_{I}$ and $A_{I}$ may be chosen arbitrarily. The one remaining parameter $\f_{I}\geqslant\f_{IR}$ may be varied to generate an entire family of solutions, and making a choice for its value amounts to choosing at which energy scale in the dual boundary theory the transition between the two CFTs occurs. However, since any one solution within this class may be obtained from any other via an additive shift of the holographic coordinate, they are all physically equivalent; we will return to this point in Sec.~\ref{Sec:RomansScaleSetting} when discussing the scale setting procedure by which we may compare the free energy of the various classes.            

\subsubsection{Confining solutions}
We introduced in Sec.~\ref{Sec:RomansEOMs} the branch of solutions which provide the geometric realisation of confinement, and which are used to compute the glueball mass spectra of the dual field theory. In this section we need only to take note of the following results, which are obtained by substitution of the $\f=0$ analytical critical point solution of Eqs.~(\ref{Eq:RomansConfChi}\,-\,\ref{Eq:RomansConfA}):
\begin{align}
\label{Eq:ConfExpAlpha}
e^{\a(\r)}=e^{4A(\r)-\c(\r)}&=\frac{1}{2}e^{4A_{I}-\c_{I}}\sinh\left(\frac{10}{3}(\r-\r_{o})\right)\, ,\\
\label{Eq:ConfExpBeta}
e^{\b(\r)}=e^{A(\r)-4\c(\r)}&=e^{A_{I}-4\c_{I}}\coth\left(\frac{5}{3}(\r-\r_{o})\right)\, ,
\end{align} 
and similarly by substituting for $\c$ and $A$ using instead the generalised IR expansions in Eqs.~(\ref{Eq:RomansChiIRExpansion}\,-\,\ref{Eq:RomansAIRExpansion}) we have
\begin{align}
\label{Eq:ConfExpAlphaIR}
e^{\a(\r)}=e^{4A(\r)-\c(\r)}&= e^{4A_{I}-\c_{I}}\,f\big(\f_{I},\,(\r-\r_{o})\big)\, ,\\
e^{\b(\r)}=e^{A(\r)-4\c(\r)}&= e^{A_{I}-4\c_{I}}\,g\big(\f_{I},\,(\r-\r_{o})\big)\, ,\label{Eq:ConfExpBetaIR}
\end{align}
where $f$ and $g$ are known (though here unspecified) numerical functions. The purpose of bringing these results to the Reader's attention will become apparent in the next subsection. 

\subsubsection{Skewed solutions}
Recall from Sec.~\ref{Sec:RomansEOMs} that the classical equations of motion Eqs.~(\ref{Eq:EOMalphaphi}\,-\,\ref{Eq:EOMbeta}), which are obtained from the five-dimensional action $\cs_{5}$, are invariant under the sign change $\b\to-\b \Leftrightarrow A-4\c\to 4\c-A$. This symmetry actually implies the existence of a distinct class of solutions to the dimensionally reduced theory which are related to the confining class, though admit a rather different background geometry: in this case the compact dimension parametrised by $\eta$ does not shrink to a point at the end of space to close off the bulk manifold, but rather increases in volume without bound as one approaches the IR region of space; the background solutions for $\c(\r)$ are all non-monotonic functions which diverge as $\r$ becomes small. As with the confining class there exist exact analytical solutions for the UV fixed point $\f=0$, given by\cite{Elander:2020ial}:
\begin{align}
\f&=0,\\
\c(\r)&=\c_{I}+\frac{1}{3}\ln\bigg[\cosh\bigg(\tfrac{{\sqrt{-5v}}}{2}(\r-\r_{o}) \bigg) \bigg]\nn\\
&\hspace{44mm}-\frac{1}{5}\ln\bigg[\sinh\bigg(\tfrac{{\sqrt{-5v}}}{2}(\r-\r_{o}) \bigg) \bigg],\label{Eq:SkewChi}\\
A(\r)&=A_{I}-\frac{4}{15}\ln(2)+\frac{4}{15}\ln\bigg[\sinh\bigg(\sqrt{-5v}(\r-\r_{o}) \bigg) \bigg]\nn\\
&\hspace{44mm}-\frac{1}{15}\ln\bigg[\tanh\bigg(\tfrac{{\sqrt{-5v}}}{2}(\r-\r_{o}) \bigg) \bigg] \,, \label{Eq:SkewA}
\end{align}
where again $\c_{I}$, $A_{I}$, and $\r_{o}$ are integration constants, and $\r_{o}$ may be freely chosen to fix the end of space (i.e. the value of the radial coordinate at which the $\c\to\infty$ singularity is located). Just as with the confining class, we can generalise these exact solutions in order to generate a family which admit arbitrary values of $\f$ by series expanding for small $(\r-\r_{o})$. The corresponding IR expansions then read as follows\cite{Elander:2020ial}:
{\small\begin{align}
	\label{Eq:RomansPhiIRskew}
	\phi(\r) &= \phi_{I}-\frac{1}{12}e^{-6\phi_{I}} \left(1-4 e^{4\phi_{I}}+3 e^{8\phi_{I}}\right)(\r-\r_{o})^2 \nn\\
	&\hspace{20mm}-\frac{1}{324} e^{-12\phi_{I}}\left(4-28 e^{4\phi_{I}}+51 e^{8\phi_{I}}-27 e^{16\phi_{I}}\right)(\r-\r_{o})^4\nn\\
	&\hspace{70mm}+\mathcal{O}\left((\r-\r_o)^6\right)\, ,\\
		\label{Eq:RomanChiIRskew}
	\chi(\r)&=\c_{I}-\frac{1}{5}\ln\left(\frac{5}{3}\right) -\frac{1}{5}\ln (\r-\r_{o})
	-\frac{1}{54} e^{-6\phi_{I}} \Big(1-12 e^{4\phi_{I}}-9 e^{8\phi_{I}}\Big)(\r-\r_{o})^2\, \nn\\
	&-\frac{1}{9720}e^{-12\phi_{I}}\Big[23+3e^{4\phi_{I}}
	\Big(-88+9 e^{4\phi_{I}}(38+24 e^{4\phi_{I}}+21 e^{8\phi_{I}})\Big)\Big](\r-\r_{o})^4\nn\\
	&\hspace{70mm}+\mathcal{O}\left((\r-\r_o)^6\right)\, ,\\
		\label{Eq:RomansAIRskew}
	A(\r)&=A_{I}+\frac{1}{5}\ln\left(\frac{5}{3}\right)+\frac{1}{5}\ln (\r-\r_{o})
	-\frac{1}{36} e^{-6\phi_{I}} \Big(1-12 e^{4\phi_{I}}-9 e^{8\phi_{I}}\Big)(\r-\r_{o})^2\, \nn\\
	&\hspace{-5mm}-\frac{1}{29160}e^{-12\phi_{I}}\Big[131+3e^{4\phi_{I}}
	\Big(-436+3 e^{4\phi_{I}}(508+84 e^{4\phi_{I}}+261 e^{8\phi_{I}})\Big)\Big](\r-\r_{o})^4\nn\\
	&\hspace{70mm}+\mathcal{O}\left((\r-\r_o)^6\right)\, , 
	\end{align}}%
where $\f_{I}$ is the free parameter which may be dialled to generate the entire family of solutions.\par 
As a slight digression, we can deduce the deep IR behaviour of the bulk geometry evaluated on the confining and skewed branches of solutions by again considering the six-dimensional metric ansatz presented in Eq.~(\ref{Eq:6Dmetric}). Substituting in for $\c$ and $A$ using the small $\r$ expansions for the confining solutions given in Eqs.~(\ref{Eq:RomansChiIRExpansion},\,\ref{Eq:RomansAIRExpansion}) implies that in the $\r\to\r_{o}$ limit the Minkowski subspace maintains a constant non-zero volume at the end of the bulk, while the size of $\eta$ dimension compactified on the $S^{1}$ vanishes (as is required). This contrasts with the behaviour of the skewed solutions: substituting instead using the IR expansions in Eqs.~(\ref{Eq:RomansPhiIRskew}\,-\,\ref{Eq:RomansAIRskew}) we find that the Minkowski dimensions scale as $(\r-\r_{o})^{2/5}$ while the compactified $\eta$ dimension scales as $(\r-\r_{o})^{-3/5}$, so that in the deep IR limit $\r\to\r_{o}$ the Minkowski volume vanishes while the size of the circle diverges. It is this characteristic scaling of the spacetime metric components, and the drastically different background geometry compared to those solutions which model confinement, that motivates our choice of the name \emph{skewed}. We conclude the aside by emphasising this rather interesting observation, namely that by simply flipping the sign of a linear combination $\b$ of the metric warp factor and the scalar parametrising the volume of the circular dimension (holding another linear combination $\a$ unchanged), one is able to construct an entirely distinct branch of new solutions to the model.\par 
We next return to the four relations which were introduced in the previous subsection, and here present the analogous results for the skewed solutions as a consistency check. From the exact analytical solutions we obtain     
\begin{align}
e^{\a(\r)}=e^{4A(\r)-\c(\r)}&=\frac{1}{2}e^{4A_{I}-\c_{I}}\sinh\left(\frac{10}{3}(\r-\r_{o})\right)\, ,\\
e^{\b(\r)}=e^{A(\r)-4\c(\r)}&=e^{A_{I}-4\c_{I}}\tanh\left(\frac{5}{3}(\r-\r_{o})\right)\, ,
\end{align}
which we thus confirm are in agreement with Eqs.~(\ref{Eq:ConfExpAlpha}) and (\ref{Eq:ConfExpBeta}) with the replacement $\b(\r)\to-\b(\r)$ (up to the contribution of an additive constant to $\b$). Likewise, we can substitute in for the skewed IR expansions to find
\begin{align}
e^{\a(\r)}=e^{4A(\r)-\c(\r)}&= e^{4A_{I}-\c_{I}}\,f\big(\f_{I},\,(\r-\r_{o})\big)\, ,\\
e^{\b(\r)}=e^{A(\r)-4\c(\r)}&= e^{A_{I}-4\c_{I}}\,\Big[g\big(\f_{I},\,(\r-\r_{o})\big)\Big]^{-1}\, ,
\end{align}
where the (unspecified) $f$ and $g$ are the same numerical functions as those which appear in Eqs.~(\ref{Eq:ConfExpAlphaIR}) and (\ref{Eq:ConfExpBetaIR}) for the confining class, so that we again find agreement after changing the sign of $\b(\r)$ while leaving $\a(\r)$ unchanged. We furthermore conclude that provided the conditions $\f_{I}^{c}=\f_{I}^{s}$ and $\r_{o}^{c}=\r_{o}^{s}$ are satisfied, where the superscripts $c$ and $s$ denote quantities associated with the confining and skewed classes of solutions respectively, then we also obtain the relation
\begin{equation}
0=\pa_{\r}\b^{c}(\r)+\pa_{\r}\b^{s}(\r)\, ,
\end{equation} 
which in turn implies that the following useful relation also holds, up to the contribution of an additive integration constant:
\begin{equation}
\label{Eq:ConfSkewBackgroundRelation}
0=\frac{3}{5}\Big[\c^{c}(\r)+A^{c}(\r)\Big]+\c^{s}(\r)-A^{s}(\r)\,.
\end{equation}
Finally then, by substituting into the above relation using the universal UV expansions (writing explicitly the superscripts which distinguish the two classes) and comparing, we extract the following parameter identities: 
\begin{align}
\label{Eq:id1}
\f_{2}^{s}&=\f_{2}^{c}\, ,\\
\label{Eq:id2}
\f_{3}^{s}&=\f_{3}^{c}\, ,\\
\label{Eq:id3}
\c_{5}^{s}&=-\c_{5}^{c}-\frac{8}{25}\f_{2}^{c}\f_{3}^{c} \,,
\end{align}
which---as we shall see---will prove to be invaluable in our analysis of the theory phase structure.

\subsubsection{General singular solutions}
As we saw when introducing the supersymmetric branch of solutions, the equations of motion derived from the action of this model admit background profiles for the scalar field $\f$ which are somewhat pathological in the deep IR region of the bulk geometry, in that $\f$ becomes divergent as one approaches the end of space. We remind the Reader that we adopt Gubser's criterion~\cite{Gubser:2000nd} in classifying these solutions, so that $\f\to\infty$ is described as a `good' singularity and the alternative $\f\to-\infty$ is labelled as a `bad' singularity; while one would otherwise disregard this latter case on the basis of them being unphysical, we will find they that play a pivotal role in our investigation of the theory phase space.\par 
A broad class of these divergent solutions, irrespective of the nature of their singularity, can be parametrised by the following expansions in proximity of the end of space at $\r=\r_{o}$:      
\begin{align}
\label{Eq:GenSingPhiExp}
\f(\r)&=\f_{I} + \f_{L}\ln(\r-\r_o) + \sum_{n=1}^{\infty} \sum_{j=0}^{2n} c_{nj}(\r-\r_o)^{2n+2n\,\f_{L}-4j\,\f_{L}}\,,\\
\label{Eq:GenSingChiExp}
\c(\r)&=\c_{I} + \c_{L}\ln(\r-\r_o) + \sum_{n=1}^{\infty} \sum_{j=0}^{2n} f_{nj}(\r-\r_{o})^{2n+2n\,\f_{L}-4j\,\f_{L}}\,,\\
\label{Eq:GenSingAExp}
A(\r)&=A_{I} + A_{L}\ln(\r-\r_o) + \sum_{n=1}^{\infty} \sum_{j=0}^{2n} g_{nj}(\r-\r_{o})^{2n+2n\,\f_{L}-4j\,\f_{L}}\,,
\end{align}
where $\f_{I}$ and $\f_{L}$ are the two free parameters which characterise the space of solutions---the latter controlling the type of logarithmic singularity encountered by $\f(\r)$ in the deep IR---and where $c_{nj}=c_{nj}(\f_{I},\f_{L})$, $f_{nj}=f_{nj}(\f_{I},\f_{L},\zeta)$, and $g_{nj}=g_{nj}(\f_{I},\f_{L},\zeta)$ are term coefficients which additionally depend on a third (discrete) parameter $\zeta=\pm1$. To leading order (see Appendix B of Ref.~\cite{Elander:2020ial} for the rather unwieldy sub-leading corrections) we therefore find the following small-$\r$ expansions: 
\begin{align}
\label{Eq:GenSingPhi}
\f(\r)&=\f_I+\f_L \ln(\r-\r_o)\,+\ldots\,,\\
\label{Eq:GenSingChi}
\c(\r)&=\c_I+\frac{1}{15}\left[4\zeta\sqrt{1-5\f_{L}^{2}}+1\right]
\ln(\r-\r_o)\,+\ldots\,,\\
\label{Eq:GenSingA}
A(\r)&=
A_I+\frac{1}{15}\left[\zeta\sqrt{1-5\f_{L}^{2}}+4\right]\ln(\rho-\rho_o)
\,+\ldots\,,
\end{align}
where the system is parametrised by the five integration constants\\ $\{\f_{I},\,\f_{L},\,\c_{I},\,A_{I},\,\r_{o}\}$, supplemented by the discrete choice of $\zeta$. It is worth noting that for the choice $\f_{L}=0$ we recover the IR expansions for the confining and skewed classes of solutions, when $\zeta=+1$ and $\zeta=-1$ respectively. We observe that the singularity parameter $\f_{L}$ is not entirely unconstrained; from the above expansions it can be seen that to ensure $\c$ and $A$ are real we must impose that $\f_{L}\geqslant-\frac{1}{\sqrt{5}}$, and by saturating this bound (and choosing $A_{I}=4\c_{I}$) one recovers the warp factor constraint $A=4\c$ satisfied by all domain-wall background solutions.\par
We furthermore observe from the complete expansion presented in Eq.~(\ref{Eq:GenSingPhiExp}) that, for any given value of $n$ and with $\f_{L}>0$, the most rapidly diverging exponent at $\r=\r_{o}$ is the sub-leading correction with $j=2n$ given by $2n(1-3\f_{L})$; we hence deduce that to ensure the IR singularity of $\f$ is logarithmic in $\r$ we require that all sub-leading exponents in the expansion are positive, or equivalently that $\f_{L}<\frac{1}{3}$. If we instead consider the case of $\f_{L}<0$ then the most singular contribution comes from the $j=0$ exponent, and the corresponding bound is $\f_{L}>-1$; this is less stringent than the requirement that $\c$ and $A$ be real, and is hence of no consequence. Combining these two bounded constraints then, we have the following allowed interval for the parameter $\f_{L}$:      
\begin{equation}
\label{Eq:GenSingBound}
-\frac{1}{\sqrt{5}}\leqslant\f_{L}<\frac{1}{3}\,.
\end{equation}
Finally, we observe that the limit  $\f_{L}\to\frac{1}{3}$ is completely pathological and renders the general series expansions for this class unusable; for every one of the infinite values of $n$ there are additive contributions to the expansions which all scale as $(\r-\r_{o})^{p}$ for $p=0,\,\frac{4}{3},\,\frac{8}{3}\ldots$, and hence no truncation is possible. This issue brings us to the one remaining branch of solutions which we shall require for our analysis, and which we will introduce in the next section.
\vspace{4mm}
\begin{table}[h!]
	\begin{center}
		\begin{tabular}{|c|c|c|c|c|}
			\hline\hline
			& $\f_{L}=-5^{-\frac{1}{2}}$ &  $-5^{-\frac{1}{2}}<\f_{L}<0$ &  $\f_{L}=0$ & $0<\f_{L}<\frac{1}{3}$ \cr
			\hline\hline
			$\zeta=+1$  & Good, DW & Good & Confining & Bad \cr
			$\zeta=-1$  & Good, DW & Good & Skewed & Bad \cr
			\hline\hline
		\end{tabular}
	\end{center}
	\vspace{-3mm}
	\caption[Parametrisation of \emph{general singular} backgrounds within the six-dimensional supergravity]{Parametrisation of the solutions obtainable from Eqs.~(\ref{Eq:GenSingPhi}\,-\,\ref{Eq:GenSingA}): here \emph{Good} and \emph{Bad} refer to which type of singularity is present at the end of space. 
		For $\f_{L}=-5^{-\frac{1}{2}}$ both choices of $\zeta=\pm1$ correspond to the same family of good singularity domain-wall backgrounds. 
	}
	\label{Tbl:GenSing}
\end{table}   

\subsubsection{Badly singular domain-wall (BSDW) solutions}
As anticipated, there exists one additional class of solutions which provide the non-trivial limiting case of the more general singular solutions presented in the previous section, and as their name suggests they satisfy the domain-wall constraint $A=4\c$ which ensures that five-dimensional Poincar\'{e} invariance is locally preserved. Their series expansions in proximity to the end of space at $\r=\r_{o}$ are given by~\cite{Elander:2020ial}   
\begin{align}
\label{Eq:BSDWPhiExpansion}
\f(\r)&=\frac{1}{3}\ln\left(\frac{3}{2}\right)+\frac{1}{3}\ln(\r-\r_o)+ \f_{b}(\r-\r_o)^{4/9}
+\sum_{j=2}^{\infty}f_j \,(\r-\r_o)^{\frac{4j}{9}}\,,\\
\c(\r)&=\c_I+\frac{1}{27}\ln(\r-\r_o)
+\frac{2}{5}\f_{b}\, (\r-\r_o)^{4/9} 
+\sum_{j=2}^{\infty}g_j \,(\r-\r_o)^{\frac{4j}{9}}\,,
\end{align}
where $\c_{I}$ is an integration constant which may be fixed arbitrarily, and $\f_{b}$ assumes the role of the one free parameter which may be dialled to generate an entire family of distinct background solutions. The sub-leading terms have as coefficients the polynomial functions $f_{j}=f_{j}(\f_{b})$ and $g_{j}=g_{j}(\f_{b})$, specific details of which may be found in Appendix C of Ref.~\cite{Elander:2020ial}.\par
\begin{figure}[h!]
	\begin{center}
		\makebox[\textwidth]{\includegraphics[width=0.44\paperwidth]
			{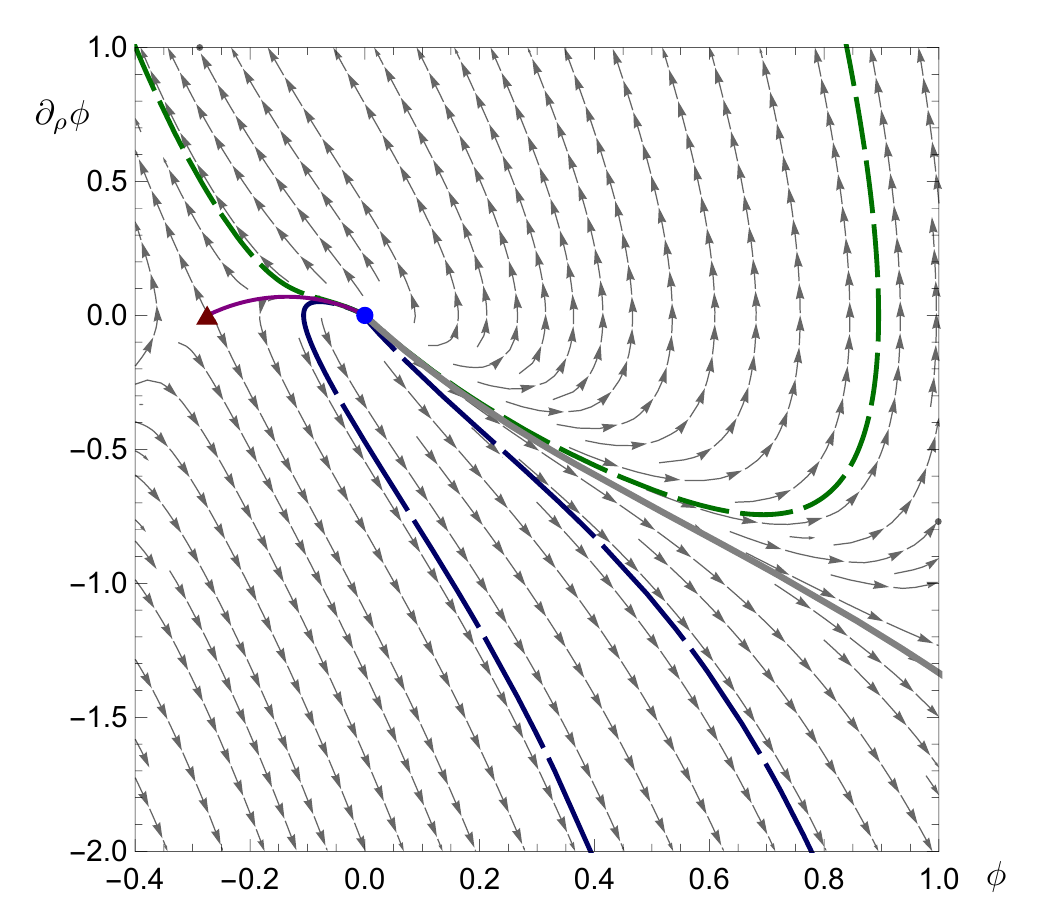}}
	\end{center}
\vspace{-5mm}
	\caption[Parametric plot of domain-wall solutions in the six-dimensional supergravity]{Parametric plot of $\pa_{\r}\f$ as a function of $\f$ for solutions 
			which satisfy the warp factor constraint $A=\frac{4}{3}\ca=4\c$. 
The blue disk and dark-red triangle respectively denote the UV and IR critical points of the
six-dimensional potential $\mathcal{V}_{6}$, the purple line represents the class of IRC solutions with duals which flow between these two critical points, 
and the light-grey line represents the analytical SUSY solutions.\ 
The dark-grey arrows exhibit the underlying vector field appearing in the first-order differential equation for $(\f, \pa_{\r} \f)$.
We also show two representative examples of the (good) singular solutions obeying the IR expansion in Eq.~(\ref{Eq:GenSingPhiExp}), for $\f_L=-1/\sqrt{5}$ and $\f_{I}=-0.3, 0.1$ (long-dashed dark blue), and two examples of the BSDW solutions with $\f_{b}=-0.06,40$ (dashed, dark green). We observe that the SUSY solutions form the separatrix between numerical backgrounds which flow to good ($\f\to\infty$) and bad ($\f\to-\infty$) singularities for positive $\f$, while the IR-conformal solutions play the same role when $\f$ is negative.}
\label{Fig:ParametricPlot}
\end{figure}

Recall from Sec.~\ref{Sec:RomansEOMs} that with some simple algebraic manipulation we derived from the equations of motion in Eq.~(\ref{Eq:EOMc1}) and (\ref{Eq:EOMc3}) a second-order differential equation dependent only on the scalar $\f$, and which is satisfied by all classes of solutions which obey the DW constraint $A=\frac{4}{3}\ca=4\c$. For convenience we reproduce this equation below:
\begin{equation*}
0=3\f''+\sqrt{5}\f'
\Big[\big(3\f'\big)^{2}+\g^{-3}\Big(9\g^{4}+12\g^{2}-1\Big)\Big]^{\frac{1}{2}}
+\g\Big(3-4\g^{-2}+\g^{-4}\Big)\,,
\end{equation*}
and we remind the Reader that here $\g\equiv e^{2\f(\r)}$. Now that we have finished our classification of admissible solution types, we find it illustrative to compare how those branches of solutions which satisfy the DW constraint flow away from the unique supersymmetric fixed point $\f=0$; these DW-type solutions are the SUSY and IRC classes, the subset of the general singular solutions which have $\f_{L}=-1/\sqrt{5}$, and the BSDW solutions of this subsection. To this end, in Fig.~\ref{Fig:ParametricPlot} we present the results of parametrically plotting $\big(\f,\pa_{\r}\f\big)$ for each of these classes, and for completeness we use the second-order differential equation above to also generate the underlying constrained vector field. 
\\
\clearpage 
We conclude Sec.~\ref{Sec:RomansClasses} by acknowledging that we are unable to state definitively that our classification of solutions as presented here is exhaustive, and that there may exist other types of singular solutions with more complicated IR behaviour which were omitted by our analysis. However, as a consistency check we carried out a complementary exercise which entails scanning the space of perturbations of the supersymmetric fixed point (controlled by the five universal UV parameters) and generating backgrounds which evolve back towards the end of space (note that this alternative method of constructing solutions is not numerically reliable as one approaches the small $\r$ region near to the end of space, and is hence not otherwise used); in so doing we did not find any evidence that additional classes of solutions exist in the theory.      


\section{Free energy derivation}
\label{Sec:FderivationRomans}
Our energetics analysis of this model and investigation into its phase space structure requires that we identify a method by which we are able to legitimately and unambiguously compare the stability of the various solution branches that we have presented; as earlier alluded to, this method will be the numerical computation of the free energy density as a function of the five universal UV expansion parameters, holographically renormalised and appropriately rescaled.\par 
Our starting point is the six-dimensional defining action of the theory provided in Eq.~(\ref{Eq:6DAction}), truncated to retain only the scalar $\f$; the other supergravity fields  do not affect the classical equations of motion and hence we set them to zero for simplicity. Regulating boundaries are introduced in the deep IR and far UV regions of the bulk geometry (at $\r_{i}$ for $i=1,2$, respectively), and we must therefore supplement the bulk action $\cs_{B}$ with two types of general boundary-localised contributions: $\cs_{\ck,i}$ are the Gibbons-Hawking-York terms and $\cs_{\l,i}$ are boundary-localised potentials. In addition to the usual UV regulator---which is a requisite in the process of holographic renormalisation~\cite{Bianchi:2001kw,Skenderis:2002wp,Papadimitriou:2004ap}---we include the IR regulator due to the fact that a subset of solutions contain a singularity at the end of space; we regulate the bulk by restricting the space to the open interval $\r\in(\r_{1},\r_{2})$ with the understanding that physical results for the free energy are obtained by removing the regulators, by taking the limits $\r_{1}\to\r_{o}$ and $\r_{2}\to\infty$ (where $\r_{o}$ is the end of the geometry).\par 
The action we shall adopt is written as follows:     
\begin{align}
\cs&=\cs_{B}+\sum_{i=1,2}\big(\cs_{\ck,i}+ \cs_{\l,i}\big)\nonumber\\
&=\int\is\text{d}^{4}x\,\text{d}\eta\,\text{d}\r\, \sqrt{-\hat{g}_{6}}\bigg(\frac{\car_{6}}{4}-\hat{g}^{\hat{M}\hat{N}}
\pa_{\hat{M}}\f\,\pa_{\hat{N}}\f-\cv_{6}(\f)\bigg)\nonumber\\
&\hspace{25mm}+\sum_{i=1,2} (-)^{i}\is\int\is\text{d}^{4}x\,\text{d}\eta\,\sqrt{-\tilde{\hat g}}
\bigg(\frac{\ck}{2}+\l_{i} \bigg)\bigg|_{\r=\r_{i}} \,,
\end{align}
where $\hat{g}_{\hat{M}\hat{N}}$ is the metric tensor for the six-dimensional line
element in Eq.~(\ref{Eq:6Dmetric}) (for $V_M=0$), $\hat{g}_{6}=-e^{8A-2\c}$ is its determinant,
$\car_{6}$ is the corresponding Ricci scalar, $\tilde{\hat g}_{\hat{M}\hat{N}}$ is the metric induced on each boundary, $\ck$ is the extrinsic curvature scalar, and $\l_{i}$ are the boundary-localised potentials. As we discussed for a generic five-dimensional system in Sec.~\ref{Subsec:HolographicFormalism}, to construct the analogous boundary-induced metric $\tilde{\hat g}_{\hat{M}\hat{N}}$ here for our six-dimensional model we must first introduce the six-vector $n_{\hat{M}}=(0,0,0,0,1,0)$ so that the following defining relations are satisfied:
\begin{align}
\label{Eq:OrthoCond1}
1&=\hat{g}_{\hat{M}\hat{N}}n^{\hat{M}}n^{\hat{N}}=n^{\hat{M}}n_{\hat{M}}\, ,\\
\label{Eq:OrthoCond2}
0&=\tilde{\hat{g}}_{\hat{M}\hat{N}}n^{\hat{M}}\,,
\end{align}
which together ensure that $n_{\hat{M}}$ is orthonormal to each five-dimensional regulating boundary. We therefore define the induced metric tensor as
\begin{equation}
\label{Eq:InducedMetric}
\tilde{\hat g}_{\hat{M}\hat{N}}\equiv \hat{g}_{\hat{M}\hat{N}}-n_{\hat{M}}n_{\hat{N}}\,.
\end{equation}
The covariant derivative acting on a generic (0,1)-tensor $f_{\hat{M}}$  is written in terms of the metric connection as
\begin{align}
\label{Eq:CovariantDeriv}
\nabla_{\hat{M}}f_{\hat{N}}&\equiv\pa_{\hat{M}}f_{\hat{N}}-\Gamma^{\hat{Q}}_{\hat{M}\hat{N}}f_{\hat{Q}}\, ,\\
\label{Eq:Connection}
\Gamma^{\hat{P}}_{\hat{M}\hat{N}}&\equiv\frac{1}{2}\hat{g}^{\hat{P}\hat{Q}}\Big(\pa_{\hat{M}}\hat{g}_{\hat{N}\hat{Q}}
+\pa_{\hat{N}}\hat{g}_{\hat{Q}\hat{M}}
-\pa_{\hat{Q}}\hat{g}_{\hat{M}\hat{N}}\Big)\,,
\end{align}
so that we have the following expression for the extrinsic curvature:
\begin{align}
\ck\equiv \hat g^{\hat{M}\hat{N}}\ck_{\hat{M}\hat{N}}
&\equiv\hat g^{\hat{M}\hat{N}}\nabla_{\hat{M}}n_{\hat{N}}\\
&=-\hat{g}^{\hat{M}\hat{N}}\Gamma^{5}_{\hat{M}\hat{N}}
=4\pa_{\r}A-\pa_{\r}\c\,.
\end{align}
As anticipated in Sec.~\ref{Sec:EnergeticsFormalism}, we must evaluate the complete on-shell action in order to obtain our result for the free energy density $\cf$. Let us first consider the bulk contribution $\cs_{B}$, which may be rewritten as a total derivative by making use of Eq.~(\ref{Eq:EOMbeta}); recalling our definitions $\a\equiv 4A-\c$ and $\b\equiv A-4\c$, the six-dimensional Ricci scalar is given by
\begin{equation}
\car_{6}=-2\Big(\a''+10(A')^{2}+7(\c')^{2}-8A'\c'\Big)\,,
\end{equation}
while Eq.~(\ref{Eq:EOMbeta}) provides us with the following relation:
\begin{equation}
17A'\c'=\b''+4\Big[(A')^{2}+(\c')^{2}\Big]\, ,
\end{equation} 
where primes denote differentiation with respect to $\r$. After some algebraic manipulation we find
\begin{equation}
\cs_{B}\equiv\cs_{B,1}+\cs_{B,2} = -\frac{3}{8}\int_{\r_{1}}^{\r_{2}}\is
\text{d}^{4}x\,\text{d}\eta\,\text{d}\r\,
\pa_{\r}\Big(e^{\a}\pa_{\r}A\Big)\, .
\end{equation}
We can also write explicitly the boundary-localised actions $\cs_{\ck,i}$ and $\cs_{\l,i}$ as follows:
\begin{align}
\cs_{\ck,1}&=
-\frac{1}{2}\int\is\text{d}^{4}x\,\text{d}\eta\, e^{\a}\Big(\pa_{\r}\a\Big)\Big\rvert_{\r=\r_{1}}
\,,\\
\cs_{\l,1}&=
-\int\is\text{d}^{4}x\,\text{d}\eta\, e^{\a}\Big(\l_{1}\Big)\Big\rvert_{\r=\r_{1}}\,,\\
\cs_{\ck,2}&=
\frac{1}{2}\int\is\text{d}^{4}x\,\text{d}\eta\, e^{\a}\Big(\pa_{\r}\a\Big)\Big\rvert_{\r=\r_{2}}
\,,\\
\cs_{\l,2}&=
\int\is\text{d}^{4}x\,\text{d}\eta\, e^{\a}\Big(\l_{2}\Big)\Big\rvert_{\r=\r_{2}}\label{Eq:Spot}\,.
\end{align}
The free energy $F$ and the free-energy density ${\cal F}$ are defined in terms of the complete action $\cs$ via the relation
\begin{equation}
\label{Eq:DefineF}
F\equiv - \lim_{\r_1\rightarrow \r_o}\lim_{\r_2\rightarrow +\infty}
{\cal S}\equiv\int\is\text{d}^{4}x\,\text{d}\eta\,\cf\, ,
\end{equation}
so that after summing the various contributions to $\cs$, we obtain the following general result:
\begin{align}
\label{Eq:RomansGeneralF}
{\cal F}&= \lim_{\r_1\rightarrow \r_o}\frac{1}{8}e^{\a}\Big(13\partial_{\r}A-4\partial_{\r}\chi+8\lambda_{1}\Big)\Big|_{\r_1}\notag\\
&-\lim_{\r_2\rightarrow +\infty}\frac{1}{8}e^{\a}\Big(13\partial_{\r}A-4\partial_{\r}\chi+8\lambda_{2}\Big)\Big|_{\r_2}\,.
\end{align}
For this expression to be of any use, it is necessary that we next specify the two boundary-localised potentials $\l_{i}$. The choice for the IR potential $\l_{1}$ is determined by the requirement that the variational principle be well defined, and that we recover the classical equations of motion (supplemented by $\r=\r_{1}$ boundary conditions for $\f$, $\c$, and $A$) when taking the variation of $S_{B}$ together with the IR boundary actions; this requisite condition selects $\l_{1}=-\frac{3}{2}\pa_{\r}A$ (see Ref.~\cite{Elander:2010wd} for details).\par 
Before we discuss the UV boundary potential, let us first make an important observation. We noted when introducing $\cs$ that we need to include a regulatory boundary in the deep IR region of the bulk geometry, despite the fact that some of the solutions we are considering (the \emph{confining} class) are completely regular and smooth at small $\r$. If we consider the sum contribution of the two IR boundary-localised actions $\cs_{\ck,1}$ and $\cs_{\l,1}$, with $\l_{1}$ now defined above, we have 
\begin{equation}
\cs_{\ck,1} + \cs_{\l,1}=-\frac{1}{2}\int\is\di^{4}x\,\di\eta
\Big( e^{\a}(\pa_{\r}A-\pa_{\r }\c)\Big)\Big|_{\r_1}\,,
\end{equation}
which, by direct substitution of the IR expansions presented in Eq.~(\ref{Eq:RomansChiIRExpansion}) and (\ref{Eq:RomansAIRExpansion}), we see has a vanishing integrand in the $\r\to\r_{o}$ limit. Hence, in this limit the free energy for the class of regular solutions is unaffected by the presence of boundary-localised terms in the deep IR, and our inclusion of an IR regulator (necessary for singular backgrounds) is justified.\par 
Our prescription for the boundary potential $\l_{2}$ is dictated by the requirements that our choice is covariant, and that it ensures the cancellation of all divergences for our asymptotically (locally) $\text{AdS}_{6}$ background solutions in the far UV. By substituting in for $\c$ and $A$ using the UV expansions, and implementing the coordinate change defined by $\r=-\frac{3}{2}\ln(z)\Rightarrow\pa_{\r}=-\frac{2}{3}z\pa_{z}$, we have the following boundary contributions:   
\begin{align}
\cs_{B,2}&=\int\is\di^{4}x\,\di\eta\, \frac{e^{\a_{U}}}{z^5}\left(-\frac{1}{3}+\frac{\f_{2}^{2}}{12}z^{4}
+\frac{1}{80} \big(4\f_{2}\f_{3} +25\c_{5}\big)z^{5} 
+\ldots \right)\bigg|_{\r_2}\,,\\
\cs_{\ck,2}&=\int\is\di^{4}x\,\di\eta\, \frac{e^{\a_{U}}}{z^5}\,\left(\frac{5}{3}-\frac{5}{12}\f_{2}^{2}z^{4}
+ 0\times z^{5} + \ldots \right)\bigg|_{\r_2}\,,\\
\cs_{\l,2}&=\int\is\di^{4}x\,\di\eta\, \frac{ e^{\a_{U}}}{z^5}\l_{2}\left(1-\frac{5}{4}\f_{2}^{2}z^{4}
-\frac{12}{5}\f_{2}\f_{3}z^{5} +\ldots \right)\bigg|_{\r_2}\, ,
\end{align}
where we have defined $\a_{U}\equiv 4A_{U}-\c_{U}$ and have truncated to show only terms up to zeroth order in $z$. We see that all three contributions contain divergences in the physical $z\to 0$ limit, and that a subset of these divergent terms are proportional to the (squared) deformation parameter $\f_{2}$ which sources the operator dual to $\f$ in the boundary field theory. As anticipated with our introduction of the SUSY branch of solutions, we notice that the superpotential expansion $\cw_2(\f)$ which was presented in Eq.~(\ref{Eq:W2}) provides exactly the necessary counter-terms to ensure that all divergences in the combined contributions of the UV-localised actions cancel; moreover it is sufficient to retain terms in $\cw_2(\f)$ up to quadratic order in $\f$, and sub-leading power corrections are not necessary to ensure that $\cf$ is appropriately renormalised. We can demonstrate this explicitly by substituting $\l_{2}=\cw_2(\f)\approx -\frac{4}{3}(1+\f^{2})$ into $\cs_{\l,2}$ to obtain:
\begin{equation}
\cs_{\l,2}=\int\is\di^{4}x\,\di\eta\, \frac{e^{\a_{U}}}{z^5}\left(-\frac{4}{3}+\frac{1}{3}\f_{2}^{2}z^{4}
+\frac{8}{15}\f_{2}\f_{3}z^{5} +\ldots \right)\bigg|_{\r_2}\,,
\end{equation}
and then summing the $\r=\r_{2}$ boundary actions. The integrand divergences exactly cancel, the $z\to 0$ limit becomes well defined, and we are left with a finite contribution to the free energy.\par
It is important to clarify, however, that although our prescription of $\l_{2}=\cw_2(\f)$ is convenient and physically motivated, it is not necessarily a unique choice. Indeed, the freedom to add finite counter-terms to the renormalised action means that the potential $\l_{2}$ and its second derivative with respect to the source $\f_{2}$ are scheme-dependent, and hence the same is also true for the free energy. While this dependence on the implemented subtraction scheme is a well-documented feature of free energy calculations in the context of holography (see for example the discussion in Ref.~\cite{Bobev:2013cja}), it does consequently mean that the concavity theorems typical of classical statistical mechanics can not intuitively be applied to our energetics analysis of this system.\par 
With our prescription for the boundary-localised potentials $\l_{i}$ specified, from Eq.~(\ref{Eq:RomansGeneralF}) we now have the following expression for $\cf$:     
\begin{equation}
\label{Eq:Flambda}
{\cal F}=\lim_{\r_1\rightarrow \r_o}\frac{e^{\a}}{8}\bigg(\pa_{\r}A-4\pa_{\r}\c\bigg)\bigg|_{\r_1}
-\lim_{\r_2\rightarrow +\infty}\frac{e^{\a}}{8}\bigg(13\pa_{\r}A-4\pa_{\r}\c+8\cw_{2}\bigg)\bigg|_{\r_2}\,.
\end{equation}
The final step in our derivation of the free energy starts with the observation that the first term of Eq.~(\ref{Eq:Flambda}) is proportional to the conserved quantity defined in Eq.~(\ref{Eq:C}), and is hence equal to some background-dependent constant which is invariant with respect to the radial coordinate. Rather conveniently, this implies that we are free to evaluate the conserved term at the UV boundary---rather than in the IR---where we find that it provides a finite contribution to $\cf$. Reformulated in this way, we have
\begin{equation}
\label{Eq:Fuv}
\cf=-\lim_{\r_2\rightarrow +\infty}e^{\a}\bigg(\frac{3}{2}\pa_{\r}A+\cw_{2}\bigg)\bigg|_{\r_2}\,.
\end{equation}
Equivalently, by treating the two terms separately and simply substituting in for the UV expansions, we find that 
\begin{align}
\label{Eq:RomansQuasiFinalF}
\cf&=\frac{1}{16}e^{\a_{U}}\Big(4\phi_{2}\phi_{3}+25\chi_{5}\Big)
-\frac{1}{48}e^{\a_{U}}\Big(28\phi_{2}\phi_{3}+15\chi_{5}\Big)\\
\label{Eq:RomansFinalF}
&=-\frac{1}{12}e^{\a_{U}}\Big(4\phi_{2}\phi_{3}-15\chi_{5}\Big)\,,
\end{align}
where the first term of Eq.~(\ref{Eq:RomansQuasiFinalF}) comes from the evaluation at the UV boundary of the conserved quantity in Eq.~(\ref{Eq:Flambda}). As we earlier anticipated, our final result for the free energy density is a function solely of the deformation parameters $\{\f_{2},\,\f_{3},\,\c_{5},\,\c_{U},\,A_{U}\}$ which characterise the asymptotic field expansions in the far UV, and is therefore universally applicable to every branch of solutions that we have discussed. For those backgrounds which locally preserve five-dimensional Poincar\'{e} invariance by satisfying the domain-wall constraint $A=4\c$, we find from the UV expansions that the deformation parameters further satisfy
\begin{align}
A_{U}&=4\c_{U}\,,\\
\label{Eq:Chi5}
\c_{5}&=-\frac{4}{25}\f_{2}\f_{3}\,,
\end{align}
so that
\begin{equation}
\label{Eq:RomansDWF}
\cf^{(DW)}=-\frac{8}{15}e^{15\c_{U}}\f_{2}\f_{3}\,.
\end{equation}\par 
We conclude this section by stating that, in order to facilitate the comparison of $\cf$ between the various classes, we will henceforth always choose to set $\c_{U}=0$ and $A_{U}\hspace{-0.3mm}=\hspace{-0.3mm}0$; the former identification may be implemented by rescaling the holographic coordinate via $z\to ze^{3\c_{U}}$, while the latter is permitted since the classical equations of motion describing the system are invariant under a simple additive shift of the warp factor $A\to A-A_{U}$.

\section{Scale setting and numerical implementation}
\label{Sec:RomansScaleSetting}
\subsubsection{Scale setting}
Having derived our expression for the free energy density $\cf$ from the holographically renormalised on-shell action, we proceed by introducing the scale setting procedure which was briefly mentioned in Sec.~\ref{Sec:EnergeticsFormalism}, and which will be a crucial component of our energetics analysis.\par 
To appreciate the necessity of this procedure, we remind the Reader that two of the branches of solutions that were discussed in Sec.~{\ref{Sec:RomansClasses}}, those which we refer to as \emph{confining} and \emph{skewed}, are related to one another via the transformation $\b\to-\b$ (where $\b$ represents the linear combination $\b=A-4\c$). As we saw, this condition is encoded by the identities of Eqs.~(\ref{Eq:id1}\,-\,\ref{Eq:id3}), however we also remind the Reader of the caveat that these parameter relations are satisfied only up to an additive constant. To demonstrate this point, for the confining solutions we can consider evaluating the conserved quantity Eq.~(\ref{Eq:C}) using the IR expansions for $\c$ and $A$ in the $\r\to\r_{o}$ limit, and using the UV expansions in the $z\to 0$ limit, and then equating the two expressions. We can repeat this exercise instead for the skewed class of solutions, and we obtain the following parameter relations:  
\begin{align}
-\frac{10}{3}&=e^{\a_{U}^{c}-\a_{I}^{c}}
\Big(4\f_{2}^{c}\f_{3}^{c}+25\c_{5}^{c}\Big)\,,\\
\frac{10}{3}&=e^{\a_{U}^{s}-\a_{I}^{s}}
\Big(4\f_{2}^{s}\f_{3}^{s}+25\c_{5}^{s}\Big)\,,
\end{align}
where we have defined $\a_{U}\equiv 4A_{U}-\c_{U}$ and $\a_{I}\equiv 4A_{I}-\c_{I}$, and the superscripts $c$ and $s$ denote evaluation using the IR expansions for the confining and skewed classes, respectively. We notice that by making the substitutions dictated by the identities Eqs.~(\ref{Eq:id1}\,-\,\ref{Eq:id3}) we are able to recover one of these relations given the other, assuming that the exponential terms are identical. As earlier stated, we can always rescale the radial coordinate to set $A_{U}=\c_{U}=0$ and we are free to choose $A_{I}=\c_{I}$, though we are still left with the requirement that $\c_{I}^{c}=\c_{I}^{s}$.\par
This is precisely the issue that we seek to address: while in the case of the confining solutions the IR parameter $\c_{I}^{c}$ takes a fixed value determined by the necessity of avoiding a conical singularity at the end of space, no such constraint is imposed on $\c_{I}^{s}$ for the skewed solutions. In the latter case the geometry does not smoothly close off at the end of space---since the size of the circle diverges---and hence $\c_{I}^{s}$ is a free parameter. We therefore deduce that the space of free parameters for the confining and skewed classes have different \emph{dimensionality}, and that in order to properly compare them we must implement an appropriate scale setting procedure with which the ambiguity in choosing $\c_{I}$ for the skewed solutions is resolved.\par
To this end, and motivated by the discussion in Ref.~\cite{Csaki:2000cx}, we introduce a universal energy scale $\Lambda$ which we conveniently define via the time taken by a massless particle to reach the end of space from the UV boundary: 
\begin{equation}
\label{Eq:Lambda}
\Lambda^{-1}\equiv t \equiv\int_{r_{o}}^{\infty}\is\text{d}\tilde{r}\,
\sqrt{\frac{g_{rr}}{|g_{tt}|}}=
\int_{r_{o}}^{\infty}\is\text{d}\tilde{r}\,e^{-A(\tilde{r})}=\int_{\r_{o}}^{\infty}\is\text{d}\tilde{\r}\,e^{\c(\tilde{\r})-A(\tilde{\r})}\, ,
\end{equation}
where the absolute value of the metric component $g_{tt}$ ensures that $\Lambda\in\mathbb{R}$ irrespective of which Minkowski metric signature is adopted, and where $\c$ and $A$ are evaluated on the numerical backgrounds. Let us now examine how this energy scale may actually be used, by first considering the trivial rescaling of the coordinates $x^{\mu}\to\s x^{\mu}$ and $\eta\to\s\eta$. We see from the six-dimensional metric in Eq.~(\ref{Eq:6Dmetric}) that this transformation is equivalent to the rigid shifts $\c\to\c+\frac{1}{3}\ln(\s)$ and $A\to A+\frac{4}{3}\ln(\s)$, and from the UV expansions in Eqs.~(\ref{Eq:RomansUVexpChi},\,\ref{Eq:RomansUVexpA}) we observe that these shifts should be supplemented by the radial rescaling $z\to \s z \-\ \Leftrightarrow \-\ \r\to\r\,-\,\frac{3}{2}\ln(\s)$ to ensure that $A_{U}=\c_{U}=0$. Consequently, the remaining non-zero UV parameters are rescaled as
\begin{align}
&\f_{2}\to\s^{2}\f_{2}\,,\\
&\f_{3}\to\s^{3}\f_{3}\,,\\
&\c_{5}\to\s^{5}\c_{5}\,,
\end{align}
while the energy scale $\Lambda$ undergoes the transformation $\Lambda\to\s\Lambda$. We may therefore construct dimensionless UV parameters by rescaling $\{\f_{2},\,\f_{3},\,\c_{5}\}$ with appropriate powers of $\Lambda$ as follows:
\begin{align}
\hat{\f}_{2}&\equiv\f_{2}\Lambda^{-2}\,,\\
\hat{\f}_{3}&\equiv\f_{3}\Lambda^{-3}\,,\\
\hat{\c}_{5}&\equiv\c_{5}\Lambda^{-5}\,,
\end{align}
and we can also define $\hat{\cf}\equiv\cf\Lambda^{-5}$. Henceforth, we shall adopt this hatted notation to denote physical quantities which have been rescaled in this manner. In the next subsection we shall provide a thorough description of the numerical procedure with which we extract the UV parameter data for the various classes of solutions, and in so doing we will also clarify how exactly this scale setting method implements the aforementioned $\c_{I}^{c}=\c_{I}^{s}$ constraint.         

\subsubsection{Numerical implementation}
At this stage we have derived a general expression for the free energy density as a function of a set of universal UV deformation parameters $\{\f_{2},\,\f_{3},\,\c_{5},\,\c_{U},\,A_{U}\}$, and have now introduced a physically motivated energy scale which will allow us to compare $\cf$ for all branches of solutions. It only remains that we provide a description of how exactly we may obtain the necessary parameter data, and the numerical techniques that we employ in the process.\par
We start by reiterating the general outline provided in Sec.~\ref{Sec:EnergeticsFormalism}, with some elaboration; the numerical routine is implemented for each class of solutions as follows.
\begin{enumerate}
	\item For any given choice of the free parameters which characterise the IR expansions, and having fixed the physical end of space by assigning $\r_{o}=0$, we construct numerical backgrounds for $\f(\r)$, $\c(\r)$, and $A(\r)$ by setting up boundary conditions near to the end of space and evolving the solutions towards the UV using the equations of motion.
	\item We match these numerical backgrounds (and their first derivatives) to the general UV expansions at some value $\r=\r_{m}$, solving for each UV parameter in turn to extract the set $\{\f_{2},\,\f_{3},\,\c_{5},\,\c_{U}\neq 0,\,A_{U}\neq 0\}$. The value of the holographic coordinate $\r_{m}$ at which the matching is performed should be chosen carefully; one must ensure that numerical noise effects are minimised, while also ensuring that $\r_{m}$ is sufficiently large that the background $\f(\r)$ has closely converged to the UV fixed point $\f=0$.
	\item We rescale the radial coordinate $z\to ze^{3\c_{U}}$ and then shift the warp factor background $A(\r)\to A(\r)-A_{U}$, using the values of $\c_{U}$ and $A_{U}$ obtained in the previous step, to set $\c_{U}=A_{U}=0$. We match the resulting rescaled background profiles to the UV expansions again to extract the new set of parameter data $\{\bar{\f}_{2},\,\bar{\f}_{3},\,\bar{\c}_{5},\,\bar{\c}_{U}=0,\,\bar{A}_{U}=0\}$, where we use bars here to emphasise that the other parameters have also been rescaled in the process.
	\item Finally, we compute the scale parameter $\Lambda$ defined in Eq.~(\ref{Eq:Lambda}) by substituting in for the rescaled background solutions $\c(\r)$ and $A(\r)$, integrating over their entire domain. For each numerical background we hence obtain the parameter data $\{\hat{\f}_{2},\,\hat{\f}_{3},\,\hat{\c}_{5}\}$ and compute $\hat{\cf}$ using Eq.~(\ref{Eq:RomansFinalF}).     
\end{enumerate}
While this schematic overview of the numerical procedure is fairly general, there are some subtle technicalities associated with some of the classes of solutions which we shall now address individually.\par 
We start with the family of supersymmetric solutions, for which we notice that the IR expansion for $\f$ in Eq.~(\ref{Eq:RomansSUSYphi}) contains no free parameters once the end of space has been fixed; for this class we find that all backgrounds have $\f_{2}=0$ when matched to the UV expansions, and hence from Eq.~(\ref{Eq:RomansDWF}) we see that they have exactly vanishing free energy. Furthermore, we observe that the integral of Eq.~(\ref{Eq:Lambda}) diverges when evaluated on these solutions ($\Lambda\to 0$), and hence the scale setting of $\hat{\cf}=\cf\Lambda^{-5}$ is poorly defined; this is not problematic however, as the vanishing free energy $\cf=0$ would remain so irrespective of any rescaling we might apply.\par 
We remind the Reader that any one numerical background within the IRC class of solutions may be used to generate any other by simply shifting the radial coordinate $\r\to\r-\d$ for some arbitrary $\d$, and that $\r$ is not bounded from below (hence the dual field theories described by this class are scale invariant at low energies). As a consequence of this, the integral defining $\Lambda^{-1}$ is also divergent for this class. We may nevertheless compute the free energy for these solutions by observing that the following parameter ratio is an invariant quantity under a generic $z\to\s z$ rescaling:
\begin{equation}
\label{Eq:KappaIRC}
\kappa \equiv \frac{\lvert\f_{3}\rvert}{\lvert\f_{2}\rvert^{\frac{3}{2}}}\,,
\end{equation}
so that we may reformulate the free energy as
\begin{equation}
\label{Eq:FreeIRC}
\cf^{(DW)}\to\cf^{(IRC)}=-\frac{8}{15}\kappa\,\f_{2}\lvert\f_{2}\rvert^{\frac{3}{2}}\,.
\end{equation}  
We then need only match a single numerical background $\f(\r)$ to the UV expansions in order to extract $\{\f_{2},\,\f_{3}\}$ (and hence $\kappa$), and then simply plot $\cf^{(IRC)}$ for $\f_{2}<0$. We determined that $\kappa\simeq 2.87979$.\par 
There are no specific numerical subtleties to mention for the class of confining solutions, and the parameter extraction is carried out as outlined above to obtain $\{\f_{2}^{c},\,\f_{3}^{c},\,\c_{5}^{c},\,\Lambda^{c}\}$. As we have discussed, the parameter data for the related skewed solutions can be obtained using the identities Eqs.~(\ref{Eq:id1}\,-\,\ref{Eq:id3}) without ever needing to match skewed numerical backgrounds to the UV expansions. Moreover, using Eq.~(\ref{Eq:ConfSkewBackgroundRelation}) we may compute $\Lambda$ for the skewed solutions by substituting instead for the confining backgrounds as follows:
\begin{equation}
(\Lambda^{s})^{-1}=\int_{\r_{o}}^{\infty}\is\text{d}\tilde{\r}\,
e^{\c^{s}(\tilde{\r})-A^{s}(\tilde{\r})}
=\int_{\r_{o}}^{\infty}\is\text{d}\tilde{\r}\,
e^{-\frac{3}{5}\big[\c^{c}(\tilde{\r})+A^{c}(\tilde{\r})\big]}\, ,
\end{equation} 
so that in fact we are not required to even generate the numerical background solutions for the skewed class; the complete set of data $\{\f_{2}^{s},\,\f_{3}^{s},\,\c_{5}^{s},\,\Lambda^{s}\}$ may be derived from the table of data for the confining class. This elucidates how the scale setting procedure effectively reduces the dimensionality of the space of free parameters for the skewed branch: by calculating $\Lambda^{s}$ in terms of the numerical background $\c^{c}(\r)$, the requirement that $\c_{I}^{s}=\c_{I}^{c}$ is manifestly satisfied.\par
The general singular numerical backgrounds are constructed using the IR expansions presented in Eqs.~(\ref{Eq:GenSingPhi}\,-\,\ref{Eq:GenSingA}), and each is uniquely identified by the choice of $\zeta$, $\f_{I}$, and $\f_{L}$; the parametrisation of this class is summarised in Table~\ref{Tbl:GenSing}. The procedure of matching to the UV expansions is just as described above, with one important caveat: we noticed that for the choice $\zeta=+1$ we were not able to reliably generate smooth numerical backgrounds, and hence the extracted parameter data could not be trusted. This issue is resolved by observing that the same relation which exists between the confining and skewed classes is more generally applicable to solutions with $\f_{L}\neq 0$ which differ by the choice of $\zeta=\pm 1$. For any given value of $\f_{L}$, we find that the following relation is satisfied: 
\begin{equation}
\label{Eq:GenSingBackgroundRelation}
0=\frac{3}{5}\Big[\c^{\mp}(\r)+A^{\mp}(\r)\Big]+\c^{\pm}(\r)-A^{\pm}(\r)\,,
\end{equation}
where the $+$ and $-$ superscript labels denote the choice $\zeta=+1$ and $\zeta=-1$ respectively. As a result, we see from the UV expansions that the parameter identities in Eqs.~(\ref{Eq:id1}\,-\,\ref{Eq:id3}) also hold true for backgrounds which differ only by the choice of $\zeta$:
\begin{align}
\f_{2}^{+}&=\f_{2}^{-}\, ,\\
\f_{3}^{+}&=\f_{3}^{-}\, ,\\
\c_{5}^{\pm}&=-\c_{5}^{\mp}-\frac{8}{25}\f_{2}^{\mp}\f_{3}^{\mp} \,.
\end{align}
Just as we obtained the parameter data for the skewed solutions from the confining ones, we may carry out the matching procedure for the $\zeta=-1$ backgrounds for various choices of $\f_{L}$, and then simply use the above identities to extract the corresponding set of data $\{\f_{2}^{+},\,\f_{3}^{+},\,\c_{5}^{+},\,\Lambda^{+}\}$.\par 
For completeness, we conclude by stating that for the branch of badly singular domain-wall (BSDW) solutions there are no specific numerical subtleties which need to be addressed, and the parameter matching process is carried out as described in the schematic outline above. A summary of the parametrisation of each class of backgrounds that we have discussed, together with any constraints which we are required to impose, is provided in Table~\ref{Tbl:ParameterSummary}; the UV parameters $\c_{U}$ and $A_{U}$ are omitted as they are always rescaled to zero. 
\begin{table}[h!]
	\begin{center}
		\begin{tabular}{|c|c|c|c|c|}
			\hline\hline
		Class & $\f_{2}$ & $\f_{3}$ & $\c_{5}$ & Scale setting \cr
			\hline\hline
SUSY & $0$ & Free & $\ca=3\c$ & None \cr
IRC & $<0$ & $\f_{3}=\kappa\f_{2}|\f_{2}|^{1/2}$ & $A=4\c$ & None \cr
Confining & Free & Curvature sing. & Conical sing. & $\Lambda$\cr
Skewed & Free & $\a^{s}=\a^{c}$ & $\b^{s}=-\b^{c}$ & $\Lambda$ \cr
Good sing. & Free & Free & Free & $\Lambda$ \cr
Bad sing. & Free & Free & Free & $\Lambda$ \cr
BSDW & Free & Free & $A=4\c$ & $\Lambda$ \cr
			\hline\hline
		\end{tabular}
	\end{center}
	\caption[Summary of parametrisation of backgrounds within the six-dimensional supergravity]{Summary of parametrisation, constraints, and scale setting procedure for each class of solutions in our energetics analysis of the circle-reduced supergravity.}
	\label{Tbl:ParameterSummary}
\end{table}\\

\section{Phase structure}
\label{Sec:RomansPhaseStruct}
\subsubsection{Free energy plots}
In computing the mass spectrum of scalar fluctuations in Sections \ref{Sec:RomansMassPlots} and \ref{Sec:RomansProbePlots} we uncovered the existence of a tachyonic state, which is symptomatic of a classical instability in the theory parameter space. As we have discussed, this observation leads us to conclude that there must by necessity exist a phase transition away from the unstable region of the branch of confining solutions to one of the other classes introduced in Sec.~\ref{Sec:RomansClasses}. Using the numerical procedure described in Sec.~\ref{Sec:RomansScaleSetting} we systematically compute the free energy for each class of solutions, and in so doing we demonstrate that our energetics analysis indeed reveals evidence of such a phase transition; we dedicate this section to presenting these results.\par
We start by reminding the Reader that each class of solutions in our analysis exhibits the same asymptotic behaviour in the far UV, and that they all correspond to deformations of the unique trivial solution $\f=0$. This solution corresponds to a background geometry that is $\text{AdS}_{6}$, and there are two such deformations which may be introduced to the dual five-dimensional theory.\par  
As shown in Eq.~(\ref{Eq:RomansDeltas}) the first of these corresponds to the insertion of a relevant ($\Delta=3$) operator $\co_{3}$; the source for this operator is identified with the leading order coefficient $\f_{2}$ in Eq.~(\ref{Eq:RomansUVexpPhi}), while its vacuum expectation value (VEV), or \emph{condensate}, is provided by the sub-leading coefficient $\f_{3}$ (see for example Ref.~\cite{Skenderis:2002wp} for a review). The second possible deformation is that of the compactification of one space-like dimension on a circle, the size of which is controlled by the additional scalar $\c$ introduced in the gravity theory; the boundary operator dual to this field is sourced by $\c_{U}$, and its VEV is encoded by the parameter $\c_{5}$.\par 
As we briefly mentioned in Sec.~\ref{Sec:EnergeticsFormalism}, the non-perturbative physics of the boundary theory manifests through non-trivial functional relations between the various UV parameters. From our numerical routine we find that $\f_{3}=\f_{3}(\f_{2})$ and $\c_{5}=\c_{5}(\f_{2})$ behave as non-linear response functions to the source $\f_{2}$, and are themselves determined within each branch of solutions, for each distinct background, by the value of $\f_{2}$. Hence, in the following plots we shall present the free energy $\hat{\cf}$ for each class as a function of the UV parameter $\hat{\f}_{2}$, and we remind the Reader that here hatted quantities have been rescaled by appropriate powers of the scale parameter $\Lambda$ defined in Eq.~(\ref{Eq:Lambda}). Given the number of solution classes which play a role in our analysis we find that the results are more conveniently portrayed using multiple plots, and we shall comment on some of their general features before discussing the phase transition itself.      
\begin{figure}[h!]
\begin{center}
	\makebox[\textwidth]{\includegraphics[width=0.54\paperwidth]
		{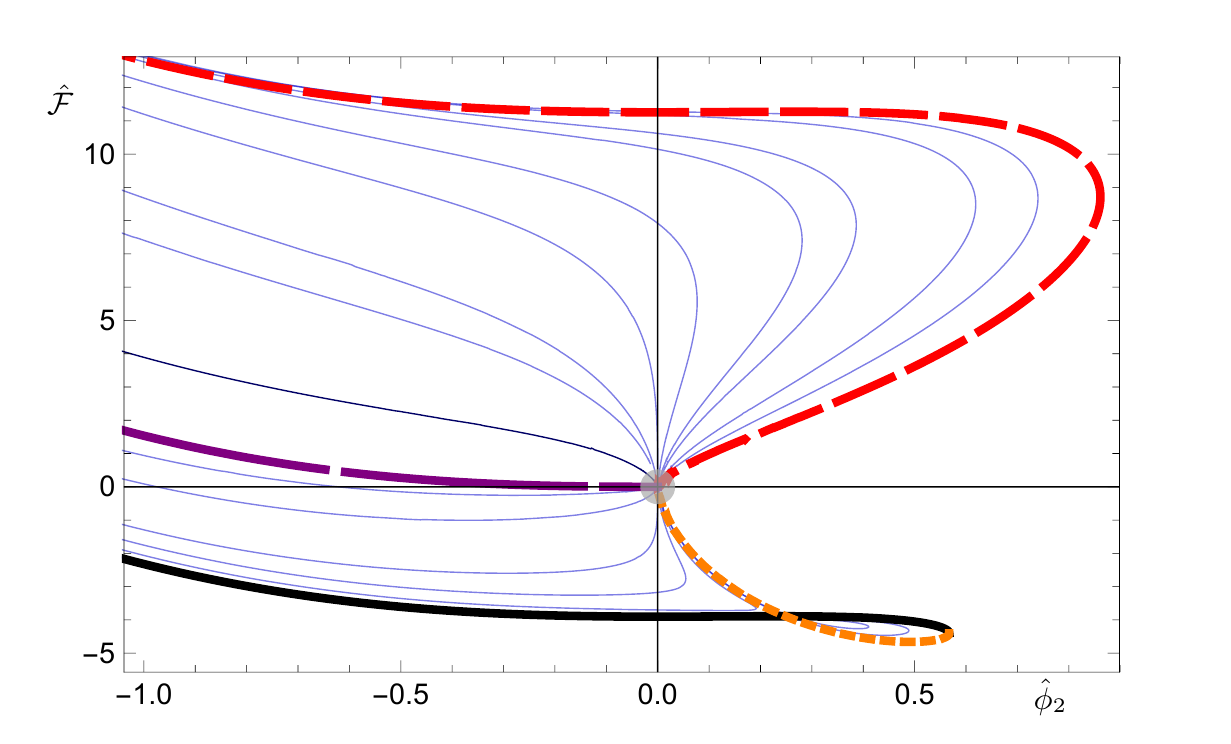}}
	\end{center}
\vspace{-5mm}
		\caption[First free energy plot for the six-dimensional supergravity]{The free energy density $\hat{\cal F}$ as a function of the deformation parameter $\hat\phi_2$ for the IR-conformal solutions (longest-dashed purple line), 
			the confining solutions (solid black and short-dashed orange lines), and the skewed solutions (dashed red line), compared 
			to a few representative choices of (good) singular solutions (thin blue lines). For the latter, we generated the numerical solutions from the IR
			expansions, by setting $(\phi_L,\zeta)=(-0.02,-1)$, $(-0.04,-1)$, $(-0.08,-1)$, $(-0.15,-1)$, $(-0.2, -1)$, $(-0.25,-1)$, $(-0.3,-1)$, $(-0.35,-1)$, $(-0.35,1)$, $(-0.3,1)$, $(-0.25,1)$, $(-0.2,1)$, $( -0.15,1)$, $(-0.04,1)$, $(-0.02,1)$, respectively (top blue line to bottom blue line), and varied the value of $\phi_I$. The darker blue line, separating the cases $\zeta = \pm 1$, corresponds to the domain-wall solutions obtained with $\phi_L = -1/\sqrt{5}$ and varying $\phi_I$. 
			The SUSY solutions are represented by a grey point at the origin. 
			The short-dashed orange line shows the region within the branch of confining solutions for which a tachyonic state appears in the scalar mass spectrum. (Note that the very top blue line crosses the red one for large negative values of $\hat \f_{2}$. We expect this to be a purely numerical artefact that could be removed with higher numerical precision.) }
		\label{Fig:FplotBlue}
\end{figure}\\
\indent Let us first consider the plot of Fig.~\ref{Fig:FplotBlue}, in which we present the free energy for all classes except for the $\f_{L}>0$ general singular solutions, and the BSDW backgrounds. We start with the simple case of the SUSY solutions, in which the scalar field $\f$ evolves monotonically towards a good singularity at the end of space, controlled by the formation of a non-zero condensate $\f_{3}$ associated with the $\Delta=3$ operator dual to $\f$. All of these solutions yield $\hat{\f}_{2}=0$ when matched to the UV expansions (with $\hat{\f}_{3}$ an unconstrained parameter), and hence their free energy $\hat{\cf}$ is always exactly vanishing. This entire class is therefore represented by a single grey point at the origin of the phase space.\par 
As we discussed when clarifying some numerical technicalities in Sec.~\ref{Sec:RomansScaleSetting}, the IR-conformal solutions are parametrised solely by the scale-invariant ratio $\k$ in Eq.~(\ref{Eq:KappaIRC}), which is defined in terms of the source $\f_{2}$ and VEV $\f_{3}$ associated with the $\co_{3}$ operator on the boundary. The free energy in this case is straightforward to compute: we determined that $\kappa\simeq 2.87979$ and then used Eq.~(\ref{Eq:FreeIRC}) to plot $\cf^{IRC}$ (note that the scale-invariant nature of these solutions guarantees that $\cf^{IRC}=\hat{\cf}^{IRC}$ would be identical for any other definition of $\Lambda$). This class is represented by the longest-dashed purple line, in the $\f_{2}<0$ region of the parameter space.\par 
The class of confining solutions, generated by varying the free IR parameter $\f_{I}$, are depicted in Fig.~\ref{Fig:FplotBlue} by the solid black line and the short-dashed orange line; the latter denotes those background solutions for which the corresponding spectrum of scalar fluctuations contains a tachyonic state (see Figs.~\ref{Fig:Spectra1} and \ref{Fig:RomansSpectrumProbe}), and hence highlights the region of the confining branch that contains an instability. The skewed solutions, which we repeat are related to the confining ones via the transformation $\b\to-\b$, are shown by the dashed red line.\par 
Lastly for this plot, the general singular solutions with $\f_{L}<0$ (those which encounter a good singularity at the end of space) are represented by the set of thin blue lines, for several representative values of $\f_{L}$ and for $\zeta=\pm 1$; these lines are generated by varying the remaining IR parameter $\f_{I}$. The domain-wall solutions obtained by setting $\f_{L}=-\frac{1}{\sqrt{5}}$ (for either choice of $\zeta=\pm 1$) are denoted by the darker blue line which lies just above the purple one. For $\zeta=+1$ we see that these solutions approach the black-orange line of the confining branch in the $\f_{L}\to 0$ limit, and instead reach the red line of the skewed solutions for $\zeta=-1$ in the same limit. Every other possible choice of $\{\f_{L} \hspace{-1mm}<\hspace{-1mm} 0,\,\zeta\}$ generates a line residing somewhere between the two extrema, and by varying these parameters we are able to completely fill the region of the parameter space which is delimited by the confining and skewed classes of solutions. \par
Before proceeding we shall highlight a few important features of Fig.~\ref{Fig:FplotBlue}. First and foremost, we observe that the free energy for each branch of solutions connects to the supersymmetric class situated at the origin of the parameter space, which corroborates our statement that all solutions correspond to different deformations of the unique fixed point $\f=0$; our prescription for introducing a universal scale via $\Lambda$ is hence sufficient for us to legitimately compare the energetics for different types of backgrounds.\par 
We next observe that the energetically favoured branch of solutions (that which minimises the free energy of the system) appears to be the class of regular solutions, for all $\hat{\f}_{2}$. Furthermore, we notice that there is a region of the parameter space with $\hat{\f}_{2}>0$ within which the system apparently prefers to realise a configuration containing a tachyonic instability; we shall soon see that this is not in fact the case, and that our investigation of the theory phase space would be incomplete had we decided to neglect the badly singular backgrounds.\par
\begin{figure}[h!]
\begin{center}
	\makebox[\textwidth]{\includegraphics[width=0.54\paperwidth]
		{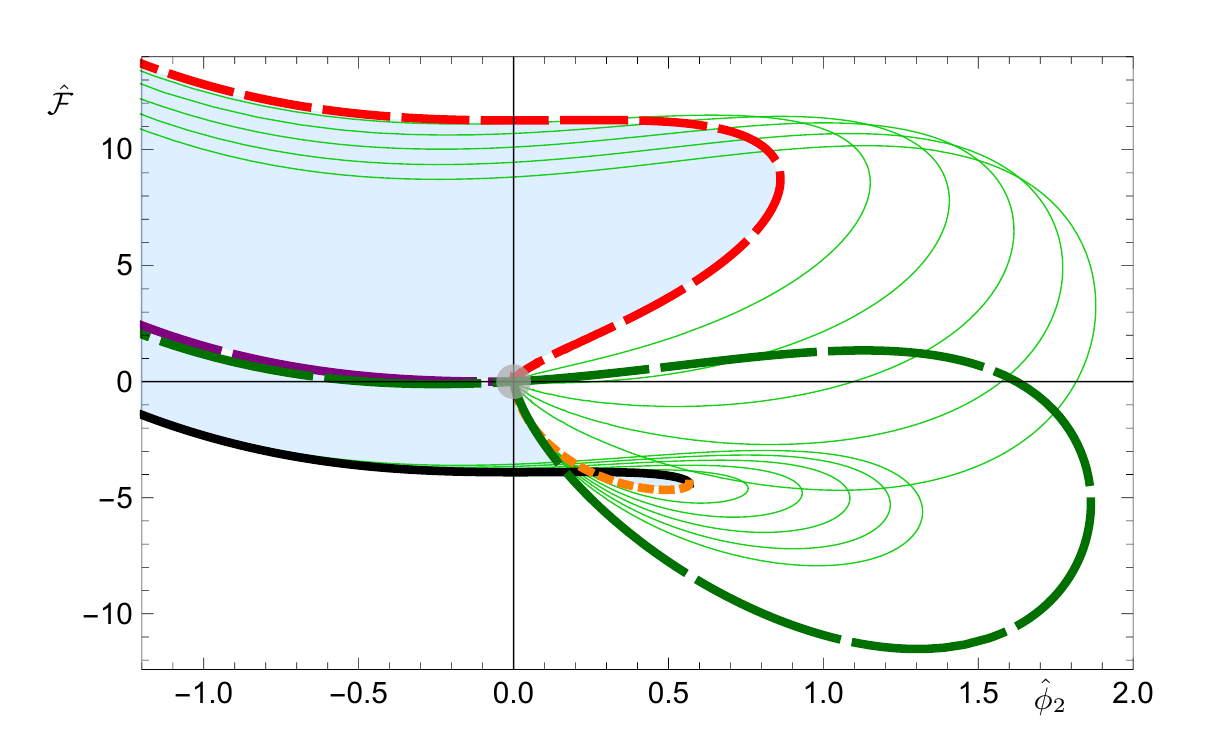}}
\end{center}
\vspace{-5mm}
		\caption[Second free energy plot for the six-dimensional supergravity]{The free energy density $\hat{\cal F}$ as a function of the deformation parameter $\hat\phi_2$ for the IR-conformal solutions (longest-dashed purple line), 
			the confining solutions (solid black and short-dashed orange lines), and the skewed solutions (dashed red line), compared 
			to a few representative choices of (badly) singular solutions (green).
			For the latter, we generated the numerical solutions from the IR
			expansions, by setting $(\phi_L,\zeta) = (0.05,-1)$, $( 0.1,-1)$, $( 0.15,-1)$, $( 0.2, -1)$, $(0.25, -1)$, 
			$( 0.25,1)$, $( 0.2,1)$, $( 0.15,1)$, 
			$( 0.1,1)$, $(  0.05,1)$, 
			respectively (lighter green lines), and varied the value of $\f_{I}$.
			The long-dashed dark-green line represents the domain-wall (badly) singular solutions, obtained by varying the parameter $\f_{b}$.
			The SUSY solutions are represented by a grey point at the origin. 
			The short-dashed orange line shows the region within the branch of confining solutions for which a tachyonic state appears in the scalar mass spectrum.
			We have shaded in light blue the region covered by the good singular solutions shown in Fig.~\ref{Fig:FplotBlue}. }
		\label{Fig:FplotGreen}
\end{figure}

 
We now turn our attention to the plot in Fig.~\ref{Fig:FplotGreen} which portrays all of the information presented in Fig.~\ref{Fig:FplotBlue}, supplemented by the free energy for the backgrounds with bad singularities: the general singular solutions with $0<\f_{L}<\frac{1}{3}$ for $\zeta=\pm 1$ (thin green lines), and the domain-wall solutions obtained in the $\f_{L}\to\frac{1}{3}$ limit (long-dashed dark-green line). We have also shaded in blue the region of the parameter space enclosed by the confining and skewed branches, which is covered by the subset of the general singular solutions that encounter a good singularity in the deep IR.                  
\\
\indent Our inclusion of the bad singularity backgrounds recontextualizes our initial observation from Fig.~\ref{Fig:FplotBlue}, that the branch of confining solutions always minimise the system free energy; crucially, we see from the complete plot in Fig.~\ref{Fig:FplotGreen} that the tachyonic region of the parameter space is \emph{never} energetically favoured, and that the system shows evidence of undergoing a first-order phase transition. We will soon provide a more thorough description of this transition by characterising it in terms of the behaviour of condensates and order parameters, but we first summarise some important features of our results so far.\par 
We re-emphasise that every class of solutions connects back to the supersymmetric ones at the origin $\big(\hat{\f}_{2}\,,\hat{\cf}\big)=(0\,,0)$, and that this still holds true also for the bad singularity backgrounds; moreover all solutions have a finite, computable free energy. If we consider only the $\hat{\f}_{2}\leqslant 0$ region of the parameter space we observe that the free energy of the system is bounded from below by the branch of regular solutions, and from above by the related class of skewed solutions; the various other classes, which include the supersymmetric solutions, the scale-invariant IRC solutions, and the other classes of singular backgrounds (for any permitted values of the parameters $\{\f_{L},\zeta\}$), lie within the region delimited by the two. In terms of energetic stability we see that the system favours background geometries in which one of the external dimensions is compactified on a circle; these solutions are free from curvature and conical singularities and have duals which can intuitively be interpreted as field theories which exhibit confinement at low energies.\par
\begin{figure}[h!]
	\begin{center}
		\makebox[\textwidth]{\includegraphics[
		    width=0.5\paperwidth
			]
			{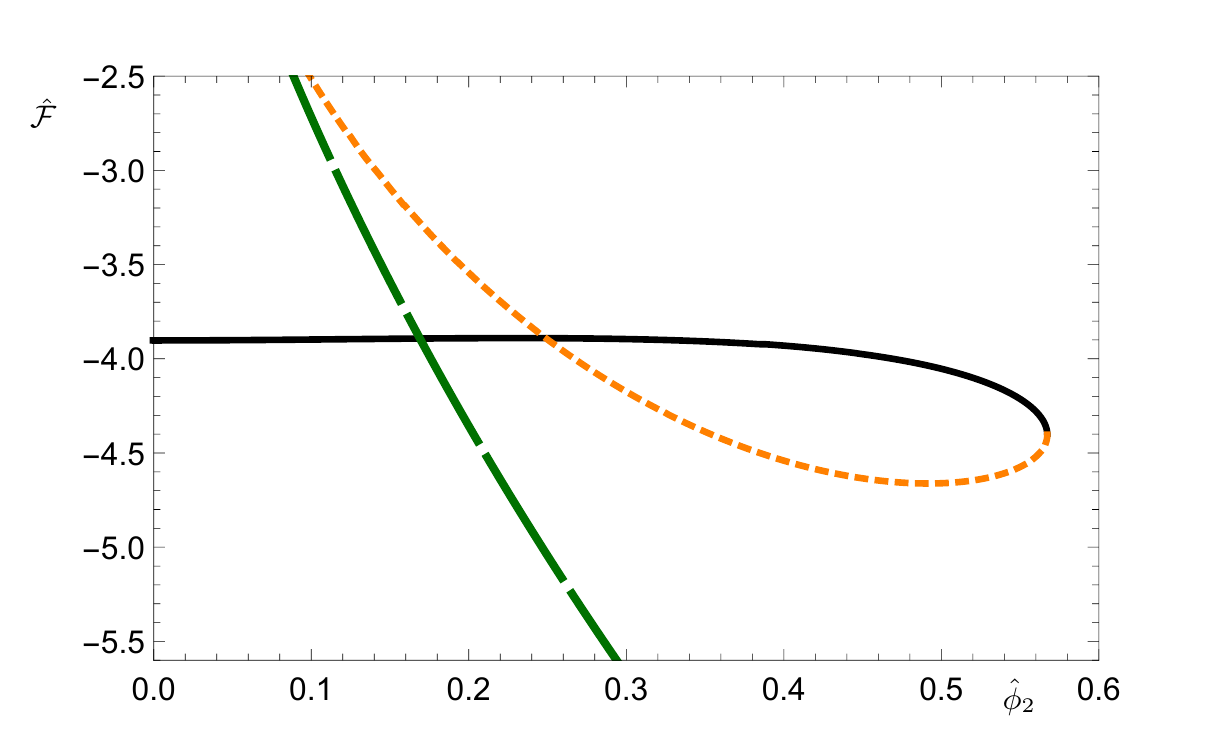}}
	\end{center}
\vspace{-5mm}
	\caption[Magnification of the phase transition of the six-dimensional supergravity]{The free energy density $\hat{\cf}$ as a function of the deformation parameter $\hat{\f_{2}}$ for the confining solutions (solid black and short-dashed orange lines), and
		the BSDW solutions (long-dashed dark-green line). We focus in particular on the region of the parameter space near to the phase transition at $\hat{\f}_{2}=\hat{\f}_{2}^{*}$. The metastable confining solutions lie along the $\hat{\f}_{2}>\hat{\f}_{2}^{*}$ portion of the solid black line.}
	\label{Fig:FplotCrossing}
\end{figure}

 
In the complementary $\hat{\f}_{2}>0$ region of the parameter space we see evidence of a first-order phase transition at some critical value of the source $\hat{\f}_{2}=\hat{\f}_{2}^{*}$. Over the interval $0\leqslant\hat{\f}_{2}<\hat{\f}_{2}^{*}$ the system prefers to maintain a compactified dimension as it did for negative $\hat{\f}_{2}$. However, as we continue to increase the source in order to approach the tachyonic instability, at the critical value $\hat{\f}_{2}^{*}$ the badly singular domain-wall solutions abruptly assume the role of the class which minimises $\hat{\cf}$, and hence are energetically favoured. This transition identifies two distinct phases of the theory: the \emph{confining phase} for $\hat{\f}_{2}<\hat{\f}_{2}^{*}$, and the \emph{domain-wall phase} for $\hat{\f}_{2}^{*}<\hat{\f}_{2}$, in which the system favours the spontaneous decompactification of the circular dimension to realise a geometry which (locally) restores the full five-dimensional Poincar\'{e} invariance. We also observe that for $\hat{\f}_{2}>\hat{\f}_{2}^{*}$ there is a finite portion of the black line which, while not energetically favoured, does not contain a tachyonic state in the spectrum of scalar fluctuations; we refer to this as the
\emph{metastable} region of the confining branch, and background solutions therein to be metastable. A magnification of these features is provided in Fig.~\ref{Fig:FplotCrossing}.    
\\
\indent The approximate coordinates of the phase transition can be extracted numerically. We find that the critical values of the source and free energy are given by
\begin{equation}
\big(\hat{\f}_{2}^{*}\,,\,\hat{\cf}^{*}\big)\simeq\big(0.169\,,-3.893\big)\,,
\end{equation}
which correspond to backgrounds generated using the IR parameter choices
\begin{equation}
\f_{I}^{*}\simeq 0.027\quad,\quad\f_{b}^{*}\simeq 98.9\,.
\end{equation}
We can furthermore determine the values of the condensate parameters $\f_{3}$ and $\c_{5}$ on either side of the transition. By introducing the subscripts `$<$' and `$>$' to denote quantities extracted in the confining phase and domain-wall phase, respectively, we obtain the following:
\begin{align}
\hat{\f}_{3\,<}^{*}&\simeq -0.092\quad,\quad
\hat{\f}_{3\,>}^{*}\simeq 43.2 \,,\notag\\
\hat{\c}_{5\,<}^{*}&\simeq -3.12\quad,\quad
\hat{\c}_{5\,>}^{*}\simeq -1.17 \,,
\end{align}
and we notice in particular the enhancement of the condensate associated with the dimension-3 operator dual to $\f$; we shall return to this observation in Sec.~\ref{Sec:RomansProbeRevisit} when we re-examine the results of our probe spectrum computation in the context of our newly acquired parameter data.\par
We conclude this subsection by observing that the results of our free energy analysis present two unexpected pathologies. The first of these is that the parameter space of the theory seems to impose an upper bound on the permitted values of the source $\hat{\f}_{2}$; despite our extensive classification of admissible background solutions, we were not able to find any class with which one is able to explore the parameter space to arbitrarily large values of $\hat{\f}_{2}$. The second pathology is that for $\hat{\f}_{2}>\hat{\f}_{2}^{*}$, when the system enters the domain-wall phase, the background solutions which minimise the free energy are (badly) singular at the end of space. The putative dual field theories living at the boundary would hence exhibit unphysical behaviour at low energies, which we would be unable to reasonably interpret. Effectively then, from this second observation we infer that a sensible holographic field theory interpretation of the system bulk dynamics only exists for $\hat{\f}_{2}$ below the upper bound given by the critical value at the phase transition  $\hat{\f}_{2}=\hat{\f}_{2}^{*}$. Crucially, this limitation does not necessarily imply the existence of a pathology in the underlying theory, but rather highlights the fact that our analysis of the phase structure using an effective supergravity description is likely to be neglecting other important contributions. We shall return to this point in Chapter~\ref{Chap:Discussion}.

\subsubsection{Characterising the phase transition}
The results of our energetics analysis have revealed the presence of a first-order phase transition, by which the system spontaneously decompactifies the periodic dimension at some critical value of the source $\hat{\f}_{2}$ in order to avoid a pathological region of the parameter space. In this section we shall attempt to provide a more detailed characterisation of the phase transition by introducing two new dynamical quantities $\hat{\cm}$ and $\hat{\Delta}_{\rm DW}$; these quantities will here assume an analogous role to that of \emph{order parameters} which are in general used to study the properties of phase transitions, and we will therefore refer to them as such.\par 
We define the first of these order parameters $\hat{\cm}$ to be the variation of the free energy density $\hat{\cf}$ with respect to the source $\hat{\f}_{2}$, holding fixed the universal scale $\Lambda$ and the leading order UV parameters $\c_{U}=A_{U}=0$; our choice of letter used to denote this quantity is motivated by its similarity to the magnetisation of a thermodynamical system, computed by differentiating the system free energy with respect to an externally sourced magnetic field (holding other quantities such as temperature fixed). The order parameter is defined for our purposes as follows:        
\begin{equation}
\label{Eq:OrderParamM}
\hat{\cm} \equiv \Lambda^{-3}\frac{\pa}{\pa\f_{2}}\cf(\f_{2},\Lambda) = \frac{\pa}{\pa\hat{\f}_{2}}\hat{\cf}(\hat{\f}_{2}) \,.
\end{equation}
We remind the Reader that the three UV parameters $\{\f_{2},\,\f_{3},\,\c_{5}\}$ which are used to compute the free energy are related by non-trivial functional dependences which are not known analytically, and hence we must resort to calculating $\hat{\cm}$ by taking the finite-difference numerical derivative of the extracted data sets. We furthermore note that, while for the IRC, confining, skewed, and BSDW classes of solutions $\hat{\cm}$ is a well-defined quantity, the same is not true for the general singular solutions; from Table~\ref{Tbl:ParameterSummary} we see that for these backgrounds the $\co_{3}$ VEV is a free parameter, and hence the derivative with respect to the source is ambiguous.\par 
Our second order parameter $\hat{\Delta}_{\rm DW}$ provides a measure of to what degree five-dimensional Poincar\'{e} invariance is broken for any given background geometry; we define this quantity as   
\begin{equation}
\label{Eq:OrderParamDelta}
\Delta_{\rm DW}\equiv\c_{5} + \frac{4}{25}\f_{2}\f_{3} 
\quad,\quad
\hat{\Delta}_{\rm DW}\equiv\hat{\c}_{5} + \frac{4}{25} \hat{\f}_{2} \hat{\f}_{3} \,.
\end{equation}
From Eq.~(\ref{Eq:Chi5}) we note that for any solution which satisfies the domain-wall constraint $\b=0$, and hence which preserves Poincar\'{e} symmetry in five dimensions, the order parameter $\hat{\Delta}_{\rm DW}$ vanishes identically. In plotting the order parameters we shall here restrict our attention to the confining and badly singular domain-wall classes of solutions only, as it is these branches which realise the phase transition and give rise to the two distinct phases of the theory (though in Fig.~\ref{Fig:DeltaPhi2} of Appendix~\ref{App:ParamPlots} we show $\hat{\Delta}_{\rm DW}$ for some other classes).\par
\begin{figure}[h!]
	\begin{center}
		\makebox[\textwidth]{\includegraphics[
			width=0.74\paperwidth
			]
			{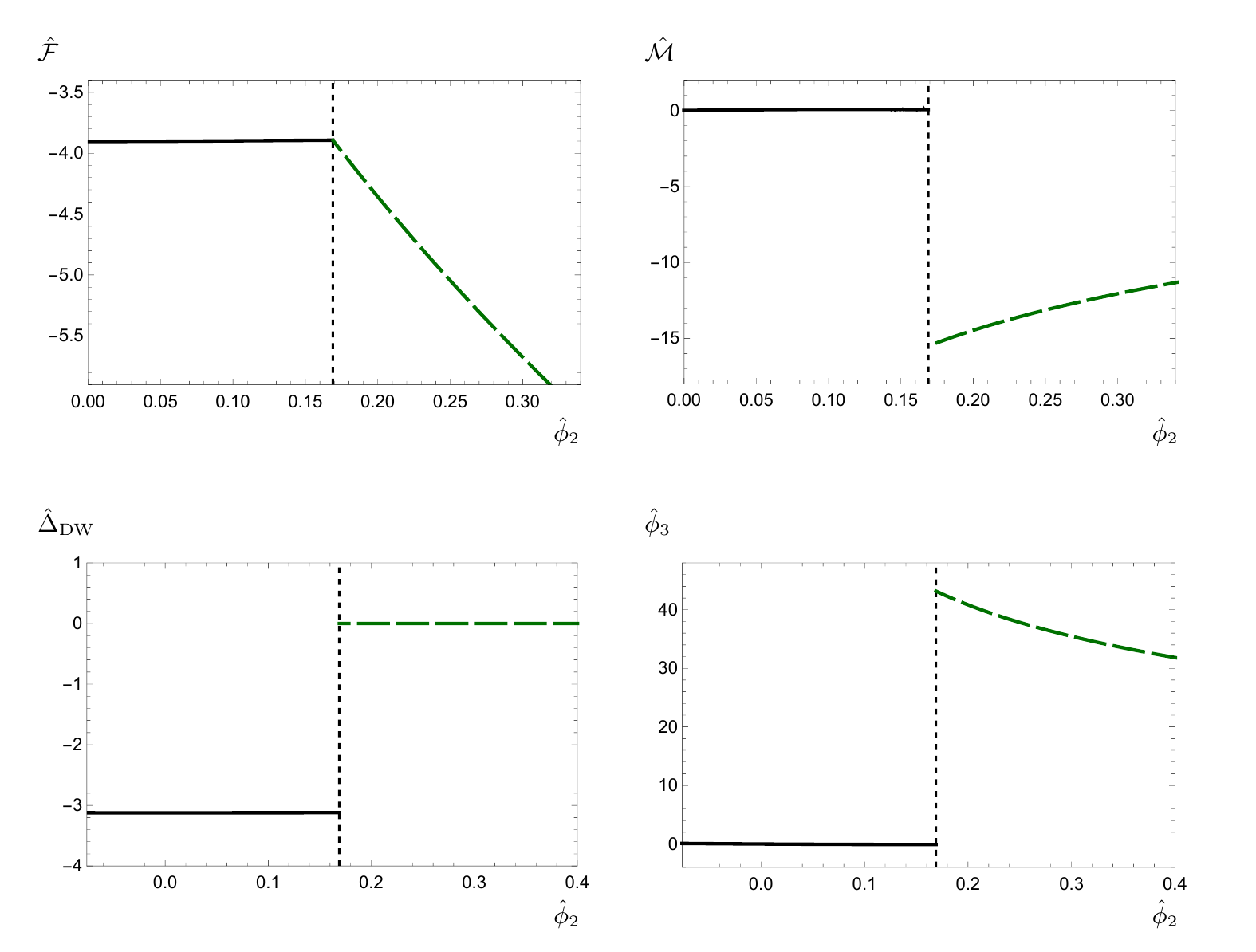}}
	\end{center}
	\vspace{-5mm}
	\caption[Characterising the phase transition of the six-dimensional supergravity]{The minimum free energy density $\hat{\cf}$ (top-left) and its derivative $\hat{\cm}$ (top-right), as a function of the deformation parameter $\hat{\f}_{2}$, for solutions within the confining (solid black), and badly singular domain-wall (dashed dark-green) classes. The bottom panels show the order parameter $\hat{\Delta}_{\rm DW}$ (bottom-left) and the condensate $\hat{\f}_{3}$ (bottom-right), for the same solutions. The vertical (short-dashed black) line in each plot denotes the critical value $\hat{\f}_{2}=\hat{\f}_{2}^{*}\simeq0.169$ at the phase transition.}
	\label{Fig:PhaseTransition}
\end{figure}

 
In Figure~\ref{Fig:PhaseTransition} we present four parameter plots, each focusing in particular on the region of the phase space in proximity of the transition at $\hat{\f}_{2}=\hat{\f}_{2}^{*}$ (denoted in each plot by the vertical dashed line). The top-left panel shows the minimum free energy density $\hat{\cf}$ as a function of the source, while the top-right panel shows its variation $\hat{\cm}$; we see that the derivative of the free energy is discontinuous at the critical value $\hat{\f}_{2}^{*}$, a feature which clearly evinces a first-order phase transition.\par 
In the bottom-left panel we plot the order parameter $\hat{\Delta}_{\rm DW}$, and we see that it too is discontinuous. In the confining phase  $\hat{\f}_{2}<\hat{\f}_{2}^{*}$ the system realises a background geometry which smoothly closes off at the end of space, breaking Poincar\'{e} invariance along the dimension parametrised by $\eta$; the non-zero value of $\hat{\Delta}_{\rm DW}$ in this phase attests to this. Conversely, in the domain-wall phase $\hat{\f}_{2}>\hat{\f}_{2}^{*}$ the system prefers to (locally) preserve the full five-dimensional Poincar\'{e} invariance, and we see that the order parameter $\hat{\Delta}_{\rm DW}$ vanishes identically.\par 
The final plot of Fig.~\ref{Fig:PhaseTransition} shows $\hat{\f}_{3}$, the VEV associated with the $\Delta=3$ boundary operator dual to $\f$, as a function of the operator's source $\hat{\f}_{2}$. From this panel we deduce that the spontaneous decompactification of the dimension parametrised by $\eta$ in the domain-wall phase is associated with the significant enhancement of the condensate $\hat{\f}_{3}$, which is almost vanishing in the confining phase. In the next section we will more closely examine the results of our probe spectrum computation by comparing the behaviour of this condensate for the other branches of solutions.           
\clearpage

\section{More about the dilaton}
\label{Sec:RomansProbeRevisit}
In Sec.~\ref{Sec:RomansMassPlots} we presented the results of our mass spectra computation for the complete set of fields comprising Romans six-dimensional supergravity compactified on a circle, and in the subsequent Sec.~\ref{Sec:RomansProbePlots} we analysed the spin-0 sector using the \emph{probe approximation} to investigate the phenomenology of the dilaton. Having now completed our calculation of the free energy density---in the process uncovering strong evidence of a first-order phase transition which energetically disfavours the branch of regular solutions within a certain region of the parameter space---we now return to the probe spectrum to contextualise our previous findings.\par 
We remind the Reader that in computing the spectra of fluctuations we restrict our attention to the class of confining backgrounds, as these are the only solutions for which the bulk geometry has a regular end of space. The entire family of backgrounds within this class is generated by varying the IR expansion parameter $\f_{I}$, and the various spectra are presented as a function of this parameter. In the following discussion we shall highlight some important features of the UV deformation parameters which were used in our energetics analysis, and compare their behaviour in certain limits to that of the spectra; plots of the functional relations between these parameters are presented in Appendix~\ref{App:ParamPlots}, though these are not crucial for our discussion and we shall not need to directly reference them here.\par
\begin{figure}[h!]
	\begin{center}
		\makebox[\textwidth]{\includegraphics[
			width=0.45\paperwidth
			]
			{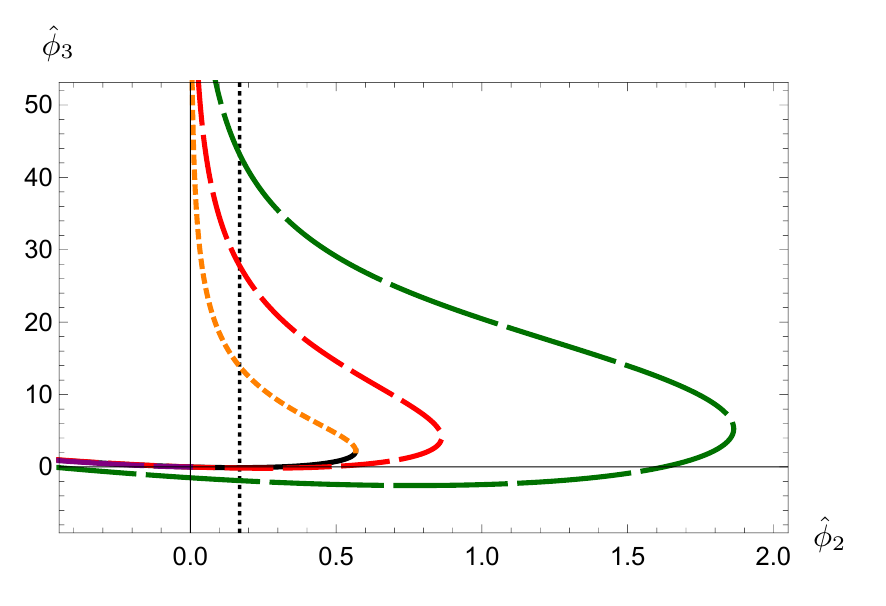}}
	\end{center}
	\vspace{-6mm}
	\caption[Condensate enhancement at the phase transition of the six-dimensional supergravity]{The UV parameter $\hat{\f}_{3}$ as a function of the deformation parameter $\hat{\f}_{2}$, for the confining (solid black and short-dashed orange), skewed (dashed red), IR-conformal (longest-dashed purple), and singular domain-wall (long-dashed dark-green) classes of background solutions. The vertical dashed line denotes the critical value $\hat{\f}_{2}=\hat{\f}_{2}^{*}\simeq0.169$ at the phase transition.}
	\label{Fig:CondensatePhi3}
\end{figure}

 
We start by considering the large-$\f_{I}$ region of the probe spectrum plot shown in Fig.~\ref{Fig:RomansSpectrumProbe}. We observe that the lightest mass eigenstate is tachyonic, and moreover that it must contain a significant contribution from the dilaton due to the fact that the probe approximation unambiguously fails to capture it. In the limit of $\f_{I}\to\infty$ this state asymptotically approaches zero from below, a feature which, for a pseudo-Nambu--Goldstone Boson, is associated with spontaneously broken scale invariance being explicitly restored. If we examine the behaviour of the UV parameters we indeed find evidence to corroborate this interpretation: the parameter $\hat{\f}_{2}$---which sources the boundary operator $\co_{3}$ and controls explicit symmetry breaking in the dual field theory---vanishes in the $\f_{I}\to\infty$ limit. Furthermore, we find that in this same limit the parameter $\hat{\f}_{3}$, the condensate associated with the spontaneous breaking of scale invariance, becomes divergently large; this contrasts, however, with the behaviour of the other VEV $\hat{\c}_{5}$, which instead approaches zero as $\f_{I}$ is dialled larger. We therefore infer that the probe approximation correctly identifies a parametrically light dilatonic state at large $\f_{I}$, a region of the theory parameter space in which scale invariance is spontaneously broken by the enhancement of the VEV $\langle\co_{3}\rangle$. In Fig.~\ref{Fig:CondensatePhi3} we show how this condensate diverges also for the skewed and BSDW classes of solutions as $\hat{\f}_{2}\to 0$, the limit in which they each converge to the supersymmetric solution denoted by the grey disk in the free energy plots; we also clearly see the enhancement of $\hat{\f}_{3}$ at the phase transition.\par
We must acknowledge, however, that the large $\f_{I}$ (or equivalently, the divergent $\hat{\f}_{3}$) region of the parameter space along the branch of confining solutions is never energetically favoured, and hence the system does not realise a regular background geometry which contains within its spectra of fluctuations the aforementioned parametrically light dilatonic state. This becomes apparent when we instead examine the small $\f_{I}$ region of the probe spectrum, where we observe that the critical value $\f_{I}^{*}$ at the phase transition (which we remind the Reader is denoted by the vertical dashed line) is reached well before we are able to dial the IR parameter $\f_{I}$ high enough to reach the tachyonic instability.\par 
As we discussed in Sec.~\ref{Sec:RomansPhaseStruct} the critical value of $\hat{\f}_{2}=\hat{\f}_{2}^{*}$ imposes an upper bound on the $\co_{3}$ source, above which the bulk dynamics of the gravity side does not admit a sensible holographic interpretation in terms of a dual field theory; the same is also true for the critical value of $\hat{\f}_{I}=\hat{\f}_{I}^{*}$. From Fig.~\ref{Fig:RomansSpectrumProbe} we see that just before the system undergoes the phase transition, the lightest state is actually very well approximated by the probe computation and is hence not to be identified with the dilaton. By contrast, we also see that the probe approximation shows an appreciable discrepancy when compared to the next-to-lightest state, despite this state not being particularly light when compared to the rest of the spectrum.\par 
This phenomenon may be understood by more closely examining the behaviour of the condensate parameters. As the $\Delta=3$ operator source $\hat{\f}_{2}$ approaches zero from below, the corresponding VEV $\hat{\f}_{3}$ is suppressed; conversely, in the same limit we find that the parameter $\hat{\c}_{5}$---associated with the condensate of the marginal operator dual to $\c$---is unsuppressed, and is ultimately responsible for spontaneously breaking conformal invariance and introducing dilaton mixing effects in this portion of the spin-0 spectrum. We hence deduce that---at least in the region of parameter space which is energetically favoured (and is thus accessible by the system)---it is the next-to-lightest scalar resonance which may be identified as an \emph{approximate} dilaton, although this state is not parametrically light.

\chapter{Seven-dimensional maximal supergravity}
\label{Chap:EnergeticsWitten}
\section{Classes of solutions}
\label{Sec:WittenClasses}
As with our exploration of the six-dimensional Romans supergravity in Chapter~\ref{Chap:EnergeticsRomans}, our investigation into the phase structure of the seven-dimensional maximal supergravity theory 
 is predicated on the numerical computation of the free energy for various distinct classes of solutions, all of which satisfy the classical equations of motion Eqs.~(\ref{Eq:WittenEOM1}\,-\,\ref{Eq:WittenEOM4}) derived from the seven-dimensional action $\cs_{7}$. We begin our energetics analysis by presenting a (non-exhaustive) catalogue of these backgrounds. 
 
\subsubsection{UV asymptotic expansions}
We categorise the branches of solutions according to their geometric properties, and their characteristics near to the end of space.\ The classes of solutions which we shall be considering each represent distinct deformations of the unique supersymmetric fixed point of the (seven-dimensional) theory $\f=0$, so that in the large-$\r$ limit all backgrounds exhibit the same convergent behaviour and the bulk geometry asymptotically approaches $\text{AdS}_{7}$.\par
This convergence of the various branches at large values of the holographic coordinate allows one to generate UV expansions for the supergravity scalar fields and the warp factor, which are universally applicable and written in terms of a small set of deformation parameters. By defining a new radial coordinate $z\equiv e^{-\r/2}$ so that the UV boundary is situated at $z=0$, these expansions are given by~\cite{Elander:2020csd,Elander:2020fmv}:
{\small\begin{align}
	\label{Eq:WittenUVexpPhi}
	\f(z)=\f_{2}z^{2} &+\left(\f_{4}
	-\frac{18\f_{2}^{2}\ln(z)}{\sqrt{5}}\right)z^{4}
	+\left(\frac{162}{5} \f_{2}^{3}\ln(z)
	-\frac{637\f_{2}^{3}}{30}
	-\frac{9\f_{2}\f_{4}}{\sqrt{5}}\right)z^{6}\nn\\
	&+\frac{1}{600}\bigg(-144\ln(z) \left(62\sqrt{5}\f_{2}^{4}
	-45\f_{2}^{2}\f_{4}\right)
	-11664\sqrt{5}\f_{2}^{4}\ln^{2}(z)\nn\\
	&\hspace{34mm}+2480\f_{2}^{2}\f_{4} 
	+11921\sqrt{5}\f_{2}^{4} -180\sqrt{5}\f_{4}^{2}\bigg)z^{8}\nn\\
	&+\frac{1}{1200}\bigg(
	-72\ln(z)\left(588\sqrt{5}\f_{2}^{3}\f_{4}
	+3335\f_{2}^{5}\right)
	+381024\f_{2}^{5}\ln^{2}(z)\nn\\
	&\hspace{30mm}+13340\sqrt{5}\f_{2}^{3}\f_{4}
	-22179\f_{2}^{5} +5880\f_{2}\f_{4}^{2}\bigg)z^{10}\nn\\
	&\hspace{50mm}+\mathcal{O}\Big(z^{12}\Big)\, ,
	\end{align} }
\vspace{-8mm}
{\small\begin{align}
	\label{Eq:WittenUVexpChi}
	\chi(z)=\c_{U}&-\frac{2\ln(z)}{3}-\frac{\f_{2}^{3}}{30}z^{4}
	+\frac{1}{675}\bigg(75\big(9\c_{6} -4\w_{6}\big)
	+144\sqrt{5}\f_{2}^{3}\ln(z)\nn\\
	&\hspace{56mm}-12\sqrt{5}\f_{2}^{3} -40\f_{2}\f_{4}\bigg)z^{6}\nn\\
	&-\frac{1}{1200}\bigg(72\ln(z)
	\left(27\f_{2}^{4} -4\sqrt{5}\f_{2}^{2}\f_{4}\right)
	+2592\f_{2}^{4}\ln^{2}(z)\nn\\
	&\hspace{40mm}-108\sqrt{5}\f_{2}^{2}\f_{4} -1355\f_{2}^{4}
	+40\f_{4}^{2}\bigg)z^{8}\nn\\
	&+\frac{1}{281250}\f_{2}\bigg(
	-720\ln(z)\left(1800\f_{2}^{2}\f_{4}+1663\sqrt{5}\f_{2}^{4}\right)\nn\\
	&\hspace{20mm}+2332800\sqrt{5}\f_{2}^{4}\ln^{2}(z)
	+332600\f_{2}^{2}\f_{4} -279784\sqrt{5}\f_{2}^{4}\nn\\
	&\hspace{40mm}+5625\f_{2}\Big(9\c_{6} -4\w_{6}\Big)
	+36000\sqrt{5}\f_{4}^{2}\bigg)z^{10}\nn\\
	&\hspace{50mm}+\mathcal{O}\Big(z^{12}\Big)\, ,
	\end{align} }
\vspace{-8mm}
{\small\begin{align}
	\label{Eq:WittenUVexpA}
	A(z)= A_{U}&-\frac{5\ln(z)}{3}-\frac{\f_{2}^{2}}{12}z^4
	+\frac{1}{540}\bigg(135\c_{6} -60\w_{6} 
	+288\sqrt{5}\f_{2}^{3}\ln(z)\nn\\ 
	&\hspace{52mm}-24\sqrt{5}\f_{2}^{3}
	-80\f_{2}\f_{4}\bigg)z^{6}\nn\\
	&-\frac{1}{480}\bigg(72\ln(z)\left(27\f_{2}^{4}
	-4\sqrt{5}\f_{2}^{2}\f_{4}\right)
	+2592\f_{2}^{4}\ln^{2}(z)\nn\\ 
	&\hspace{47mm}-108\sqrt{5}\f_{2}^{2}\f_{4}
	-1355\f_{2}^{4} +40\f_{4}^{2}\bigg)z^{8}\nn\\
	&+\frac{1}{225000}\f_{2}\bigg(-1440\ln(z)\left(
	1800\f_{2}^{2}\f_{4} +1663\sqrt{5}\f_{2}^{4}\right)\nn\\
	&\hspace{20mm}+4665600\sqrt{5}\f_{2}^{4}\ln^{2}(z)
	+665200\f_{2}^{2}\f_{4} -559568\sqrt{5}\f_{2}^{4}\nn\\
	&\hspace{44mm} +1125\f_{2}\Big(9\c_{6}-4\w_{6}\Big)
	+72000\sqrt{5}\f_{4}^{2}\bigg)z^{10}\nn\\
		&\hspace{50mm}+\mathcal{O}\Big(z^{12}\Big)\, ,
	\end{align} }
\vspace{-8mm}
{\small\begin{align}
	\label{Eq:WittenUVexpOmega}
\w(z)&=\w_{U}+\w_{6}z^{6} 
	+\frac{9}{50}\f_{2}^{2}\w_{6}z^{10}+\ldots \nn\\
	&=A_{U}-\frac{5}{2}\c_{U} +\left(\w_{6}-\frac{9}{4}\c_{6}\right)z^{6} -\frac{9}{200}\f_{2}^{2}\Big(9\c_{6} -4\w_{6}\Big)z^{10}+\ldots\, ,
	\end{align} }%
which are governed by the set of seven parameters $\{\f_{2},\f_{4},\chi_{6},\w_{6},\c_{U},\w_{U},A_{U}\}$. We remind the Reader that we choose to impose the constraint $A=\frac{5}{2}\c+\w$ on \emph{all} background solutions, locally preserving five-dimensional Poincar\'{e} invariance within the $\{x^{\mu},\zeta\}$ subspace; the second line of Eq.~(\ref{Eq:WittenUVexpOmega}) follows from the substitution of this constraint, supplemented by the identification $\w_{U}=A_{U}-\frac{5}{2}\c_{U}$. We hence deduce that the parameter $\c_{6}$ must always be exactly zero for consistency, though we shall nevertheless leave any dependence on $\c_{6}$ explicitly visible in equations henceforth unless otherwise stated.

\subsubsection{Supersymmetric (SUSY) solutions}
In $D=7$ dimensions the scalar potential $\cv_{7}$ presented in Eq.~(\ref{Eq:V7}) is a function only of $\f$, and hence the defining equation of the superpotential formalism given in Eq.~(\ref{Eq:Superpotential}) of Sec.~\ref{Subsec:HolographicFormalism} becomes:
\begin{equation}
\cv_{7}(\f)=\big(\pa_{\f}\cw\big)^{2}-\frac{6}{5}\cw^{2}\,,
\end{equation}
which admits an exact superpotential solution $\cw=\cw_{1}$ given by 
\begin{equation}
\cw_{1}=-\frac{1}{4}e^{-\frac{4\f}{\sqrt{5}}}-e^{\frac{\f}{\sqrt{5}}}\,.
\label{Eq:WittenFirstOrder1}
\end{equation}
The first-order differential equations presented in Eqs.~(\ref{Eq:FirstOrderPhi}) and (\ref{Eq:FirstOrderA}) reduce to
\begin{align}
\label{Eq:WittenSUSYeom1}
\pa_{\r}\f&=G^{\f\f}\pa_{\f}\cw=2\pa_{\f}\cw\,,\\
\label{Eq:WittenSUSYeom2}
\pa_{\r}\ca&=-\frac{2}{5}\cw\,,
\end{align}
from which we see that, for the supersymmetric fixed point solution $\f=0$ which realises an exactly $\text{AdS}_{7}$ background geometry, the metric warp factor is a linear function of the holographic coordinate $\ca=\half\r$. There also exists a class of more general solutions to this system of first-order equations for which $\f=\f(\r)$ evolves monotonically from the trivial fixed point in the UV towards a good singularity at the end of space in the deep IR. After substituting in for $\cw=\cw_{1}$ we obtain the following simplified EOMs:
\begin{align}
\pa_{\r}\f&=\frac{2}{\sqrt{5}}\Big(e^{-\frac{4\f}{\sqrt{5}}}
-e^{\frac{\f}{\sqrt{5}}}\Big)\,,
\label{Eq:WittenFirstOrder2}\\
\pa_{\r}\ca&=\frac{1}{10}\Big(e^{-\frac{4\f}{\sqrt{5}}}
+4e^{\frac{\f}{\sqrt{5}}}\Big)\,,
\label{Eq:WittenFirstOrder3}
\end{align}
from which we may construct a family of exact solutions, formulated in terms of a new radial coordinate defined via $\pa_{\r}\equiv e^{-\frac{3\f}{2\sqrt{5}}}\pa_{\t}$. These are given by 
\begin{align}
\f(\t)&=\frac{4}{\sqrt{5}}{\rm arctanh}
\left(e^{-2(\t-\t_{o})}\right)\\
\ca(\t)&=\ca_{o}+\frac{1}{10}\ln\Big[
\cosh(\t-\t_{o})\sinh^{4}(\t-\t_{o}) \Big]\,,
\end{align}
where $\t_{o}$ and $\ca_{o}$ are (real) integration constants, the former being used to fix the end of space. By series expanding the above analytical solutions for small $\t$ we find
\begin{align}
\f(\r)&=
-\frac{2}{\sqrt{5}}\ln(\t-\t_{o}) +\frac{2}{3\sqrt{5}}(\t-\t_{o})^{2} -\frac{7}{45\sqrt{5}}(\t-\t_{o})^{4} +\ldots\,,\\
\ca(\r)&=
\ca_{o} +\frac{2}{5}\ln(\t-\t_{o}) +\frac{7}{60}(\t-\t_{o})^{2} -\frac{19}{1800}(\t-\t_{o})^{4} +\ldots\,,
\end{align} 
or equivalently, in terms of $\r$ (with $\r_{o}$ fixing the end of space):
\begin{align}
\f(\r)&=
-\sqrt{5}\ln\Big(\tfrac{2}{5}(\r-\r_o)\Big)
+\frac{16}{1875\sqrt{5}}(\r-\r_o)^{5} +\ldots\,,\label{Eq:WittenSUSYphi}\\
\ca(\r)&=
\ca_{I}+\ln(\r-\r_o)+\frac{8}{9375}(\r-\r_o)^{5} +\ldots\,,
\end{align}
with $\ca_{I}=\ca_{o} +\ln\left(\frac{2}{5}\right)$. Recalling that $\ca=\frac{3}{5}A=\frac{3}{2}\c$, we therefore also have the following for $\c$ and $A$:
\begin{align}
\c(\r)&=\c_{I}+\frac{2}{3}\ln(\r-\r_o)+\frac{16}{28125}(\r-\r_o)^{5} +\ldots\,,\\
A(\r)&=A_{I}+\frac{5}{3}\ln(\r-\r_o)+\frac{8}{5625}(\r-\r_o)^{5} +\ldots\,,
\end{align}
where $\c_{I}=\frac{2}{3}\ca_{I}$ and $A_{I}=\frac{5}{3}\ca_{I}$.\par
We conclude by observing that the first-order equations of motion in Eqs.~(\ref{Eq:WittenSUSYeom1}) and (\ref{Eq:WittenSUSYeom2}) admit another superpotential solution $\cw=\cw_{2}$, albeit one which may only be generated term-by-term as a perturbative expansion in $\f$; this additional solution may be written as 
\begin{equation}
\label{Eq:WittenW2}
\cw_{2}= -\frac{1}{4}\left(
5 +\f^{2} -\frac{3}{\sqrt{5}}\f^{3}\ln\left(\tfrac{\f^{2}}{\k}\right)\right)
+\ldots\,,
\end{equation}
where here $\k$ is a free parameter introduced to ensure that the logarithm argument is dimensionless. This second superpotential will later prove to be a crucial component of our energetics analysis, by providing the counter-terms required to cancel the divergent contributions to the UV boundary action; the variable $\k$ thus takes the role of a scheme-dependence parameter in the process of holographic renormalisation.

\subsubsection{IR-conformal (IRC) solutions}
There also exist non-singular backgrounds within the seven-dimensional supergravity which also locally preserve the extended Poincar\'{e} invariance by realising a domain-wall geometry. These backgrounds smoothly interpolate as a function of the holographic coordinate between the two critical point solutions of the potential $\cv_{7}$, corresponding in the dual field theory to an RG flow between two distinct six-dimensional CFTs. The circle-compactified dimensions internal to the torus $T^{2}$ maintain a non-zero volume for all (finite) values of the radial coordinate, in contrast to the class of confining solutions wherein the $S^{1}$ parametrised by $\eta$ eventually shrinks to a point and the bulk geometry closes off; as a result, the boundary field theories dual to solutions within this class do not exhibit a low-energy limit. Once one of these interpolating backgrounds has reached the constant IR critical point solution $\f(\r)=\f_{IR}$, further evolving it towards yet lower values of the radial coordinate leaves $\f(\r)$ unaffected, which hence motivates the name \emph{IR-conformal}. As an aside, it is perhaps useful to note that these solutions are physically equivalent to the class of confining backgrounds in the limit at which the scale of confinement in the latter is sent to infinitesimal energies ($\r_{o}\to-\infty$).\par
Recalling from Eq.~(\ref{Eq:WittenDeltas}) that the scaling dimension of the operator dual to $\f$ at the IR critical point satisfies $6-\Delta_{IR}<0$, the small-$\r$ expansions for this branch of solutions may be formulated in terms of the quantity $e^{-(6-\Delta_{IR})\frac{\r}{R_{IR}} }$ which is vanishingly small in the $\r\to-\infty$ limit. These expansions are given by~\cite{Elander:2013jqa}
\begin{align}
\f(\r)&=\f_{IR} + \big(\f_{I}-\f_{IR} \big)e^{-(6-\Delta_{IR})\frac{\r}{R_{IR}} }\,+\ldots\, ,
\label{Eq:WittenIRCphi}\\
\c(\r)&=\c_I+\frac{2\r}{3R_{IR}} - \frac{1}{30} \big(\f_{I}-\f_{IR} \big)^{2} e^{-2(6-\Delta_{IR})\frac{\r}{R_{IR}} }\,+\ldots\, ,
\label{Eq:WittenIRCchi}\\
A(\r)&=A_I+\frac{5\r}{3R_{IR}} - \frac{1}{12} \big(\f_{I}-\f_{IR} \big)^{2} e^{-2(6-\Delta_{IR})\frac{\r}{R_{IR}} }\,+\ldots\, ,
\label{Eq:WittenIRCA}
\end{align}
where $R_{IR}$ as defined in Eq.~(\ref{Eq:WittenRIR}) is the curvature radius of the $\text{AdS}_{7}$ geometry associated with the IR critical point solution, and the integration constants $\c_{I}$ and $A_{I}$ may be chosen arbitrarily. The remaining free parameter $\f_{I}\geqslant\f_{IR}$ sets the scale at which the transition between the two CFTs occurs in the dual field theory, and its variation generates an entire family of backgrounds. Notice however that this is the only tunable scale within the class due to the fact that any one IRC background solution may be shifted by some $\r\to\r-\d$ in order to obtain any other, and hence they are all physically equivalent.\par
We shall revisit this class of solutions when we present a parametric plot of all backgrounds which realise a domain-wall geometry and preserve Poincar\'{e} symmetry within the $\{x^{\mu},\eta,\zeta\}$ subspace, analogous to that shown in Fig.~\ref{Fig:ParametricPlot} for the six-dimensional supergravity. This will enable us to better visualise how the various DW solution classes are related to each other, and moreover will show clearly the interpolating nature of the IRC solutions.

\subsubsection{Confining solutions}
We have already encountered the branch of solutions which we refer to as \emph{confining} in Sec.~\ref{Sec:WittenEOMs} and which were used to compute the spectra of bosonic fluctuations in the toroidally reduced supergravity, the results of which are discussed in Sec.~\ref{Sec:WittenMassPlots}. In this brief section we bring to the Reader's attention the following results, which are obtained by substituting in for $\c$ and $A=\frac{5}{2}\c+\w$ using the small-$\r$ expansions presented in Eqs.~(\ref{Eq:WittenChiIRExpansion}\,-\,\ref{Eq:WittenOmegaIRExpansion}):
\begin{align}
\label{Eq:WittenNotUpsilonExp}
e^{\a(\r)}=e^{4A(\r)-\c(\r)}&= e^{4A_{I}-\c_{I}}\,\tilde{f}\big(\f_{I},\,(\r-\r_{o})\big)\,,\\
\label{Eq:WittenUpsilonExp}
e^{\U(\r)}=e^{A(\r)-\frac{5}{2}\c(\r)}&= e^{A_{I}-\frac{5}{2}\c_{I}}\,\tilde{g}\big(\f_{I},\,(\r-\r_{o})\big)\,,
\end{align}
where $\tilde{f}$ and $\tilde{g}$ are known numerical functions which we neglect to write explicitly for the sake of simplicity. We shall evaluate these same quantities using a related (but geometrically distinct) class of solutions in the next subsection, and compare the two sets of results.  

\subsubsection{Skewed solutions}
The existence of a fourth class of solutions, which we shall refer to as \emph{skewed}, may be inferred from the observation that the classical equations of motion derived from the seven-dimensional supergravity action are left invariant under the sign change $\U\to-\U \Leftrightarrow A-\frac{5}{2}\c\to \frac{5}{2}\c-A$ (or equivalently $\w\to-\w$), with the linear combination $\a=4A-\c$ left unchanged. This family of backgrounds can be generated by solving the EOMs subject to boundary conditions imposed on each bulk field using the following IR expansions:
\begin{align}
\label{Eq:WittenSkewPhi}
\f(\r)&=\f_{I}-\tfrac{1}{2\sqrt{5}}e^{-\frac{8\f_{I}}{\sqrt{5}}} 
\left(1-3 e^{\sqrt{5}\f_{I}}+2 e^{2\sqrt{5}\f_{I}}\right)(\r-\r_{o})^{2}\nn\\
-&\tfrac{1}{80\sqrt{5}}e^{-\frac{16\f_{I}}{\sqrt{5}}} 
\left(9 -44 e^{\sqrt{5}\f_{I}} +57e^{2 \sqrt{5}\f_{I}} 
+2e^{3\sqrt{5}\f_{I}} -24e^{4\sqrt{5}\f_{I}}\right)(\r-\r_{o})^{4}\nn\\
&\hspace{20mm}+\co\left((\r-\r_o)^6\right)\,, \\
\label{Eq:WittenSkewChi}
\c(\r)&=\c_{I}-\tfrac{1}{9}\ln(\r-\r_{o})
-\tfrac{1}{45}e^{-\frac{8\f_{I}}{\sqrt{5}}}\left(1-8e^{\sqrt{5}\f_{I}}
-8e^{2\sqrt{5}\f_{I}}\right)(\r-\r_{o})^{2}\nn\\
-&\tfrac{1}{375}e^{-\frac{16\f_{I}}{\sqrt{5}}}\left(
\tfrac{83}{48}-\tfrac{38}{3}e^{\sqrt{5}\f_{I}}
+\tfrac{61}{2}e^{2\sqrt{5}\f_{I}}+\tfrac{34}{3}e^{3\sqrt{5}\f_{I}}
+\tfrac{62}{3}e^{4\sqrt{5}\f_{I}}\right)(\r-\r_{o})^{4}\nn\\
&\hspace{20mm}+\co\left((\r-\r_o)^6\right)\,,\\
\label{Eq:WittenSkewOmega}
\w(\r)&=\w_{I}+\tfrac{1}{2}\ln(\r-\r_{o})
+\tfrac{1}{40}e^{-\frac{8\f_{I}}{\sqrt{5}}} 
\left(1 -8e^{\sqrt{5}\f_{I}} -8e^{2\sqrt{5}\f_{I}}\right)(\r-\r_{o})^{2}\nn\\
+&\tfrac{1}{8000}e^{-\frac{16\f_{I}}{\sqrt{5}}} 
\left(
31-8\Big(32e^{\sqrt{5}\f_{I}} -81e^{2\sqrt{5}\f_{I}}
-76e^{3\sqrt{5}\f_{I}} -68e^{4\sqrt{5}\f_{I}}\Big)
\right)(\r-\r_{o})^{4}\nn\\
&\hspace{20mm}+\co\left((\r-\r_o)^6\right)\,,
\end{align}
where $\r_{o}$ again fixes the end of space, while $\f_{I}$ is the free parameter which is varied to generate the entire family. As with the branch of regular solutions, by direct substitution of these expansions we may compute the exponential quantities from the previous subsection to obtain the analogous results for this new class; we find  
\begin{align}
e^{\a(\r)}=e^{4A(\r)-\c(\r)}&= e^{4A_{I}-\c_{I}}\,\tilde{f}\big(\f_{I},\,(\r-\r_{o})\big)\,,\\
e^{\U(\r)}=e^{A(\r)-\frac{5}{2}\c(\r)}&= e^{A_{I}-\frac{5}{2}\c_{I}}\,
\Big[\tilde{g}\big(\f_{I},\,(\r-\r_{o})\big)\Big]^{-1}\,,
\end{align}
where $\tilde{f}$ and $\tilde{g}$ are exactly the same numerical functions as those appearing in Eqs.~(\ref{Eq:WittenNotUpsilonExp}) and (\ref{Eq:WittenUpsilonExp}). This verifies that the branch of solutions obtained using the expansions in Eqs.~(\ref{Eq:WittenSkewChi}) and (\ref{Eq:WittenSkewOmega}) are related to the class of solutions which holographically realise confinement (differing only by the sign of $\U$, provided that $\f_{I}$ and $\r_{o}$ are chosen to be the same), though they nevertheless exhibit a completely different background geometry.\par
To demonstrate this point explicitly, it is instructive to consider the small-$\r$ behaviour of the seven-dimensional metric provided in Eq.~(\ref{Eq:7DmetricConstrained}) by substituting in for $\c$, $\w$, and $A$ using the above expansions. Let us first consider the case of the confining solutions: we deduce that in the $\r\to\r_{o}$ limit the Minkowski dimensions maintain a constant non-zero volume at the end of space, and the same is true also for the $S^{1}$ parametrised by the $\zeta$ coordinate; as is to be expected, the volume of the other circle (parametrised by $\eta$) within the torus instead vanishes in the same limit. By contrast, for the branch of skewed solutions we observe that the Minkowski dimensions and the $\zeta$-circle both scale as $(\r-\r_{o})^{1/3}$, while the $\eta$-circle scales as $(\r-\r_{o})^{-2/3}$; as one approaches the end of space in the deep IR, the volume of the subspace spanned by $\{x^{\mu},\zeta\}$ therefore shrinks to a point while the other $S^{1}$ increases in size without bound. Although the confining and skewed solutions are related by the simple relation $\U\to-\U$, it is evident that the two classes realise dissimilar geometries; this behaviour of the two internal torus dimensions in the case of the latter compared to the former motivates our choice of the name \emph{skewed}.\par
On a slight digression we here make an important clarification: as earlier stated we have chosen to restrict the classes of solutions that we consider by adopting the constraint $A=\frac{5}{2}\c+\w$, which is motivated solely by the convenience that in doing so we ensure that Poincar\'{e} invariance is locally preserved within the five-dimensional subspace parametrised by $x^{\mu}$ and $\zeta$; it otherwise has no physical significance. This constraint in turn leaves the equations of motion presented in Eqs.~(\ref{Eq:WittenEOM1}\,-\,\ref{Eq:WittenEOM4}) invariant under the transformation $\U\to-\U$, from which follows our discussion on the class of skewed solutions. We emphasise the fact that, were this constraint to be relaxed, other admissible classes of solutions (possibly including other singular backgrounds) may be discovered. We postpone this extended exploration of the theory to a potential future study, and here remark that the skewed solutions presented in this section may prove to be merely a subset of a wider branch of singular solutions in a more comprehensive investigation of the model.\par
To conclude this subsection, let us observe---from their respective IR expansions---that the confining and skewed classes are related via the following useful relation, which is satisfied up to an additive constant:
\begin{equation}
\label{Eq:WittenConfSkewBackgroundRelation}
0=\frac{1}{3}\Big[2\c^{c}(\r)+A^{c}(\r)\Big]+\c^{s}(\r)-A^{s}(\r)\,,
\end{equation}
where the superscript labels $c$ and $s$ denote confining and skewed background solutions, respectively. By substituting into the above relation using the UV expansions presented in Eqs.~(\ref{Eq:WittenUVexpChi}) and (\ref{Eq:WittenUVexpA}), and keeping written explicitly the superscript labels which distinguish the two classes of solutions, we derive the following parameter identities: 
\begin{align}
\label{Eq:Wid1}
\f_{2}^{s}&=\f_{2}^{c}\, ,\\
\label{Eq:Wid2}
\f_{4}^{s}&=\f_{4}^{c}\, ,\\
\label{Eq:Wid3}
\w_{6}^{s}-\tfrac{9}{4}\c_{6}^{s}&=\tfrac{9}{4}\c_{6}^{c}-\w_{6}^{c}
\quad\Rightarrow\quad 
\w_{6}^{s}=-\w_{6}^{c} \,,
\end{align}
which will later prove useful in our phase structure exploration of this theory.

\subsubsection{General singular solutions}
In Sec.~\ref{Sec:RomansClasses} we provided a classification of solutions within the circle-compactified six-dimensional supergravity system, which included $\f(\r)$ backgrounds that monotonically interpolate towards a singularity in the deep IR region of the bulk geometry; as we have already seen with our earlier introduction of the supersymmetric solutions, similar divergent backgrounds are also admitted by the $T^{2}$-compactified seven-dimensional supergravity. It is possible to construct generalised IR expansions---analogous to those presented in Eqs.~(\ref{Eq:GenSingPhiExp}\,-\,\ref{Eq:GenSingA})---from which a broad class of these solutions may be obtained, encompassing both \emph{good} and \emph{bad} types of singularities (according to Gubser's criterion~\cite{Gubser:2000nd}). The expansions, valid near to the end of space at $\r=\r_{o}$, may be formulated as follows:    
\begin{align}
\label{Eq:WittGenSingPhiExp}
\f(\r)&=\f_{I} + \sqrt{5}\,\f_{L}\ln(\r-\r_o) + \sum_{n=1}^{\infty} \sum_{j=0}^{2n} c_{nj}(\r-\r_{o})^{2n+2n\,\f_{L}-5j\,\f_{L}}\,,\\
\label{Eq:WittGenSingChiExp}
\c(\r)&=\c_{I} + \c_{L}\ln(\r-\r_o) + \sum_{n=1}^{\infty} \sum_{j=0}^{2n} f_{nj}(\r-\r_{o})^{2n+2n\,\f_{L}-5j\,\f_{L}}\,,\\
\label{Eq:WittGenSingAExp}
A(\r)&=A_{I} + A_{L}\ln(\r-\r_o) + \sum_{n=1}^{\infty} \sum_{j=0}^{2n} g_{nj}(\r-\r_{o})^{2n+2n\,\f_{L}-5j\,\f_{L}}\,,
\end{align}
where $\f_{I}$ and $\f_{L}$ are the two free parameters which characterise the space of solutions, with $\f_{L}$ controlling the type of logarithmic singularity present in the deep IR. An additional (discrete) parameter $\Omega=\pm1$ is hidden within the various terms of these expressions, and to leading order we obtain: 
\begin{align}
\label{Eq:WittGenSingPhi}
\f(\r)&=\f_{I}+\sqrt{5}\,\f_{L}\ln(\r-\r_o)\,+\ldots\,,\\
\label{Eq:WittGenSingChi}
\c(\r)&=\c_{I}+\frac{1}{18}\left[4\Omega\sqrt{1-6\f_{L}^{2}}+2\right]
\ln(\r-\r_o)\,+\ldots\,,\\
\label{Eq:WittGenSingA}
A(\r)&=
A_I+\frac{1}{18}\left[\Omega\sqrt{1-6\f_{L}^{2}}+5\right]\ln(\r-\r_o)
\,+\ldots\,.
\end{align}
The summation coefficients
$c_{nj}(\f_{I},\f_{L})$, $f_{nj}(\f_{I},\f_{L},\Omega)$, and $g_{nj}(\f_{I},\f_{L},\Omega)$ can be systematically determined order-by-order, by direct substitution of the expansions into the equations of motion. We see that the complete space of solutions accessible to these general expansions is parametrised by the five integration constants $\{\f_{I},\,\f_{L},\,\c_{I},\,A_{I},\,\r_{o}\}$, supplemented by the choice of $\Omega$. Notice that the logarithm coefficients $\c_{L}$ and $A_{L}$, shown explicitly in Eqs.~(\ref{Eq:WittGenSingChi}) and (\ref{Eq:WittGenSingA}), actually extend the applicability of these expansions to backgrounds which do not encounter $\f$ singularities at the end of space; for the unique choice $\f_{L}=0$ we also recover the IR expansions for the confining and skewed classes of solutions, when $\Omega=1$ and $\Omega=-1$ respectively.\par
Let us furthermore note that $\f_{L}$ is a constrained parameter; the requirement that $\c$ and $A$ both be real functions necessitates that we impose $\f_{L}\geqslant-\frac{1}{\sqrt{6}}$, so that by saturating this bound (and fixing $A_{I}=\frac{5}{2}\c_{I}$) one recovers the geometric constraint $A-\frac{5}{2}\c=\w=0$ satisfied by all domain-wall backgrounds within this model. From the general expansions in Eqs.~(\ref{Eq:WittGenSingPhiExp}\,-\,\ref{Eq:WittGenSingAExp}) we also observe that, for any given value of $n$ when $\f_{L}>0$, the most rapidly diverging exponent as $\r\to\r_{o}$ is the sub-leading correction which maximises $j$, given by $2n(1-4\f_{L})$. To ensure that the singular behaviour of $\f$ is governed by the leading-order logarithmic term we hence require that all sub-leading exponents in the expansions are positive, which yields our second constraint $\f_{L}<\frac{1}{4}$. If we instead consider the complementary $\f_{L}<0$ case, then the most rapidly diverging contribution for any given $n$ comes from the minimum $j$ exponent and the corresponding bound is $\f_{L}>-1$; this is less stringent than the requirement that $\c$ and $A$ be real, and is hence of no consequence. The combination of our two bound constraints therefore leaves the following allowed interval for the singularity parameter $\f_{L}$:      
\begin{equation}
\label{Eq:WitGenSingBound}
-\frac{1}{\sqrt{6}}\leqslant\f_{L}<\frac{1}{4}\,.
\end{equation}
As with the analogous class of general solutions presented in Sec.~\ref{Sec:RomansClasses}, we observe that the upper bound on $\f_{L}$ represents a pathological limiting case; the general series expansions are rendered unusable as $\f_{L}\to\frac{1}{4}$ since every one of the infinite values of $n$ generates additive contributions which all scale as $(\r-\r_{o})^{p}$ for $p=0,\,\frac{5}{4},\,\frac{5}{2}\ldots$, and hence no truncation is possible. This limiting case actually represents the distinct branch of singular solutions that are introduced in the next subsection, and which will manifest an important feature of the theory phase structure.\par 
We conclude by clarifying that---unlike with our exploration of Romans six-dimensional supergravity---we shall neglect to compute the free energy density for the class of generalised singular backgrounds described in this section. The schematic IR expansions are presented for the sake of completeness, and to demonstrate explicitly that admissible \emph{good} singularity backgrounds do also exist within this system. Both the task of determining the expansion coefficients $c_{nj}$, $f_{nj}$, and $g_{nj}$, and the subsequent process of numerically extracting the parameter data required to compute $\hat{\cf}$, are rather laborious and not necessary for our current purposes. We therefore postpone this line of investigation to a potential future study, wherein a more comprehensive survey of the space of solutions (including the relaxation of our self-imposed constraint $\U=\w$) can be undertaken.
\vspace{4mm}
\begin{table}[h!]
	\begin{center}
		\begin{tabular}{|c|c|c|c|c|}
			\hline\hline
			& $\f_{L}=-6^{-\frac{1}{2}}$ &  $-6^{-\frac{1}{2}}<\f_{L}<0$ &  $\f_{L}=0$ & $0<\f_{L}<\frac{1}{4}$ \cr
			\hline\hline
			$\Omega=+1$  & Good, DW & Good & Confining & Bad \cr
			$\Omega=-1$  & Good, DW & Good & Skewed & Bad \cr
			\hline\hline
		\end{tabular}
	\end{center}
	\vspace{-3mm}
	\caption[Parametrisation of \emph{general singular} backgrounds within the seven-dimensional supergravity]{Parametrisation of the solutions obtainable from Eqs.~(\ref{Eq:WittGenSingPhi}\,-\,\ref{Eq:WittGenSingA}): here \emph{Good} and \emph{Bad} refer to which type of singularity is present at the end of space. For $\f_{L}=-6^{-\frac{1}{2}}$ both choices of $\Omega=\pm1$ correspond to the same family of good singularity domain-wall backgrounds. 
	}
	\label{Tbl:WittGenSing}
\end{table}

\subsubsection{Badly singular domain-wall (BSDW) solutions}
The final class of backgrounds which we shall consider admit the same domain-wall geometry as with the supersymmetric solutions, however as their name suggests they are characterised by the scalar field $\f$ approaching a \emph{bad} ($\f\to\infty$) singularity at the end of space; as earlier mentioned we adopt the same terminology as Gubser~\cite{Gubser:2000nd} to describe singularities, with evaluation on a badly singular background solution having the undesirable quality of leaving the scalar potential unbounded from above. Nevertheless, as with our exploration of the six-dimensional theory we will find that these otherwise pathological solutions play a pivotal role in the phase structure of the theory.\par
The scalar fields and warp factor are described by the following small-$\r$ expansions in the deep IR~\cite{Elander:2020fmv}:
\begin{align}
\f(\r)&=\tfrac{\sqrt{5}}{4}\ln\Big(\tfrac{8}{5}(\r-\r_{o})\Big) +\f_{b}(\r-\r_{o})^{5/8}
+\tfrac{\sqrt{5}}{135}\left(37\f_{b}^{2}-3\ 2^{\frac{15}{4}}\ 5^{-\frac{1}{4}}\right)(\r-\r_{o})^{5/4}\nn\\
+&\tfrac{1}{31050}\left(18192\ 10^{3/4}\f_{b} -162595\f_{b}^{3}\right)(\r-\r_{o})^{15/8} \nn\\
+&\tfrac{1}{11736900}\left(90754487\sqrt{5}\f_{b}^{4}-59745768\ 2^{3/4} \sqrt[4]{5}\ \f_{b}^{2} +900864 \sqrt{2}\right)
(\r-\r_{o})^{5/2}\nn\\
&\hspace{20mm}+\co\left((\r-\r_o)^{25/8}\right)\,,\\
\ca(\r)&=\ca_{I}+\tfrac{1}{16}\ln(\r-\r_{o})
+\tfrac{4}{3\sqrt{5}}\f_{b}(\r-\r_{o})^{5/8}\nn\\
&\hspace{20mm}+\tfrac{1}{2700}\left(192\ 10^{3/4}-2155 \f_{b}^{2}\right)(\r-\r_{o})^{5/4}\nn\\
&\hspace{20mm}-\tfrac{2\sqrt{5}}{46575}\left(7008\ 2^{\frac{3}{4}}\ 5^{-\frac{1}{4}} \f_{b}-12721\f_{b}^{3}\right)(\r-\r_{o})^{15/8}\nn\\
-&\tfrac{1}{234738000}\left(458724605\ \f_{b}^{4} -64375824\ 10^{3/4} \f_{b}^{2} +2437632\sqrt{10}\right)(\r-\r_{o})^{5/2}\nn\\
&\hspace{20mm}+\co\left((\r-\r_o)^{25/8}\right)\,,
\end{align}
where $\ca_{I}$ and $\r_{o}$ are integration constants (the latter fixing the end of space), and $\f_{b}$ is the free parameter that is varied to generate the family of backgrounds. We remind the Reader that solutions which preserve six-dimensional Poincar\'{e} invariance, which include this class, satisfy $\ca=\frac{3}{5}A=\frac{3}{2}\c \Leftrightarrow \U=0$.\par
\begin{figure}[h!]
	\begin{center}
	\makebox[\textwidth]{\includegraphics[width=0.67\paperwidth]
			{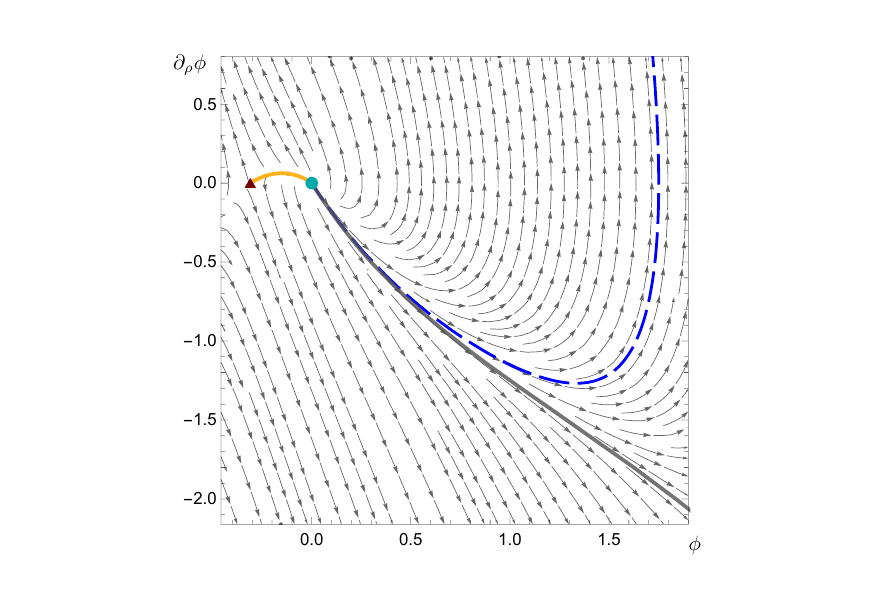}}
	\end{center}
	\vspace{-8mm}
	\caption[Parametric plot of domain-wall solutions in the seven-dimensional supergravity]{Parametric plot of $\pa_{\r}\f$ as a function of $\f$ for solutions 
		which satisfy the warp factor constraint $A=\frac{5}{3}\ca=\frac{5}{2}\c$. 
		The cyan disk and dark-red triangle respectively denote the UV and IR critical points of the
		seven-dimensional potential $\cv_{7}$, the orange line represents the class of IRC solutions with duals which flow between these two critical points, 
		and the grey line represents the class of good-singularity SUSY solutions.\ 
		The arrows exhibit the underlying vector field defined via the second-order differential equation for $\f$ shown in Eq.~(\ref{Eq:WittenParametricPhi}).
		A representative example of the BSDW solutions for the special choice $\f_{b}=\f_{b}^{*}\simeq 33.54$ (the critical value at the phase transition, to be discussed in Sec.~\ref{Sec:WittenPhaseStruct}) is shown with the dashed blue line.\ We observe that the SUSY solutions form the separatrix between numerical backgrounds which flow to good ($\f\to\infty$) and bad ($\f\to-\infty$) singularities for positive $\f$, while the IR-conformal solutions play the same role when $\f$ is negative.}
	\label{Fig:WittenParametricPlot}
\end{figure}

Having now introduced our catalogue of solutions to be analysed as part of our supergravity phase structure investigation, we conclude this section by presenting the analogous parametric plot to that shown in Fig.~\ref{Fig:ParametricPlot} for the six-dimensional theory. Recall that in Sec.~\ref{Sec:WittenEOMs} we derived the second-order differential equation Eq.~(\ref{Eq:WittenParametricPhi}) in terms only of $\f$, which is satisfied by all backgrounds which realise a domain-wall geometry by obeying the constraint $\U=A-\frac{5}{2}\c=0$; this equation is reproduced below for convenience:
\begin{align*}
0=5\f'' &+\sqrt{15}\f'
\Big[2\big(\f'\big)^{2}+\g^{-\frac{8}{5}}\Big(8\g +8\g^{2}-1\Big)\Big]^{\half}\nn\\
&+\sqrt{20}\g^{-\frac{8}{5}}\Big(1-3\g +2\g^{2}\Big) \,,
\end{align*}
where primes denote derivatives with respect to $\r$, and we remind the Reader that in this expression we have defined $\g\equiv e^{\sqrt{5}\f(\r)}$. Each class discussed in this section represents a distinct deformation of the unique supersymmetric critical point solution $\f=\f_{UV}=0$, with the subset of solutions which locally preserve six-dimensional Poincar\'{e} invariance admitting a $\f(\r)$ profile that satisfies the above differential equation. For illustrative purposes it is useful to parametrically plot a representative background from each of these classes, in order to visualise how each one flows away from the trivial solution; the results of this exercise are shown in Fig.~\ref{Fig:WittenParametricPlot}.\par 
As a penultimate remark, we re-emphasise the fact that our classification of solutions in this section is by no means exhaustive.\ As discussed when introducing the branch of skewed solutions, our adopted constraint $\U=A-\frac{5}{2}\c=\w$ to locally preserve five-dimensional Poincar\'{e} invariance restricts our phase structure analysis to encompass only the subset of all backgrounds which satisfy it (and hence for which $\c_{6}=0$); were this constraint to be relaxed it is possible that other types of admissible solutions, which fail to preserve Poincar\'{e} symmetry within the $\{x^{\mu},\zeta\}$ subspace, may be discovered.\par    
We furthermore remind the Reader that although we have successfully identified \emph{good} singularity backgrounds within the toroidally compactified seven-dimensional supergravity---these are given by the subset of the general singular solutions for which $\f_{L}<0$ (see Table.~\ref{Tbl:WittGenSing})---we make the decision \emph{not} to include them in our energetics analysis of the theory phase space. Based on the corresponding results of our analogous investigation in Chapter~\ref{Chap:EnergeticsRomans} we anticipate that these backgrounds would most likely simply fill out the plot region delimited by the branches of confining and skewed solutions, though we postpone the testing of this hypothesis to a future investigation. Likewise, we opt not to compute $\hat{\cf}$ for the complementary badly singular solutions (with general $\f_{L}>0$) either.

\section{Free energy derivation}
\label{Sec:FderivationWitten}
As with our investigation into the phase structure of Romans six-dimensional supergravity in Chapter~\ref{Chap:EnergeticsRomans}, the quantity of interest to us is the free energy density of the system as a function of the universal expansion parameters which govern the asymptotic UV behaviour of the various bulk fields. The general procedure is similar to as before; we employ a numerical routine to extract sets of physical parameter data, and then explore the phase space of the theory by plotting the appropriately renormalised and rescaled free energy as a function of the parameter which sources the $\Delta=4$ boundary operator $\co_{4}$.\par
We begin by defining the seven-dimensional action which we shall adopt, noting again that---since a subset of the background solutions we are considering exhibit singular behaviour at the end of space---we include an IR regulating boundary in addition to the UV regulator necessary for holographic renormalisation. We are therefore required to supplement the bulk action of Eq.~(\ref{Eq:7DAction}) with GHY terms and boundary-localised potentials, so that our complete action is given by        
\begin{align}
\label{Eq:7DActionComplete}
\cs&=\cs_{B}+\sum_{i=1,2}\big(\cs_{\ck,i}+ \cs_{\l,i}\big)\nn\\
&=\int\is\di^{4}x\,\di\eta\,\di\zeta\,\di\r\, \sqrt{-\hat{g}_{7}}\bigg(\frac{\car_{7}}{4}-\hat{g}^{\hat{M}\hat{N}}
\pa_{\hat{M}}\f\,\pa_{\hat{N}}\f-\cv_{7}(\f)\bigg)\nn\\
&\hspace{25mm}+\sum_{i=1,2} (-)^{i}\is\int\is\di^{4}x\,\di\eta\,\di\zeta\,\sqrt{-\tilde{\hat g}}
\bigg(\frac{\ck}{2}+\l_{i} \bigg)\bigg|_{\r=\r_{i}} \,,
\end{align}
where $\hat{g}_{\hat{M}\hat{N}}$ is the seven-dimensional metric tensor for the seven-dimensional line
element in Eq.~(\ref{Eq:7Dmetric}), $\hat{g}_{7}=-e^{8A-2\c}$ is its determinant,
$\car_{7}$ is the corresponding Ricci scalar as provided in Eq.~(\ref{Eq:7DRicci}), and $\tilde{\hat g}_{\hat{M}\hat{N}}$ is the metric induced on each six-dimensional boundary. The extrinsic curvature scalar coming from each GHY term is $\ck$, while $\l_{i}$ are the boundary-localised potentials.\par
To construct the boundary-induced metric $\tilde{\hat g}_{\hat{M}\hat{N}}$ we introduce the covariant seven-vector $n_{\hat{M}}=(0,0,0,0,1,0,0)$, so that the orthonormality conditions presented in Eqs.~(\ref{Eq:OrthoCond1}) and (\ref{Eq:OrthoCond2}) are satisfied (though note for the purposes of this model the hatted uppercase Latin indices instead take values $\hat{M}\in\{0,1,2,3,5,6,7\}$). We reproduce these defining conditions here for convenience:
\begin{align*}
1&=\hat{g}_{\hat{M}\hat{N}}n^{\hat{M}}n^{\hat{N}}=n^{\hat{M}}n_{\hat{M}}\, ,\\
0&=\tilde{\hat{g}}_{\hat{M}\hat{N}}n^{\hat{M}}\,,
\end{align*}
so that the induced metric tensor is again defined as 
\begin{equation*}
\tilde{\hat g}_{\hat{M}\hat{N}}\equiv \hat{g}_{\hat{M}\hat{N}}-n_{\hat{M}}n_{\hat{N}}\,.
\end{equation*}
Our definitions for the covariant derivative and the metric connection are identical to the expressions provided in Eqs.~(\ref{Eq:CovariantDeriv}) and (\ref{Eq:Connection}), though we also present them here for reference:
\begin{align*}
\nabla_{\hat{M}}f_{\hat{N}}&\equiv\pa_{\hat{M}}f_{\hat{N}}-\Gamma^{\hat{Q}}_{\hat{M}\hat{N}}f_{\hat{Q}}\, ,\\
\Gamma^{\hat{P}}_{\hat{M}\hat{N}}&\equiv\frac{1}{2}\hat{g}^{\hat{P}\hat{Q}}\Big(\pa_{\hat{M}}\hat{g}_{\hat{N}\hat{Q}}
+\pa_{\hat{N}}\hat{g}_{\hat{Q}\hat{M}}
-\pa_{\hat{Q}}\hat{g}_{\hat{M}\hat{N}}\Big)\,,
\end{align*}
while our result for the extrinsic curvature scalar is again given by
\begin{align*}
\ck\equiv \hat g^{\hat{M}\hat{N}}\ck_{\hat{M}\hat{N}}
&\equiv\hat g^{\hat{M}\hat{N}}\nabla_{\hat{M}}n_{\hat{N}}\\
&=-\hat{g}^{\hat{M}\hat{N}}\Gamma^{5}_{\hat{M}\hat{N}}
=4\pa_{\r}A-\pa_{\r}\c \,.
\end{align*} 
Our derivation of the free energy density starts with a reformulation of the bulk contribution $\cs_{B}$ to the complete action of Eq.~(\ref{Eq:7DActionComplete}) as a total derivative; by making use of the result for the seven-dimensional Ricci scalar in Eq.~(\ref{Eq:7DRicci}), and of the conserved quantity presented in Eq.~(\ref{Eq:WittenVanish}), we observe that $\cs_{B}$ may be conveniently rewritten as 
\begin{equation}
\cs_{B}\equiv\cs_{B,1}+\cs_{B,2} = -\frac{3}{10}\int_{\r_{1}}^{\r_{2}}\is\di^{4}x\,\di\eta\,\di\zeta\,\di\r\,
\pa_{\r}\Big(e^{\a}\pa_{\r}A\Big)\,,
\end{equation}
where we have reintroduced $\a\equiv 4A-\c$.
For the sake of clarity we may also write explicitly the boundary-localised actions $\cs_{\ck,i}$ and $\cs_{\l,i}$, as follows:
\begin{align}
\cs_{\ck,1}&=
-\frac{1}{2}\int\is\di^{4}x\,\di\eta\,\di\zeta\, e^{\a}\Big(\pa_{\r}\a\Big)\Big\rvert_{\r=\r_{1}}
\,,\\
\cs_{\l,1}&=
-\int\is\di^{4}x\,\di\eta\,\di\zeta\, e^{\a}\Big(\l_{1}\Big)\Big\rvert_{\r=\r_{1}}\,,\\
\cs_{\ck,2}&=
\frac{1}{2}\int\is\di^{4}x\,\di\eta\,\di\zeta\, e^{\a}\Big(\pa_{\r}\a\Big)\Big\rvert_{\r=\r_{2}}
\,,\\
\cs_{\l,2}&=
\int\is\text{d}^{4}x\,\di\eta\,\di\zeta\, e^{\a}\Big(\l_{2}\Big)\Big\rvert_{\r=\r_{2}}\label{Eq:Spot7D}\,.
\end{align}
We adopt an analogous definition of the free energy density $\cf$ to that presented in Eq.~(\ref{Eq:DefineF}) for the case of the six-dimensional theory,
\begin{equation}
F\equiv - \lim_{\r_1\rightarrow \r_o}\lim_{\r_2\rightarrow +\infty}
{\cal S}\equiv\int\is\di^{4}x\,\di\eta\,\di\zeta\,\cf\, ,
\end{equation}
so that, by summing the contributions to the complete action $\cs$, we obtain the following universally applicable result: 
\begin{align}
\label{Eq:WittenGeneralF}
{\cal F}&= \lim_{\r_1\rightarrow \r_o}
\frac{1}{10}e^{\a}\Big(17\pa_{\r}A-5\pa_{\r}\c+10\lambda_{1}\Big)
\Big|_{\r_1}\notag\\
&-\lim_{\r_2\rightarrow +\infty}
\frac{1}{10}e^{\a}\Big(17\pa_{\r}A-5\pa_{\r}\c+10\lambda_{2}\Big)
\Big|_{\r_2}\,.
\end{align}
As before, the value assumed by the IR boundary-localised potential $\l_{1}=-\frac{3}{2}\pa_{\r}A$ is fixed by the requirement that the variation of the complete action is well defined, and we again direct the Reader's attention to Ref.~\cite{Elander:2010wd} for further details on this point. We notice that the sum of the two IR boundary-localised terms $\cs_{\ck,1}$ and $\cs_{\l,1}$, with $\l_{1}$ defined above, gives the following contribution to the complete action:  
\begin{align}
\cs_{\ck,1} + \cs_{\l,1}&=-\half\int\is\di^{4}x\,\di\eta\,\di\zeta\,
\Big(e^{\a}(\pa_{\r}A-\pa_{\r}\c)\Big)\Big|_{\r_1}\\
&=-\half\int\is\di^{4}x\,\di\eta\,\di\zeta\,
\Big(e^{\a}\big(\tfrac{3}{2}\pa_{\r}\c+\pa_{\r}\w\big)\Big)
\Big|_{\r_1}\,,
\end{align}
which, by substituting in for the scalars $\c$ and $\w$ using the small-$\r$ expansions presented in Eqs.~(\ref{Eq:WittenChiIRExpansion}) and (\ref{Eq:WittenOmegaIRExpansion}), we see vanishes in the $\r\to\r_{o}$ limit. Hence, the free energy for the class of regular confining solutions is again unaffected by our inclusion of boundary-localised terms in the deep IR, as is to be expected.\par
Let us now turn our attention to the UV boundary potential $\l_{2}$ which, as with the analogous derivation of $\cf$ for the six-dimensional model, must be chosen carefully to ensure the cancellation of all divergences in the far UV. We proceed by substituting in for the scalar $\c$ and the warp factor $A$ using the small-$z$ expansions presented in Sec.~\ref{Sec:WittenClasses}, and implementing the radial coordinate change $\r=-2\ln(z)\Rightarrow\pa_{\r}=-\half z\pa_{z}$, to obtain the following terms localised at the UV boundary:   
\begin{align}
\cs_{B,2}&=\frac{1}{200}\int\is\di^{4}x\,\di\eta\,\di\zeta\, \frac{e^{\a_{U}}}{z^6}\Big(-50 +5\f_{2}^{2}z^{4}\nn\\
&\hspace{46mm}+\big(\xi +45\c_{6} -20\w_{6}\big)z^6 +\ldots\Big)\Big|_{\r_2}\,,\\
\cs_{\ck,2}&=\frac{3}{100}\int\is\di^{4}x\,\di\eta\,\di\zeta\, \frac{e^{\a_{U}}}{z^6}\,\Big(50 -5\f_{2}^{2}z^{4} -\xi z^6 +\ldots \Big)\Big|_{\r_2}\,,\\
\cs_{\l,2}&=\frac{1}{150}\int\is\di^{4}x\,\di\eta\,\di\zeta\,
\frac{e^{\a_{U}}}{z^6}\l_{2}\bigg(150-45\f_{2}^{2}z^{4}\nn\\
&\hspace{30mm}-\left(\frac{3}{2}\xi +80\f_{2}\f_{4} -18\xi\ln(z)\right)z^{6} +\ldots \bigg)\bigg|_{\r_2}\, ,
\end{align}
where we have defined $\xi\equiv 16\sqrt{5}\f_{2}^{3}$, and have reintroduced $\a_{U}\equiv 4A_{U}-\c_{U}$. As anticipated, we see that all three contributions contain multiple types of divergences in the $z\to 0$ limit, including terms proportional to the (squared) UV deformation parameter $\f_{2}$ which sources the $\Delta=4$ boundary operator dual to $\f$, and a term which is logarithmic in $z$. We furthermore observe that, just as with Romans supergravity, there exists a convenient choice for $\l_{2}$ which provides the exact counter-terms required to cancel all of these divergences, and allows us to define our properly renormalised free energy; for this model the appropriate identification is $\l_{2}=\cw_{2}(\f)$, with $\cw_{2}(\f)$ the small-$\f$ superpotential expansion of Eq.~(\ref{Eq:WittenW2}). By substituting in for the UV potential, and noting that the sub-leading terms in the power expansion of $\cw_{2}$ are inconsequential in the renormalisation procedure, we obtain 
\begin{align}
\cs_{\l,2}&=\frac{1}{4}\int\is\di^{4}x\,\di\eta\,\di\zeta\, \frac{e^{\a_{U}}}{z^6}\bigg(-5+\half\f_{2}^{2}z^{4}\nn\\ 
&\hspace{30mm}+\bigg(\frac{\xi}{20} +\frac{2}{3}\f_{2}\f_{4}
+\frac{3}{80}\xi\ln\left(\tfrac{\f_{2}^{2}}{\k}\right)\bigg)z^{6} +\ldots \bigg)\bigg|_{\r_2}\,,
\end{align}
with which one may verify that the total UV contribution $\cs_{B,2}+\cs_{\ck,2}+\cs_{\l,2}$ is finite in the $z\to 0$ physical limit.\par
As a brief aside, let us here make two observations. Firstly, we notice that the divergences present in the UV-localised actions render the free energy density $\cf$ and its second derivative with respect to the source $\f_{2}$ scheme-dependent; consequently, the familiar concavity theorems for classical thermodynamical systems are not applicable to this holographic model. Secondly, we note that our choice of $\cw_{2}$ to cancel divergences at the UV boundary has the effect of introducing an additional scheme-dependence in the form of the free parameter $\k$; we shall specify our assigned value for this parameter soon.\par
Having established our prescriptions for the two boundary-localised potentials, we may substitute directly for $\l_{1}$ and $\l_{2}$ into Eq.~(\ref{Eq:WittenGeneralF}) to obtain the following expression for $\cf$:     
\begin{align}
\label{Eq:FlambdaWitten}
{\cal F}&=\lim_{\r_1\rightarrow \r_o}\frac{e^{\a}}{10}\bigg(2\pa_{\r}A-5\pa_{\r}\c\bigg)\bigg|_{\r_1}\nn\\
&-\lim_{\r_2\rightarrow +\infty}\frac{e^{\a}}{10}\bigg(17\pa_{\r}A-5\pa_{\r}\c+10\cw_{2}\bigg)\bigg|_{\r_2}\,.
\end{align}
As was previously observed with our analysis of Romans supergravity, we here too notice that the IR contribution to the free energy density is proportional to a conserved quantity; the expression in the first line of Eq.~(\ref{Eq:FlambdaWitten}) is equal to the $\r$-invariant quantity defined in Eq.~(\ref{Eq:CC}) (ignoring an immaterial factor of ten), and hence we may evaluate this IR-localised term at the UV boundary instead without affecting our results. Gathering together terms, we therefore obtain
\begin{equation}
\cf=-\lim_{\r_2\rightarrow +\infty}e^{4A-\c}\bigg(\frac{3}{2}\pa_{\r}A+\cw_{2}\bigg)\bigg|_{\r_2}\,,
\end{equation}
which we note is identical to the analogous result for Romans supergravity, shown in Eq.~(\ref{Eq:Fuv}). If we instead keep the two UV contributions separate and proceed to substitute in for $\c$ and $A$ using their respective UV expansions, we find 
\begin{align}
\label{Eq:WittenQuasiFinalF}
\cf&=\frac{1}{20}e^{\a_{U}}\Big(27\c_{6}-12\w_{6}\Big)\nn\\
&\hspace{16mm}-\frac{1}{120}e^{\a_{U}}\bigg(
20\f_{2}\f_{4}+27\c_{6}-12\w_{6}
-\frac{\xi}{8}\bigg(12-9\ln\left(\tfrac{\f_{2}^{2}}{\k}\right)\bigg)\bigg)\\
\label{Eq:WittenFinalF}
&=-\frac{1}{120}e^{\a_{U}}\bigg(
20\f_{2}\f_{4}-135\c_{6}+60\w_{6}
-\frac{\xi}{8}\bigg(12-9\ln\left(\tfrac{\f_{2}^{2}}{\k}\right)\bigg)\bigg)\,,
\end{align}
where the first line of Eq.~(\ref{Eq:WittenQuasiFinalF}) is obtained by evaluating the conserved quantity at the UV boundary. Our adopted constraint $\U=A-\frac{5}{2}\c=\w$ (which is satisfied by all classes of solutions) imposes that $\c_{6}=0$, and we select a renormalisation scheme by making the convenient assignment $\k=e^{-\frac{4}{3}}$, so that our final result for the free energy density is
\begin{equation}
\label{Eq:WittenFinalFinalF}
\cf=-\frac{1}{120}e^{\a_{U}}\bigg(
20\f_{2}\f_{4}+60\w_{6}
+\frac{9}{8}\xi\ln\left(\f_{2}^{2}\right)\bigg)\,.
\end{equation} 
Backgrounds which realise a domain-wall geometry and locally preserve six-dimensional Poincar\'{e} invariance furthermore require that $\w=0\Rightarrow \w_{U}=\w_{6}=0$, and the same expression then becomes
\begin{equation}
\label{Eq:WittenDWF}
\cf^{(DW)}=-\frac{1}{120}e^{9\chi_{U}}\bigg(
20\f_{2}\f_{4}+\frac{9}{8}\xi\ln\left(\f_{2}^{2}\right)\bigg)
\bigg)\,.
\end{equation}
As with our analysis of the six-dimensional supergravity, we shall here too choose to always set $\c_{U}=A_{U}=0$ in order to simplify the comparison of $\cf$ between different backgrounds. The parameter $\c_{U}$ vanishes if we implement a rescaling of the holographic coordinate via $z\to ze^{\frac{3}{2}\c_{U}}$, while the parameter $A_{U}$ may be cancelled by a simple additive shift of any given background solution for the warp factor, $A\to A-A_{U}$.

\section{Scale setting and numerical implementation}
\label{Sec:WittenScaleSetting}
\subsubsection{Scale setting}
We have classified the various solution branches of interest to our investigation, and have now derived a general expression for the free energy density $\cf$ as a function of the deformation parameters which characterise all solutions in the far UV. As with our energetics analysis of the six-dimensional supergravity in Chapter~\ref{Chap:EnergeticsRomans}, it is convenient to introduce a universal energy scale in order to facilitate comparison between the different classes; this also ensures that their respective parameter spaces have the same dimensionality. In Eqs.~(\ref{Eq:Wid1}\,-\,\ref{Eq:Wid3}) we presented identities which relate the UV parameters for the confining and skewed classes of solutions, which are obtained as a consequence of the relation shown in Eq.~(\ref{Eq:WittenConfSkewBackgroundRelation}). This relation is satisfied only up to an additive constant, which is a constrained parameter in the case of the confining solutions (fixed by the requirement that no conical singularity exists at the end of space) though may be freely chosen for the skewed solutions.\par
Recall from Sec.~\ref{Sec:WittenEOMs} that the combination defined in Eq.~(\ref{Eq:CC}) represents a conserved quantity which is invariant with respect to the radial coordinate; for any given background within any class, this quantity may be evaluated at any value of $\r$ and will yield the same result. We may therefore consider substituting in for the IR expansions of any one branch of solutions and taking the $\r\to\r_{o}$ limit, and then equating this result to the same expression evaluated instead using the UV expansions presented in Eqs.~(\ref{Eq:WittenUVexpChi},\,\ref{Eq:WittenUVexpA}) in the $z\to 0$ limit; this exercise would hence provide us with a scale-independent relation between the IR and UV expansion parameters, unique to each class of backgrounds. We focus in particular on the confining and skewed solutions, for which we derive the following:  
\begin{align}
-\frac{2}{3}&=e^{\a_{U}^{c}-\a_{I}^{c}}
\Big(9\c_{6}^{c}-4\w_{6}^{c}\Big)\,,\\
\frac{2}{3}&=e^{\a_{U}^{s}-\a_{I}^{s}}
\Big(9\c_{6}^{s}-4\w_{6}^{s}\Big)\,,
\end{align}
where we have reintroduced $\a_{U}\equiv 4A_{U}-\c_{U}$ and $\a_{I}\equiv 4A_{I}-\c_{I}$, and where the superscripts $c$ and $s$ denote evaluation using the IR expansions for the confining and skewed classes, respectively. As anticipated, these expressions are identical after making the parameter replacement $\w_{6}\to-\w_{6}$ provided that $\c_{I}^{c}=\c_{I}^{s}$ (recall that $\c_{6}^{c}=\c_{6}^{s}=0$ by necessity). The parameter $\c_{I}$ is constrained for the class of backgrounds which holographically model confinement, while it may be freely chosen for the related branch of skewed solutions; as with our exploration of the $D=6$ supergravity, we can reduce the dimensionality of the space of free parameters for the latter by using the former to derive the free energy density for both classes.\par
To facilitate this, and again inspired by the discussion in Ref.~\cite{Csaki:2000cx}, we reintroduce the universal energy scale $\Lambda$ which allows us to legitimately compare the energetics for the various branches of solutions listed in Sec.~\ref{Sec:WittenClasses}.
We adopt the same prescription for this universal scale as in Eq.~(\ref{Eq:Lambda}), defining $\Lambda$ to be the reciprocal of the time taken by a massless particle to reach the end of space at $\r=\r_{o}$ from the UV boundary. This equation is reproduced below: 
\begin{equation*}
\Lambda^{-1}\equiv t \equiv\int_{r_{o}}^{\infty}\is\text{d}\tilde{r}\,
\sqrt{\frac{g_{rr}}{|g_{tt}|}}=
\int_{r_{o}}^{\infty}\is\text{d}\tilde{r}\,e^{-A(\tilde{r})}=
\int_{\r_{o}}^{\infty}\is\text{d}\tilde{\r}\,
e^{\c(\tilde{\r})-A(\tilde{\r})}\, ,
\end{equation*}
and we remind the Reader that taking the absolute value of the metric component $g_{tt}$ ensures that $\Lambda$ is a real quantity, and that $\c$ and $A$ are evaluated on the numerical backgrounds. Let us consider a simple coordinate rescaling of the form $x^{\mu}\to\s x^{\mu}$, $\eta\to\s\eta$, and $\zeta\to\s\zeta$, which we see from the constrained seven-dimensional metric in Eq.~(\ref{Eq:7DmetricConstrained}) is equivalent to the linear field shifts $\c\to\c+\frac{2}{3}\ln(\s)$ and $A\to A+\frac{5}{3}\ln(\s)$ (so that $\ca\to\ca+2\ln(\s)$). From the UV asymptotic expansions presented in Eqs.~(\ref{Eq:WittenUVexpChi},\,\ref{Eq:WittenUVexpA}) we furthermore observe that these shifts should be supplemented by the rescaling of the holographic coordinate $z\to\s z \-\ \Leftrightarrow \-\ \r\to\r-2\ln(\s)$ to ensure that $A_{U}=\c_{U}=0$. Under such a transformation the remaining UV parameters are rescaled as
\begin{align}
&\f_{2}\to\s^{2}\f_{2}\,,\\
&\f_{4}\to\s^{4}\Big[\f_{4}-\tfrac{18}{\sqrt{5}}\f_{2}^{2}\ln(\s)\Big]\,,
\label{Eq:Phi4Rescale}\\
&\w_{6}\to\s^{6}\w_{6}\,,
\end{align}
while the energy scale satisfies $\Lambda\to\s\Lambda$. The more complicated scaling transformation of $\f_{4}$ is due to the presence of a logarithmic term at order $z^{4}$ in the UV expansion for $\f$ in Eq.~(\ref{Eq:WittenUVexpChi}), and we must account for this extra contribution when extracting data for this parameter. By inspection of Eq.~(\ref{Eq:WittenFinalFinalF}) we see that dimensional analysis demands that the combination $\f_{2}\f_{4}$ has the same units as $\w_{6}$, and hence we see that the following combinations represent dimensionless (scaling-invariant) quantities:
\begin{align}
\hat{\f}_{2}&\equiv\f_{2}\Lambda^{-2}\,,\\
\hat{\w}_{6}&\equiv\w_{6}\Lambda^{-6}\,,\\
\hat{\cf}&\equiv\cf\Lambda^{-6}\,.
\end{align}
In Sec.~\ref{Sec:WittenPhaseStruct} we will present the results of our energetics analysis in terms of these rescaled parameters, which we shall henceforth distinguish with hats.\par
We conclude this subsection with a brief but important clarification. From the non-trivial scaling behaviour of $\f_{4}$ shown in Eq.~(\ref{Eq:Phi4Rescale})---which is a consequence of the additional logarithmic term present in the sub-leading coefficient of the expansion in Eq.~(\ref{Eq:WittenUVexpPhi})---we infer that the UV parameter $\f_{4}$ does not directly correspond to the VEV of the $\Delta=4$ boundary operator $\co_{4}$ dual to $\f$. Were we to conduct a more careful analysis, we would compute the operator one-point function $\langle\co_{4}\rangle$ by functionally differentiating the holographically renormalised on-shell action with respect to the source $\f_{2}$ (see for example Refs.~\cite{Bianchi:2001kw,Skenderis:2002wp,Papadimitriou:2004ap} for details). Nevertheless, for our purposes it is sufficient to know that the deformation parameter $\f_{4}$ is \emph{associated} with the $\co_{4}$ condensate, and with some abuse of terminology we shall refer to it as such from this point forward.

\subsubsection{Numerical implementation}
Our derived expression for the free energy density $\cf$, which plays a foundational role in our energetics analysis of the theory phase structure, is formulated as a function of the universal UV deformation parameters $\{\f_{2},\,\f_{4},\,\w_{6},\,\w_{U},\,\c_{U},\,A_{U}\}$. To plot $\cf$ we are therefore required to employ a numerical routine in order to extract physical values for this set of parameters for each class of backgrounds, and we now turn our attention to detailing this process.\par
Our numerical method is essentially the same as that described in Sec.~\ref{Sec:RomansScaleSetting}, though we nevertheless provide a separate outline here to highlight any differences compared to the circle-compactified theory:  
\begin{enumerate}
	\item For any given choice of the free parameters which characterise the IR field expansions of the class in question, and having fixed the end of space by assigning $\r_{o}=0$, we construct numerical backgrounds for $\f$, $\c$, $\w$, and $A$ by setting up boundary conditions in the deep IR and evolving the solutions towards the UV using the equations of motion. Note that our adopted constraint $\U=\w$ ensures that any one of the backgrounds $\{\c,\w,A\}$ may be obtained as a linear combination of the other two.
	\item We match the constructed backgrounds to the general UV expansions at some choice of $\r=\r_{m}$, solving for each UV parameter in turn to extract the set
	$\{\f_{2},\,\f_{4},\,\w_{6},\,
	\w_{U}\neq 0,\,\c_{U}\neq 0,\,A_{U}\neq 0\}$.
	 The value of the radial coordinate at which the matching is performed must be chosen to ensure that any numerical noise is minimised, and should be sufficiently far into the UV region of the geometry that the background $\f(\r)$ has had sufficient time to reach the UV fixed point $\f=0$ (or as close as is numerically feasible).
	\item Using the values of $\c_{U}$ and $A_{U}$ obtained in the previous step, the holographic coordinate is rescaled according to $z\to ze^{\frac{3}{2}\c_{U}}$ and then the warp factor background is shifted by $A(\r)\to A(\r)-A_{U}$, to set $\c_{U}=A_{U}=0$ (note that this consequently also sets $\w_{U}=0$). We match these rescaled background profiles to the UV expansions again to extract the new set of parameter data 
	$\{\bar{\f}_{2},\,\bar{\f}_{4},\,\bar{\w}_{6},\,
	\bar{\w}_{U}=0,\,\bar{\c}_{U}=0,\,\bar{A}_{U}=0\}$,
	 where we use bars here to emphasise that the other parameters have also been rescaled as a result. We remind the Reader that $\f_{4}$ (associated with the VEV of the boundary operator $\co_{4}$ dual to $\f$) exhibits non-trivial rescaling behaviour under the transformation of the radial coordinate shown above, so that $\bar{\f}_{4}=\Big(\f_{4}-\tfrac{27}{\sqrt{5}}\c_{U}\f_{2}^{2}\Big)e^{6\c_{U}}$. 
	\item Finally we compute the universal scale $\Lambda$ as defined in Eq.~(\ref{Eq:Lambda}), by substituting in for the rescaled background solutions $\c(\r)$ and $A(\r)$ and integrating over their entire domain. For each numerical background we are therefore able to extract the parameter data $\{\hat{\f}_{2},\,\hat{\f}_{4},\,\hat{\w}_{6}\}$, and can compute $\hat{\cf}$ using Eq.~(\ref{Eq:WittenFinalFinalF}).     
\end{enumerate}
It is instructive to supplement the above schematic overview with a more specific description of the numerical process for each class of solutions individually, to clarify any numerical technicalities case-by-case. We now proceed to address each class in the order that they were introduced in Sec.~\ref{Sec:WittenClasses}.\par
As with the analogous class for the six-dimensional Romans supergravity, the supersymmetric background solutions all yield $\f_{2}=0$ when matched to the UV expansions since the IR expansion for $\f$ shown in Eq.~(\ref{Eq:WittenSUSYphi}) contains no free parameters once the end of space has been fixed; from Eq.~(\ref{Eq:WittenDWF}) we therefore deduce that this class has identically vanishing free energy. The integral which defines the universal scale $\Lambda$ is a divergent quantity when evaluated in the deep IR for these backgrounds (verified by simply substituting in using the IR expansions for $\c=\frac{2}{3}\ca$ and $A=\frac{5}{3}\ca$), though this is inconsequential for our purposes.\par 
From Eq.~(\ref{Eq:Phi4Rescale}) we see that the existence of an additional logarithmic term at order $z^{4}$ in the $\f$ UV expansion induces non-trivial scaling behaviour in the parameter $\f_{4}$. This subtlety slightly complicates our treatment of the IR-conformal solutions in comparison to the $D=6$ supergravity, due to the fact that it is not obvious how one should define an appropriate scale-invariant ratio $\kappa$ of the source $\f_{2}$ and condensate parameter $\f_{4}$ (analogously to that of Eq.~(\ref{Eq:KappaIRC})), or even if such a ratio exists. Irrespective of this issue, we may still extract parameter data for this class of backgrounds by taking a different approach: a set of numerical backgrounds is generated using the IR expansions presented in Eqs.~(\ref{Eq:WittenIRCphi}\,-\,\ref{Eq:WittenIRCA}) by dialling the free parameter $\f_{I}\geqslant\f_{IR}$, and each background is matched in turn to the UV expansions at some value $\r=\r_{m}$. The point at which the matching is performed should be sufficiently high so as to ensure that $\f(\r)$ has properly converged at the trivial fixed point $\f=0$ in the UV, and moreover that this is the case for \emph{all} of the generated backgrounds (recall that any one profile in this class may be shifted by $\r\to\r-\d$ to produce another completely equivalent profile). The leading-order UV parameters $A_{U}$ and $\c_{U}$ may simply be set to zero by hand, and the extracted data $\{\f_{2},\f_{4}\}$ does not require any further manipulation. We remind the Reader that no end of space exists for this branch of solutions (i.e.\ $\r$ is not physically bounded from below), and hence the integral defining $\Lambda$ diverges; since these backgrounds exhibit scale-invariance this observation is of no real importance, as $\cf$ would be identical had we adopted any other definition for $\Lambda$.\par 
Parameter data for the class of confining solutions $\{\f_{2}^{c},\,\f_{4}^{c},\,\w_{6}^{c},\,\Lambda^{c}\}$ is obtained by simply matching backgrounds to the UV expansions, and the numerical procedure for this class does not present any technical issues which must be specifically addressed. The identities presented in Eqs.~(\ref{Eq:Wid1}\,-\,\ref{Eq:Wid3}) enable us to simultaneously extract the corresponding sets of UV parameter data $\{\f_{2}^{s},\,\f_{4}^{s},\,\w_{6}^{s}\}$ for the related class of skewed solutions; moreover, the universal scale $\Lambda$ may be computed by substituting instead for the confining backgrounds according to Eq.~(\ref{Eq:WittenConfSkewBackgroundRelation}):
\begin{equation}
(\Lambda^{s})^{-1}=\int_{\r_{o}}^{\infty}\is\text{d}\tilde{\r}\,
e^{\c^{s}(\tilde{\r})-A^{s}(\tilde{\r})}
=\int_{\r_{o}}^{\infty}\is\text{d}\tilde{\r}\,
e^{-\frac{1}{3}\big[2\c^{c}(\tilde{\r})+A^{c}(\tilde{\r})\big]}\, .
\end{equation} 
By computing $\Lambda^{s}$ for the skewed solutions in terms of the numerical backgrounds which manifest a smoothly tapered geometry, we guarantee that $\c_{I}^{s}=\c_{I}^{c}$ and hence ensure that the parameter identities Eqs.~(\ref{Eq:Wid1}\,-\,\ref{Eq:Wid3}) are satisfied. We therefore find that it is unnecessary to numerically generate backgrounds for the branch of skewed solutions using Eqs.~(\ref{Eq:WittenSkewPhi}\,-\,\ref{Eq:WittenSkewOmega}), and that the complete set of data $\{\f_{2}^{s},\,\f_{4}^{s},\,\w_{6}^{s},\,\Lambda^{s}\}$ is obtainable from that of the regular backgrounds.\par 
Finally, the required UV parameter data for the BSDW solutions is extracted according to the numerical process outlined above, with no particular class-specific subtleties to mention. In Table~\ref{Tbl:WittenParameterSummary} we present a summary of how the UV deformation parameters which characterise each branch of solutions are constrained, omitting $\{\c_{U},\,\w_{U},\,A_{U}\}$ since they are always rescaled to zero.
\vspace{4mm}   
\begin{table}[h!]
	\begin{center}
		\begin{tabular}{|c|c|c|c|c|}
			\hline\hline
			Class & $\f_{2}$ & $\f_{4}$ & $\w_{6}$ & Scale setting \cr
			\hline\hline
			SUSY & $0$ & Free & $2\ca=3\c$ ($\w_{6}=0$) & None \cr
			IRC & $<0$ & $\f_{4}=\f_{4}(\f_{2})$ & $2A=5\c$ ($\w_{6}=0$) & None \cr
			Confining & Free & Curvature sing. & Conical sing. & $\Lambda$\cr
			Skewed & Free & $\a^{s}=\a^{c}$ & $\U^{s}=-\U^{c}$ & $\Lambda$ \cr
			BSDW & Free & Free & $2A=5\c$ ($\w_{6}=0$) & $\Lambda$ \cr
			\hline\hline
		\end{tabular}
	\end{center}
	\caption[Summary of parametrisation of backgrounds within the seven-dimensional supergravity]{Summary of parametrisation, constraints, and scale setting procedure for each class of solutions in our energetics analysis of the torus-reduced supergravity. For the IRC solutions $\f_{4}$ has a functional dependence on the source $\f_{2}$ and is not a free parameter, though as previously discussed this dependence is not known analytically.}
	\label{Tbl:WittenParameterSummary}
\end{table}

\section{Phase structure}
\label{Sec:WittenPhaseStruct}
\subsubsection{Free energy plots}
In Sections~\ref{Sec:WittenMassPlots} and~\ref{Sec:WittenProbePlots} we computed the spectra of composite states for the field theory living on the boundary of the $D=7$ bulk spacetime, by considering fluctuations about background solutions which realise a tapered geometry in the deep IR. We uncovered the existence of a tachyonic state in a certain region of the parameter space, which is indicative of an instability in the theory; such background profiles for $\f$ correspond holographically to unstable RG trajectories in the dual field theory. We therefore anticipate the existence of a phase transition by necessity, whereby the system would prevent these unstable background configurations from ever being energetically favoured, and our analysis mirrors that of the six-dimensional model: we systematically compute the free energy density $\hat{\cf}$ for the various branches of solutions discussed in Sec.~\ref{Sec:WittenClasses} using the numerical procedure detailed in the previous section, and present the results of our investigation here.\par
Let us start by recalling that all distinct classes of backgrounds which are admitted as solutions by the compactified supergravity theory exhibit the same asymptotically convergent behaviour in the far UV, and are obtainable as deformations of the unique supersymmetric fixed point solution $\f=0$ by the set of UV parameters $\{\f_{2},\,\f_{4},\,\w_{6},\,\w_{U},\,\c_{U},\,A_{U}\}$. This fixed point solution realises an $\text{AdS}_{7}$ background geometry, and for the dual six-dimensional field theory living on the boundary these deformations fall into one of two categories.\\ 
From Eq.~(\ref{Eq:WittenDeltas}) we see that the first of these corresponds to the insertion of a relevant $\Delta=4$ operator $\co_{4}$, the source for which is identified as the leading order coefficient $\f_{2}$ in the asymptotic expansion Eq.~(\ref{Eq:WittenUVexpPhi}). The vacuum expectation value for this operator is associated with the sub-leading parameter $\f_{4}$. The second type of possible deformation is the compactification of an external space-like dimension on a circle, the size of which is governed by an additional scalar field introduced in the sigma-model coupled to gravity; as previously discussed, for the toroidal compactification on $T^{2}=S^{1}\times S^{1}$ of the $D=7$ supergravity we extend the scalar manifold to include $\c$ and $\w$. These fields are dual to marginal boundary operators sourced by the leading-order UV parameters $\c_{U}$ and $\w_{U}$, respectively, and their VEVs are associated with the sub-leading parameters $\c_{6}$ and $\w_{6}$. However, we remind the Reader that by choosing to impose the constraint $A-\frac{5}{2}\c=\w$ in order to (locally) preserve five-dimensional Poincar\'{e} invariance within the $\{x^{\mu},\zeta\}$ subspace, we consequently completely suppress the $\c_{6}$ condensate for \emph{all} backgrounds.\par 
Although the asymptotic behaviour of the various classes in the far UV may be described in terms of a finite set of deformation parameters, the non-perturbative dynamics of the dual field theory is encoded by the non-trivial functional relations between them; as with the corresponding quantities of the six-dimensional theory, the operator condensate parameters $\f_{4}$ and $\w_{6}$ behave as non-linear response functions of the $\Delta=4$ operator source $\f_{2}$. For this reason we will show the free energy plotted as a function of $\hat{\f}_{2}$, so that the dependence of $\hat{\cf}$ on the other UV parameters is implicitly accounted for. With these preambulatory comments out of the way, we shall now turn our attention to discussing the results of our investigation into the theory phase space.\par
\begin{figure}[h!]
	\begin{center}
		\makebox[\textwidth]{\includegraphics[width=0.52\paperwidth]
			{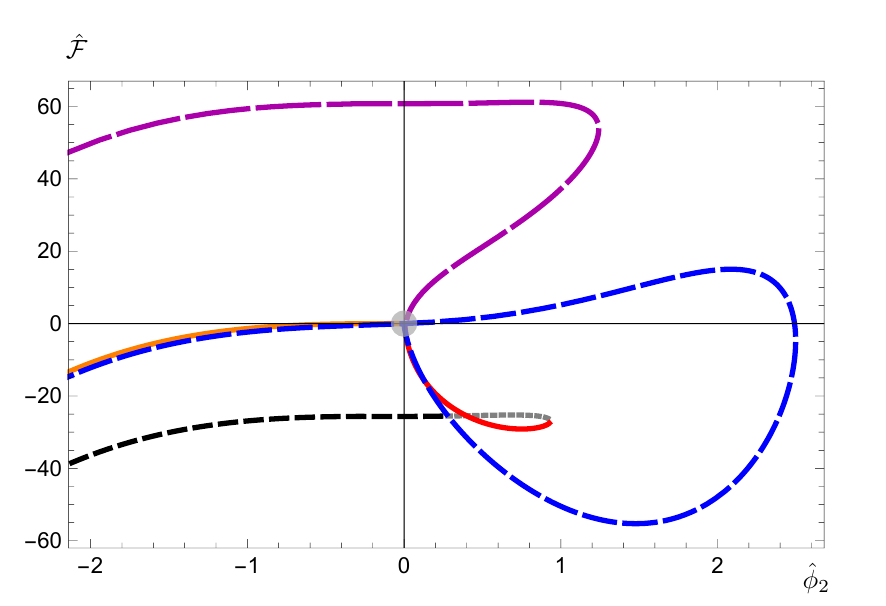}}
	\end{center}
	\vspace{-5mm}
	\caption[Free energy plot for the seven-dimensional supergravity]{The free energy density $\hat{\cf}$ as a function of the deformation parameter $\hat{\f}_{2}$ for the IR-conformal solutions (solid orange line), the skewed solutions (long-dashed magenta line), and the badly-singular domain-wall (BSDW) solutions (dashed blue line). The SUSY solutions are denoted by the grey point at the origin. The confining solutions are separated into three regions: the stable portion of the branch (short-dashed black line), the metastable portion (shortest-dashed grey line), and the unstable portion (solid red line).}
	\label{Fig:FplotWitten}
\end{figure}

 
In Figure~\ref{Fig:FplotWitten} we present the holographically renormalised free energy density $\hat{\cf}$ as a function of the source $\hat{\f}_{2}$, rescaled with appropriate powers of $\Lambda$ as defined in Eq.~(\ref{Eq:Lambda}), for five of the distinct classes of solutions listed in Sec.~\ref{Sec:WittenClasses}. Starting with the simplest case, the branch of supersymmetric backgrounds corresponds to deforming the $\f=0$ critical point solution with the development of a non-zero condensate---associated with $\hat{\f}_{4}$---for the operator dual to $\f$ (referred to as a \emph{VEV deformation}, see Ref.~\cite{Skenderis:2002wp}), which on the gravity side of the duality drives the solutions monotonically towards a good singularity ($\f\to\infty$) at the end of space. Every background within this family yields $\f_{2}=0$ when matched to the UV expansions, and hence according to Eq.~(\ref{Eq:WittenDWF}) this class always has identically vanishing free energy; they are denoted by the grey point at the origin of the phase space (enlarged for visibility, they do not fill a disk).  
\\
\indent The backgrounds which we refer to as \emph{IR-conformal} preserve local six-dimensional \pin within the subspace spanned by the coordinates $\{x^{\mu},\eta,\zeta\}$, and their interpolation between the two critical point solutions of the gravitational theory corresponds holographically to a stable RG flow between two distinct $D=6$ CFTs. Technically there is only a single physically distinct solution within this branch, as any one background may be shifted by $\r\to\r-\d$ to generate any other; the specific choice of the only tunable parameter $\f_{I}$ simply determines at which energy scale the RG trajectory transitions from the supersymmetric CFT to the other. Unlike with our treatment of the analogous IRC class in the six-dimensional supergravity, a scale-invariant ratio $\k$ of the source and VEV of the $\Delta=4$ operator $\co_{4}$ dual to $\f$ was not identified, and hence the $\{\f_{2},\f_{4}\}$ parameter data for this class was extracted manually using the numerical procedure described in Sec.~\ref{Sec:WittenScaleSetting}. The results of this exercise are shown with the solid orange line in the $\hat{\f}_{2}<0$ region of Fig.~\ref{Fig:FplotWitten}.\par
\begin{figure}[h!]
	\begin{center}
		\makebox[\textwidth]{\includegraphics[
			width=0.5\paperwidth
			]
			{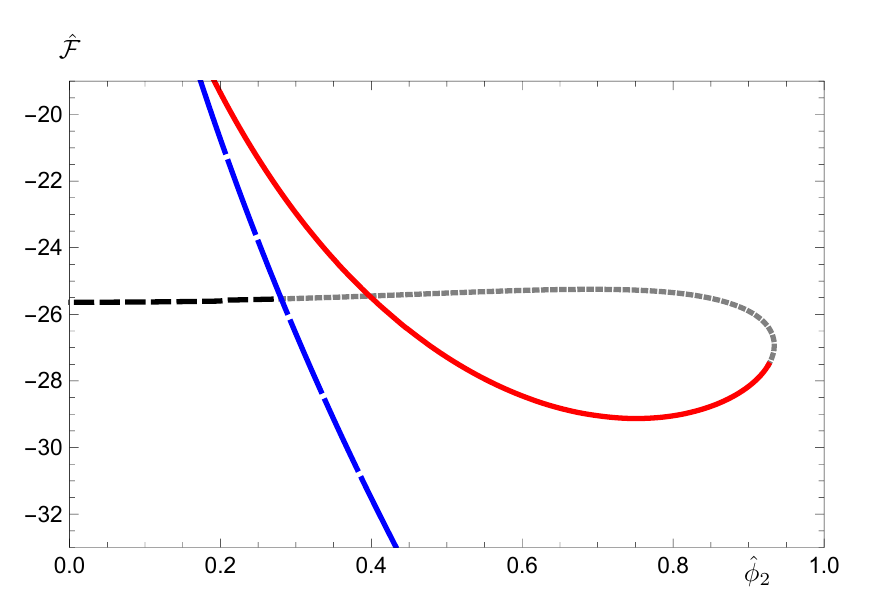}}
	\end{center}
	\vspace{-5mm}
	\caption[Magnification of the phase transition of the seven-dimensional supergravity]{The free energy density $\hat{\cf}$ as a function of the deformation parameter $\hat{\f}_{2}$ for the confining solutions, and
		the BSDW solutions (long-dashed blue line). We focus in particular on the region of the parameter space near to the phase transition at $\hat{\f}_{2}=\hat{\f}_{2}^{*}$; we denote the stable portion of the confining branch by the dashed black line, the metastable portion by the short-dashed grey line, and highlight the unstable tachyonic region in solid red.}
	\label{Fig:FplotCrossingWitten}
\end{figure}

 
The regular backgrounds that holographically realise confinement---for which one of the circles internal to the torus shrinks to a point in the deep IR and the geometry smoothly closes off---are represented in Figures~\ref{Fig:FplotWitten} and \ref{Fig:FplotCrossingWitten} by the short-dashed black, shortest-dashed grey, and solid red lines; this segmentation denotes solution stability, as explained in the captions. Finally, the class of \emph{skewed} solutions which are related to the confining backgrounds via the transformation $\U\to-\U$ are denoted by the long-dashed magenta line, and the badly singular domain-wall (BSDW) solutions are represented by the dashed blue line.\par 
Before proceeding to discuss the evident first-order phase transition in these plots, let us first emphasise some other important features of Fig.~\ref{Fig:FplotWitten}. We start by observing that all branches of solutions share a common point in the theory phase space and that, as expected, they each connect to the supersymmetric solutions at the origin; this corroborates our claim that each class of backgrounds, irrespective of their dissimilar geometric properties, are obtainable as deformations of the trivial critical point solution. Furthermore all classes have a finite, computable free energy density.\par
Within the $\hat{\f}_{2}\leqslant0$ region of the parameter space we find that $\hat{\cf}$ is bounded by the only two branches of solutions which fail to (locally) preserve the maximum six-dimensional Poincar\'{e} invariance. The confining backgrounds, which admit a sensible field theory interpretation at all scales and are free from singularities, minimise the free energy of the system and hence provide the energetically favoured geometric configuration; the skewed backgrounds, for which the $\eta$-parametrised $S^{1}$ diverges in volume rather than shrinking in the deep IR, instead maximise it. The various other branches of solutions, which all realise a domain-wall geometry with \pin extended to include the circle-compactified dimension parametrised by $\eta$, lie within the region of the phase space delimited by the $\U\neq0$ backgrounds.
\\
\indent Turning our attention to the complementary $\hat{\f}_{2}>0$ region of the plots, we observe similar evidence of a first-order phase transition as was encountered in our investigation of Romans $D=6$ supergravity. For small positive values of $\hat{\f}_{2}$ we see that the confining backgrounds continue to be energetically favoured, and the system prefers to maintain geometries which smoothly close off in the deep IR. As the source is dialled higher in order to approach the tachyonic instability, however, we notice the existence of a critical value $\hat{\f}_{2}=\hat{\f}_{2}^{*}$ at which the badly singular solutions intersect the confining branch and the two classes briefly have identical free energy. Beyond this critical value it is instead the BSDW solutions which minimise $\hat{\cf}$, and it becomes energetically favourable for the system to locally restore six-dimensional Poincar\'{e} invariance by allowing the $\eta$-parametrised $S^{1}$ to maintain a non-zero volume at all scales.\par
This feature is reminiscent of that identified for the six-dimensional supergravity in Sec.~\ref{Sec:RomansPhaseStruct}, and we again infer the existence of a first-order phase transition which identifies two distinct phases of the theory: when the source of the $\Delta=4$ operator is sufficiently small $\big(\hat{\f}_{2}<\hat{\f}_{2}^{*}\big)$ the system is in the \emph{confining phase}, and beyond the critical value $\big(\hat{\f}_{2}>\hat{\f}_{2}^{*}\big)$ it enters the \emph{domain-wall phase}. The region of the parameter space which contains the tachyonic instability is rendered inaccessible as a consequence of the system transitioning from the former phase to the latter, wherein energetic stability necessitates the spontaneous decompactification of the $\eta$-parametrised $S^{1}$ internal to the torus. We furthermore notice that there are confining branch backgrounds within the $\hat{\f}_{2}>\hat{\f}_{2}^{*}$ region of the plot which---while not energetically favoured---nevertheless do not contain a tachyonic state within their spectra of fluctuations; we refer to these backgrounds as \emph{metastable}, and represent them by the shortest-dashed grey line. In Fig.~\ref{Fig:FplotCrossingWitten} we present a magnified view of Fig.~\ref{Fig:FplotWitten}, focusing in particular on the region of the parameter space near to the phase transition.\par
We can be slightly more quantitative in our examination of the theory phase space, by numerically extracting the values of the various parameters at the phase transition. The coordinates of the point at which the confining and BSDW classes of solutions intersect in Figures~\ref{Fig:FplotWitten} and~\ref{Fig:FplotCrossingWitten}, corresponding to the critical values of the source and free energy, are given by
\begin{equation}
\big(\hat{\f}_{2}^{*}\,,\,\hat{\cf}^{*}\big)
\simeq\big(0.281\,,-25.54\big)\,,
\end{equation}
and the numerical backgrounds situated at this special point are generated using the IR parameter choices
\begin{equation}
\f_{I}^{*}\simeq 0.039\quad,\quad\f_{b}^{*}\simeq 33.54\,.
\end{equation}
The values of the two condensate parameters $\f_{4}$ and $\w_{6}$ on either side of the transition may also be determined numerically, though we remind the Reader that $\w_{6}$ is identically zero for all backgrounds which satisfy the domain-wall constraint $A=\frac{5}{2}\c$. By reintroducing the notation whereby subscripts `$<$' and `$>$' are used to denote quantities extracted in the confining and domain-wall phases, respectively, we obtain the following:
\begin{align}
\hat{\f}_{4\,<}^{*}&\simeq -0.347\quad,\quad
\hat{\f}_{4\,>}^{*}\simeq 546 \,,\notag\\
\hat{\w}_{6\,<}^{*}&\simeq 51.21\quad,\quad
\hat{\w}_{6\,>}^{*}= 0 \,,
\end{align}
and we note the significant enhancement of the parameter $\hat{\f}_{4}$ associated with condensate of the dimension-4 operator $\co_{4}$ dual to $\f$. This point will be of particular interest in Sec.~\ref{Sec:WittenProbeRevisit} when we come to re-examine the results of our probe spectrum analysis in the context of the theory phase space.\par
We conclude by observing that the free energy plots presented and discussed in this section show many similarities to the analogous plots in Figures~\ref{Fig:FplotBlue}, \ref{Fig:FplotGreen}, and \ref{Fig:FplotCrossing} for the six-dimensional theory, although unfortunately this resemblance also includes the two pathologies that were identified at the end of Sec.~\ref{Sec:RomansPhaseStruct}. The first is that there once again appears to be a maximum admissible value $\hat{\f}_{2}\simeq 2.50\-\ (\f_{2}\simeq 0.55)$ to which the source may be dialled, and no branch of solutions listed in our catalogue is able to explore the phase space beyond this point; there is no physical reason a priori to predict that such an upper bound should be imposed on this deformation parameter. The other reoccurring pathology presents itself once the system has transitioned to the domain-wall phase of the theory at the critical value $\hat{\f}_{2}=\hat{\f}_{2}^{*}$. The backgrounds which minimise $\hat{\cf}$ in this region of the parameter space, and hence which would be energetically favoured and physically realised, are those which evolve $\f$ towards a bad singularity at the end of space. These backgrounds do not admit a sensible dual description in terms of a lower-dimensional field theory, and our ability to interpret them holographically breaks down. As a consequence of this second observation we infer that the phase transition identifies an upper bound on the source at $\hat{\f}_{2}=\hat{\f}_{2}^{*}$. For deformations of the supersymmetric boundary CFT which correspond to the insertion of the $\Delta=4$ operator $\co_{4}$ dual to $\f$, and furthermore for which the source of this operator is dialled beyond the aforementioned upper bound, our dual formulation in terms of a sigma-model coupled to gravity is ineffective. We shall return to this discussion in Chapter~\ref{Chap:Discussion}.

\subsubsection{Characterising the phase transition}
At this stage we have conducted a systematic exploration of the phase space for the toroidally reduced seven-dimensional supergravity, by numerically computing the free energy density $\hat{\cf}$ as a function of the universal deformation parameters which characterise the asymptotic behaviour of the bulk fields in the far UV. The results of our analysis are presented in Figures~\ref{Fig:FplotWitten} and~\ref{Fig:FplotCrossingWitten}, and they reveal clear evidence of a first-order phase transition which prevents the system from accessing an unstable region of the theory parameter space as the $\co_{4}$ source is dialled beyond the critical value $\hat{\f}_{2}=\hat{\f}_{2}^{*}$. With the existence of this transition established, we proceed in this subsection to provide a more rigorous characterisation by studying the properties of some convenient order parameters within each of the two phases of the theory.\par
We refer to the first of these order parameters as the \emph{magnetisation} $\hat{\cm}$, here defined analogously to Eq.~(\ref{Eq:OrderParamM}) in Sec.~\ref{Sec:RomansPhaseStruct} for the six-dimensional supergravity:
\begin{equation}
\label{Eq:OrderParamM(Witten)}
\hat{\cm} \equiv \Lambda^{-4}\frac{\pa}{\pa\f_{2}}\cf(\f_{2},\Lambda) = \frac{\pa}{\pa\hat{\f}_{2}}\hat{\cf}(\hat{\f}_{2}) \,.
\end{equation}
That is, we consider the variation of $\cf$ with respect to the source $\f_{2}$ (measured in appropriate units of $\Lambda$) while holding fixed $\c_{U}=A_{U}=0$. We remind the Reader that our final expression for the free energy density as shown in Eq.~(\ref{Eq:WittenFinalFinalF}) is a function of the three UV parameters $\{\f_{2},\,\f_{4},\,\w_{6}\}$, and that those associated with the two operator VEVs ($\f_{4}$ and $\w_{6}$) are themselves implicitly dependent on $\f_{2}$; these functional dependences are not known in closed form, and hence the derivative in Eq.~(\ref{Eq:OrderParamM(Witten)}) cannot be evaluated analytically. Nevertheless, we can instead compute the derivative numerically by plotting the ratio of finite differences $\Delta\hat{\cf}$ and $\Delta\hat{\f}_{2}$ for the extracted data.\par 
Similarly to $\hat{\Delta}_{\rm DW}$ defined in Eq.~(\ref{Eq:OrderParamDelta}), we require that our second order parameter provides a convenient measure of to what degree any given background solution fails to locally preserve \pin along the $S^{1}$-compact dimension parametrised by $\eta$. For the seven-dimensional supergravity we find that the condensate $\w_{6}$ of the boundary operator dual to the sigma-model scalar $\w$ is a suitable candidate for such an order parameter; from the constrained metric ansatz in Eq.~(\ref{Eq:7DmetricConstrained}) we see that it is this VEV which is ultimately responsible for controlling the volume of the $S^{1}$ parametrised by $\eta$. Those backgrounds which realise a DW geometry by satisfying $\U=A-\frac{5}{2}=\w=0$ (and hence which preserve the full six-dimensional Poincar\'{e} invariance) have $\w_{6}$ vanishing identically, while we find that $\w_{6}(\f_{2})$ is some non-trivial function when determined for the confining and skewed branches of backgrounds.
\begin{figure}[h!]
	\begin{center}
		\makebox[\textwidth]{\includegraphics[
			width=0.43\paperwidth
			]
			{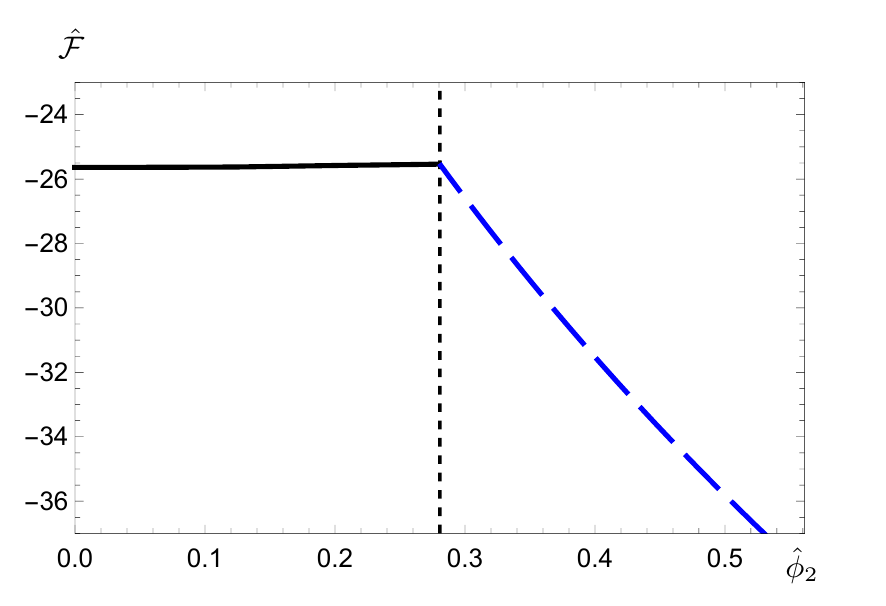}}
	\end{center}
	\vspace{-6mm}
	\caption[Minimum free energy at the phase transition of the seven-dimensional supergravity]{The minimum free energy density $\hat{\cf}$ as a function of the deformation parameter $\hat{\f}_{2}$, for the confining (solid black) and badly singular domain-wall (dashed blue) classes of background solutions. The vertical dashed line denotes the critical value $\hat{\f}_{2}=\hat{\f}_{2}^{*}\simeq0.281$ at the phase transition.}
	\label{Fig:minimumF}
\end{figure}\\
\indent In Figures~\ref{Fig:minimumF} and~\ref{Fig:dF} we present plots of the (minimum) free energy density $\hat{\cf}$ and magnetisation $\hat{\cm}$ as functions of $\hat{\f}_{2}$, restricting our attention to the confining and BSDW classes of solutions only and focusing in particular on the region of the parameter space near to the phase transition. Figure~\ref{Fig:minimumF} shows that although the minimum free energy of the system is continuous, it is evidently not differentiable at the critical value $\hat{\f}_{2}=\hat{\f}_{2}^{*}$ (denoted by the vertical dashed line). This observation is also clearly demonstrated in Figure~\ref{Fig:dF}, where the two phases are demarcated by a discontinuity in the order parameter $\hat{\cm}$ at the critical value of the source.
\begin{figure}[h!]
	\begin{center}
		\makebox[\textwidth]{\includegraphics[
			width=0.43\paperwidth
			]
			{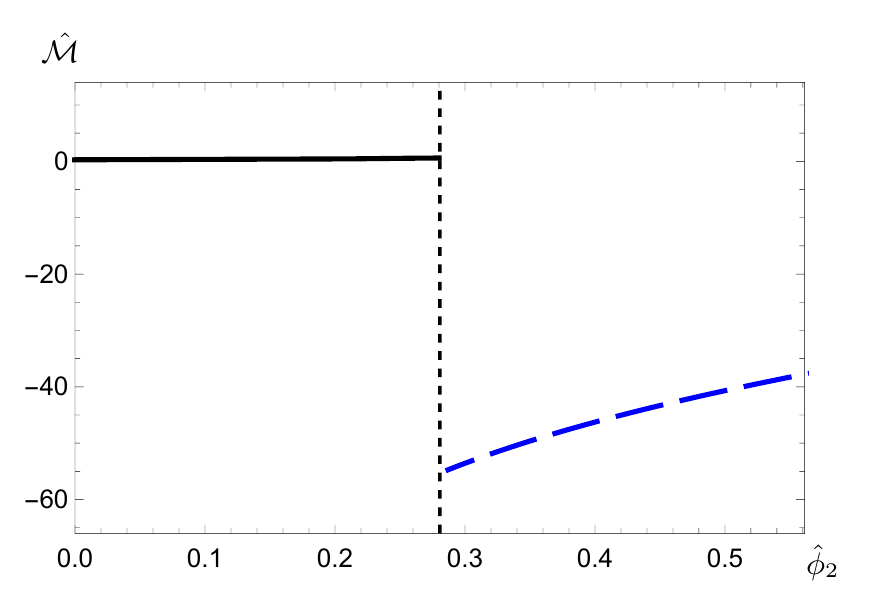}}
	\end{center}
	\vspace{-6mm}
	\caption[Magnetisation at the phase transition of the seven-dimensional supergravity]{The magnetisation $\hat{\cm}$ as a function of the deformation parameter $\hat{\f}_{2}$, for the confining (solid black) and badly singular domain-wall (dashed blue) classes of background solutions. The vertical dashed line denotes the critical value $\hat{\f}_{2}=\hat{\f}_{2}^{*}\simeq0.281$ at the phase transition.}
	\label{Fig:dF}
\end{figure}\\
\indent The second of our two order parameters $\hat{\w}_{6}$ is plotted in Fig.~\ref{Fig:omega6phi2}, and as predicted it too shows discontinuous behaviour as the theory transitions from the confining phase to the domain-wall phase. In the former case, the energetically favoured backgrounds geometrically realise confinement by shrinking the $\eta$-parametrised circle to a point in the deep IR; Poincar\'{e} invariance is preserved only within the $\{x^{\mu},\zeta\}$ subspace, and the operator dual to $\w$ acquires a non-zero VEV. Conversely, in the domain-wall phase it is the class of BSDW solutions which instead minimise $\hat{\cf}$, and the full six-dimensional Poincar\'{e} invariance is restored; this is reflected by the fact that the condensate $\hat{\w}_{6}$ is completely suppressed once the $\co_{4}$ source is dialled beyond the critical value $\hat{\f}_{2}>\hat{\f}_{2}^{*}$.     
\begin{figure}[h!]
	\begin{center}
		\makebox[\textwidth]{\hspace{4mm}\includegraphics[
			width=0.43\paperwidth
			]
			{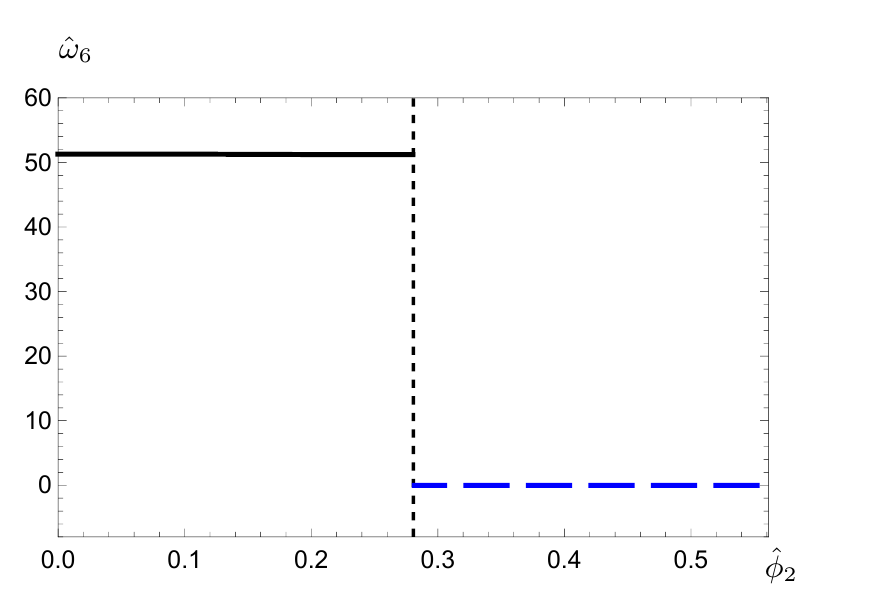}}
	\end{center}
	\vspace{-6mm}
	\caption[An order parameter at the phase transition of the seven-dimensional supergravity]{The UV parameter $\hat{\w}_{6}$ as a function of the deformation parameter $\hat{\f}_{2}$, for the confining (solid black) and badly singular domain-wall (dashed blue) classes of background solutions. The vertical dashed line denotes the critical value $\hat{\f}_{2}=\hat{\f}_{2}^{*}\simeq0.281$ at the phase transition.}
	\label{Fig:omega6phi2}
\end{figure}\\ 
\indent Finally for this subsection, in Fig.~\ref{Fig:phi4phi2} we present the parameter $\hat{\f}_{4}$ associated with the VEV of the operator dual to $\f$ as a function of its source $\hat{\f}_{2}$. We observe that the transition of the theory from the confining phase to the domain-wall phase, and consequently the spontaneous decompactification of the circular dimension parametrised by $\eta$, is associated with the significant enhancement of the condensate $\langle\co_{4}\rangle$. As previously discussed, the supersymmetric CFT dual to the trivial critical point solution $\f(\r)=0$ admits a deformation corresponding to the insertion of a $\Delta=4$ operator $\co_{4}$; this operator is sourced by the UV parameter $\f_{2}$ (which controls the explicit breaking of scale invariance) and has a VEV associated with $\f_{4}$ (governing the spontaneous breaking of scale invariance). We shall examine this phenomenon more closely in the next section, when we revisit the results of our probe spectrum computation from Sec.~\ref{Sec:WittenProbePlots}.    
\begin{figure}[h!]
	\begin{center}
		\makebox[\textwidth]{\includegraphics[
			width=0.43\paperwidth
			]
			{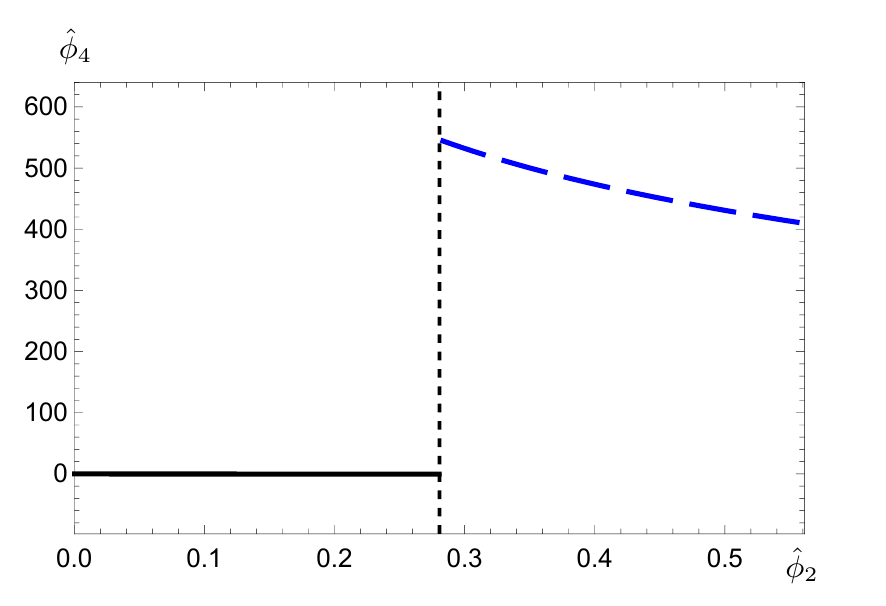}}
	\end{center}
	\vspace{-6mm}
	\caption[First plot of condensate enhancement at the phase transition of the seven-dimensional supergravity]{The UV parameter $\hat{\f}_{4}$ as a function of the deformation parameter $\hat{\f}_{2}$, for the confining (solid black) and badly singular domain-wall (dashed blue) classes of background solutions. The vertical dashed line denotes the critical value $\hat{\f}_{2}=\hat{\f}_{2}^{*}\simeq0.281$ at the phase transition.}
	\label{Fig:phi4phi2}
\end{figure}\\

\section{More about the dilaton }
\label{Sec:WittenProbeRevisit}
In Sec.~\ref{Sec:WittenMassPlots} we presented the numerical results of our spectra computation for the gauge-invariant modes which descend from the maximal seven-dimensional supergravity, truncated to retain only a sigma-model scalar coupled to gravity. Subsequently, in Sec.~\ref{Sec:WittenProbePlots} we conducted a probe state analysis of the spin-0 sector---implemented by `switching off' the scalar fluctuation of the metric---in order to determine how appreciably the excitations exhibit dilaton mixing. Motivated by our discovery of a tachyonic instability within the class of confining solutions, we have furthermore computed the free energy density as a function of a set of universal deformation parameters for several geometrically distinct backgrounds; in the process we uncovered evidence of a first-order phase transition which prevents the theory from accessing the unstable region of the parameter space. 
In this section we shall use the extracted data to revisit and contextualise the results of our probe state analysis, by examining the functional relations between the various UV parameters and comparing their behaviour to that of the spin-0 spectrum shown in Fig.~\ref{Fig:WittenSpectrumProbe}; some supplementary plots are presented in Appendix~\ref{App:ParamPlots}, though we shall not find it necessary to refer to these plots directly for our discussion here.\par 
Let us start by considering the rightmost region of the plot in Fig.~\ref{Fig:WittenSpectrumProbe}, where we observe that the lightest gauge-invariant resonance---which we remind the Reader is tachyonic---is gradually becoming massless in the large-$\f_{I}$ limit. There is a significant discrepancy between this state and the probe approximation, and we hence infer that it contains a substantial dilatonic component; since the dilaton is the pNGB associated with the spontaneous breaking of scale invariance, we additionally infer that its vanishing mass is indicative of scale invariance being explicitly restored. We find evidence to support this inference upon examining the behaviour of the UV parameter $\hat{\f}_{2}$---which sources the operator dual to $\f$, and which is responsible for governing the explicit breaking of conformal invariance---as a function of the tunable IR parameter $\f_{I}$. In the limit $\f_{I}\to\infty$ we find that $\hat{\f}_{2}\to0$, signalling that the deformation of the supersymmetric CFT which explicitly breaks dilatation invariance is being suppressed and that the corresponding pNGB should asymptotically become massless.\par 
Similarly, we may consider the behaviour of the other UV parameters in this same limit; we deduce that $\hat{\f}_{4}$ diverges as the source $\hat{\f}_{2}$ asymptotically approaches zero, while the parameter $\hat{\w}_{6}$ instead vanishes. Since $\w_{6}$ is identified as the condensate of the marginal operator dual to the sigma-model scalar $\w$---and moreover is ultimately responsible for governing the volume of the $S^{1}$ parametrised by $\eta$---we should expect to find that it is suppressed in the limit $\f_{I}\to\infty$, for which $\hat{\cf}$ along all branches of backgrounds converges to the supersymmetric solutions (realising a domain-wall geometry). The enhancement of the UV parameter $\hat{\f}_{4}$ associated with the condensate $\langle\co_{4}\rangle$ of the $\Delta=4$ operator instead implies that conformal invariance is spontaneously broken in the dual field theory as $\f_{I}\to\infty$.          
\begin{figure}[h!]
	\begin{center}
		\makebox[\textwidth]{\includegraphics[
			width=0.45\paperwidth
			]
			{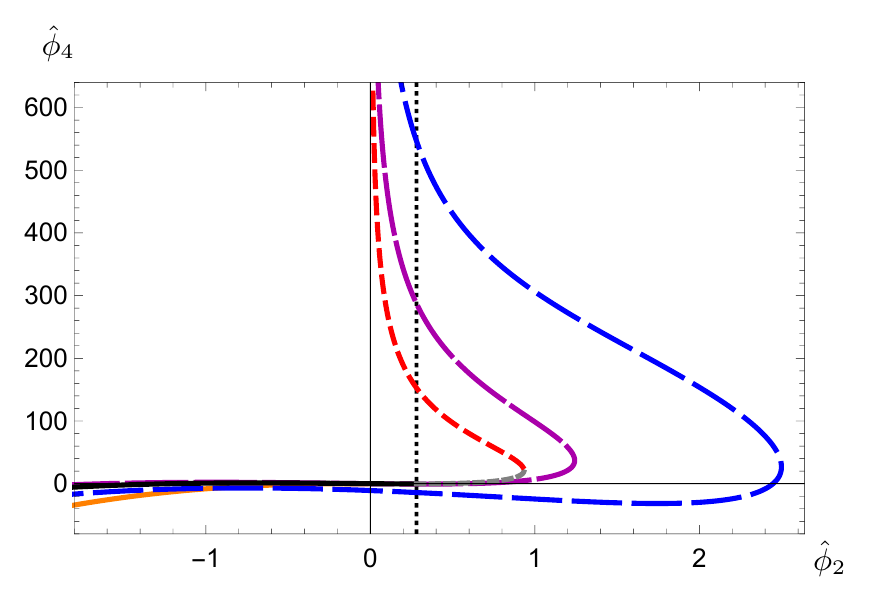}}
	\end{center}
	\vspace{-6mm}
	\caption[Second plot of condensate enhancement at the phase transition of the seven-dimensional supergravity]{The UV parameter $\hat{\f}_{4}$ as a function of the deformation parameter $\hat{\f}_{2}$, for the confining (solid black, short-dashed grey, and dashed red), skewed (longest-dashed magenta), and badly singular domain-wall (long-dashed blue) classes of background solutions. The vertical dashed line denotes the critical value $\hat{\f}_{2}=\hat{\f}_{2}^{*}\simeq0.281$ at the phase transition.}
	\label{Fig:CondensatePhi4}
\end{figure}\\
\indent At large values of the tunable IR parameter $\f_{I}$ we therefore have that the explicit breaking of scale invariance is suppressed (since $\hat{\f}_{2}\to0$), the spontaneous breaking of scale invariance is enhanced (as $\hat{\f}_{4}$ becomes large), and an asymptotically massless scalar resonance is unambiguously missed by our probe comparison; considered altogether, we may infer that this parametrically light spin-0 state experiences significant mixing with the dilaton. In Fig.~\ref{Fig:CondensatePhi4} we also show $\hat{\f}_{4}(\hat{\f}_{2})$ for the branch of badly singular DW solutions, emphasising in particular the enhancement of $\hat{\f}_{4}$ at the phase transition.\par
As with the six-dimensional theory discussed in Chapter~\ref{Chap:EnergeticsRomans} however, we should interpret this interesting observation cautiously; our energetics analysis of the theory phase structure uncovered the existence of a first-order phase transition, which prevents the confining backgrounds within the large-$\f_{I}$ region of the parameter space (and hence also the aforementioned light dilatonic state) from ever being energetically favoured. To clarify this point, we remind the Reader of our discussion in Sec.~\ref{Sec:WittenPhaseStruct} where we identified two distinct phases within the torus-reduced theory; it is the \emph{domain-wall phase}---which does not admit a sensible dual description in terms of a lower-dimensional QFT---that is physically realised in the $\f_{I}\to\infty$ limit, and we must therefore concede that our results can only be interpreted holographically for confining backgrounds that are generated with $\f_{I}$ below the critical value $\hat{\f}_{I}^{*}$ at the phase transition. As we have previously mentioned in Sec.~\ref{Sec:RomansPhaseStruct} this limitation is not necessarily indicative of a pathology in the theory, but rather highlights the fact that our analysis of the phase structure using an effective supergravity approximation is most likely incomplete; we shall elaborate on this issue in Chapter~\ref{Chap:Discussion}. \par 
Turning our attention now to the small-$\f_{I}$ region of Fig.~\ref{Fig:WittenSpectrumProbe} we notice that---in contrast to the analogous probe plot in Fig.~\ref{Fig:RomansSpectrumProbe}, neither the lightest nor next-to-lightest gauge-invariant resonances in the spectrum show any indication of containing a significant dilatonic component; they are well approximated by the probe computation (refer back to Sec.~\ref{Sec:WittenProbePlots} for a brief discussion of this phenomenon as it pertains to the results of Chapter.~\ref{Chap:nTorus}). We instead observe that in proximity of the phase transition, the lightest state which exhibits dilaton mixing is a relatively heavy $\mathfrak{a}^{\c}$ excitation with mass $M\approx1.45$. As with the corresponding region of the plot in Fig.~\ref{Fig:RomansSpectrumProbe}, the reason for this interesting phenomenon may be attributed to the behaviour of the marginal operator's VEV. As the source of the $\Delta=4$ operator $\hat{\f}_{2}$ approaches zero from below, we find that the UV deformation parameter $\hat{\f}_{4}$ (associated with the $\co_{4}$ operator condensate) is suppressed. In the same limit, we also observe that the condensate of the marginal operator dual to $\w$---governed by the parameter $\hat{\w}_{6}$---instead takes a comparatively large non-zero value; it is hence this VEV which is responsible for the spontaneous breaking of scale invariance in the dual theory, and for the mixing of a subset of the massive scalar states with the dilaton.


\chapter{Discussion}
\label{Chap:Discussion}
We dedicate this short chapter to summarising the key results of this Thesis---based on work contained within Refs.~\cite{Elander:2018aub,Elander:2020csd,Elander:2020ial,Elander:2020fmv}---and emphasising the novel research which it contributes to the literature on top-down holography.

\subsubsection{Summary of research}
In Chapters~\ref{Chap:SpectraRomans} and~\ref{Chap:SpectraWitten} we computed the spectra of massive excitations (dually corresponding to composite states) for two well-established and rigorously defined supergravities, which are known to provide the low-energy effective description of superstring theory and M-theory; these are the six-dimensional half-maximal supergravity originally written by Romans~\cite{Romans:1985tw} and the seven-dimensional maximal supergravity first constructed in Refs.~\cite{Pernici:1984xx,Pernici:1984zw}, respectively. For both models we considered the field fluctuations of a sigma-model coupled to five-dimensional gravity, on background geometries which holographically realise confinement by smoothly shrinking a circle-compactified dimension to zero volume at the end of space.\par 
Our numerical computations for these modes extend those which have previously been carried out in the literature by considering background which interpolate between the two critical point solutions of each theory, in addition to those which permit $\f$ to explore a runaway direction of the scalar potentials. In the case of the $D=6$ theory we furthermore considered the physical excitations which descend from the $1$- and $2$-forms of the complete action which defines the theory, an exercise that had not previously been attempted prior to the work in Ref.~\cite{Elander:2018aub}. To tackle this issue we considered the fluctuations of generic $p$-forms (for $p=1,2$), and demonstrated explicitly the derivation of their corresponding fluctuation equations using appropriate $R_{\xi}$ gauge-fixing terms; in the process we addressed some technical subtleties associated with gauge-invariance and Higgsing effects.\par 
The other important result of these spectra computations was our discovery---for both of the supergravity theories---of a tachyonic instability within their respective scalar sectors. We found that in each case a tunable IR parameter $\f_{I}$ could be dialled in such a way that the mass of the lightest resonance is parametrically suppressed, until eventually the system is brought close enough to the aforementioned instability that this state becomes tachyonic at some value $\f_{I}=\f_{I}^{\t}$.\par 
Motivated by previous work in Ref.~\cite{Kaplan:2009kr} and Ref.~\cite{Pomarol:2019aae}, we tested a diagnostic tool (discussed in Ref.~\cite{Elander:2013jqa} and first implemented in Ref.~\cite{Elander:2020csd}) designed to detect mixing effects between scalar resonances and the pseudo-Nambu--Goldstone boson associated with the spontaneous breaking of conformal invariance: the dilaton. This test consists of neglecting the contribution $h$ to the gauge-invariant variables $\mathfrak{a}^{a}$---which corresponds to the spin-0 fluctuations of the decomposed metric---(hence `switching off' the back-reaction these states might otherwise induce in the underlying geometry), and comparing the resultant \emph{probe} spectra to the proper complete computation; where appreciable discrepancies arise we infer the presence of dilaton mixing effects.\par 
We applied this probe analysis to the scalar sectors of the two supergravity theories and found that---in proximity of their unique trivial solutions---the lightest spin-0 resonance is \emph{not} parametrically light, and nor does it exhibit the features associated with being a dilaton admixture; some heavier states in this region of the parameter space, however, did appear to contain a significant dilatonic component, and we referred to these states as \emph{approximate} dilatons. By dialling an IR parameter in order to explore beyond the trivial fixed point solution---specifically, in proximity of the aforementioned tachyonic instability---we furthermore found that the (parametrically) lightest scalar excitation in each theory is \emph{not} effectively approximated by its corresponding probe state $\mathfrak{p}^{a}$, and hence \emph{is} identifiable with the dilaton. Finally, in the $\f_{I}\to\infty$ limit (well beyond the appearance of an instability) we noticed that in both supergravity theories the mass of the tachyonic state asymptotically converges to zero from below, a phenomenon which we again attributed to dilaton mixing effects.\par
Having uncovered an instability in the spectra of the two theories we proposed to investigate their respective phase structures, with the understanding that there must necessarily exist some mechanism by which the branch of pathological confining backgrounds would be prohibited from being physically realised. With this motivation established, we proceeded to compile a catalogue of geometrically-distinct backgrounds which are admissible within each compactified supergravity; several of these branches of solutions were unknown in the literature before Refs.~\cite{Elander:2020ial,Elander:2020fmv}, and some within the seven-dimensional theory (the \emph{skewed} and \emph{general singular} classes) are new to this Thesis. Interestingly, these catalogues of solutions---while not exhaustive---show a remarkable degree of commonality between the two supergravity theories.\par
Further to our proposed investigation, we proceeded to derive a general expression for the free energy density $\cf$ from the holographically renormalised on-shell action of each theory, as a function of a set of universal (class-independent) deformation parameters.\ We additionally introduced a physically motivated common energy scale $\Lambda$ to facilitate comparison between the various types of solutions, effectively ensuring that the space of free parameters within each class has the same dimensionality. By implementing a numerical routine in order to extract the required data for the aforementioned UV deformation parameters, we were therefore able to explore the phase structure of the two supergravity theories in Chapters~\ref{Chap:EnergeticsRomans} and~\ref{Chap:EnergeticsWitten}.\par
Our next important result---and perhaps our most significant---came from this energetics analysis: we uncovered strong evidence within both models of a first-order phase transition which prevents the theory from ever reaching the tachyonic instability by moving in parameter space along the branch of regular solutions. We found that beyond a certain critical value of a deformation parameter $\hat{\f}_{2}^{*}$ (which is associated in each case with the source of a relevant boundary operator) the system energetically favours a branch of singular domain-wall backgrounds, and the parameter space is divided into two distinct phases: the \emph{confining} phase for small source deformations $\hat{\f}_{2}<\hat{\f}_{2}^{*}$, and the \emph{domain-wall} phase for $\hat{\f}_{2}>\hat{\f}_{2}^{*}$. This crucial observation furthermore identifies three separate regions along the branch of confining backgrounds: \emph{stable} ($\f_{I}<\f_{I}^{*}$), \emph{metastable} ($\f_{I}^{*}<\f_{I}<\f_{I}^{\t}$), and \emph{unstable} ($\f_{I}>\f_{I}^{\t}$).\par
Finally, by examining the numerical parameter data extracted from our phase structure investigation we were able to contextualise the results of our probe state analysis. Crucially, we deduced that the parametrically light dilatonic excitation---which appears in proximity of the tachyonic instability---actually lies within the metastable portion of the confining branch, and is hence not physically realised.\ We furthermore discovered that the \emph{approximate} dilaton states within the stable portion of the branch arise due to the complicated interplay between enhanced operator condensates and the nearby instability. The parametrically light dilaton---which emerges in the $\f_{I}\to\infty$ limit due to the eventual suppression of the source deformation which explicitly breaks conformal invariance---is rendered inaccessible, as it resides within the region of the parameter space well beyond a phase transition.        

\subsubsection{General observations}
We next briefly comment on some interesting general phenomena which have proven to be recurrent within both of the toroidally compactified supergravities that we have studied, and which we expect may be more widely applicable to other similar models.\par
The first such observation is regarding the spectra of a subset of the resonances which descend from the sigma-model coupled to gravity: specifically, the spin-2 fluctuations $\mathfrak{e}^{\mu}_{\ \nu}$ of the metric, and the spin-0 fluctuations $\mathfrak{a}^{a}$ associated with $\c$. As discussed in Chapters~\ref{Chap:SpectraRomans} and~\ref{Chap:SpectraWitten}, we noticed that for certain choices of the IR parameter $\f_{I}$---corresponding to solutions which interpolate between the two critical points of the scalar potential---these excitations exhibit an interesting \emph{universality} feature; the spectra show no dependence on specific details of the background fields being fluctuated within this region of the parameter space, and appear sensitive only to the confinement mechanism. For the six-dimensional supergravity we demonstrated that this is also the case for the spin-1 excitations corresponding to the graviphoton. While it is not clear whether this feature of the spectra is indicative of some deeper underlying physical phenomenon, it is nonetheless noteworthy.\par
Our second general observation also pertains to the mass spectra of each supergravity, and we have already discussed it several times throughout this Thesis; we have found that both of the theories encounter a tachyonic instability in their spin-0 sector for choices of an IR parameter which dials $\f$ too far beyond their respective trivial fixed point solutions, along a runaway direction of the potential. Since we were investigating the dimensional-reduction of two well-defined and established supergravities, this discovery could have been rather problematic. However, as we have demonstrated, each theory is protected by the fact that this instability always resides within an inaccessible region of the parameter space---owing to the presence of a first-order phase transition which energetically disfavours the branch of regular solutions---and the system is prohibited from approaching arbitrarily close to the instability.\par
The third general feature that we have observed to be common to both the six- and seven-dimensional theories, and which follows from our probe state analysis, is that dilaton mixing effects are always present to some degree within the scalar mass spectra; for all values of the tunable IR parameter $\f_{I}$, at least some of the resonances are missed by the probe approximation. Furthermore we find that the lightest spin-0 resonance can always be dialled in such a way that its mass is parametrically suppressed compared to all other states within the tower, and in proximity of the point at which this state becomes massless our probe approximation unambiguously fails to capture it. The important caveat to this observation is that this phenomenon occurs only along the metastable portion of the confining branch, and no such parametrically light dilaton was found to exist within the stable region of the parameter space.\par
For our final general observation, we notice that the catalogue of admissible backgrounds compiled for each of the two supergravity theories---while not necessarily exhaustive---shows rather surprising similarities in both cases; we find that every class of solutions within one model has an analogue exhibiting the same geometric properties within the other. Moreover, each class of backgrounds assumes a similar role within their respective phase structures: we find that two related branches of solutions---those which we referred to as \emph{confining} and \emph{skewed}---provide the delimitations of the free energy density $\hat{\cf}$ within the $\f\leqslant0$ region of the parameter space (the former minimising $\hat{\cf}$ and hence being energetically favoured, the latter instead maximising it), while for both theories it is a branch of singular domain-wall backgrounds which instigate their respective phase transitions, rendering an instability inaccessible. There does not seem to be any obvious reason that such remarkably similar phase structures should have been predicted a priori, nor is it clear whether this compelling phenomenon is attributable to some deeper underlying physical mechanism. What is evident however, is that our study of these two theories must be incomplete; in both cases we have uncovered a region of parameter space in which a class of badly singular solutions provide the energetically stable background geometries, and for which our ability to interpret the dynamics of the dual boundary theory breaks down. We shall discuss this issue further in the next subsection.

\subsubsection{Outlook and open questions}
We conclude this discussion by addressing some of the issues that our work has left unresolved, and posing some interesting questions which future efforts may seek to address. Furthermore, we provide a brief summary of potential avenues for future research which would build upon the foundational work presented within this Thesis.\par
We remind the Reader that our investigation into the phase structure of each supergravity yielded two rather unsatisfying results. The first of these was the unexpected upper bound which was uncovered for the UV deformation parameter associated with the source of the operator dual to $\f$. Within both models that we considered, the existence of a maximum permitted value for $\hat{\f}_{2}$ limited our ability to fully explore the theory phase space. There is no obvious reason that such an upper bound should be imposed on each supergravity, and we must hence consider the possibility that other branches of solutions not listed in our catalogue may yet be identified. Perhaps a more thorough classification of admissible backgrounds---which may require that we either extend the model field content, or consider more general ans\"{a}tze for the backgrounds---would yield solutions which permit one to explore regions of the phase space corresponding to arbitrarily large \emph{operator deformations}~\cite{Skenderis:2002wp} of the holographically dual supersymmetric CFT.\par 
The other pathology which our work uncovered was the existence of a so-called \emph{domain-wall phase} within each compactified supergravity. As we have discussed, one of our major new findings is that a tachyonic instability is, in both theories, rendered physically inaccessible due to the presence of a first-order phase transition; a branch of \emph{badly} singular domain-wall backgrounds are energetically favoured for choices of the source parameter $\hat{\f}_{2}>\hat{\f}_{2}^{*}$ which would otherwise drive the system towards this instability. These badly singular solutions do not admit a sensible dual interpretation in terms of a lower-dimensional QFT, and hence we must concede that our holographic description of boundary dynamics is applicable only to the stable portion of the confining branch, before the phase transition. It is not clear whether the gravitational model may be improved to overcome this limitation; perhaps a more complete analysis would require that we also retain the Kaluza-Klein modes of the compactified spaces in each case. Moreover it is possible that an effective supergravity formulation is insufficient for the purposes of such a phase structure investigation, and that we should instead seek to explore the two theories in terms of extended objects to capture physical effects which are omitted by the supergravity approximation. Either way, we defer this challenging problem to future studies.\par   
Let us conclude by briefly discussing some potential avenues for further research, which might entail either extending the work contained in this Thesis or applying the tools and methods we have developed to other interesting models.\par
Firstly, a natural extension to our investigation of the seven-dimensional maximal supergravity would be to relax our self-imposed geometrical constraint $\U=A-\frac{5}{2}\c=\w$, and hence to conduct a similar exploration of the theory's phase structure for a potentially much richer space of solutions. Although we found it convenient to restrict our attention to a subset of backgrounds for the purposes of this preliminary work, allowing for other---perhaps more exotic---solutions might go some way to addressing the two pathologies that we have encountered. Further building on this Thesis, a more complete study would also include the branch of general singular backgrounds which were identified in Sec.~\ref{Sec:WittenClasses}, but which we neglected to include in our numerical free energy analysis. We predict that these solutions would fill the parameter space region delimited by the confining and skewed branches---analogously to the case of the six-dimensional theory---but nonetheless it would be worthwhile to test this hypothesis.\par
Secondly, while our study of the two distinct supergravities has in both cases uncovered evidence of a parametrically light dilaton, we have furthermore demonstrated that this resonance exists only along a metastable portion of a branch of confining solutions (beyond a phase transition). It is possible that other models exist---perhaps involving alternative compactifications---which realise a similar mechanism to avoid pathological regions of their parameter space. Moreover, there is no reason to assume that such phase transitions should be as strong as those discovered in Refs.~\cite{Elander:2020ial,Elander:2020fmv}; it would be of significant interest to find examples of models wherein such a phase transition is sufficiently weak that---by dialling appropriate deformation parameters---the theory yields a dilatonic excitation with a parametrically suppressed mass, while still realising a stable field configuration.\par 
The third and final potential line of research that we shall discuss is based upon Nahm's classification~\cite{Nahm:1977tg} of supersymmetric $\text{AdS}_{D}$ solutions within supergravity. This Thesis has focused primarily on two of these cases: the six-dimensional theory---first constructed by Romans~\cite{Romans:1985tw}---which is obtainable by compactifying and reducing ten-dimensional massive type-IIA supergravity on a warped four-sphere $\cm^{10}\to \text{AdS}_{6}\times S^{4}$, and the seven-dimensional theory first constructed in Refs.~\cite{Pernici:1984xx,Pernici:1984zw} by compactifying eleven-dimensional supergravity on a four-sphere $\cm^{11}\to\text{AdS}_{7}\times S^{4}$. In a recent paper~\cite{Elander:2021wkc} we conducted a similar investigation for the five-dimensional theory---first constructed in Refs.~\cite{Pernici:1985ju,Gunaydin:1984qu,Gunaydin:1985cu}---which is obtainable by compactifying ten-dimensional type-IIB supergravity on a five-sphere~\cite{Gunaydin:1984fk,Kim:1985ez} $\cm^{10}\to \text{AdS}_{5}\times S^{5}$ and (consistently) truncating the Kaluza-Klein modes of the compactified space~\cite{Lee:2014mla,Baguet:2015sma}. The study restricted attention to a sub-truncation which preserves certain subgroups of the five-sphere isometry group $SO(6)$ (isomorphic to the $SU(4)_{R}$ symmetry of the dual theory), and we refer the interested Reader to the paper itself for further details.\par 
Intriguingly, although this study was conducted within a lower-dimensional gravitational model---including backgrounds which geometrically realise a dual description of confinement in a \emph{three}-dimensional boundary theory---we nevertheless found that remarkably similar phenomena to those discussed in this Thesis emerged. A tachyonic instability was uncovered along a branch of regular backgrounds, and furthermore we demonstrated the existence of yet another phase transition which rendered the pathological region of the parameter space energetically inaccessible.\par
Based on these fascinating recurrent physical features, a final avenue for potential future work---which promises to be fruitful and provides a natural extension of this Thesis---would be to conduct an analogous investigation for the last remaining theory encompassed by Nahm's classification: the four-dimensional maximal supergravity~\cite{Freund:1980xh,deWit:1986oxb,deWit:1981sst,deWit:1982bul,Page:1983mke,Duff:1990xz,Nicolai:2011cy} obtainable via the reduction on a seven-sphere of eleven-dimensional supergravity $\cm^{11}\to\text{AdS}_{4}\times S^{7}$, retaining an $SO(8)$ gauge group.\ It is reasonable to predict that similar phenomenological results might be uncovered: a tachyonic instability along a branch of regular backgrounds which geometrically realise confinement, a first-order phase transition which protects the theory by rendering the pathology inaccessible, and a parametrically light scalar resonance which is identifiable with the dilaton, though which exists only along a metastable portion of the solution branch. Observing these same physical features within otherwise distinct theories is rather exciting, and provides us with the motivation to continue conducting similar investigations within other models---the hope being that there is perhaps some deeper underlying physics yet to be uncovered.


\appendix
\chapter{Formulating 2-forms in four dimensions}
\label{App:2-Forms}
In this appendix we shall derive some convenient expressions for the Lagrangian of a $U(1)$-invariant 2-form. These equivalent formulations naturally generalise to higher dimensions, which we exploit in our derivation of the $p$-form fluctuation equations in Sec.~\ref{Sec:MoreBosons}.\par 
Let us start by considering the following generic action describing a four-dimensional theory with a spontaneously-broken $U(1)$ gauge symmetry, and with Minkowski signature $\{-,+,+,+\}$:
\begin{equation}
\label{Eq:4dtheory}
\cs_{o}=\int\is\di^{4}x\,\cl_{o}=
\int\is\di^4x\,\bigg\{ 
-\frac{1}{4}F_{\mu\nu}F^{\mu\nu}
-\half\big(\pa_{\mu}\pi + mA_{\mu}\big)\big(\pa^{\mu}\pi + mA^{\mu}\big)
\bigg\}\,,
\end{equation}
where $F_{\mu\nu}\equiv 2\pa_{[\mu}A_{\nu]}$ is the field strength of the four-vector $A_{\mu}$, $\pi$ is a pseudo-scalar field, and the mass $m$ is a symmetry-breaking parameter; gauge-invariance of the term $\pa_\mu \pi +mA_{\mu}$ is guaranteed $\forall m$ under the transformations $A_{\mu}\rightarrow A_{\mu}-\pa_{\mu}\a$ and $\pi\rightarrow\pi+m\a$, where $\a=\a(x^{\mu})$.\par 
We can remove unphysical mixing between vector and scalar terms by supplementing this action with the following gauge-fixing term:
\begin{equation}
\cl_{\xi}=-\frac{1}{2\xi}\big(\pa^{\mu}A_{\mu} + \xi m\pi\big)^{2}\,,
\end{equation}
so that we have
\begin{align}
\cl_{o}+\cl_{\xi}=&-\frac{1}{4}F_{\mu\nu}F^{\mu\nu}-\half m^{2}A_{\mu}A^{\mu}-\frac{1}{2\xi}\big(\pa^{\mu}A_{\mu}\big)^{2}\nn\\
&-\half\pa_{\mu}\pi\pa^{\mu}\pi -\frac{\xi}{2} m^{2}\pi^{2} -\pa^{\mu}\big(m\pi A_{\mu}\big)\,.
\end{align}
The total derivative term can be neglected, and hence we find that the classical equations of motion for the two fields $\{A_{\mu},\,\pi\}$ decouple. The unspecified gauge-fixing parameter $\xi$ may be freely chosen. As usual, the computation of correlation functions requires that we write the corresponding generating functional; we introduce the partition function (or \emph{path integral}) as follows:
\begin{equation}
\label{Eq:PartFunc}
\cz[J]\equiv\cn_{o}\is\int\hspace{-1mm}\cd A_{\mu}\cd\pi\,e^{i\int\text{d}^{4}x\, 
	\big(\cl_{o} + \cl_{\xi} + \cl_{J}\big)} 
\quad \text{with}\quad \cd A_{\mu}\equiv\underset{x}{\prod}\,\di A_{\mu}(x)\,,
\end{equation}
and the integrand measure for $\pi(x)$ is similarly defined. The prefactor $\cn_{o}$ is some generic constant, while the supplementary Lagrangian density $\cl_{J}$ collectively represents source terms, which we do not specify. From this point onwards we shall follow closely the derivation in Appendix~A.2 of Ref.~\cite{Elander:2018aub}, which itself generalises the discussion of Ref.~\cite{Bijnens:1995ii} (see also Refs.~\cite{Ecker:1989yg,Bruns:2004tj}).\par
In four dimensions a massless 2-form is dual to a massless scalar field (both propagating a single degree of freedom), while a massive 2-form is equivalent to a massive vector field (each carrying 3 degrees of freedom). Motivated by this observation, we shall demonstrate that the same physical theory described by the action in Eq.~(\ref{Eq:4dtheory}) can be equivalently reformulated in terms of 2-forms only, though differing by which gauge symmetries are manifested. To proceed, we introduce the 2-forms $B_{\mu\nu}$ and $\tilde{B}_{\mu\nu}$ by defining: 
\begin{equation}
\label{Eq:Define2forms}
\pa_{\mu}\pi +m A_{\mu} \equiv \frac{1}{2}\epsilon_{\mu\nu\r\s} \pa^{\nu}B^{\r\s}\equiv\pa^{\nu}\tilde{B}_{\mu\nu}\,,
\end{equation}
where $\epsilon_{\mu\nu\r\s}$ is the four-index Levi-Civita tensor, and we note that this expression is invariant under the following vectorial gauge transformation:
\begin{equation}
\label{Eq:2formTransform}
B_{\mu\nu}\rightarrow B_{\mu\nu}-{2}\partial_{[\mu}\alpha_{\nu]}\,,
\end{equation}
with $\a_{\nu}$ dependent on the Minkowski coordinates $x^{\mu}$, leaving both the combination $\pa_{\mu}\pi +m A_{\mu}$ and the 2-form $\tilde{B}_{\mu\nu}$ unaffected; we shall return to this observation later.\par 
To rewrite our partition function $\cz[J]$ in terms of the new 2-form field $\tilde{B}_{\mu\nu}$ we exploit the ``insertion of one'' functional identity introduced within the Faddeev--Popov procedure~\cite{Faddeev:1967fc}, which is used to factor out divergent contributions to a path integral coming from gauge redundancies. Schematically, this identity reads:
\begin{equation}
1=\int\is\cd\a\, 
\d\big(\mathscr{G}\big)\Delta_{\mathscr{G}}^{\a }
=\int\is\cd\a\, 
\d\big(\mathscr{G}(A^{\a})\big)
\bigg|\frac{\d}{\d\a}\mathscr{G}(A^{\a})\bigg|\,,
\end{equation}
which would be inserted into a theory that admits a gauge transformation sending $A\to A^{\a}$ (where $\a$ denotes the transformed field) for some redundant gauge-parameter $\a$. The expression $\mathscr{G}$ appearing in the delta function argument is chosen to enforce a gauge-fixing condition that constrains the functional integral to cover only physically distinct field configurations, by selecting a single representative configuration along each \emph{gauge orbit}. The Faddeev--Popov determinant $\Delta_{\mathscr{G}}^{\a}$ is obtained by functionally differentiating $\mathscr{G}$; for Abelian theories (such as our Eq.~(\ref{Eq:4dtheory}) theory) this quantity is independent of the gauge field $A$, and may hence be factored out of the complete path integral. The non-Abelian case is slightly more subtle as $\Delta_{\mathscr{G}}^{\a}$ picks up a dependence on the gauge field and the group structure constants, and one typically proceeds by representing the determinant as a functional integral over a set of anti-commuting \emph{ghost} fields; this will not be of any further relevance to our discussion, however.\par 
With this brief aside out of the way, we introduce the following functional identities:
\begin{align}
1&=\cn_{\ci}\is\int\is\cd\bar{\ci}_{\mu\nu}\,  e^{i\int\text{d}^{4}x\, 
\bar{\ci}_{\mu\nu}\bar{\ci}^{\mu\nu}}\,,\\
1&=\cn_{B}\is\int\is\cd\tilde{B}_{\mu\nu}\, 
\d\left(A_{\mu}+\tfrac{1}{m}\pa_{\mu}\pi
-\tfrac{1}{m}\pa^{\nu}\tilde{B}_{\mu\nu}\right)\,, \label{Eq:FunctIdent}
\end{align}
for some unspecified prefactor constants $\cn_{\ci}$ and $\cn_{B}$. The new 2-form $\bar{\ci}_{\mu\nu}$ is a generic auxiliary field, and the delta function in Eq.~(\ref{Eq:FunctIdent}) is chosen to enforce the equivalence from Eq.~(\ref{Eq:Define2forms}). By inserting these identities into Eq.~(\ref{Eq:PartFunc}) we therefore have:
\begin{align}
\cz[J]&=\cn\is\int\is
\cd A_{\mu}\cd\pi\cd\bar{\ci}_{\mu\nu}\cd\tilde{B}_{\mu\nu}\, \d\left(A_{\mu}+\tfrac{1}{m}\pa_{\mu}\pi
-\tfrac{1}{m}\pa^{\nu}\tilde{B}_{\mu\nu}\right)\nn\\
&\hspace{10mm}
\times\,\,  e^{i\int\text{d}^{4}x\,\bar{\ci}_{\mu\nu}\bar{\ci}^{\mu\nu}}
\,e^{i\int\text{d}^{4}x
\,\big(\cl_{o} +\cl_{\xi} +\cl_{J}  \big)}\,,
\end{align}
where now $\cn=\cn_{o}\cn_{\ci}\cn_{B}$. Next we perform the integration over $A_{\mu}$, using the $\d$-function to rewrite the vector field in terms of the pseudo-scalar $\pi$ and the 2-form $\tilde{B}_{\mu\nu}$. All dependence of the Lagrangian density $\cl_{o}$ on $\pi$ disappears in the process, so that $\cl_{o}(A_{\mu},\pi)=\cl_{o}(\tilde{B}_{\mu\nu})$:
{\small
\begin{align}
\cl_{o}(\tilde{B}_{\mu\nu})&=-\frac{1}{4m^2}
\left[\pa_{\mu}\pa^{\s}\tilde{B}_{\nu\s}
-\pa_{\nu}\pa^{\s}\tilde{B}_{\mu\s}\right]
\left[\pa^{\mu}\pa^{\bar{\s}}\tilde{B}^{\nu}_{\,\,\,\,\bar{\s}}
-\pa^{\nu}\pa^{\bar{\s}}\tilde{B}^{\mu}_{\,\,\,\,\bar{\s}}\right]
-\frac{1}{2}\pa^{\s}\tilde{B}_{\mu\s}
\pa^{\bar{\s}}\tilde{B}^{\mu}_{\,\,\,\,\bar{\s}}
\nn\\
&\equiv \label{Eq:BadKinetic} -F_{\mu\nu}\big[\tilde{B}\big]^2-\frac{1}{2}\pa^{\s}\tilde{B}_{\mu\s}
\pa^{\bar{\s}}\tilde{B}^{\mu}_{\,\,\,\,\bar{\s}}\,.
\end{align} 
}%
Conversely, the antisymmetry condition $\tilde{B}_{\mu\nu}=-\tilde{B}_{\nu\mu}$ implies that $\pa^{\mu}\pa^{\nu}\tilde{B}_{\mu\nu}$ is vanishing; we therefore find that the gauge-fixing term retains a dependence on $\pi$ only, with $\cl_{\xi}(A_{\mu},\pi)=\cl_{\xi}(\pi)$ given by:
\begin{equation}
\cl_{\xi}(\pi)=-\frac{1}{2\xi}\left(-\frac{1}{m}\pa^{\mu}\pa_{\mu}\pi
+\xi m \pi\right)^2\,.
\end{equation}
This further simplifies our expression for the partition function $\cz[J]$, since the integral over $\pi$ simply yields another constant $\cn_{\pi}\equiv\int\is\cd\pi\,e^{i\int\text{d}^{4}x\,\cl_{\xi}}$, which we may absorb into the overall normalisation with yet another redefinition:
\begin{equation}
\cz[J]=\cn'\is\int\is \cd\bar{\ci}_{\mu\nu}\cd\tilde{B}_{\mu\nu}\, 
e^{i\int\text{d}^{4}x\,
\big(\bar{\ci}_{\mu\nu}\bar{\ci}^{\mu\nu} +\cl_{o} +\cl_{J}\big)}\,.
\end{equation}
where $\cn'=\cn_{\pi}\cn$. At this stage we have reformulated the path integral in terms only of a 2-form field $\tilde{B}_{\mu\nu}$, though we are still left with the pathology that our Lagrangian density Eq.~(\ref{Eq:BadKinetic}) contains kinetic terms with four spacetime derivatives; it is for this reason---and following Ref.~\cite{Bijnens:1995ii}---that we earlier introduced the auxiliary field $\bar{\ci}_{\mu\nu}$, which we shall now exploit to address this problem. By defining the change of variable:
\begin{equation}
\bar{\ci}_{\mu\nu}\equiv\hat{\mu}\ci_{\mu\nu} +F_{\mu\nu}\big[\tilde{B}\big]\,,
\end{equation}
one finds that the partition function may be rewritten as
\begin{equation}
\cz[J]=\cn'\is\int\is\cd\tilde{B}_{\mu\nu}
\cd\big(\hat{\mu}\ci_{\mu\nu}\big)\,
e^{i\int\text{d}^{4}x \, \big(\bar{\cl}_{o} +\cl_{J}\big)}\,,
\end{equation}
where we have now $\bar{\cl}_{o}\equiv\cl_{o} +\bar{\ci}_{\mu\nu}\bar{\ci}^{\mu\nu}$, which (neglecting an inconsequential total derivative term) is given by the following:
\begin{equation}
\bar{\cl}_{o}=-\frac{1}{2}\pa^{\nu}\tilde{B}_{\mu\nu}
\pa^{\s}\tilde{B}^{\mu}_{\,\,\,\,\s}
+\hat{\mu}^{2}\ci_{\mu\nu}\ci^{\mu\nu}
+\frac{2\hat{\mu}}{m}
\pa^{\nu}\ci_{\mu\nu}\pa^{\s}\tilde{B}^{\mu}_{\,\,\,\,\s}\,.
\end{equation}
The $F_{\mu\nu}^{2}$ terms have cancelled, and we are left with a Lagrangian density describing two (mass) dimension-1 fields $\tilde{B}_{\mu\nu}$ and $\ci_{\mu\nu}$. Hence we see that removing the four-derivative kinetic terms in Eq.~(\ref{Eq:BadKinetic}) comes at the expense of introducing an additional dynamical field. Next, we diagonalise the kinetic terms of $\bar{\cl}_{o}$ by implementing a rotation according to: 
\begin{equation}
\begin{pmatrix}
\tilde{B}_{\mu\nu}\\
\ci_{\mu\nu}
\end{pmatrix} \equiv 
\begin{pmatrix}
\cos\th & \sin\th\\
-\sin\th & \cos\th
\end{pmatrix} \begin{pmatrix}
G_{\mu\nu}\\
H_{\mu\nu}
\end{pmatrix}\,,
\end{equation}
with the rotation angle defined via the dimensionless ratio $\tan(2\th)=\frac{4\hat{\mu}}{m}$. The resulting fields are then further rescaled using the following redefinitions:
\begin{align}
\tilde{G}_{\mu\nu}&\equiv \tfrac{\cos(\th)}{\sqrt{\cos(2\th)}} G_{\mu\nu}\,,\\
\tilde{H}_{\mu\nu}&\equiv - \tfrac{\sin(\th)}{\sqrt{\cos(2\th)}} H_{\mu\nu}\,,
\end{align}
after which we find that the reparametrised Lagrangian density is given by
\begin{align}
\bar{\cl}_{o}=&-\tfrac{1}{2}\pa^{\nu}\tilde{G}_{\mu\nu}
\pa^{\s}\tilde{G}^{\mu}_{\,\,\,\,\s}
+\tfrac{1}{2}\pa^{\nu}\tilde{H}_{\mu\nu}
\pa^{\s}\tilde{H}^{\mu}_{\,\,\,\,\s}\\
&\nn
+\hat{\mu}^2\cos(2\theta)
\begin{pmatrix}
\tilde{G}_{\mu\nu} \\ 
\tilde{H}_{\mu\nu}\end{pmatrix}^{\is T}
\is\begin{pmatrix}
	\tan^{2}\th & 1 \\
	1 & \frac{1}{\tan^{2}\th}
\end{pmatrix}
\begin{pmatrix}
	\tilde{G}^{\mu\nu} \\ 
	\tilde{H}^{\mu\nu}\end{pmatrix}\,.
\end{align}
At this point we make two observations. Firstly, we see that the kinetic terms of the 2-form fields $\tilde{G}_{\mu\nu}$ and $\tilde{H}_{\mu\nu}$ have opposite signs. The correct (i.e. physical) choice for the relative sign between the kinetic and mass terms is determined according to which Minkowski metric signature has been adopted; since we are using the ``mostly plus'' convention, we require that kinetic and mass terms have the same sign. Hence, it is the kinetic term for $\tilde{H}_{\mu\nu}$ which is compatible with causal propagation.\par 
Our second observation is that the mass terms of $\bar{\cl}_{o}$ must now also be diagonalised, though we further require that the relative sign difference between the kinetic terms for the two fields be preserved in the process. To this end, we next introduce another field rotation:  
\begin{equation}
\begin{pmatrix}
\tilde{G}_{\mu\nu}\\
\tilde{H}_{\mu\nu}
\end{pmatrix} \equiv 
\begin{pmatrix}
\cosh\b & \sinh\b\\
\sinh\b & \cosh\b
\end{pmatrix} \begin{pmatrix}
W_{\mu\nu}\\
K_{\mu\nu}
\end{pmatrix}\,,
\end{equation}
and we determine that the aforementioned condition for diagonalised mass terms is satisfied by demanding that the rotation angle $\b$ is related to $\th$ via
\begin{equation}
e^{2\b}=\cos(2\th)\,,
\end{equation}
with which we obtain the following useful trigonometric identities:
{\small	
\begin{equation*}
\forall\th,\,0=
\begin{cases}
\tan^{2}(\th)\cosh^{2}(\b) \,+\, 2\cosh(\b)\sinh(\b) \,+\, \tan^{-2}(\th)\sinh^{2}(\b) \,,\\[-1mm]
\tan^{2}(\th)\cosh(\b)\sinh(\b) \,+\, \cosh(2\b) \,+\, \tan^{-2}(\th)\cosh(\b)\sinh(\b) \,,\\[-1mm]
\tan^{2}(\th)\sinh^{2}(\b) \,+\, 2\cosh(\b)\sinh(\b) \,+\, \tan^{-2}(\th)\cosh^{2}(\b) \,-\, 4\cot(2\th)\csc(2\th) \,.
\end{cases}
\end{equation*} 
}%
After some algebraic manipulation, we find that $\bar{\cl}_{o}$ may be written as follows: 
\begin{equation}
\bar{\cl}_{o}=
-\frac{1}{2}\pa^{\s}{W}_{\mu\s}\pa^{\bar{\s}}{W}^{\mu}_{\,\,\,\,\bar{\s}}
+\frac{1}{2}\pa^{\s}{K}_{\mu\s}\pa^{\bar{\s}}{K}^{\mu}_{\,\,\,\,\bar{\s}}
+\frac{1}{4}m^{2}K_{\mu\nu}K^{\mu\nu}\,.
\end{equation}
Notice that any trace of the auxiliary mass parameter $\hat{\mu}$ has vanished, and the two antisymmetric fields have completely decoupled. The relative sign difference between the two kinetic terms has been preserved, and the unstable $W_{\mu\nu}$ remains as a massless artefact of the reformulation; nevertheless, we may once again simply absorb its contribution to the path integral into a  redefinition of the overall constant prefactor. The partition function is therefore given by
\begin{equation}
\cz[J]=\cn''\is\int\is\cd K_{\mu\nu}\, 
e^{i\int\text{d}^{4}x \big(\cl_{K} +\cl_{J}\big) }\,,
\end{equation}
for some new constant $\cn''$, and where $\cl_{K}$ describes the massive 2-form $K_{\mu\nu}$:
\begin{equation}
\label{Eq:LagrK}
\cl_{K}=\frac{1}{2}\pa^{\s}{K}_{\mu\s}
\pa^{\bar{\s}}{K}^{\mu}_{\,\,\,\,\bar{\s}}
+\frac{1}{4}m^{2}K_{\mu\nu}K^{\mu\nu}\,.
\end{equation}
This Lagrangian is indeed physically equivalent to our original $U(1)$ theory from Eq.~(\ref{Eq:4dtheory}), though it does not manifest any gauge invariance. We can proceed to construct another equivalent theory which does admit vectorial gauge transformations akin to those of Eq.~(\ref{Eq:2formTransform}), starting from the following field redefinition:
\begin{equation}
K_{\mu\nu}\equiv\tfrac{1}{2}\e_{\mu\nu\r\s}
\left(\cb^{\r\s}+\tfrac{1}{m}\cf^{\r\s}\right)
\equiv\tfrac{1}{2m}\e_{\mu\nu\r\s}\ch^{\r\s}
\,,
\end{equation}
where $\cf^{\r\s}$ is the field strength of an Abelian gauge field $\ca^{\r}$:
\begin{equation}
\cf^{\r\s}\equiv2\pa^{[\r}\ca^{\s]}
=\pa^{\r}\ca^{\s} -\pa^{\s}\ca^{\r}\,.
\end{equation}
Since this 2-form is exact, it is necessarily also closed. Hence we also find that the following identity is satisfied:
\begin{equation}
\pa^{\a}K_{\mu\a}=\frac{1}{2}\epsilon_{\mu\a\r\s}\pa^{\a}\cb^{\r\s}\,.
\end{equation}
By making use of the Levi-Civita tensor identities:
\begin{align}
\e_{\mu\nu\r\s}\e^{\mu\nu}_{\,\,\,\,\,\,\bar{\r}\bar{\s}}=
&-2\big(\eta_{\r\bar{\r}}\eta_{\s\bar{\s}}
-\eta_{\r\bar{\s}}\eta_{\s\bar{\r}}\big)\,,\\
\e_{\mu\nu\r\s}\e^{\mu}_{\,\,\,\,\bar{\nu}\bar{\r}\bar{\s}}=
&-\eta_{\nu\bar{\nu}}\eta_{\r\bar{\r}}\eta_{\s\bar{\s}}
-\eta_{\nu\bar{\r}}\eta_{\r\bar{\s}}\eta_{\s\bar{\nu}}
-\eta_{\nu\bar{\s}}\eta_{\r\bar{\nu}}\eta_{\s\bar{\r}}\nn\\
&+\eta_{\nu\bar{\r}}\eta_{\r\bar{\nu}}\eta_{\s\bar{\s}}
+\eta_{\nu\bar{\s}}\eta_{\r\bar{\r}}\eta_{\s\bar{\nu}}
+\eta_{\nu\bar{\nu}}\eta_{\r\bar{\s}}\eta_{\s\bar{\r}}
\,,
\end{align}
and furthermore by defining the completely anti-symmetrised 3-form field strength:
\begin{equation}
{G}_{\mu\nu\r}\equiv
3\pa_{[\mu}\cb_{\nu\r]} =
\pa_{\mu}\cb_{\nu\r} +\pa_{\r}\cb_{\mu\nu} +\pa_{\nu}\cb_{\r\mu}\,,
\end{equation}
we obtain our third and final reformulation of the Lagrangian density:
\begin{equation}
\label{Eq:LagrGH}
\cl_{K}(\ca_{\mu},\,\cb_{\mu\nu})=-\frac{1}{12}{G}_{\mu\nu\r}G^{\mu\nu\r}
-\frac{1}{4}\ch_{\mu\nu}\ch^{\mu\nu}\,,
\end{equation}
which is invariant under the simultaneous gauge transformations
\begin{align}
\ca_{\mu}&\to \ca_{\mu} +m\a_{\mu}\,,\\
\cb_{\mu\nu}&\to \cb_{\mu\nu}-2\pa_{[\mu}\a_{\nu]}\,,
\end{align}
for some arbitrary four-vector $\a_{\mu}$ that depends on the Minkowski coordinates. As a consistency check, we notice that for $m=0$ the Lagrangian $\cl_{K}$ reduces to kinetic terms for a massless 2-form $\cb_{\mu\nu}$ (dual to a massless scalar $\pi$) and a massless $U(1)$ gauge boson $\ca_{\mu}$. Conversely, in the case that $m\neq 0$, the Lagrangian describes a massive 2-form $\cb_{\mu\nu}$ (dual to a massive 1-form $A_{\mu}$).\par  
To summarise then, we have derived three completely equivalent formulations of a Lagrangian density to describe the same underlying physical theory, differing only by their gauge invariances; these are Eq.~(\ref{Eq:4dtheory}) which admits a $U(1)$ gauge transformation parametrised by the scalar $\a$, Eq.~(\ref{Eq:LagrK}) which does not contain a gauge redundancy, and Eq.~(\ref{Eq:LagrGH}) which is invariant under transformations parametrised by the vector $\a_{\mu}$. We adapted this third formulation in Sec.~\ref{Sec:MoreBosons}, to describe a $U(1)$ theory of 2-forms in $D=5$ dimensions. For a brief discussion regarding the generalisation of this procedure to higher dimensions, the interested Reader is directed to Appendix~A.2 of Ref.~\cite{Elander:2018aub}.

\chapter{Spectra from critical point solutions}
\label{App:CritSpectra}
\subsubsection{Circle-compactified $D=6$ supergravity}
Tabulated overleaf are the masses extracted from our numerical spectra computation, obtained by fluctuating the bosonic fields of the $S^{1}$-reduced six-dimensional maximal supergravity discussed in Chapter~\ref{Chap:SpectraRomans}. We show only the massive excitations of the backgrounds which correspond to the critical points of the $D=6$ scalar potential $\cv_{6}(\f)$ defined in Eq.~(\ref{Eq:V6}). Both Tables~\ref{Tbl:RomansUV} and~\ref{Tbl:RomansIR} have been adapted from those in Ref.~\cite{Elander:2018aub}.
\begin{table}[h!]
	\begin{center}
		\begin{small}
			\begin{tabular}{|ccccccccc|}
				\hline\hline
				Spin-0 & Spin-1 & Spin-2 & Spin-0 & Spin-1 & Spin-0 & Spin-1 & Spin-1 & Spin-1 \cr
				$\mathfrak{a}^a$ &  $V_{\mu}$ & $\mathfrak{e}^{\mu}_{\ \nu}$ & $\pi^i$ & $A^{i}_{\mu}$ & $X$&$B_{6\nu}$ & $X_{\mu}$ &$B_{\mu\nu}$\cr
				\hline
				0.54\,\small{($\mathfrak{p}$)} & 1.23 & 1.00 & 1.00 & 0.73 & 0.60 & 0.40 & 1.02 & 0.66 \\
				
				{\bf 0.62}\,\textcolor{white}{\small{($\mathfrak{p}$)}} & 1.91 & 1.65 & 1.65 & 1.38 & 1.35 & 1.07 & 1.66 & 1.34 \\
				
				1.15\,\small{($\mathfrak{p}$)} & 2.55 & 2.28 & 2.28 & 2.00 & 2.00 & 1.72 & 2.29 & 1.98 \\
				
				{\bf 1.53}\,\textcolor{white}{\small{($\mathfrak{p}$)}} & 3.18 & 2.90 & 2.90 & 2.63 & 2.64 & 2.35 & 2.91 & 2.60 \\
				
				1.77\,\small{($\mathfrak{p}$)} & 3.81 & 3.53 & 3.53 & 3.25 & 3.27 & 2.97 & 3.53 & 3.22 \\
				
				{\bf 2.20}\,\textcolor{white}{\small{($\mathfrak{p}$)}} &  \text{} & \text{} & \text{} & 3.87 & 3.89 & 3.60 &  \text{}  &3.84   \\
				
				2.39\,\small{($\mathfrak{p}$)} & \text{} & \text{} & \text{} & \text{} & \text{} & \text{} & \text{} & \text{} \\
				
				{\bf 2.84}\,\textcolor{white}{\small{($\mathfrak{p}$)}} & \text{} & \text{} & \text{} & \text{} & \text{} & \text{} & \text{} & \text{} \\
				
				3.01\,\small{($\mathfrak{p}$)} & \text{} & \text{} & \text{} & \text{} & \text{} & \text{} & \text{} & \text{} \\
				
				{\bf 3.48}\,\textcolor{white}{\small{($\mathfrak{p}$)}} & \text{} & \text{} & \text{} & \text{} & \text{} & \text{} & \text{} & \text{} \\
				
				3.64\,\small{($\mathfrak{p}$)} & \text{} & \text{} & \text{} & \text{} & \text{} & \text{} & \text{} & \text{} \\
				\hline\hline
			\end{tabular}
		\end{small}
	\end{center}
	\caption[Numerical masses from the UV critical point solution within the six-dimensional supergravity]{Numerical masses $M$ of the lightest excitations within the ten ($a\in\{1,2\}$) towers of bosonic modes of the $S^{1}$-compactified $D=6$ supergravity, computed on backgrounds with $\f=\f_{UV}=0$. All states are normalised in units of the lightest tensor mass, and the spectra were computed using regulators $\r_{1}=10^{-3}$ and $\r_{2}=8$. Our implementation of the midpoint determinant method used the intermediate value $\r_{\ast}=4$. We use bold font to denote $\mathfrak{a}^{a}$ excitations which exhibit background-independence in Fig.~\ref{Fig:Spectra1}, while those states labelled with a $\mathfrak{p}$ are captured effectively by the probe approximation.   }
	\label{Tbl:RomansUV}
\end{table}

\begin{table}[h!]
	\begin{center}
		\begin{small}
			\begin{tabular}{|ccccccccc|}
				\hline\hline
				Spin-0 & Spin-1 & Spin-2 & Spin-0 & Spin-1 & Spin-0 & Spin-1 & Spin-1 & Spin-1 \cr
				$\mathfrak{a}^a$ &  $V_{\mu}$ & $\mathfrak{e}^{\mu}_{\ \nu}$ & $\pi^i$ & $A^{i}_{\mu}$ & $X$&$B_{6\nu}$ & $X_{\mu}$ &$B_{\mu\nu}$\cr
				\hline
				{\bf 0.62}\,\textcolor{white}{\small{($\mathfrak{p}$)}} & 1.23 & 1.00 & 1.00 & 0.73 & 1.08 & 0.82 & 1.48 & 1.10 \\
				
				1.44\,\small{($\mathfrak{p}$)} & 1.90 & 1.65 & 1.65 & 1.37 & 1.82 & 1.54 & 2.13 & 1.80 \\
				
				{\bf 1.53}\,\textcolor{white}{\small{($\mathfrak{p}$)}} & 2.55 & 2.28 & 2.28 & 2.00 & 2.49 & 2.19 & 2.77 & 2.45 \\
				
				2.11\,\small{($\mathfrak{p}$)} & 3.18 & 2.90 & 2.90 & 2.62 & 3.13 & 2.83 & 3.40 & 3.08 \\
				
				{\bf 2.20}\,\textcolor{white}{\small{($\mathfrak{p}$)}} & 3.81 & 3.53 & 3.53 & 3.25 & 3.76 & 3.46 & \text{}& 3.71  \\
				
				2.76\,\small{($\mathfrak{p}$)}  & \text{} & \text{} & \text{} & 3.87 & \text{} & \text{} & \text{} & \text{}\\
				
				{\bf 2.84}\,\textcolor{white}{\small{($\mathfrak{p}$)}} & \text{} & \text{} & \text{} & \text{} & \text{} & \text{} & \text{} & \text{} \\
				
				3.39\,\small{($\mathfrak{p}$)} & \text{} & \text{} & \text{} & \text{} & \text{} & \text{} & \text{} & \text{} \\
				
				{\bf 3.48}\,\textcolor{white}{\small{($\mathfrak{p}$)}} & \text{} & \text{} & \text{} & \text{} & \text{} & \text{} & \text{} & \text{} \\
				\hline\hline
			\end{tabular}
		\end{small}
	\end{center}
	\caption[Numerical masses from the IR critical point solution within the six-dimensional supergravity]{Numerical masses $M$ of the lightest excitations within the ten ($a\in\{1,2\}$) towers of bosonic modes of the $S^{1}$-compactified $D=6$ supergravity, computed on backgrounds with $\f=\f_{IR}=-\frac{1}{4}\ln(3)$. All states are normalised in units of the lightest tensor mass, and the spectra were computed using regulators $\r_{1}=10^{-3}$ and $\r_{2}=8$. Our implementation of the midpoint determinant method used the intermediate value $\r_{\ast}=4$. We use bold font to denote $\mathfrak{a}^{a}$ excitations which exhibit background-independence in Fig.~\ref{Fig:Spectra1}, while those states labelled with a $\mathfrak{p}$ are captured effectively by the probe approximation.   }
	\label{Tbl:RomansIR}
\end{table}
\clearpage

\subsubsection{Torus-compactified $D=7$ supergravity}
We present below the numerical masses extracted from our spectra computation for the $T^{2}$-reduced seven-dimensional half-maximal supergravity discussed in Chapter~\ref{Chap:SpectraWitten}. We restrict attention to the massive excitations of the backgrounds which correspond to the critical points of the $D=7$ scalar potential $\cv_{7}(\f)$ defined in Eq.~(\ref{Eq:V7}).
\begin{table}[h!]
	\begin{center}
		\begin{small}
			\begin{tabular}{|cccc|cccc|}
				\hline\hline
				Spin-0 & \text{} & Spin-2 & \text{} & Spin-0 & \text{} & Spin-2 & \text{}\cr
				$\mathfrak{a}^a$ & $\cq_{\mathfrak{a} }$ & $\mathfrak{e}^{\mu}_{\ \nu}$ & $\cq_{\mathfrak{e} }$ & $\mathfrak{a}^a$ & $\cq_{\mathfrak{a} }$ & $\mathfrak{e}^{\mu}_{\ \nu}$ & $\cq_{\mathfrak{e} }$ \cr
				\hline
				{\bf 0.58}\,\small{($\mathfrak{p}$)} &  -   & 1.00 & - & {\bf 0.58}\,\small{($\mathfrak{p}$)} & - & 1.00 & - \\
				
				0.59\,\small{($\mathfrak{p}$)} & 1.02 & 1.58 & 1.58 & {\bf 1.03}\,\small{($\mathfrak{p}$)} & 1.78 & 1.58 & 1.58\\
				
				{\bf 1.03}\,\small{($\mathfrak{p}$)} & 1.75 & 2.15 & 1.36 & 1.37\,\small{($\mathfrak{p}$)} & 1.33 & 2.15 & 1.36\\
				
				1.14\,\textcolor{white}{\small{($\mathfrak{p}$)}} & 1.11 & 2.71 & 1.26 & {\bf 1.46}\,\textcolor{white}{\small{($\mathfrak{p}$)}} & 1.07 & 2.70 & 1.26\\
				
				{\bf 1.45}\,\textcolor{white}{\small{($\mathfrak{p}$)}} & 1.27 & 3.26 & 1.20 & {\bf 1.61}\,\small{($\mathfrak{p}$)} & 1.10 & 3.26 & 1.21\\
				
				{\bf 1.61}\,\small{($\mathfrak{p}$)} & 1.11 & \text{} & \text{} & 1.98\,\small{($\mathfrak{p}$)} & 1.23 & \text{} & \text{}\\
				
				1.69\,\small{($\mathfrak{p}$)} & 1.05 & \text{} & \text{} & {\bf 2.06}\,\textcolor{white}{\small{($\mathfrak{p}$)}} & 1.04 & \text{} & \text{}\\
				
				{\bf 2.06}\,\textcolor{white}{\small{($\mathfrak{p}$)}} & 1.22 & \text{} & \text{} & {\bf 2.18}\,\small{($\mathfrak{p}$)} & 1.06 & \text{} & \text{}\\
				
				{\bf 2.18}\,\small{($\mathfrak{p}$)} & 1.06 & \text{} & \text{} & 2.56\,\small{($\mathfrak{p}$)} & 1.17 & \text{} & \text{}\\
				
				2.23\,\small{($\mathfrak{p}$)} & 1.02 & \text{} & \text{} & {\bf 2.64}\,\textcolor{white}{\small{($\mathfrak{p}$)}} & 1.03 & \text{} & \text{}\\
				
				{\bf 2.64}\,\textcolor{white}{\small{($\mathfrak{p}$)}} & 1.18 & \text{} & \text{} & \text{} & \text{} & \text{} & \text{}\\
				\hline\hline
			\end{tabular}
		\end{small}
	\end{center}
	\caption[Numerical masses from the critical point solutions within the seven-dimensional supergravity]{Numerical masses $M$ of the lightest excitations within the four ($a\in\{1,2,3\}$) towers of bosonic modes of the (truncated) $T^{2}$-compactified $D=7$ supergravity, computed on backgrounds with $\f=\f_{UV}=0$ (left) and $\f=\f_{IR}=-\frac{1}{\sqrt{5}}\ln(2)$ (right). All states are normalised in units of the lightest tensor mass, and the spectra were computed using regulators $\r_{1}=10^{-4}$ and $\r_{2}=12$. Our implementation of the midpoint determinant method used the intermediate value $\r_{\ast}=4$. We use bold font to denote $\mathfrak{a}^{a}$ excitations which exhibit background-independence in Fig.~\ref{Fig:WittenSpectrum}, while those states labelled with a $\mathfrak{p}$ are captured effectively by the probe approximation. We additionally provide the mass ratio of each state with its predecessor, for the spin-0 modes ($\cq_{\mathfrak{a} }$) and spin-2 modes ($\cq_{\mathfrak{e} }$).   }
	\label{Tbl:WittenUVandIR}
\end{table}

\chapter{An alternative normalisation of spectra}
\label{App:LambdaSpectra}
\subsubsection{Circle-compactified $D=6$ supergravity}
\begin{figure}[h]
	\begin{center}
		\makebox[\textwidth]{\includegraphics[width=0.76\paperwidth]
			{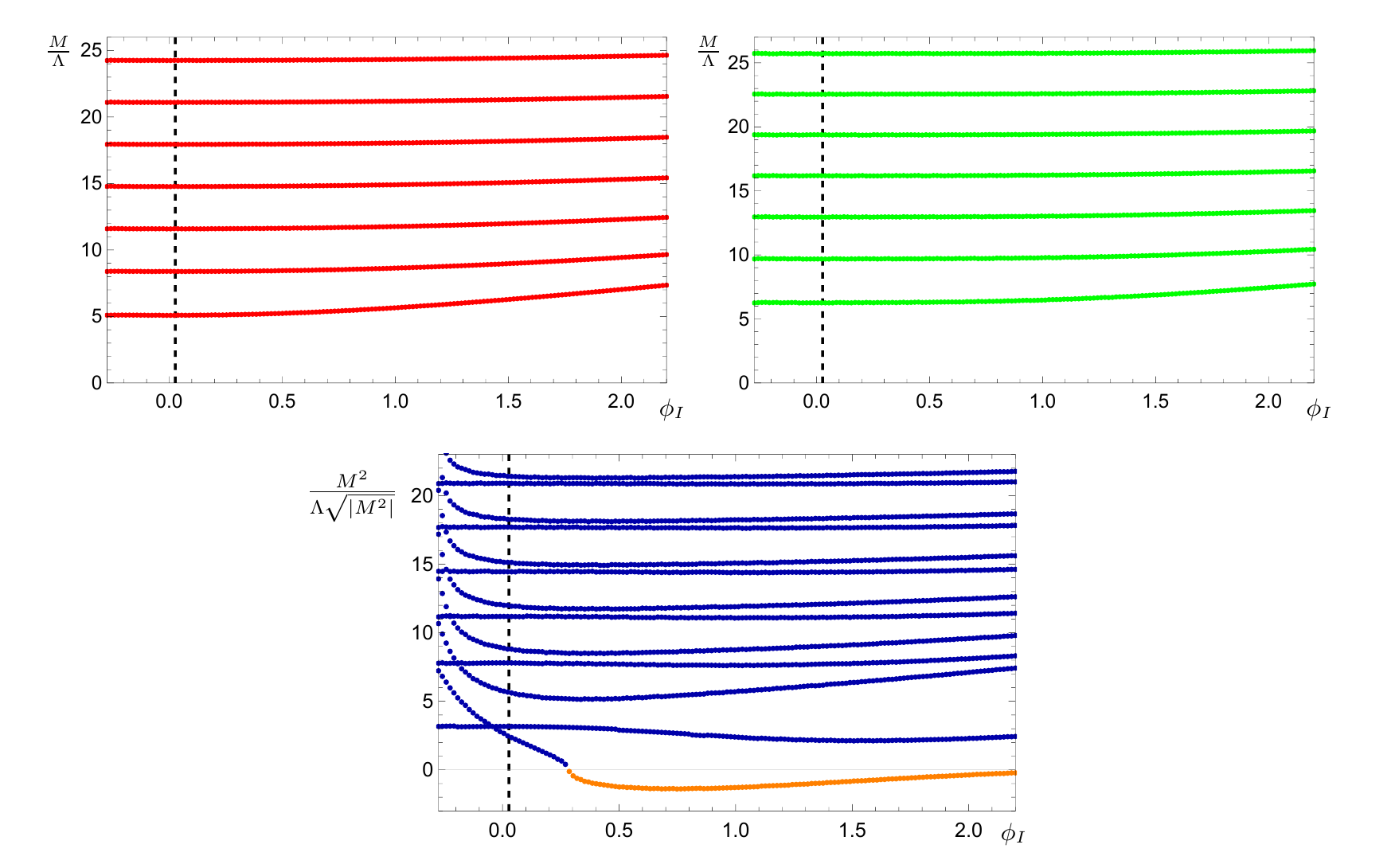}}
	\end{center}
	\caption[An alternative normalisation of the mass spectra of graviton resonances for the six-dimensional supergravity]{The spectra of masses $M$ as a function of the one free parameter which characterises the class of confining solutions, $\f_{I}\in[\f_{IR},2.2]$, normalised in units of the universal scale $\Lambda$. From top to bottom, left to right: 
		the spectra of fluctuations for the tensors 
		$\mathfrak{e}^{\mu}_{\ \nu}$ (red), 
		the graviphoton $V_{\mu}$ (green), and the two scalars 
		$\mathfrak{a}^{a}$ (blue). The orange disks in the scalar 
		spectrum represent masses for which $M^{2}<0$, and hence denote a tachyonic state. The vertical dashed lines mark the critical value of the IR parameter $\f_{I}= \f_{I}^{*}>0$ at the phase transition, discussed in Sec.~\ref{Sec:RomansPhaseStruct}. All states were computed using regulators $\r_{1}=10^{-4}$ and $\r_{2}=12$.}
	\label{Fig:LambdaSpectra1}
\end{figure}

\begin{figure}[h]
	\begin{center}
		\makebox[\textwidth]{\includegraphics[width=0.76\paperwidth]
			{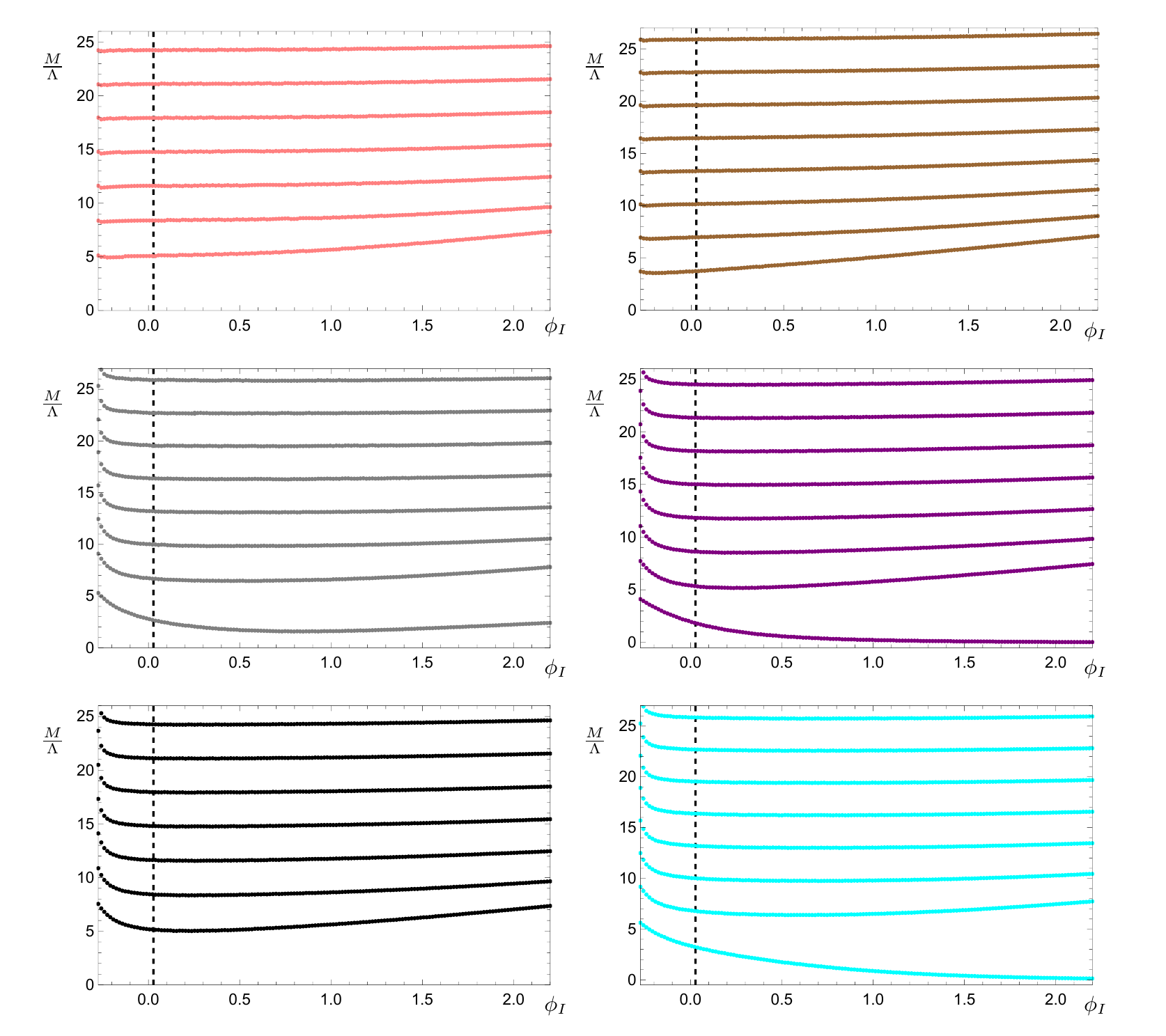}}
	\end{center}
	\vspace{-5mm}
	\caption[An alternative normalisation of the mass spectra of $p$-form resonances for the six-dimensional supergravity]{The spectra of masses $M$ as a function of the one free parameter which characterises the class of confining solutions, $\f_{I}\in[\f_{IR},2.2]$, normalised in units of the universal scale $\Lambda$. From top to bottom, left to right: the spectra of fluctuations 
		of the $SU(2)$ adjoint (pseudo-)scalars $\pi^{i}$ (pink), $SU(2)$ adjoint vectors $A^{i}_{\mu}$ (brown), 
		$U(1)$ scalar combination $X$ (grey), 
		$U(1)$ transverse vector $B_{6\mu}$ (purple), 
		$U(1)$ transverse vector combination $X_{\mu}$ (black), and 
		the $U(1)$ 2-form $B_{\mu\nu}$ (cyan). The vertical dashed lines mark the critical value of the IR parameter $\f_{I}= \f_{I}^{*}>0$ at the phase transition, discussed in Sec.~\ref{Sec:RomansPhaseStruct}. The spectra were computed using regulators $\r_{1}=10^{-4}$ and $\r_{2}=12$ with the exception of the $U(1)$ scalar combination $X$, for which the choice $\r_{1}= 10^{-7}$ was used instead to minimise numerical cutoff effects which were present for the lightest state at large values of $\f_{I}$. }
	\label{Fig:LambdaSpectra2}
\end{figure}

In Figures~\ref{Fig:LambdaSpectra1} and~\ref{Fig:LambdaSpectra2} below we provide an alternative normalisation---in units of the universal scale $\Lambda$ introduced in Eq.~(\ref{Eq:Lambda})---for the spectra of bosonic modes which descend from the six-dimensional maximal supergravity discussed in Chapter~\ref{Chap:SpectraRomans}. The previously applied normalisation which measures the spectra in units of the lightest spin-2 mass is removed.  

\clearpage

\subsubsection{Torus-compactified $D=7$ supergravity}
\begin{figure}[!h]
	\centering
	\subfloat{\makebox[0.2\paperwidth]
		{\includegraphics[width=0.38\paperwidth]
			{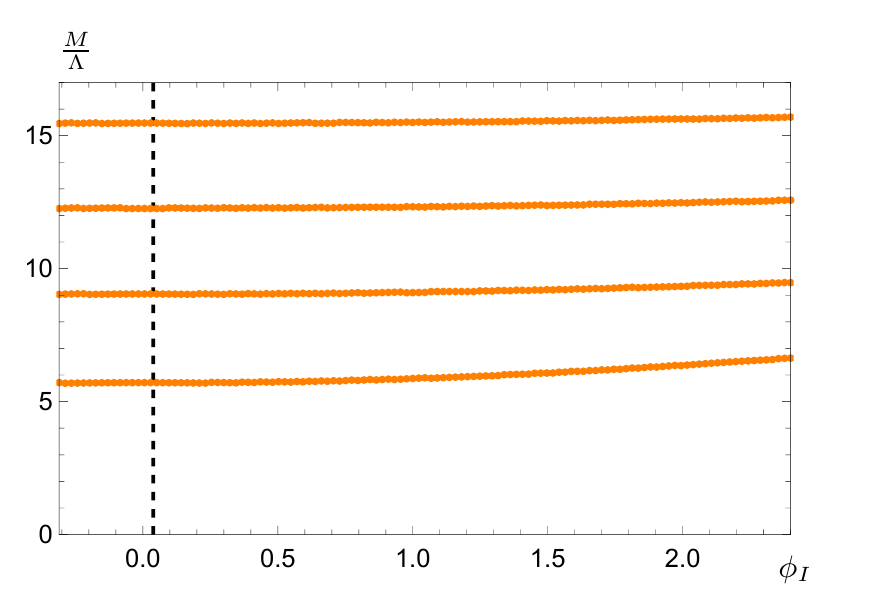}}  }
	\hfill
	\subfloat{\makebox[0.2\paperwidth]
		{\includegraphics[width=0.38\paperwidth]
			{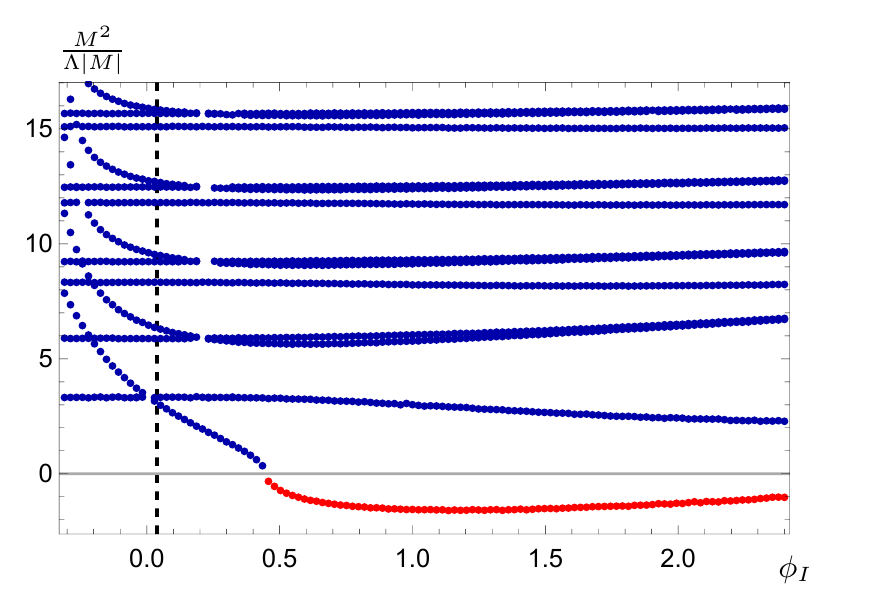}}  }
	\caption[An alternative normalisation of the mass spectra of graviton resonances for the seven-dimensional supergravity]{The spectra of masses $M$ as a function of the one free parameter which characterises the class of confining solutions, $\f_{I}\in[\f_{IR},2.4]$, normalised in units of the universal scale $\Lambda$. The left plot shows 
		the spectra of tensor fluctuations 
		$\mathfrak{e}^{\mu}_{\ \nu}$ (orange), while the right plot shows the mass eigenstates of the scalar fluctuations associated with
		$\{\f,\c,\w\}$ (blue). The red disks in the scalar 
		spectrum represent masses for which $M^{2}<0$, and hence denote a tachyonic state. The vertical dashed lines represent the critical value of the IR parameter $\f_{I}= \f_{I}^{*}>0$ at a first-order phase transition, discussed in Sec.~\ref{Sec:WittenPhaseStruct}. All states were computed using regulators $\r_{1}=10^{-4}$ and $\r_{2}=12$. We acknowledge the existence of some small gaps in the scalar spectrum; these are regions where the eigenstates were so close to degenerate in mass that the numerical routine was unable to resolve and identify them separately, and are hence not of any physical significance. }
		\label{Fig:LambdaSpectraWitten}
\end{figure}

We present in Figure~\ref{Fig:LambdaSpectraWitten} an alternative normalisation---in units of the universal scale $\Lambda$ introduced in Eq.~(\ref{Eq:Lambda})---for the spectra of bosonic modes which descend from the seven-dimensional maximal supergravity discussed in Chapter~\ref{Chap:SpectraWitten}. The previously applied normalisation which measures the spectra in units of the lightest spin-2 mass is removed.  

\chapter{Gravitational invariants}
\label{App:GravInvariants}
We dedicate this Appendix to presenting some simplified expressions for the gravitational invariants of the two supergravity theories, and subsequently plotting these quantities as a function of the holographic coordinate to demonstrate explicitly the differing background geometries which are realised by the confining, skewed, and badly singular domain-wall solutions. In addition to the Ricci curvature scalar we shall also consider in each case the squared Ricci tensor, and the squared Riemann tensor (also known as the Kretschmann scalar), defined to be:
\begin{align}
\car_{\text{(2)}}^{2}&\equiv
\car_{\hat{M}\hat{N}}\car^{\hat{M}\hat{N}}\\
\car_{\text{(4)}}^{2}&\equiv
\car_{\hat{M}\hat{N}\hat{R}\hat{S}}\car^{\hat{M}\hat{N}\hat{R}\hat{S}}
\,.
\end{align}

\subsubsection{Circle-compactified $D=6$ supergravity}
The following expressions are derived using the metric ansatz introduced in Eq.~(\ref{Eq:6Dmetric}), using the equations of motion presented in Eqs.~(\ref{Eq:RomansEOM1}\,-\,\ref{Eq:RomansEOM4}). After some algebraic manipulation we obtain:
\begin{align}
\car_{6}&=6\cv_{6} +4\big(\f'\big)^{2}\,\\
\car_{\text{(2)}}^{2}&=
\tfrac{1}{6}\Big[\car_{6}^{2} +80\big(\f'\big)^{4}\Big]\,,\\
\car_{\text{(4)}}^{2}&=
\tfrac{1}{108}\Big(139\car_{\text{(2)}}^{2} -13\Big(\car_{6}^{2} +2\car_{6}\cv_{6} -6\cv_{6}^{2}\Big)\Big)\nn\\
&\quad+\tfrac{4}{3}\Big[\Big(5\car_{6} -28\cv_{6} +282\big(\c'\big)^{2}\Big)\c''
+\Big(35\car_{6} -256\cv_{6}\Big)\big(\c'\big)^{2}\Big]\nn\\
&\quad+4\Big[11\big(\c''\big)^{2} +199\big(\c'\big)^{4}
-16\c'\big(A'\big)^{3}\Big]\,,
\end{align}
where primes denote differentiation with respect to $\r$. Notice that the curvature scalar and the (squared) Ricci tensor may be formulated in terms solely of the potential $\cv_{6}(\f)$ and the one sigma-model scalar of the theory $\f$. From this observation we infer that both of these gravitational invariants remain finite at all scales, provided $\f$ is non-divergent; this condition is satisfied by the confining, skewed, and IR-conformal classes of solutions in our catalogue. For the skewed backgrounds---in which the size of the $\eta$-circle blows up in the deep-IR region of the geometry---we see that the volume divergence manifests in the Kretschmann scalar $\car_{\text{(4)}}^{2}$, which retains a dependence on the field $\c$ and the metric warp factor. 
\vspace{-10mm}\begin{figure}[!h]
	\centering
	\subfloat{\makebox[0.2\paperwidth]
		{\includegraphics[width=0.38\paperwidth]
			{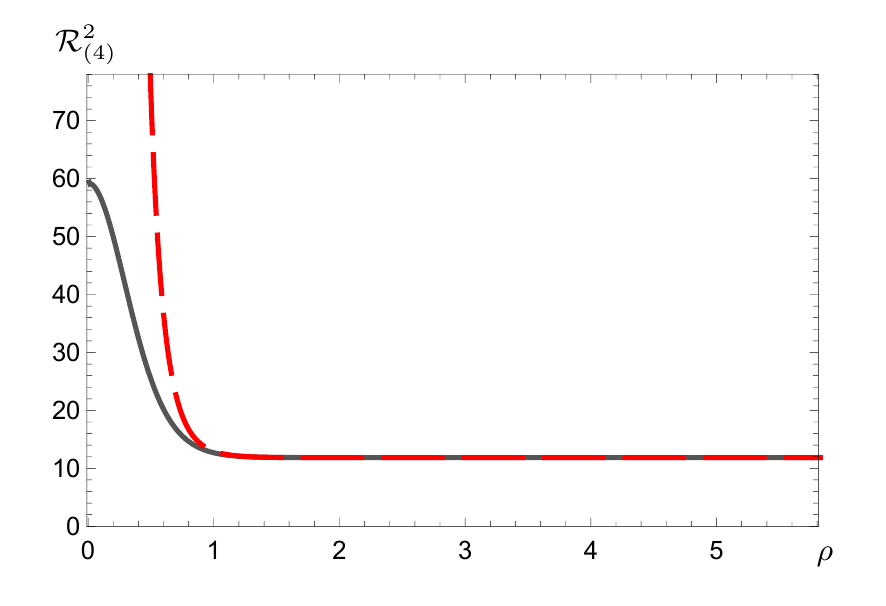}}  }
	\hfill
	\subfloat{\makebox[0.2\paperwidth]
		{\includegraphics[width=0.38\paperwidth]
			{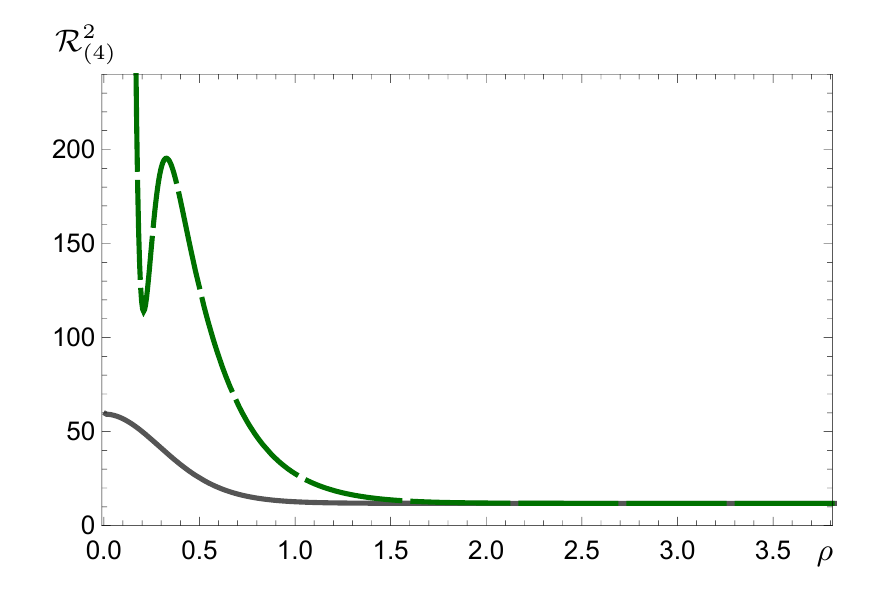}}  }
	\caption[Kretschmann scalar for some special backgrounds in the six-dimensional supergravity]{Plots of the Kretschmann scalar $\car_{\text{(4)}}^{2}$ as a function of the holographic coordinate $\r$, for the six-dimensional metric. The left panel is evaluated on the trivial background $\f=0$, for the confining (solid dark grey line) and skewed (dashed red line) classes of solutions. The right panel shows the invariant evaluated instead on the critical backgrounds at the phase transition: the $\f_{I}=\f_{I}^{*}\simeq 0.027$ confining background (solid dark grey line), and the $\f_{b}=\f_{b}^{*}\simeq 98.9$ BSDW background (dashed dark green line). }
	\label{Fig:GravInvRomans}
\end{figure}

\newpage
\subsubsection{Torus-compactified $D=7$ supergravity}
The following expressions are derived using the metric ansatz introduced in Eq.~(\ref{Eq:7Dmetric}), using the equations of motion presented in Eqs.~(\ref{Eq:1}\,-\,\ref{Eq:5}). After some algebraic manipulation we obtain:
\begin{align}
\car_{7}&=\tfrac{28}{5}\cv_{7} +\big(\f'\big)^{2}\,\\
\car_{\text{(2)}}^{2}&=
\tfrac{1}{7}\Big[\car_{7}^{2} +6\big(\f'\big)^{4}\Big]\,,\\
\car_{\text{(4)}}^{2}&=
\tfrac{7}{6}\car_{\text{(2)}}^{2}
+\tfrac{8}{25}\Big(94\cv_{7}^{2} -15\car_{7}\cv_{7}\Big)\nn\\
&\quad+\tfrac{2}{15}
\Big(55\car_{7}-528\cv_{7} +960\big(A'\big)^{2} +165\big(\U'\big)^{2}
\Big)\big(\U'\big)^{2}\nn\\
&\quad+\Big(\tfrac{221}{30}\Big(5\car_{7} -48\cv_{7}\Big)
+416\big(A'\big)^{2} +\tfrac{3515}{8}\big(\c'\big)^{2} +\tfrac{887}{3}\big(\U'\big)^{2}
\Big)\big(\c'\big)^{2}\nn\\
&\quad-\tfrac{16}{3}\Big(5\car_{7} -48\cv_{7}
+12\big(A'\big)^{2} +150\big(\c'\big)^{2} +88\big(\U'\big)^{2}
\Big)A'\c'  \,,
\end{align}
where primes denote differentiation with respect to $\r$. Notice again that the curvature scalar $\car_{7}$ and the squared Ricci tensor may be formulated in terms solely of the potential $\cv_{7}(\f)$ and the one sigma-model scalar field of the theory $\f$. Hence both of these gravitational invariants remain finite at all scales, provided $\f$ is non-divergent. As with the six-dimensional supergravity this condition is satisfied by the confining, skewed, and IR-conformal classes of solutions in our catalogue. Again, for the skewed backgrounds we see that the volume divergence first appears at the level of the squared Riemann tensor $\car_{\text{(4)}}^{2}$, which retains a dependence on the field $\c$ and the metric warp factor $A$.\par
In the case of the six-dimensional theory we verified that in lifting to ten-dimensional massive type-IIA supergravity, both the confining and skewed classes of solutions yield a finite ten-dimensional Ricci scalar $\car_{10}$.\ Conversely the badly singular domain-wall solutions cause $\car_{10}$ to become divergent, verifying that they retain their singular nature even in the uplifted theory; the Reader is directed to Appendix~D of Ref.~\cite{Elander:2020ial} for details.
\begin{figure}[!h]
	\centering
	\subfloat{\makebox[0.2\paperwidth]
		{\includegraphics[width=0.38\paperwidth]
			{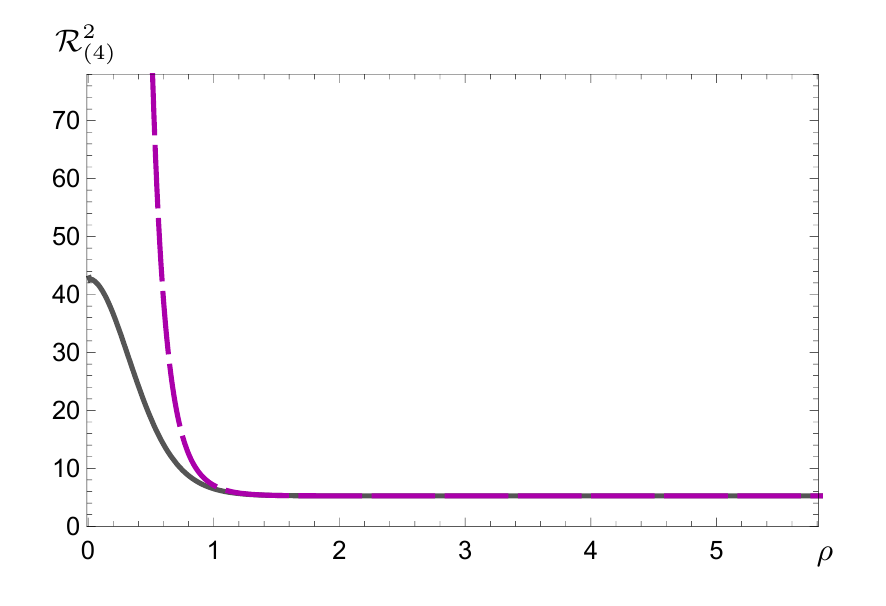}}  }
	\hfill
	\subfloat{\makebox[0.2\paperwidth]
		{\includegraphics[width=0.38\paperwidth]
			{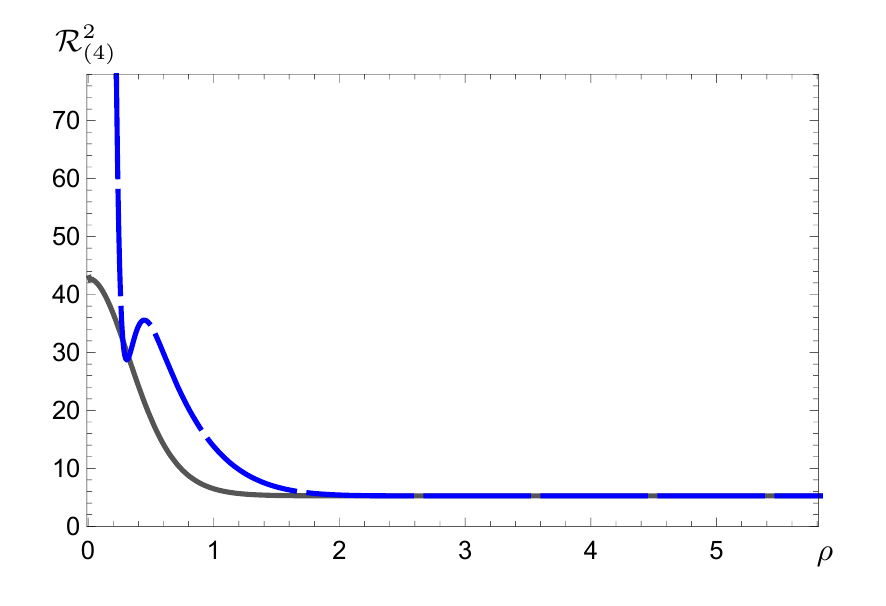}}  }
	\caption[Kretschmann scalar for some special backgrounds in the seven-dimensional supergravity]{Plots of the Kretschmann scalar $\car_{\text{(4)}}^{2}$ as a function of the holographic coordinate $\r$, for the seven-dimensional metric. The left panel is evaluated on the trivial background $\f=0$, for the confining (solid dark grey line) and skewed (dashed magenta line) classes of solutions. The right panel shows the invariant evaluated instead on the critical backgrounds at the phase transition: the $\f_{I}=\f_{I}^{*}\simeq 0.039$ confining background (solid dark grey line), and the $\f_{b}=\f_{b}^{*}\simeq 33.54$ BSDW background (dashed blue line).}
	\label{Fig:GravInvWitten}
\end{figure}

\chapter{Additional parameter plots }
\label{App:ParamPlots}
\subsubsection{Circle-compactified $D=6$ supergravity}
\begin{figure}[h!]
	\begin{center}
		\makebox[\textwidth]{\includegraphics[
			width=0.7\paperwidth
			]
			{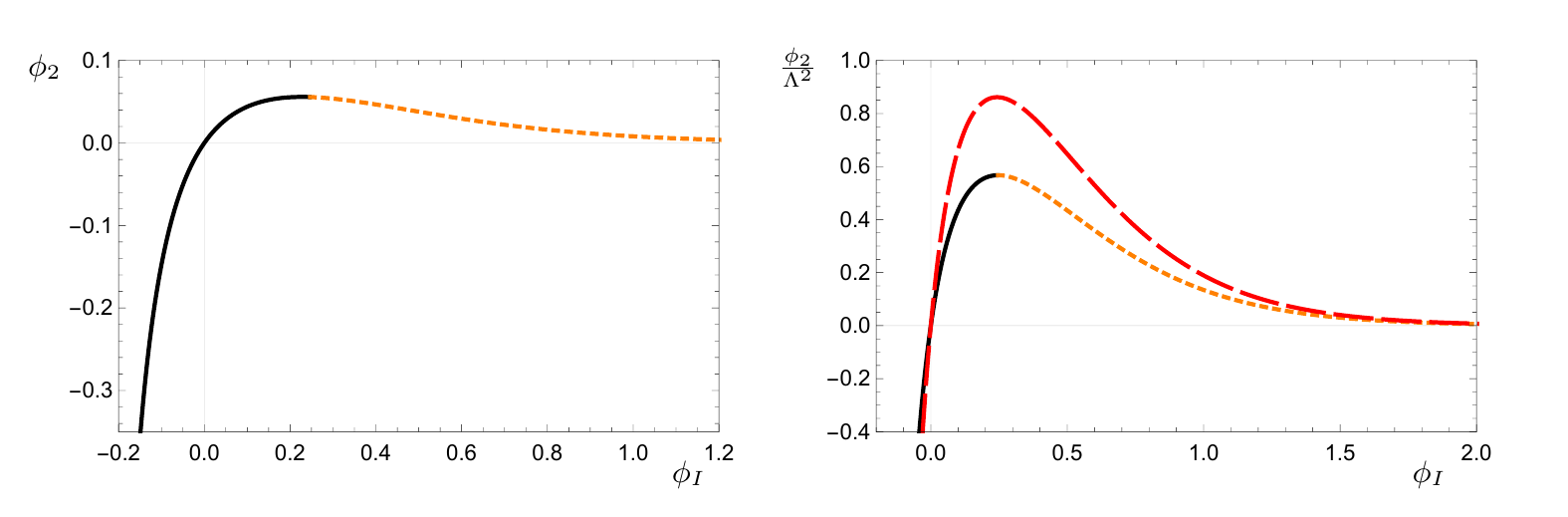}}
	\end{center}
	\vspace{-5mm}
	\caption[First additional parameter plot for the six-dimensional supergravity]{Plots showing the relationship between the UV expansion parameter 
		$\f_{2}$ and the IR parameter $\f_{I}$, for the confining (solid black and short-dashed orange) and skewed (dashed red) branches of solutions.
		The left plot shows the bare extracted parameters: the confining and skewed branches conincide, as $\f_{2}^{c}=\f_{2}^{s}$, $\f_{I}^{c}=\f_{I}^{s}$. The right plot shows the same parameters
		after rescaling with the appropriate powers of $\L$.}
	\label{Fig:Phi2PhiI}
\end{figure}

In Figures~\ref{Fig:Phi2PhiI}\,-\,\ref{Fig:DeltaPhi2} below we present some additional plots which help to elucidate the non-trivial implicit relations between the various parameters which characterise the confining, skewed, and IR-conformal branches of backgrounds within the $S^{1}$-reduced six-dimensional supergravity discussed in Chapter~\ref{Chap:SpectraRomans}. These plots are supplementary to our discussion of dilaton phenomenology in Sec.~\ref{Sec:RomansProbeRevisit}.  

\begin{figure}[h!]
	\begin{center}
		\makebox[\textwidth]{\includegraphics[
			width=0.7\paperwidth
			]
			{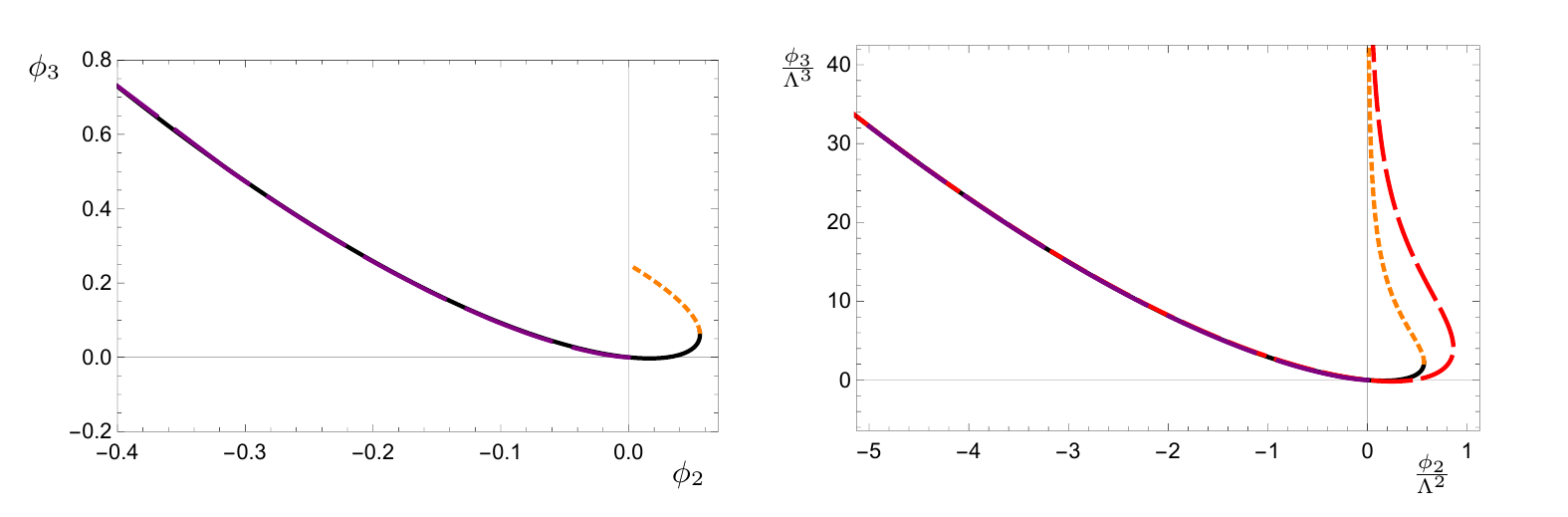}}
	\end{center}
	\vspace{-5mm}
	\caption[Second additional parameter plot for the six-dimensional supergravity]{Plots showing the relationship between the two UV expansion parameters $\f_{2}$ and $\f_{3}$ for solutions belonging to the confining (solid black and short-dashed orange), skewed (dashed red), and IR-conformal (longest-dashed purple) classes. 
		The left plot shows the base parameters extracted by matching to the UV expansions, 
		with $\f_{2}^{c}=\f_{2}^{s}$, $\f_{3}^{c}=\f_{3}^{s}$. The right panel shows the same parameters
		after rescaling with the appropriate powers of $\L$. (For $\f_{2} \leqslant 0$, although the confining, skewed, and IR-conformal classes are not in complete agreement, they are close enough that in these plots the black and red lines are hidden behind the purple one.)}
	\label{Fig:Phi3Phi2}
\end{figure}%

\begin{figure}[h!]
	\begin{center}
		\makebox[\textwidth]{\includegraphics[
			width=0.7\paperwidth
			]
			{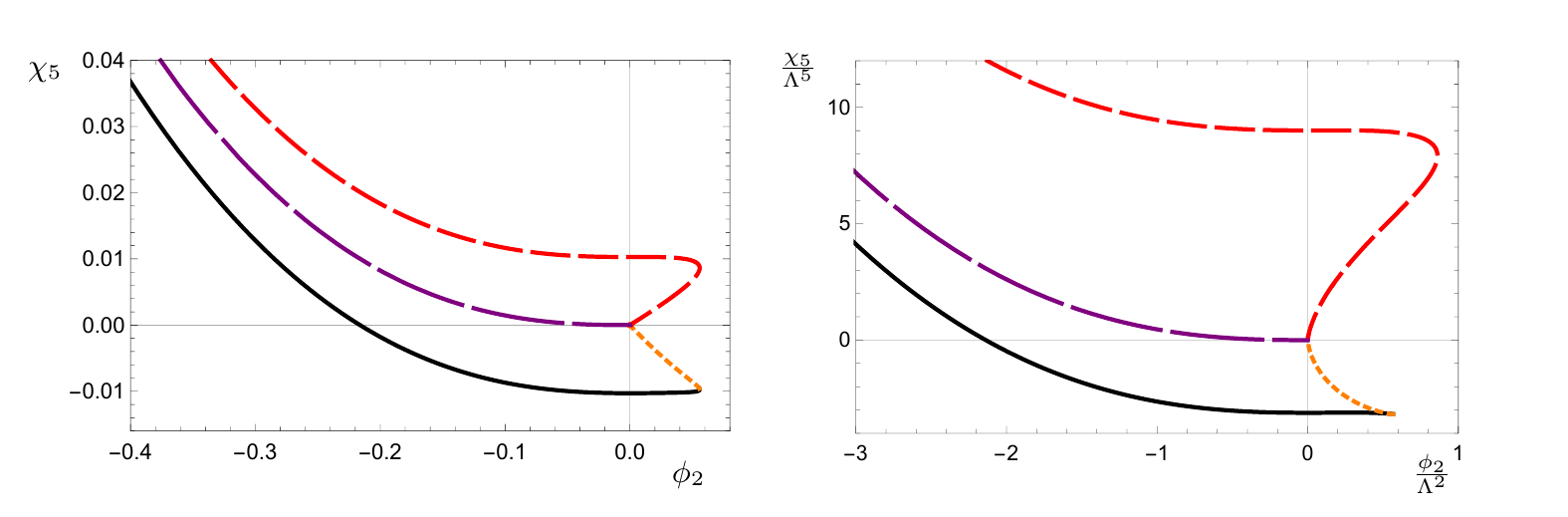}}
	\end{center}
	\vspace{-5mm}
	\caption[Third additional parameter plot for the six-dimensional supergravity]{Plots showing the relationship between the two UV expansion parameters $\f_{2}$ and $\c_{5}$ for solutions within the confining (solid black and short-dashed orange), skewed (dashed red), 
		and IR-conformal (longest-dashed purple) classes. The left plot shows the parameters extracted by matching to the UV expansions, 
		with $\f_{2}^{c}=\f_{2}^{s}$, $\f_{3}^{c}=\f_{3}^{s}$, and $\c_{5}^{s}=-\c_{5}^{c}-\frac{8}{25}\f_{2}^{c}\f_{3}^{c}$. The right panel shows the same parameters after rescaling with the appropriate powers of $\L$.}
	\label{Fig:Chi5Phi2}
\end{figure}%

\begin{figure}[t!]
	\begin{center}
		\makebox[\textwidth]{\includegraphics[
			width=0.7\paperwidth
			]
			{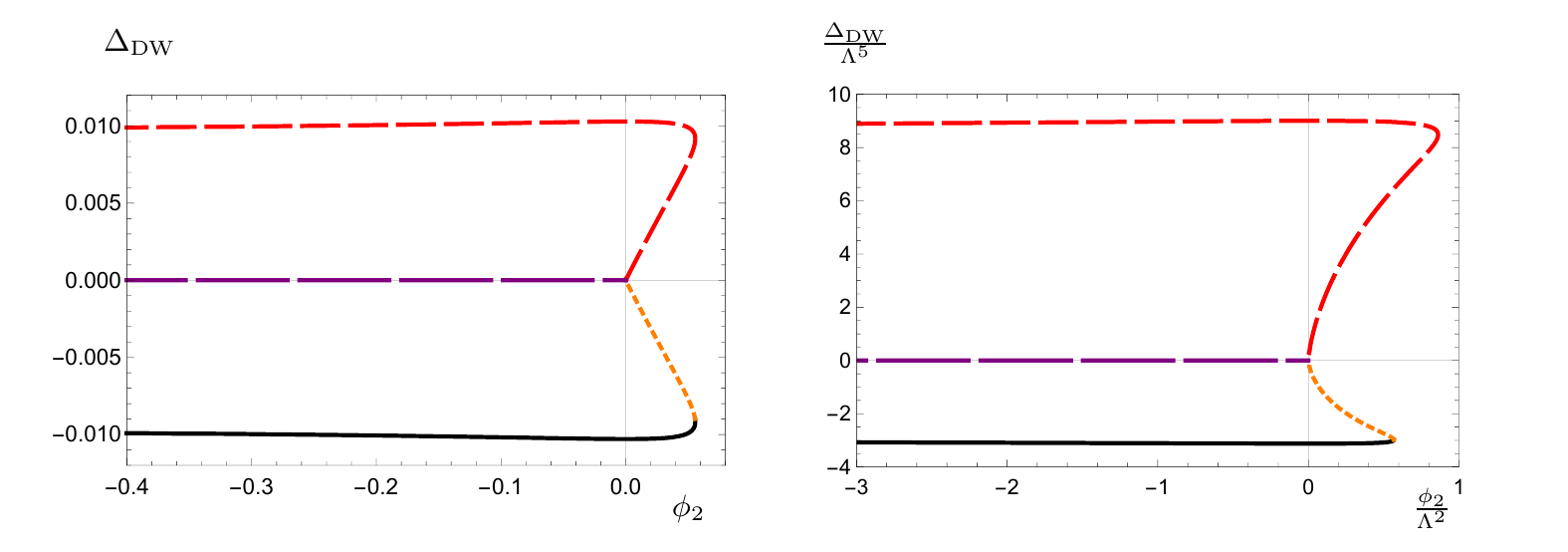}}
	\end{center}
	\vspace{-5mm}
	\caption[Fourth additional parameter plot for the six-dimensional supergravity]{The order parameter $\Delta_{\rm DW}$ as defined in Eq.~(\ref{Eq:OrderParamDelta}), for solutions within the confining (solid black and short-dashed orange), skewed (dashed red), and IR-conformal (longest-dashed purple) classes. The left plot shows the parameters extracted by matching to the UV expansions. The right panel shows the same parameters after rescaling with the appropriate powers of $\L$.}
	\label{Fig:DeltaPhi2}
\end{figure}
\clearpage

\subsubsection{Torus-compactified $D=7$ supergravity} 
\begin{figure}[!h]
	\centering
	\subfloat{\makebox[0.2\paperwidth]
		{\includegraphics[width=0.36\paperwidth]
			{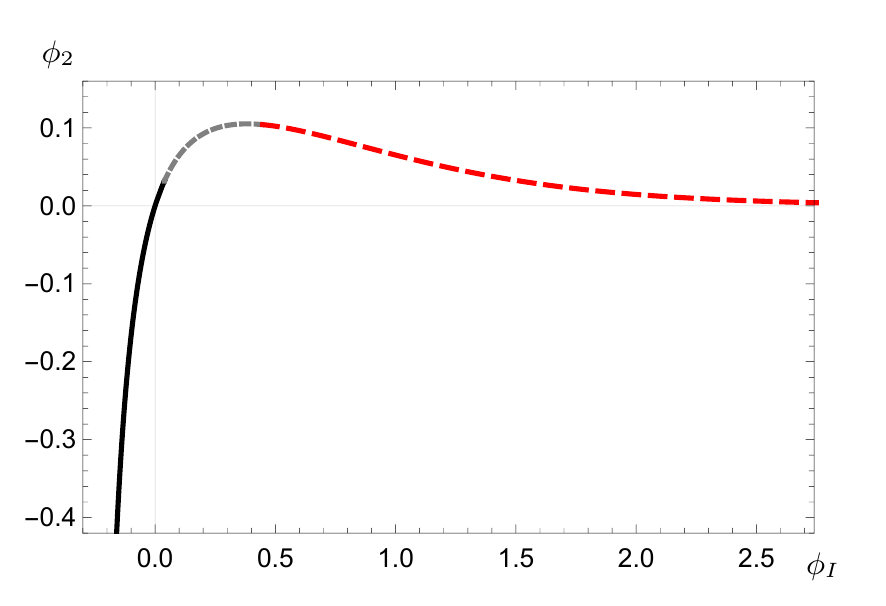}}  }
	\hfill
	\subfloat{\makebox[0.2\paperwidth]
		{\includegraphics[width=0.36\paperwidth]
			{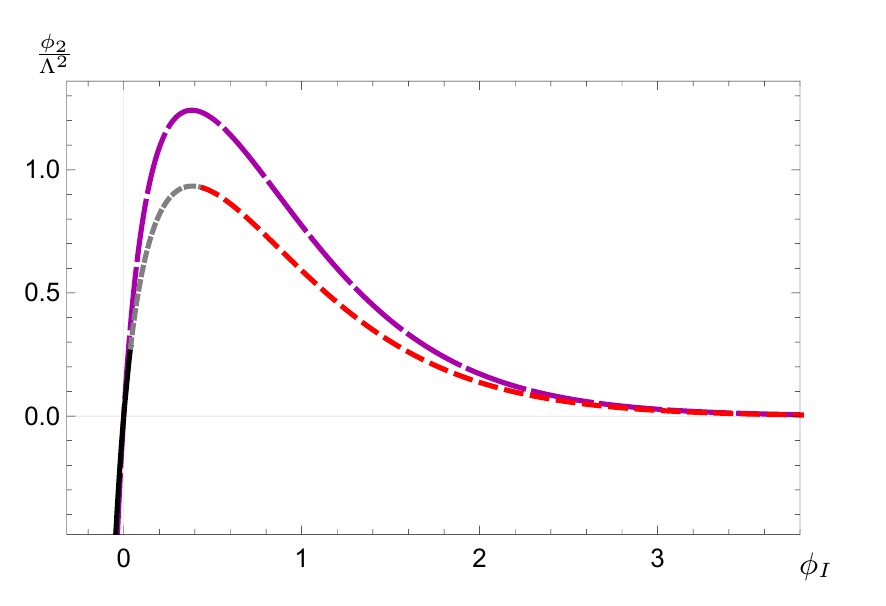}}  }
	\caption[First additional parameter plot for the seven-dimensional supergravity]{Plots showing the relationship between the UV expansion parameter 
		$\f_{2}$ and the IR parameter $\f_{I}$, for the confining (solid black, short-dashed grey, and dashed red) and skewed (long-dashed magenta) branches of solutions.
		The left plot shows the bare extracted parameters: the confining and skewed branches coincide, as $\f_{2}^{c}=\f_{2}^{s}$, $\f_{I}^{c}=\f_{I}^{s}$. The right plot shows the same parameters
		after rescaling with the appropriate powers of $\L$.}
	\label{Fig:WittenPhi2PhiI}
\end{figure}

In Figures~\ref{Fig:WittenPhi2PhiI}\,-\,\ref{Fig:omega6phi2Appendix} below we present some additional plots which help to elucidate the non-trivial implicit relations between the various parameters which characterise the confining, skewed, IR-conformal, and badly singular domain-wall branches of backgrounds within the $T^{2}$-reduced seven-dimensional supergravity discussed in Chapter~\ref{Chap:SpectraWitten}. These plots are supplementary to our discussion of dilaton phenomenology in Sec.~\ref{Sec:WittenProbeRevisit}.

\begin{figure}[!h]
	\centering
	\subfloat{\makebox[0.21\paperwidth]
		{\includegraphics[width=0.364\paperwidth]
			{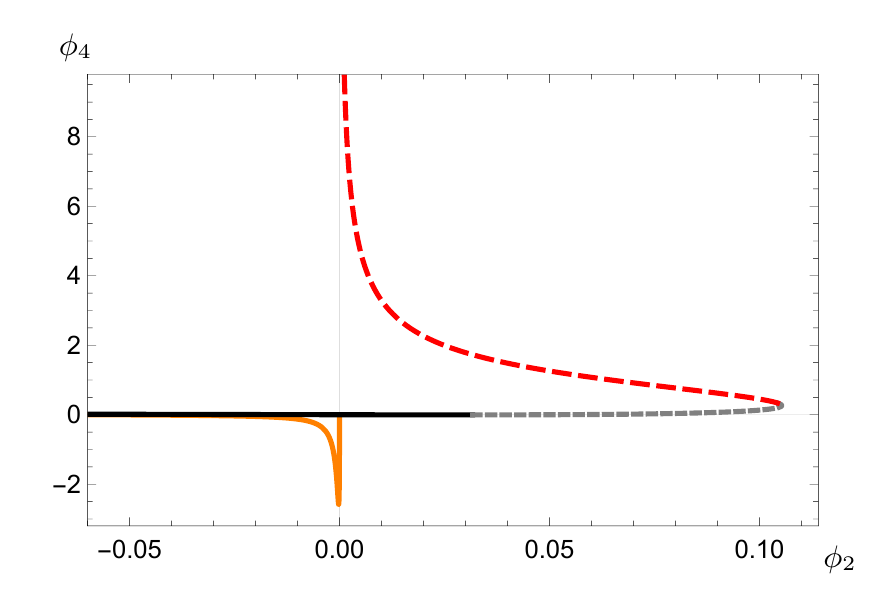}}  }
	\hfill
	\subfloat{\makebox[0.21\paperwidth]
		{\includegraphics[width=0.364\paperwidth]
			{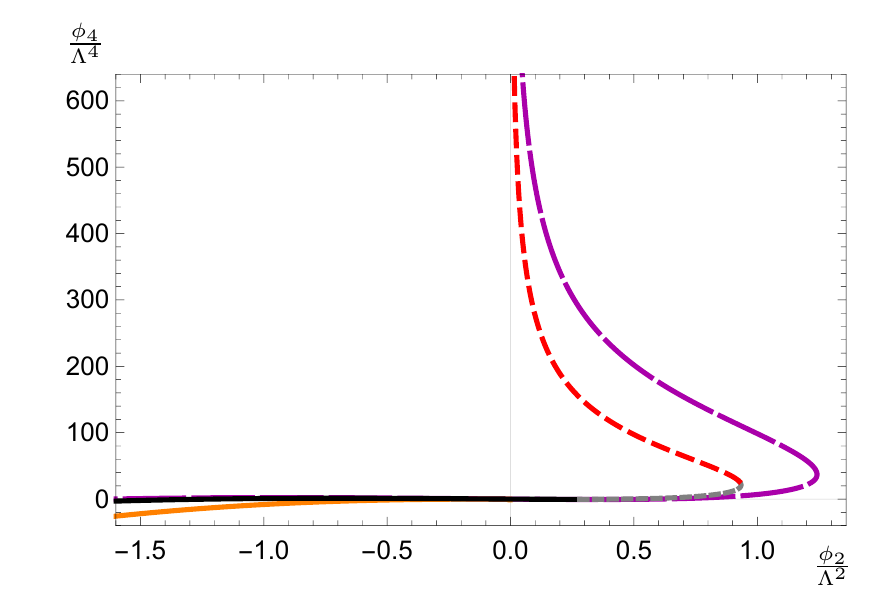}}  }
	\caption[Second additional parameter plot for the seven-dimensional supergravity]{Plots showing the relationship between the two UV expansion parameters $\f_{2}$ and $\f_{4}$ for solutions belonging to the confining (solid black, short-dashed grey, and dashed red), skewed (long-dashed magenta), and IR-conformal (solid orange) classes. 
	The left plot shows the bare parameters extracted by matching to the UV expansions, 
	with $\f_{2}^{c}=\f_{2}^{s}$, $\f_{4}^{c}=\f_{4}^{s}$. The right panel shows the same parameters
	after rescaling with the appropriate powers of $\L$.  }
\label{Fig:Phi4Phi2}
\end{figure}%

\begin{figure}[t!]
	\centering
	\subfloat{\makebox[0.2\paperwidth]
		{\includegraphics[width=0.36\paperwidth]
			{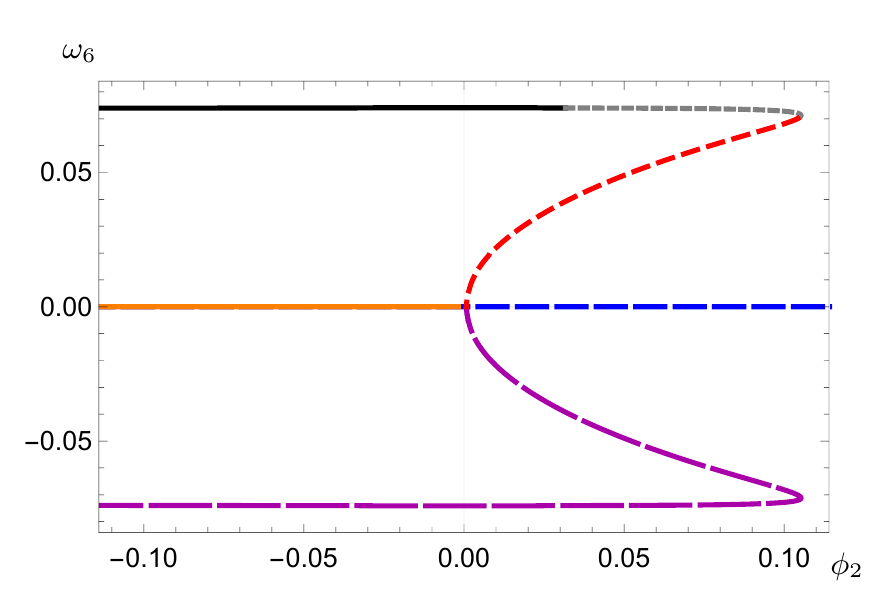}}  }
	\hfill
	\subfloat{\makebox[0.2\paperwidth]
		{\includegraphics[width=0.36\paperwidth]
			{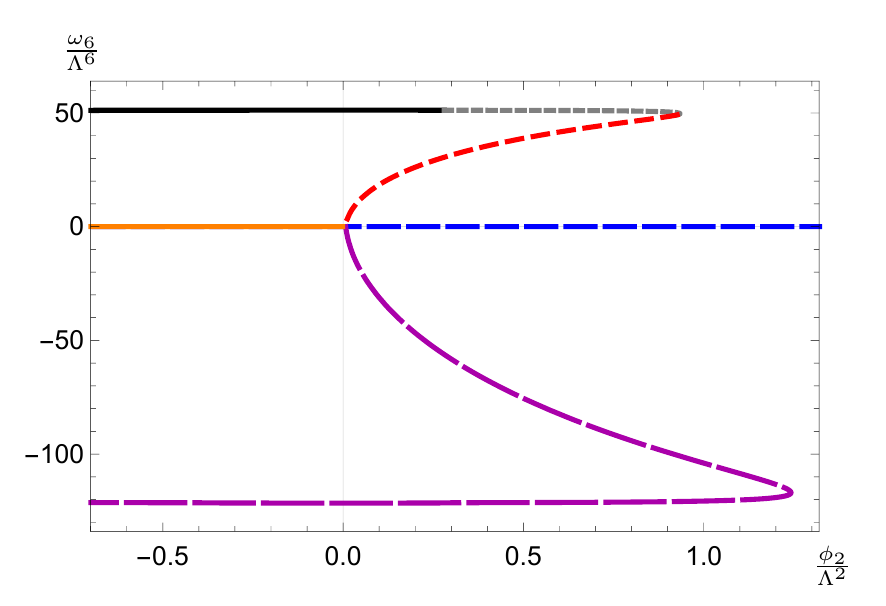}}  }
	\caption[Third additional parameter plot for the seven-dimensional supergravity]{The UV expansion parameter $\w_{6}$, for solutions within the confining (solid black, shortest-dashed grey, and short-dashed red), skewed (long-dashed magenta), IR-conformal (solid orange), and badly singular domain-wall (dashed blue) classes. The left plot shows the parameters extracted by matching to the UV expansions. The right panel shows the same parameters after rescaling with the appropriate powers of $\L$. Although not evident from these plots, we remind the Reader that there exists an upper bound on the $\Delta=4$ operator source at approximately $\hat{\f}_{2}\simeq 2.50\-\ (\f_{2}\simeq 0.55)$ and hence the blue line will eventually terminate.}
	\label{Fig:omega6phi2Appendix}
\end{figure}

\begin{singlespace}

\end{singlespace}

\end{document}